\documentclass[lefteqn]{elsart_modifiedbyrs}

 \usepackage{graphicx}
 \usepackage{epsfig}

\usepackage{amssymb}

\begin{document}

\begin{frontmatter}


\title{Rigorous solution of a Hubbard model extended by
nearest-neighbour Coulomb and isotropic exchange interaction on a triangle and tetrahedron}
\author[a]{Rolf Schumann\corauthref{cor1}}
 \corauth[cor1]{R. Schumann schumann@theory.phy.tu-dresden.de}
\address[a]{Institut f\"ur Theoretische Physik, TU Dresden, 
D-01062 Dresden, Germany}
\begin{abstract}
In the preprint condmat0701060 we detected a factor two error in the coding of the
Heisenberg term of the Hamiltonian. In the result the exchange term used in the paper 
was that of an anisotropic Heisenberg model with $J_x=J_y=J$ and $J_z=J/2$. Thus all results given for which $J=0$ was chosen,
remain unchanged, whereas all results containing the exchange parameter remain true for
the extended Hubbard model with anisotropic exchange. That model is of some interest by itself,
due to the relation to the XXZ model. The differing results for 
isotropic exchange are given here.
\end{abstract}
\begin{keyword}
Hubbard model, rigorous solution, nearest-neighbour interaction, cluster
\PACS 85.80.+n, 73.22.-f, 71.27.+a
\end{keyword}
\end{frontmatter}
\newcommand{\beq}{\begin{eqnarray} }
\newcommand{\eeq}{\end{eqnarray}}
\newcommand{\eee}{\mathrm e}
\newcommand{\punkt}{\, .}
\newcommand{\komma}{\, ,}
\newcommand{\summe}[1]{\sum\limits_{#1}}
\newcommand{\sumijs}{ \sum\limits_{i\neq j\sigma}}
\newcommand{\product}[1]{\prod\limits_{#1}}
%
\newcommand{\cplus}[1]{{\bf c}^+_{#1}}
\newcommand{\cminus}[1]{{\bf c}_{#1}}
\newcommand{\cpup}[1]{{\bf c}^+_{#1 \uparrow}}
\renewcommand{\cup}[1]{{\bf c}_{#1 \uparrow}}
\newcommand{\cpdown}[1]{{\bf c}^+_{#1 \downarrow}}
\newcommand{\cdown}[1]{{\bf c}_{#1 \downarrow}}
\newcommand{\cpsigma}[1]{{\bf c}^+_{#1 \sigma}}
\newcommand{\csigma}[1]{{\bf c}_{#1 \sigma}}
\newcommand{\cpis}{{\bf c}^+_{i \sigma}}
\newcommand{\cis}{{\bf c}_{i \sigma}}
\newcommand{\cpjs}{{\bf c}_{j \sigma}}
\newcommand{\cjs}{{\bf c}_{j \sigma}}
\newcommand{\apnull}{{\bf a}^+_{0}}
\newcommand{\anull}{{\bf a}_{0}}
\newcommand{\apup}{{\bf a}^+_{\uparrow}}
\newcommand{\aup}{{\bf a}_{\uparrow}}
\newcommand{\apdown}{{\bf a}^+_{\downarrow}}
\newcommand{\adown}{{\bf a}_{\downarrow}}
\newcommand{\apzwei}{{\bf a}^+_{2}}
\newcommand{\azwei}{{\bf a}_{2}}
\newcommand{\apsigma}[1]{{\bf a}^+_{#1 \sigma}}
\newcommand{\asigma}[1]{{\bf a}_{#1 \sigma}}
\newcommand{\aplus}[1]{{\bf a}^+_{#1 }}
\newcommand{\aminus}[1]{{\bf a}_{#1 }}
%
%
\newcommand{\nup}[1]{{\bf n}_{#1 \uparrow}}
\newcommand{\ndown}[1]{{\bf n}_{#1 \downarrow}}
\newcommand{\nsigma}[1]{{\bf n}_{#1 \sigma}}
\newcommand{\noperator}[1]{{\bf n}_{#1 }}
\newcommand{\dyade}[2]{| #1 \rangle \langle #2|}
\newcommand{\hamilton}{{\bf H}}
\newcommand{\gesamtn}{\mbox{\boldmath $N$}}
\newcommand{\gesamtr}{\mbox{\boldmath $R$}}
\newcommand{\gesamts}{\mbox{\boldmath $S$}}
\newcommand{\liouville}{\mbox{\boldmath $L$}}
\newcommand{\interaction}{\mbox{\boldmath $V$}}
\newcommand{\koperator}{\mbox{\boldmath $K$}}
\newcommand{\operator}[1]{{\bf #1}}
\newcommand{\opP}{\mbox{\boldmath$ P $}}
\newcommand{\opQ}{\mbox{\boldmath $ Q $}}
\newcommand{\opA}{\mbox{\boldmath $ A $}}
\newcommand{\opB}{\mbox{\boldmath $ B $}}
\newcommand{\opC}{\mbox{\boldmath $ C $}}
\newcommand{\opD}{\mbox{\boldmath $ D $}}
\newcommand{\opE}{\mbox{\boldmath $ E $}}
\newcommand{\opL}{\mbox{\boldmath $ L $}}
\newcommand{\opF}{\mbox{\boldmath $ F $}}
\newcommand{\opH}{\mbox{\boldmath $ H $}}
\newcommand{\opM}{\mbox{\boldmath $ M $}}
\newcommand{\opT}{\mbox{\boldmath $ T $}}
\newcommand{\opU}{\mbox{\boldmath $ U $}}
\newcommand{\opX}{\mbox{\boldmath $ X $}}
\newcommand{\ops}{\mbox{\boldmath $ s $}}
\newcommand{\opS}{\mbox{\boldmath $ S $}}
\newcommand{\opR}{\mbox{\boldmath $ R $}}
\newcommand{\ri}[1]{{\bf R}_{i #1}} 
\newcommand{\si}[1]{{\bf S}_{i #1}}
\newcommand{\eins}{\mbox{\boldmath$I$ }}
\newcommand{\einsa}{\mbox{\boldmath$I\!_2$ }}
\newcommand{\laplace}{\mbox{\boldmath$\Delta$ }}
\newcommand{\opbeta}{\mbox{\boldmath $ \beta $ }}
\newcommand{\Sigmaz}{\mbox{\boldmath $ \Sigma_z$}}
\newcommand{\sigmax}{\mbox{\boldmath $ \sigma_x$}}
\newcommand{\sigmay}{\mbox{\boldmath $ \sigma_y$}}
\newcommand{\sigmaz}{\mbox{\boldmath $ \sigma_z$}}
\newcommand{\opsigma}{\vec{\mbox{\boldmath $ \sigma $}}}
\newcommand{\opalpha}{\vec{\mbox{\boldmath $ \alpha $}}}
\newcommand{\tket}[1]{$ | #1 \rangle$}
\newcommand{\tbra}[1]{$\langle #1 |$}
\newcommand{\mket}[1]{| #1 \rangle}
\newcommand{\mbra}[1]{\langle #1|}
\newcommand{\ket}[1]{| #1 \rangle}
\newcommand{\bra}[1]{\langle #1|}
\newcommand{\scalar}[2]{\langle #1 \, | \, #2 \rangle}
\newcommand{\expect}[1]{\langle #1 \rangle}
\newcommand{\texpect}[1]{$\langle #1 \rangle$}
\newcommand{\mean}[1]{\langle #1 \rangle}
\newcommand{\tmean}[1]{$\langle #1 \rangle$}
\newcommand{\ketvacuum}{\mket{\mbox{vac}}}
\newcommand{\bravacuum}{\mbra{\mbox{vac}}}
\newcommand{\tpket}[1]{$ \left| #1 \right )$}
\newcommand{\tpbra}[1]{$ \left ( #1 \right|$}
\newcommand{\pket}[1]{\left | #1 \right )}
\newcommand{\pbra}[1]{\left ( #1 \right |}
\newcommand{\pscalar}[2]{\left ( #1 
\,\left | #2 \right . \right )}
\newcommand{\ikr}{{\rm e}^{i\vec{k}\vec{R}_i}}
\newcommand{\mikr}{{\rm e}^{-i\vec{k}\vec{R}_i}}
\newcommand{\ikvonr}[1]{{\rm e}^{i\vec{k}\vec{R}_{#1}}}
\newcommand{\mikvonr}[1]{{\rm e}^{-i\vec{k}\vec{R}_{#1}}}
\newcommand{\ryd}{\mbox{ryd}}

\newcommand{\sumviervier}{\summe{k \alpha} \summe{k' \alpha'} 
                   \summe{q \beta} \summe{q' \beta'}}
\newcommand{\sumvier}{\summe{k \alpha} \summe{k' } 
                   \summe{q \beta}  \summe{q' }}                     
\newcommand{\sumsechs}{\summe{k \sigma} \summe{k' } 
                   \summe{q}  \summe{q' }
                   \summe{p}  \summe{p' }}                     
\newcommand{\sumkkqq}{\summe{k } \summe{k' } 
                   \summe{q }  \summe{q' }}                     
\newcommand{\sumijlm}{\summe{i \sigma} \summe{j } 
                   \summe{l}  \summe{m }}                     
\newcommand{\sumij}{\summe{i \sigma} \summe{j }} 
\newcommand{\ddelta}[1]{\delta_{#1,0}}
\newcommand{\expo}[1]{e^{\displaystyle #1}}
\newcommand{\difk}{\left(\frac{\vec{p}}{|\vec{p}|} \frac{\partial}{\partial \vec{k}}\right)}
\newcommand{\difq}{\left(\frac{\vec{p}}{|\vec{p}|} \frac{\partial}{\partial \vec{q}}\right)}
\newcommand{\diffk}{\mbox{\bf D}}
\newcommand{\difkx}{p_x \frac{\partial}{\partial k_x}}
\newcommand{\difky}{p_y \frac{\partial}{\partial k_y}}
\newcommand{\diffkx}{\mbox{\bf D}_x}
\newcommand{\diffky}{\mbox{\bf D}_y}
\newcommand{\rtot}{{\bf R}}
\newcommand{\stot}{{\bf S}}
\newcommand{\imag}{i}
\newcommand{\Aindex}[1]{A_{#1}}
\newcommand{\Bindex}[1]{B_{#1}}
\newcommand{\Cindex}[2]{C_{#1-#2}}
\newcommand{\Psiindex}[1]{\Psi_{#1}}
\newcommand{\Thetaindex}[1]{\Theta_{#1}}
\newcommand{\multsum}[1]{\summe{i_{#1} \neq j_{#1}} \summe{\alpha {#1} \neq \alpha'_{#1}}
                         \summe{\eta_{#1} \neq \eta'_{#1}}} 
\newcommand{\Anumbered}[1]{A_{i_{#1}j_{#1},\alpha_{#1}\alpha'_{#1} \eta_{#1}\eta'_{#1}}}
\newcommand{\XXnumbered}[1]{X_{i_{#1}}^{\alpha_{#1}\alpha'_{#1}}                        X_{j_{#1}}^{\eta_{#1}\eta'_{#1}}} 
\newpage
\begin{figure}
\unitlength=1mm
\begin{picture}(160,165)
\put(0,115){\epsfig{file=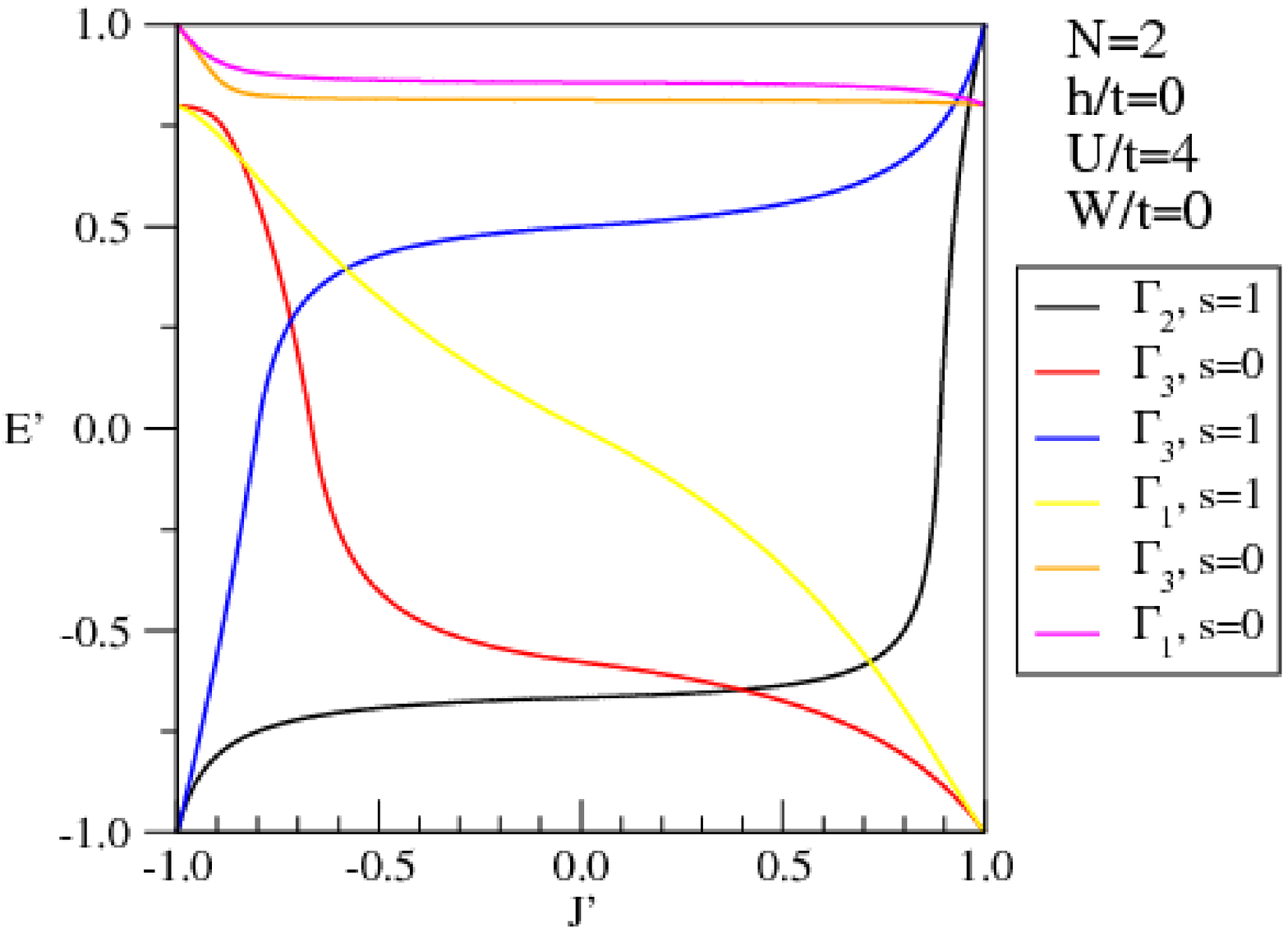,height=50mm}}
\put(70,110){\epsfig{file=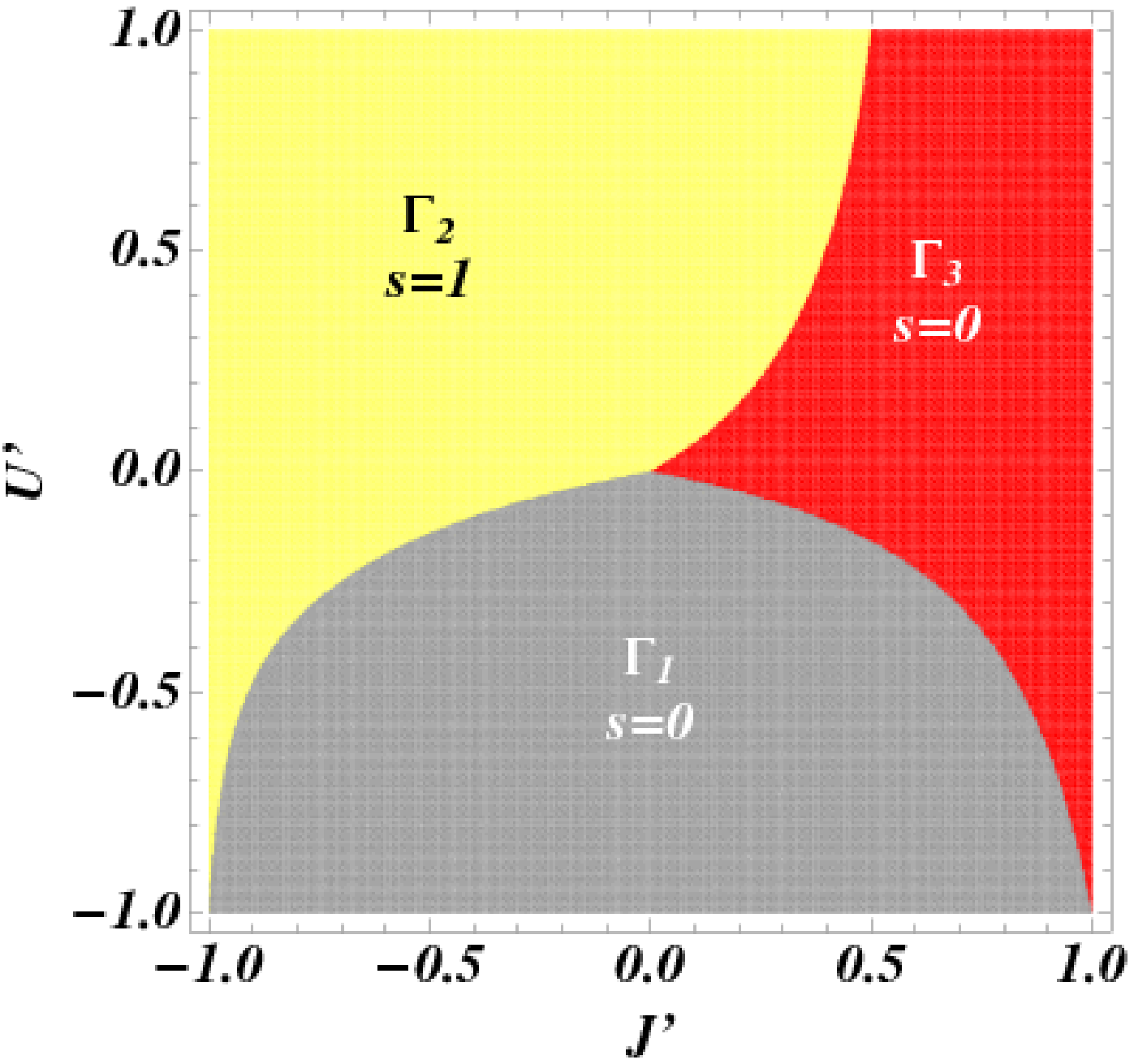,height=55mm}}
\put(0,60){\epsfig{file=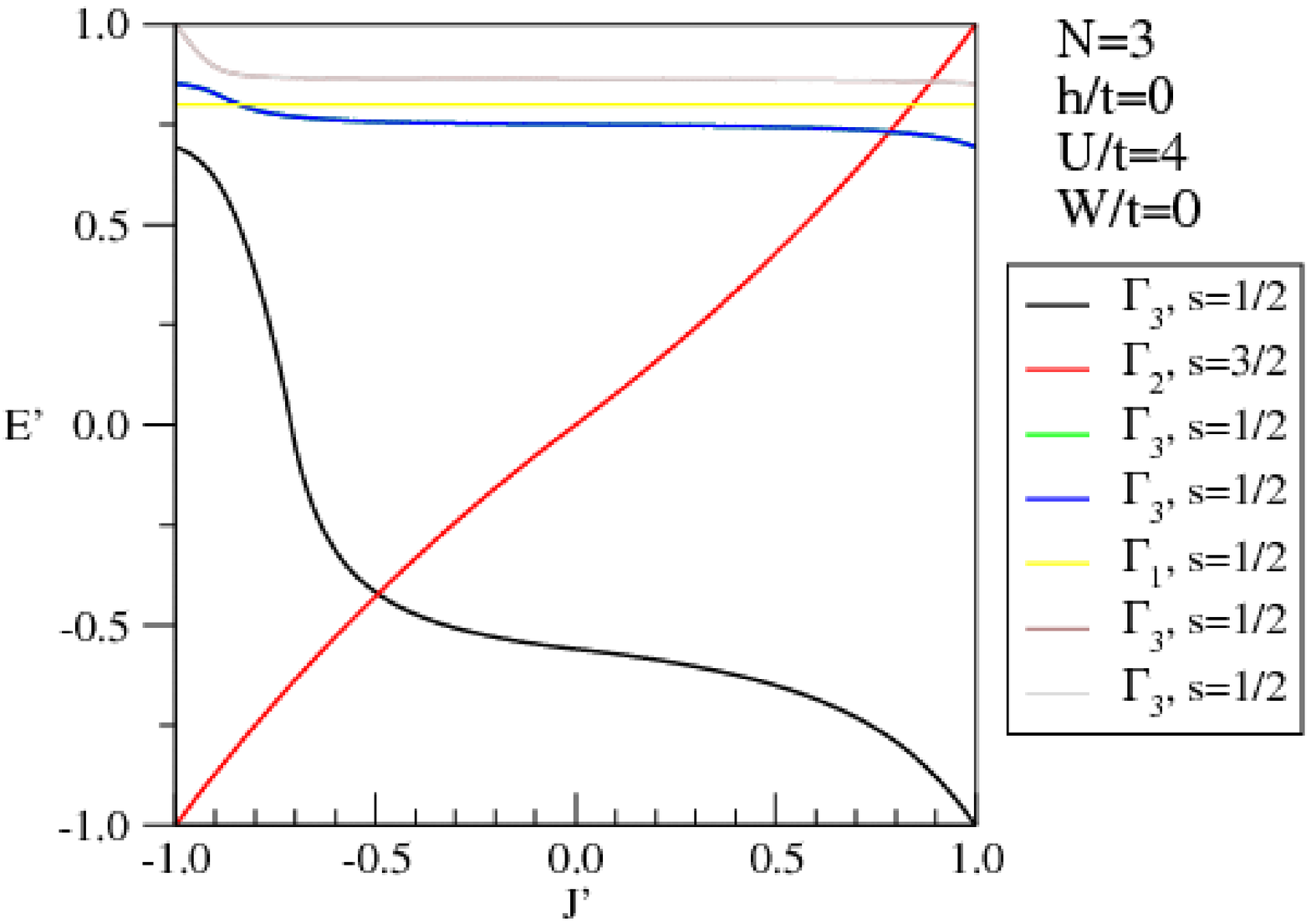,height=50mm}}
\put(70,55){\epsfig{file=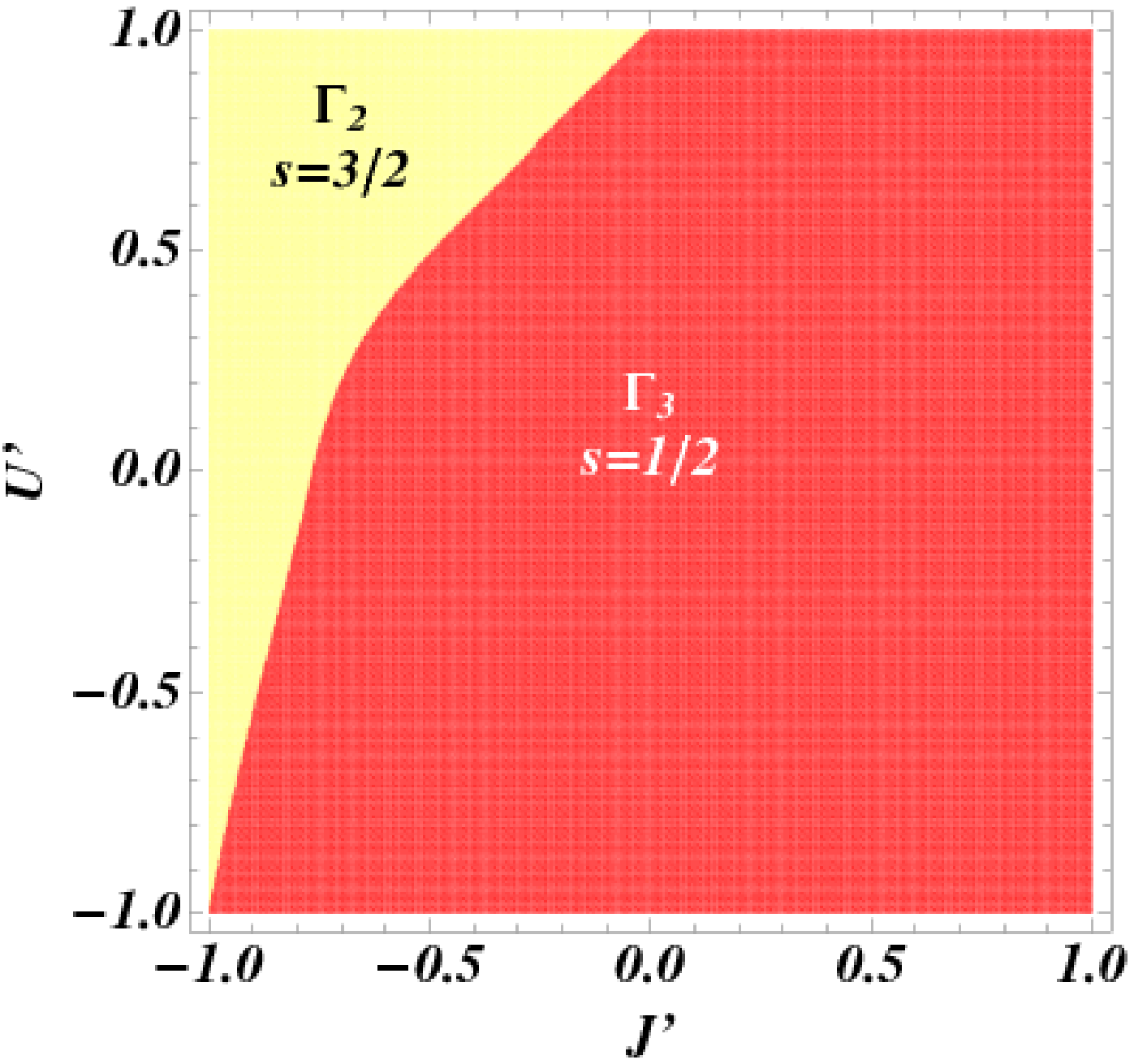,height=55mm}}
\put(0,5){\epsfig{file=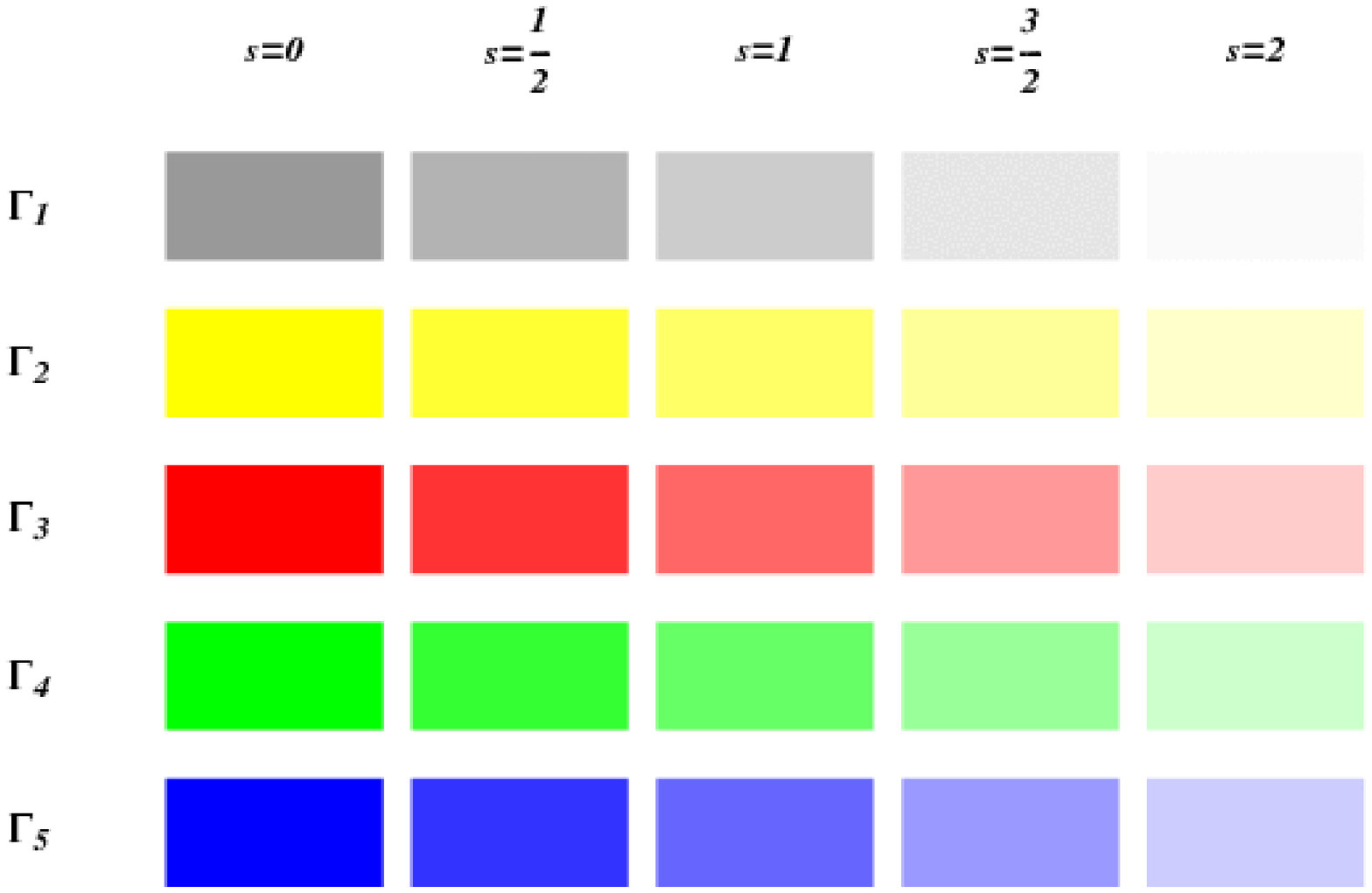,height=50mm,width=60mm}}
\put(70,0){\epsfig{file=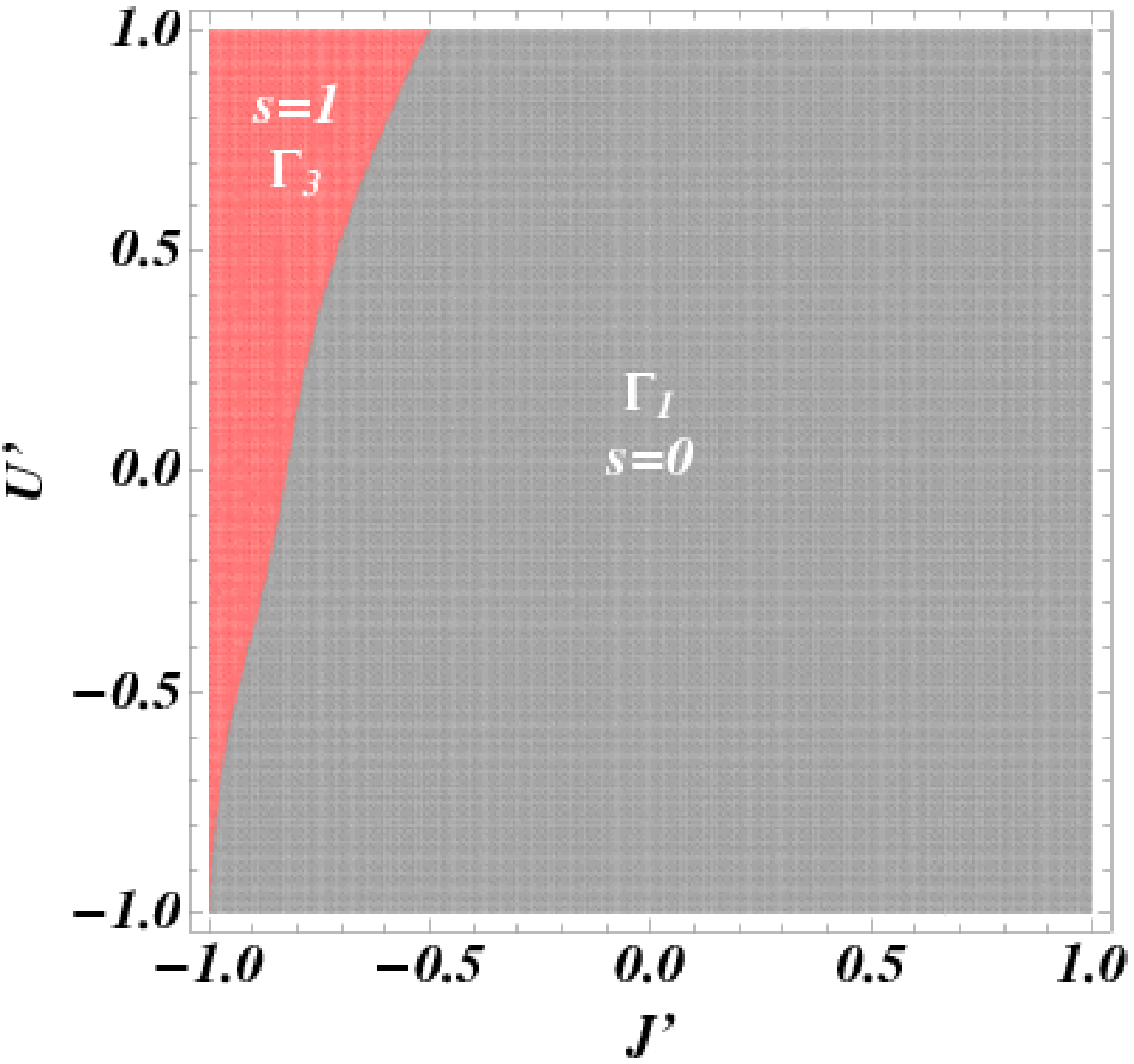,height=55mm}}
\end{picture}
\caption{
  Left: The complete canonical spectrum for the triangle
  (scaled to primed values) in dependence on $J'$ for $h/t=0$, $U/t=4$ 
  and $W/t=0$. Right: The ground states for the complete $J$-$U$-plane 
  (scaled to primed values)  for $W/t=0$ and $h/t=0$ and electron occupation 
  n=2 (top),  n=3, and n=4 (bottom).
  Down left: The palette used here and 
  in Fig. \ref{primedewsTetraJ_isotrop}.}
\label{primedewsTriangleJ_isotrop}
\end{figure}
%
\newpage
\begin{figure}[b]
 \begin{minipage}[c]{0.330\textwidth}
 \centering 
 \includegraphics[width=45mm]{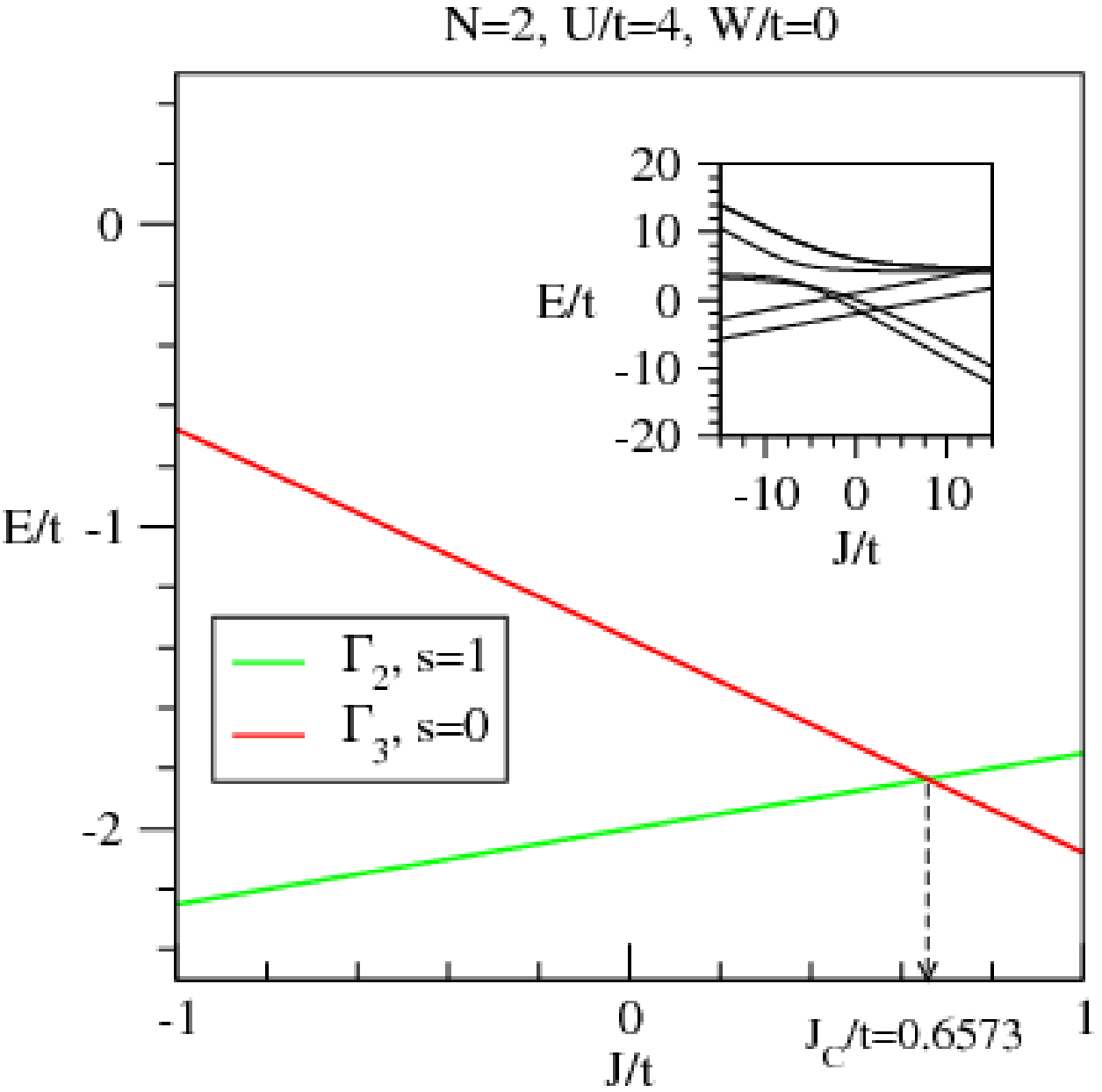}
 \end{minipage}%
 \hfill
 \begin{minipage}[c]{0.33\textwidth}
 \centering 
 \includegraphics[width=45mm]{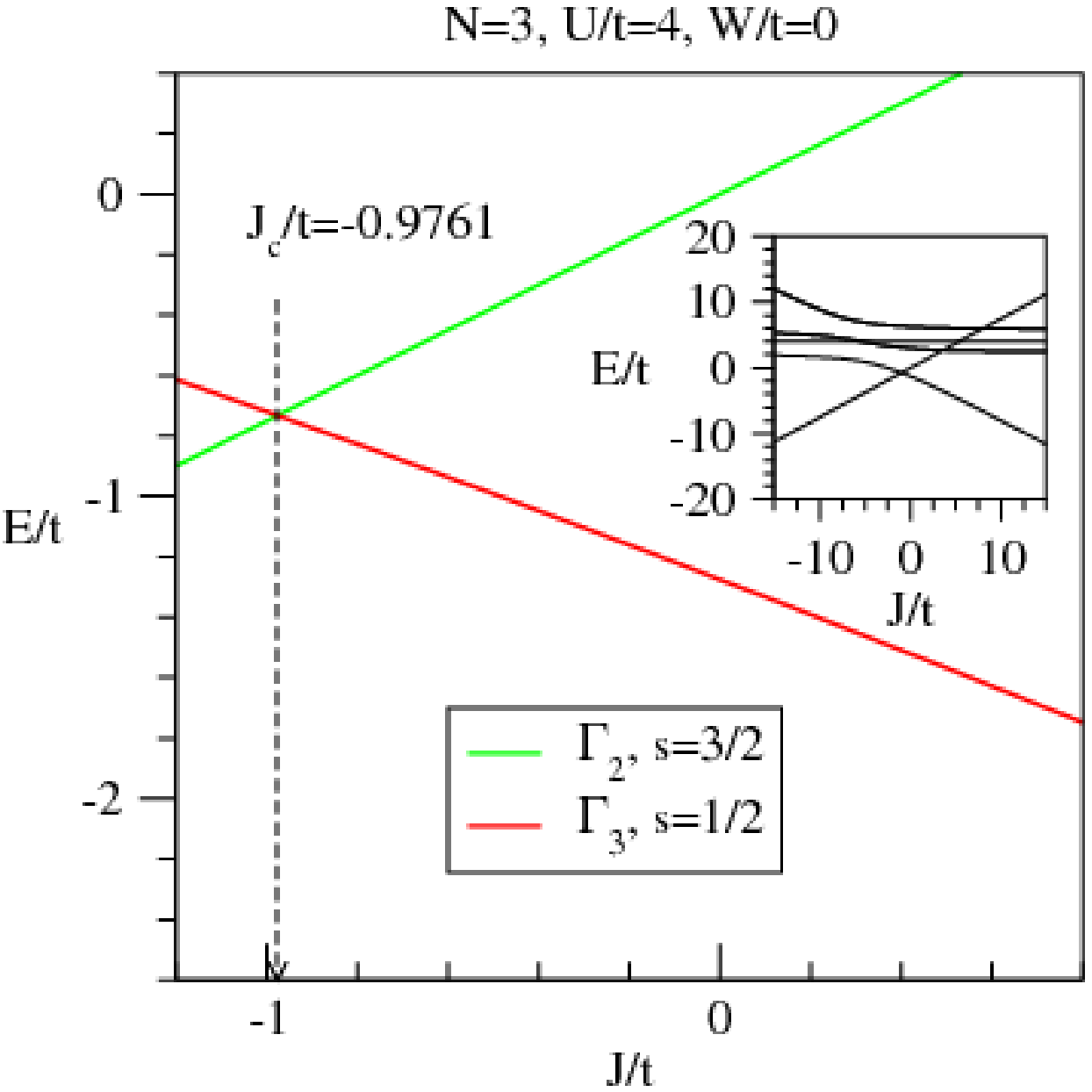}
\end{minipage}%
 \hfill
 \begin{minipage}[c]{0.33\textwidth}
 \centering 
 \includegraphics[width=45mm]{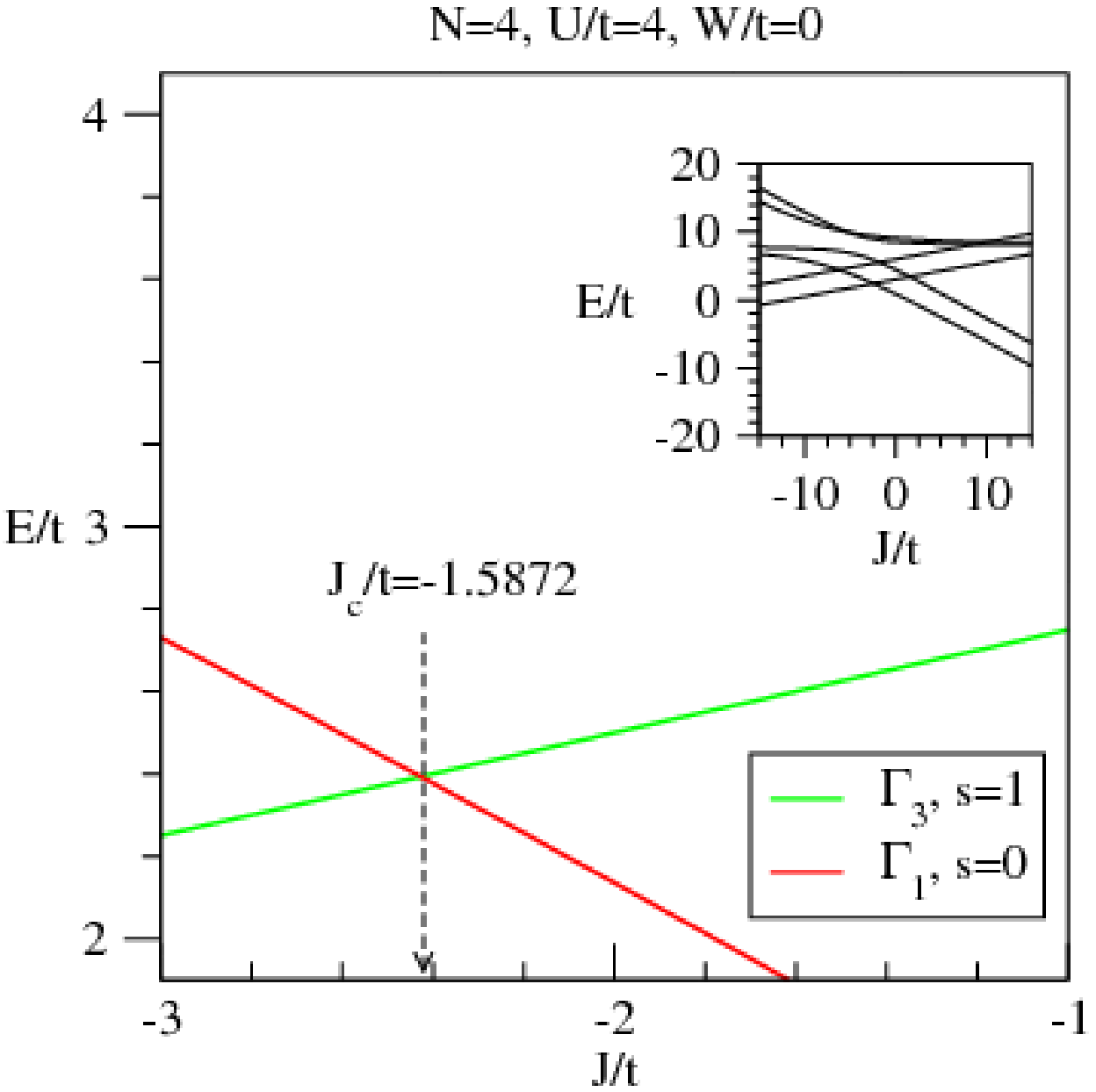}
\end{minipage}%
\caption{The crossings of the canonical ground states in dependence of the exchange parameter $J$ for the triangular cluster with
isotropic exchange for $N=2$, $N=3$, and $N=4$ respectively. The insets sketch the $J$-dependencies for the complete canonical spectra.}
\label{triangleLevelCrossings_isotrop}
\end{figure}

\newpage
\begin{figure}
\centering
\unitlength=1mm
\begin{picture}(160,200)
\put(0,155){\epsfig{file=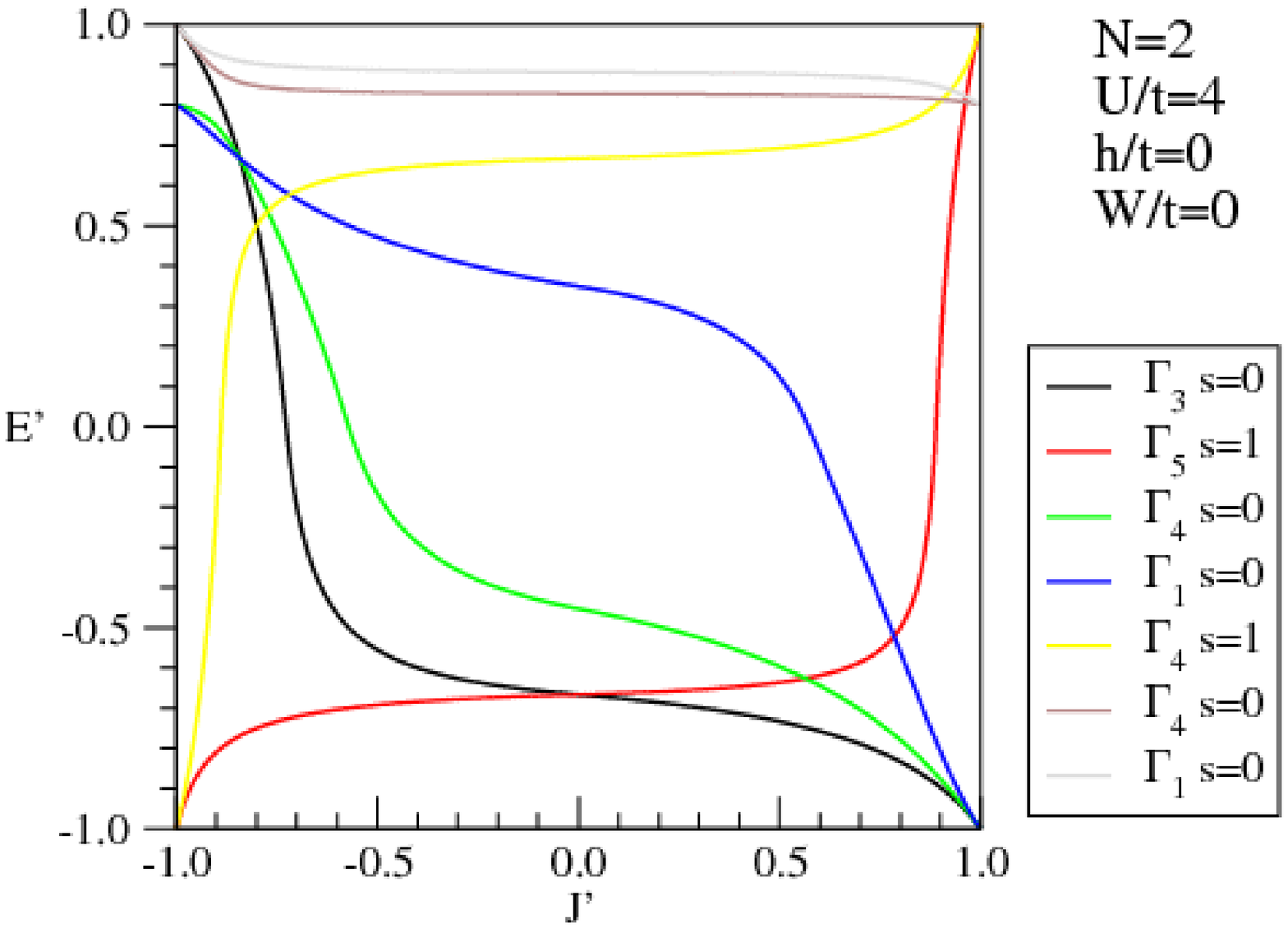,height=45mm}}
\put(0,105){\epsfig{file=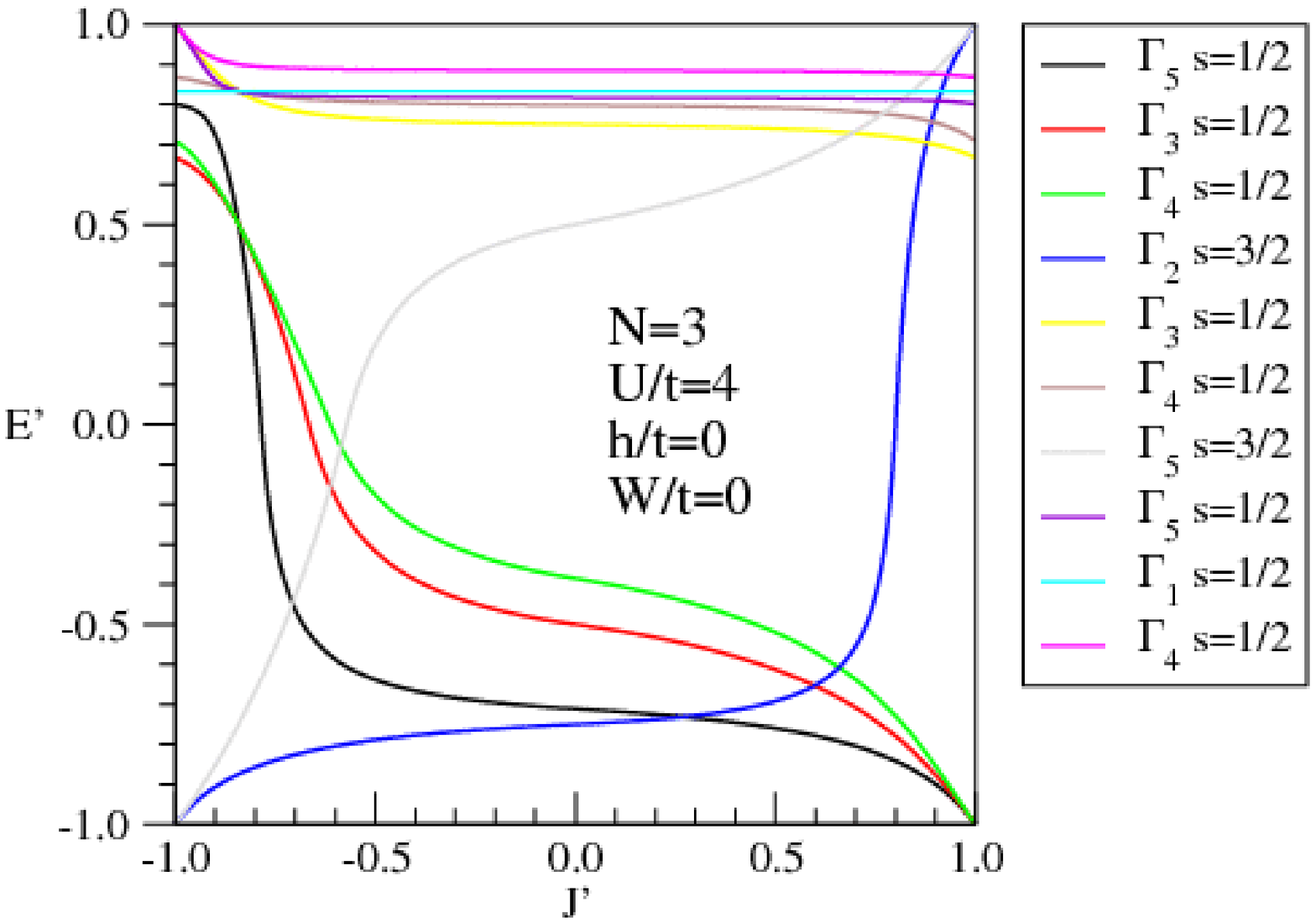,height=45mm}}
\put(0,55){\epsfig{file=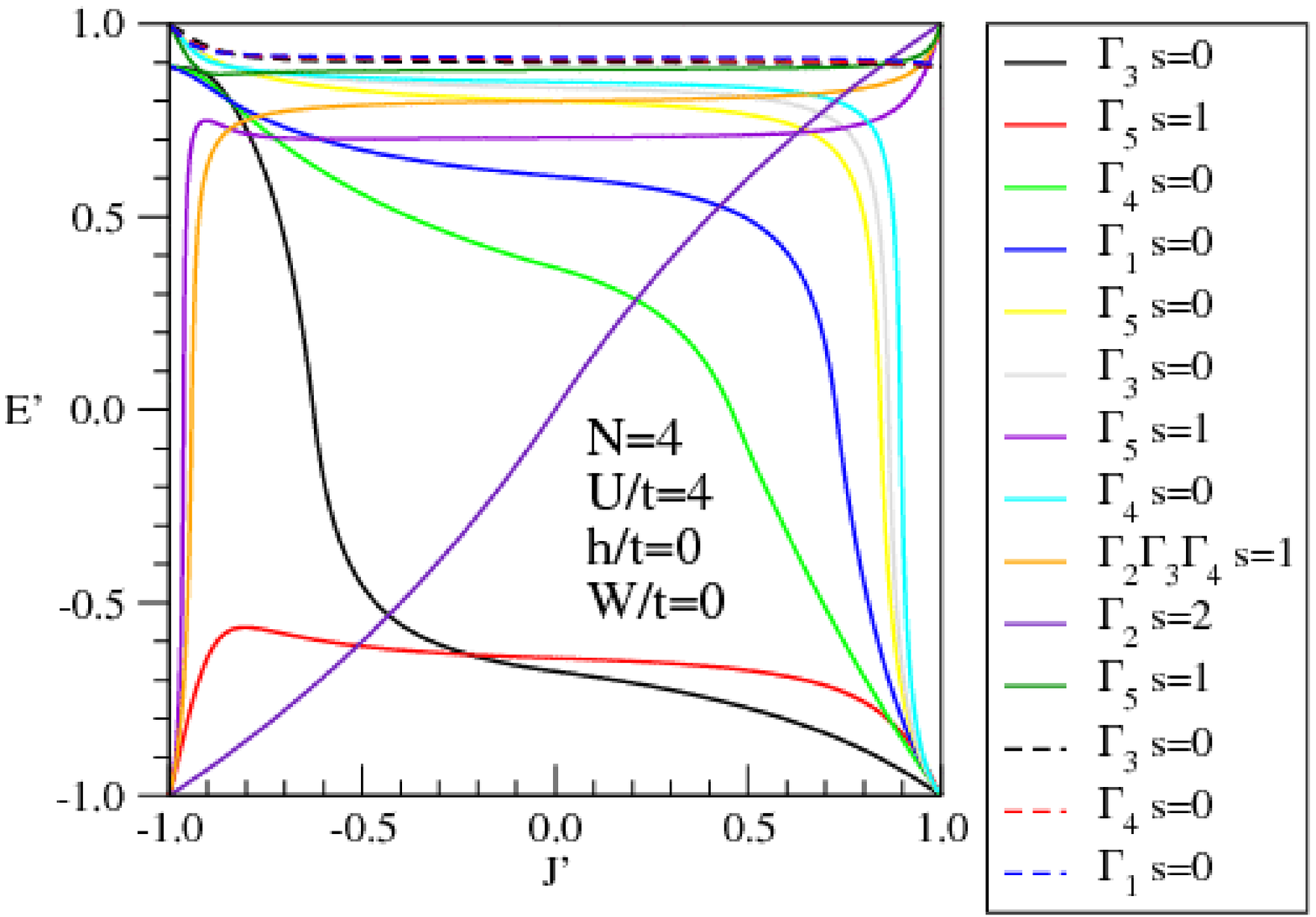,height=45mm}}
\put(0,05){\epsfig{file=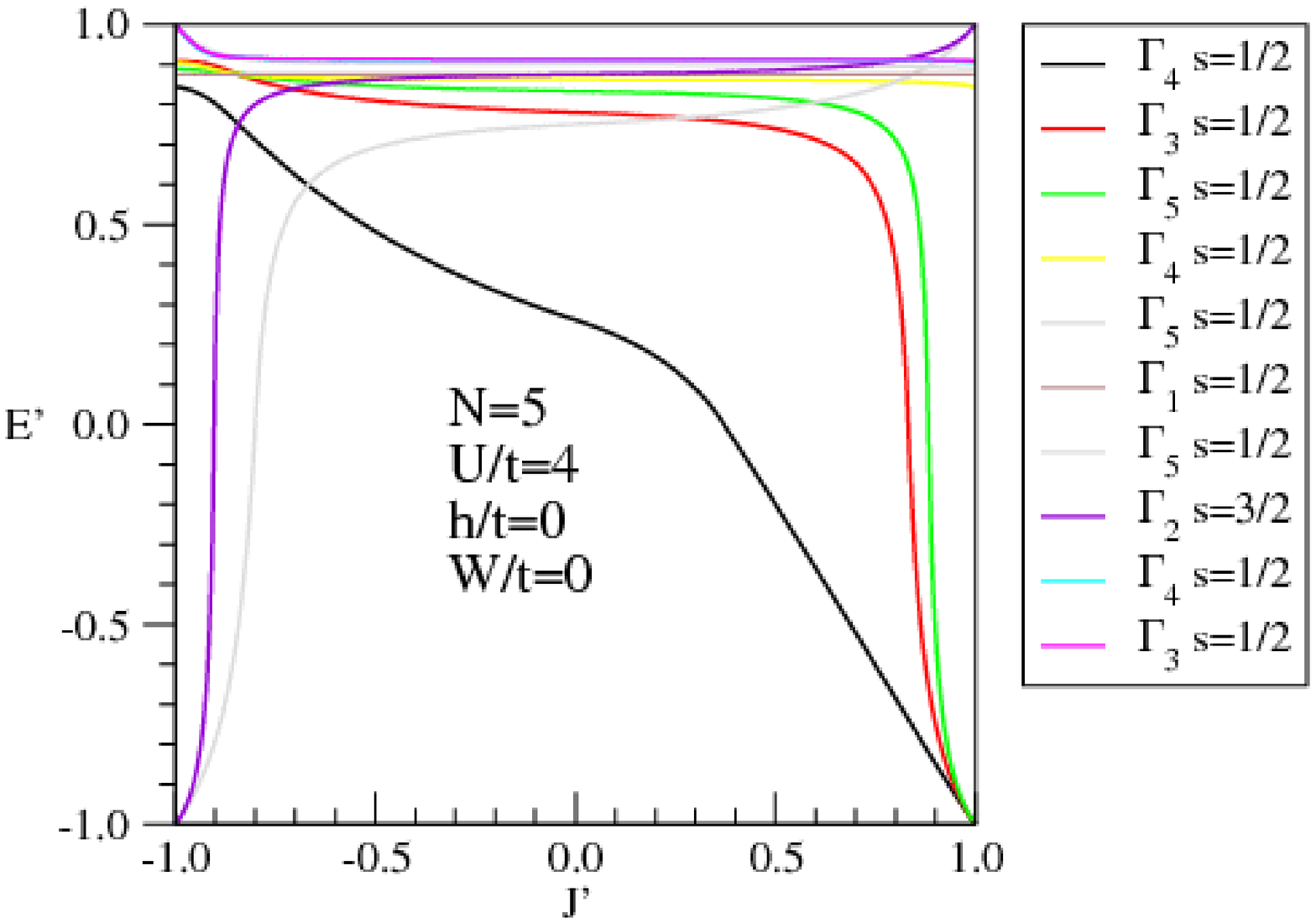,height=45mm}} 
 \put(70,150){\epsfig{file=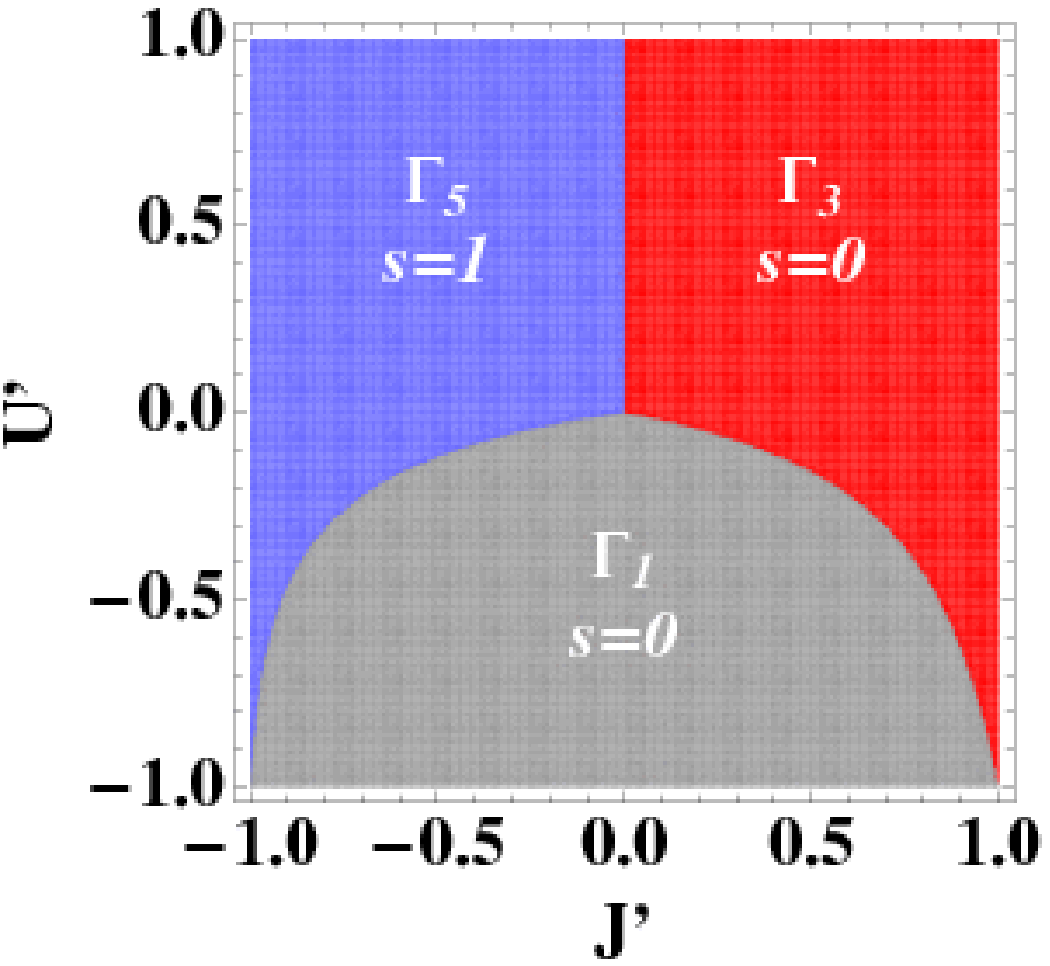,height=50mm}}
 \put(70,100){\epsfig{file=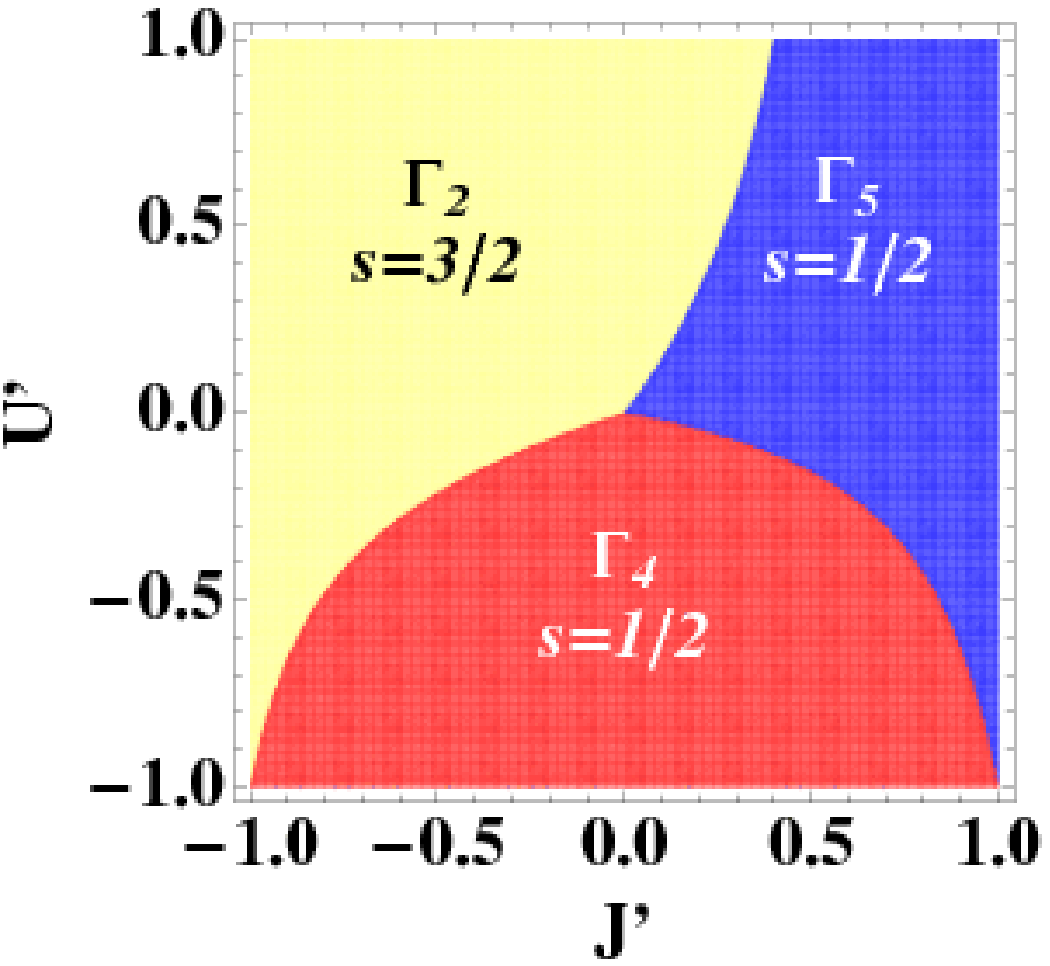,height=50mm}}
 \put(70,50){\epsfig{file=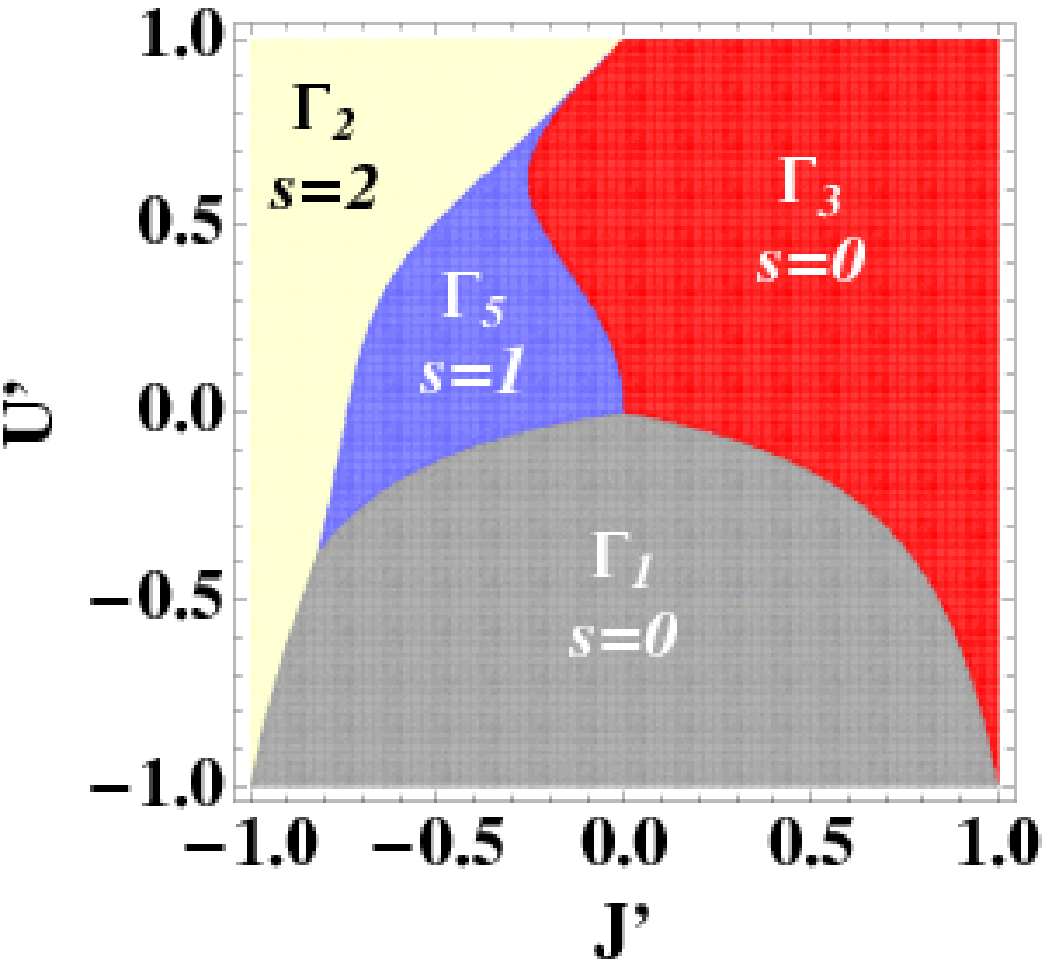,height=50mm}}
 \put(70,0){\epsfig{file=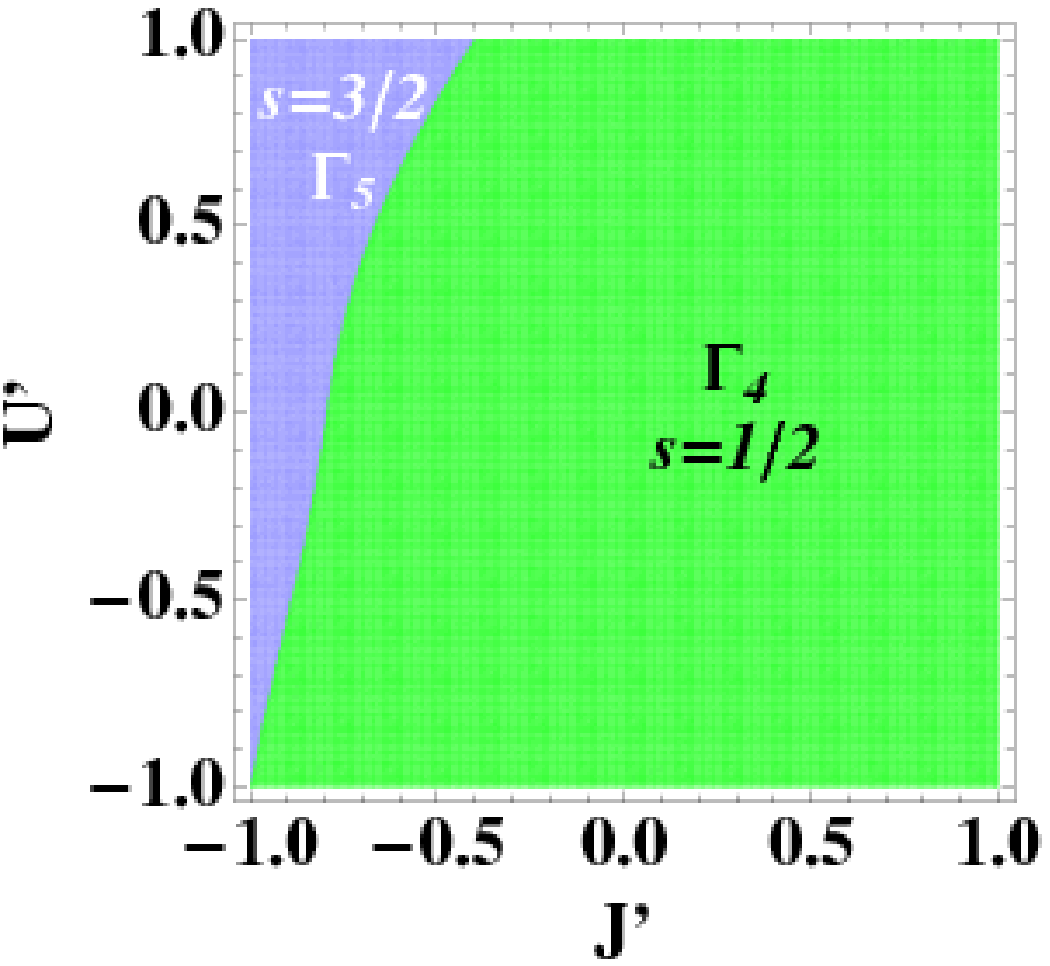,height=50mm}}
\end{picture}
\caption{Left: The complete canonical spectrum for the tetrahedron
  (scaled to primed values) in dependence on $J'$ for $h/t=0$, $U/t=4$ 
  and $W/t=0$. Right: The ground states for the complete $J$-$U$-plane 
  (scaled to primed values) and 
  for $W/t=0$ and $h/t=0$. The colours are according to the palette given in
  Figs. \ref{primedewsTriangleJ_isotrop}.
 }
\label{primedewsTetraJ_isotrop}
\end{figure}

\newpage
\begin{figure}
 \centering 
\unitlength=1mm
\begin{picture}(160,120)
\put(0,60){\epsfig{file=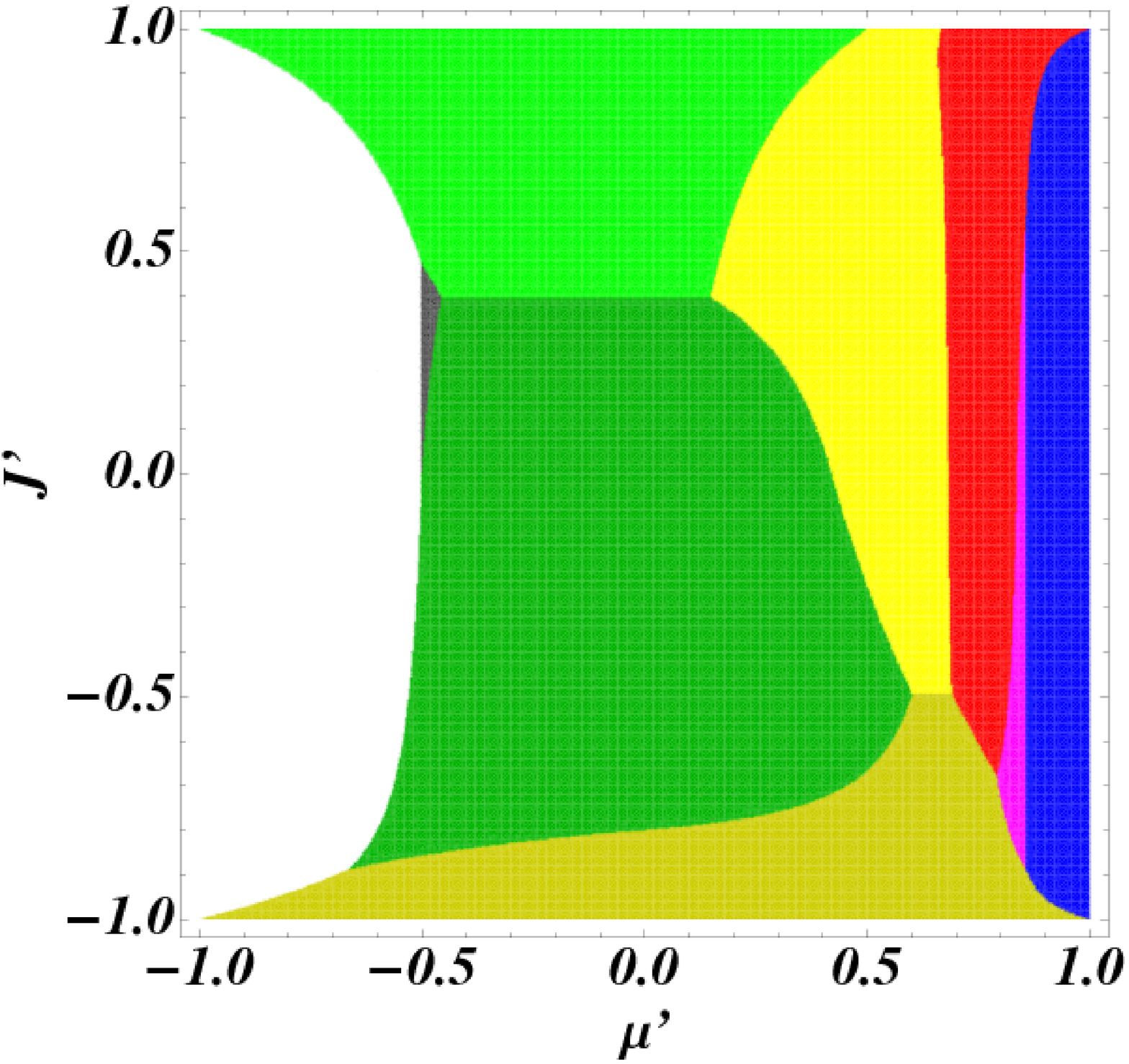,height=60mm}}
\put(70,60){\epsfig{file=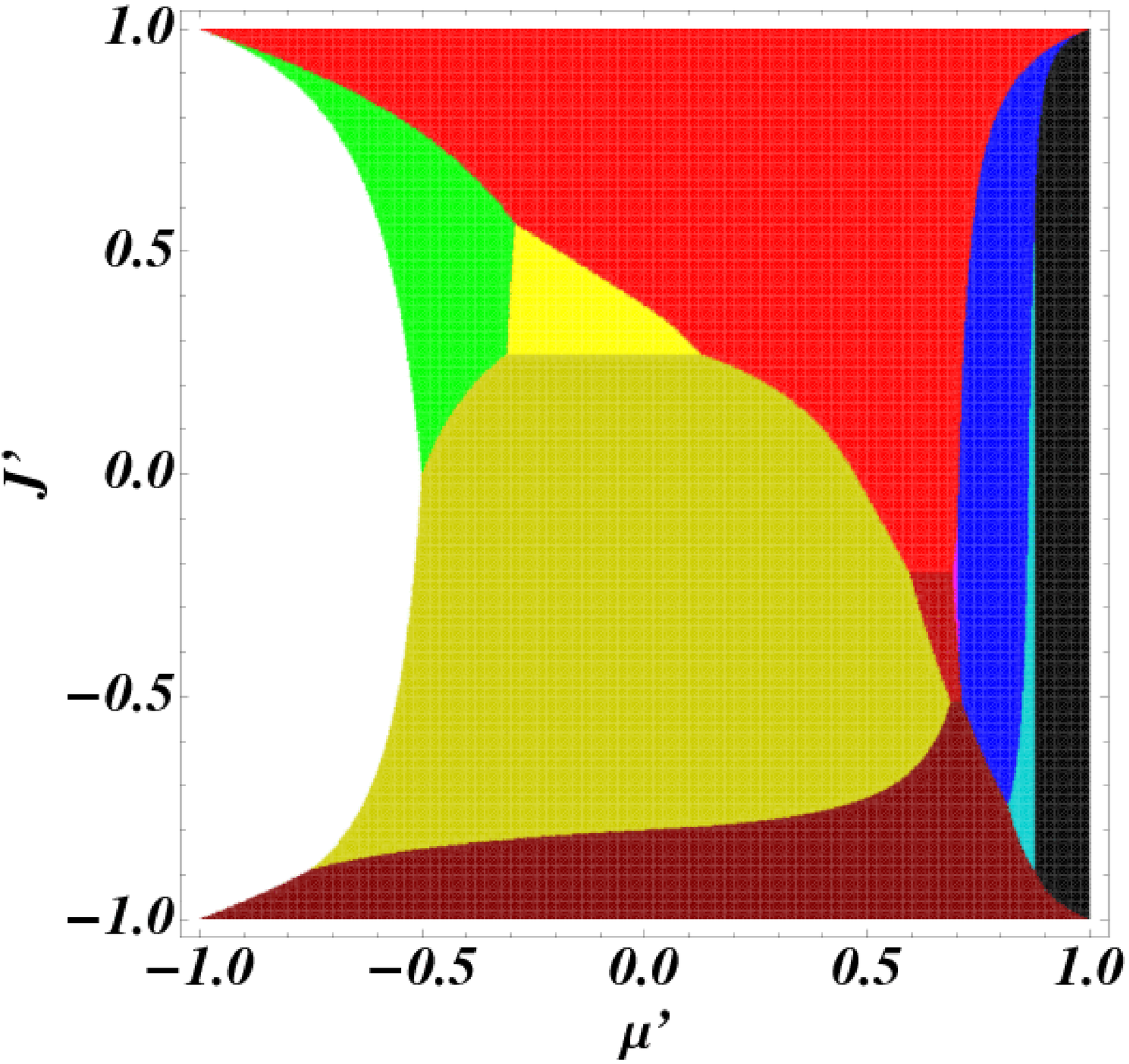,height=60mm}}
\put(0,0){\epsfig{file=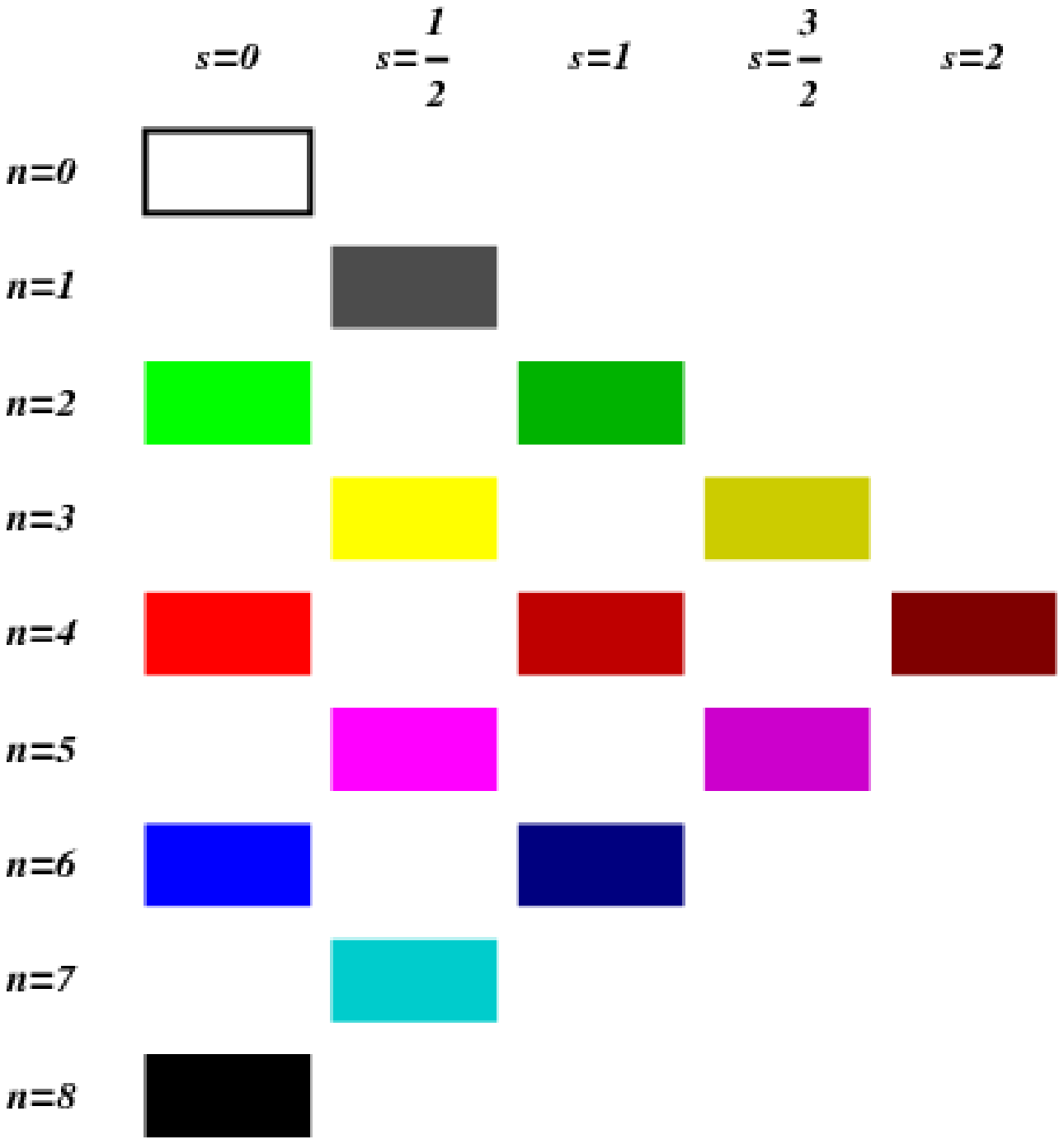,height=60mm}}
\put(70,5){\parbox[b]{65mm}{
\caption{
 The electron occupation $N$ and the spin eigenvalue $s$ of the ground state in dependence on the chemical potential and exchange parameter $J$
(both scaled to primed values)
for the triangle (upper left panel) and  tetrahedron (right panel) resp. and $U/t=4$, $h/t=0$, and $W/t=0$.
The lower left panel shows the palette, which was used to indicate the quantum numbers of the
different groundstates.
}}}
\end{picture}
%
\label{tetraNvonMueprimeJprimeU4_isotrop}
\end{figure}

This paper is concerned with the article \cite{Schumann08}.
In connection with the study of the extended Hubbard model
on an isosceles triangular cluster, we discovered a factor two error in 
the coding of the Ising part  of the Heisenberg term in the Hamiltonian.
In consequence the rigorous results remain valid, but not for
the isotropic Heisenberg model, but for an anisotropic Heisenberg model. 
In detail:  The eq. (4) of Ref. \cite{Schumann08} containing the additional Heisenberg term has to be changed to
\beq
\opH_J&=&\frac{J}{4}\,\sum_{\langle i \neq j \rangle} 
\left ( \opS_i^+ \opS_j^- +\opS_i^- \opS_j^+  +\opS_i^z \opS_j^z \right ) \nonumber \komma
\label{anisotropicexchange}
\eeq
where the double sum is  symmetrical in  $i$  and $j$, i.e. it counts every pair twice
what is corrected by the factor one half in front of it.
This term can be rewritten into the form
\beq
\opH_J&=&
\frac{J}{2}\,\sum_{\langle i \neq j \rangle} \left ( \opS_i^x \opS_j^x +\opS_i^y \opS_j^y \right )  
+ 
\frac{J}{4}\,\sum_{\langle i \neq j \rangle} \opS_i^z \opS_j^z        \nonumber \\
         &=:&
\frac{J_{\bot}}{2}\,\sum_{\langle i \neq j \rangle} 
\left ( \opS_i^x \opS_j^x +\opS_i^y \opS_j^y \right )  
+ 
\frac{J_{\|}}{2} \,\sum_{\langle i \neq j \rangle} \opS_i^z \opS_j^z   
\mbox{ with } J_{\bot}=2 J_{\|}=:J
\komma 
\eeq
what makes explicite, that this term describes an anisotropic Heisenberg exchange which is
in the $x$-$y$-plane twice as big in $z$-direction.
Thus, all results with $J=0$ of Ref. \cite{Schumann08} remain unchanged and all 
results with $J\ne0$ remain  true for the model with an anisotropic Heisenberg exchange
in the form of eq.  (\ref{anisotropicexchange}). 
Regarding the case $J\neq0$, results for 
the extended Hubbard model with anisotropic Heisenberg interaction are in general not too
different from the results with isotropic exchange 
Since the comparison of both cases allow to study the influence of
an anisotropy in the Heisenberg exchange, we add here the main differing results for the  isotropic case. 
The underlying closed analytic expressions of the eigenvalues and eigenvectors 
for the isotropic case are given in the appendix.
\section{Triangle with isotropic exchange interaction}
In Fig. \ref{primedewsTriangleJ_isotrop} we recalculated the main differing part of Fig. 2 of Ref. \cite{Schumann08}, where
we have used the same  scaling function for the primed values. For an isotropic exchange interaction the energy levels become independent on the magnetic quantum number, therefore we used the eigenvalues of the total spin 
for the intensity in the palette. Anisotropy lifts this degeneracy, resulting in more possible groundstates.
In Fig. \ref{triangleLevelCrossings_isotrop} we show  all the groundstate levelcrossings for
the isotropic case.. 
The insets show the $J$ dependence of the complete canonical spectra. 
The levels are denoted in the legend by their quantum numbers.
If the cluster is occupied with two electrons we find  crossings from a 
threefold degenerate groundstate with spin one and representation $\Gamma_2$.
 to two eigenstates of spin zero and representation $\Gamma_3$ for $U>0$ and to
a nondegenerate state of symmetry $\Gamma_1$.
The critical $J_c$, where this transition happens, is dependent on the on-site correlation according to 
\beq
J_c&=&
\frac{1}{2} \left(9 t+4 U-\sqrt{81 t^2+56 U t+16 U^2}\right)  \qquad U>0\\
J_c&=&
2 \left(3 t+U-\sqrt{9 t^2+2 U t+U^2}\right) \qquad U<0
\punkt
\eeq
For three electrons there is only one ground state level crossing, changing from
fourfold degenerate level (no. 27, 30, 36, 39)  with spin one half 
and irred. representation $\Gamma_3$ to 
the fourfold degenerate states (no. 23, 26,35,46) with spin three halves 
and irred. representation $\Gamma_2$. 
The crossing point $J_c$ is implicitely given as function on $U$ by
\beq
4 U-6 J_c&=&
\sqrt{9 J_c^2+24 U J_c+16 \left(27 t^2+U^2\right)} \times \nonumber \\
&&\hspace{1cm} \cos \left(\frac{1}{3} \cos ^{-1}\left(\frac{\left(4 U+3 J_c\right){}^3}{\left(9 J_c^2+24 U
 J_c+16 \left(27 t^2+U^2\right)\right){}^{3/2}}\right)\right)
\punkt
\eeq
The solution of the above equation is the  yellow-red borderline in the related groundstate diagram of Fig. \ref{primedewsTriangleJ_isotrop}.
If the electron occupation is four, a crossover from a sixfold degenerated groundstate (no. 44, 45, 51, 54, 56, 57) 
of symmetry $\Gamma_3$ with spin one  to a singlet (no. 46) with $\Gamma_1$ and 
spin zero takes place for negative values of $J$ (remember: $J<0$ favorites parallel spins in our notation).
The $U$ dependence of the critical exchange parameter is given by
\beq
J_c &=& \frac{1}{2} \left(3 t+4 U-\sqrt{153 t^2+40 U t+16 U^2}\right)
\punkt
\eeq
\section{Tetrahedron with isotropic exchange Interaction}
The groundstate phase diagrams and the complete spectra are given in Fig. \ref{primedewsTetraJ_isotrop}
which corresponds to Fig. 8 of Ref. \cite{Schumann08}. 
Instead, we indicate the irreducible representation and the total spin value.
The most interesting phase diagram we  find again for $N=4$. 
For the other occupation numbers the critical lines for the groundstate crossings can be given 
explicitely. We find for 
$N=2$ 
\beq
J_c&=& 
\left \{
\begin{array}{ll}
0&U>0\\[1ex]
7 t+2 U-\sqrt{49 t^2+4 U t+4 U^2}& J<0 \quad U<0\\[1ex]
-\frac{4}{3}U&0<J \quad U<0
\end{array} 
\right .
\komma.
\eeq
for 
$N=3$ 
\beq
J_c&=& 
\left \{
\begin{array}{ll}
\frac{4}{3}U&U>0\\[1ex]
\frac{1}{3} \left(5 t+2 U-\sqrt{25 t^2-4 U t+4 U^2} \right) \qquad & J<0 \quad U<0\\[1ex]
-\frac{4}{3}U&0<J \quad U<0
\end{array} 
\right .
\komma
\eeq
$N=6$
\beq
J_c&=& 3 t+2 U-\sqrt{73 t^2+20 U t+4 U^2}
\punkt
\eeq
 For the occupation numbers 4 and 5  the  "phase boundaries " can be
given implicitely from equating the adjacent groundstate levels. 
Due to the length of the formula
we refer for that case to the eigenvalues given in the appendix.
\section{$J'$-$\mu'$ phase diagram for the cluster gases }
In Fig. \ref{tetraNvonMueprimeJprimeU4_isotrop} we show the ground state phase diagram
for the triangular and tetragonal resp. cluster gas which correspond to the right panel of Fig. 17 and
Fig. 26 resp. of Ref. \cite{Schumann08}.
For the case of isotropic exchange we find a very small area where an $N=1$ state is lowest in the triangular cluster gas, which is absent in the anisotropic case.

In the appendix I give the complete eigensystem for the triangular and tetrahedral
cluster  with isotropic exchange. I am not going to publish these data elsewhere, since
it is to lengthy for any print media. I decided to do so, since I can not guarantee
the survival of my webpage \cite{webpage} in the future, where at present these results 
can be found in a more useable form.


\appendix
\newpage
\appendix
\section{Eigensystem of the extended Hubbard model with isotropic exchange interaction}
In the following tables we give the eigenvalues
for the extended model with isotropic exchange interaction.
The first column numbers the state.
Please note that there are slide
differences between the numbering in this paper and the numbers in Ref. \cite{Schumann08},
due to a different order in applying the symmetry operators. Thus, one has
to relate the states of  Ref. \cite{Schumann08} to the states given here by their quantum numbers instead, which are the eigenvalues of the $U$-independent
symmetry operators, i.e. the electron occupation number  $N_e$, the spin-projection 
$\opS_z$ in z-direction $m_s$, the eigenvalues $s(s+1)$ of $\opS^2$ and the spatial
symmetry. The latter is indicated by $\Gamma_{i,j}$, where the first index labels the irreducible
representation of the point group and the second numbers the partner.
The notation is based on Ref. \cite{CornwellBook}.
The third column gives the enery eigenvalues in abbreviated form. The abbreviations
are listed subsequently to the tables. In the last column a numerical value is given for example.
For comparison we have chosen the same parameters as in Ref. \cite{Schumann02}
corresponding to a pure Hubbard-model with $U=5t$, $W=J=0t$. If the grand-canonical 
energy levels in an applied magnetic field are needed one has to substract $\mu \, N_e+h\,m_s$.
\subsection{The triangular cluster}
\begin{tabular}[t]{|r|l|c|c|}
\multicolumn{4}{c}{\large \bf \boldmath Eigenkets and eigenvalues for ${\rm  N_e}$=0 and   ${\rm m_s}$= $0$. } \\ \hline
\parbox[c]{1cm}{No}  & \parbox[c]{2.5cm}{\begin{center} Eigenstate \end{center}}  & \parbox[c]{7.5cm}{ \begin{center}   Energy \end{center}}  & \parbox[c]{2cm}{ \begin{center} Example \end{center}}  \\ \hline 
\hline 
1 & $\ket{0,0,0,\Gamma_1}$  & $0$   & 0. \\ 
\hline
\end{tabular} \\[2ex]
\begin{tabular}[t]{|r|l|c|c|}
\multicolumn{4}{c}{\large \bf \boldmath Eigenkets and eigenvalues for ${\rm  N_e}$=1 and   ${\rm m_s}$= $- \frac{1}{2} $. } \\ \hline
\parbox[c]{1cm}{No}  & \parbox[c]{2.5cm}{\begin{center} Eigenstate \end{center}}  & \parbox[c]{7.5cm}{ \begin{center}   Energy \end{center}}  & \parbox[c]{2cm}{ \begin{center} Example \end{center}}  \\ \hline 
\hline 
2 & $\ket{1,- \frac{1}{2} , \frac{3}{4} ,\Gamma_1}$  & $2 t$   & 2. \\ 
3 & $\ket{1,- \frac{1}{2} , \frac{3}{4} ,\Gamma_{3,1}}$  & $-t$   & -1. \\ 
4 & $\ket{1,- \frac{1}{2} , \frac{3}{4} ,\Gamma_{3,2}}$  & $-t$   & -1. \\ 
\hline
\end{tabular} \\[2ex]
\begin{tabular}[t]{|r|l|c|c|}
\multicolumn{4}{c}{\large \bf \boldmath Eigenkets and eigenvalues for ${\rm  N_e}$=1 and   ${\rm m_s}$= $\frac{1}{2} $. } \\ \hline
\parbox[c]{1cm}{No}  & \parbox[c]{2.5cm}{\begin{center} Eigenstate \end{center}}  & \parbox[c]{7.5cm}{ \begin{center}   Energy \end{center}}  & \parbox[c]{2cm}{ \begin{center} Example \end{center}}  \\ \hline 
\hline 
5 & $\ket{1,\frac{1}{2} , \frac{3}{4} ,\Gamma_1}$  & $2 t$   & 2. \\ 
6 & $\ket{1,\frac{1}{2} , \frac{3}{4} ,\Gamma_{3,1}}$  & $-t$   & -1. \\ 
7 & $\ket{1,\frac{1}{2} , \frac{3}{4} ,\Gamma_{3,2}}$  & $-t$   & -1. \\ 
\hline
\end{tabular} \\[2ex]
\begin{tabular}[t]{|r|l|c|c|}
\multicolumn{4}{c}{\large \bf \boldmath Eigenkets and eigenvalues for ${\rm  N_e}$=2 and   ${\rm m_s}$= $-1$. } \\ \hline
\parbox[c]{1cm}{No}  & \parbox[c]{2.5cm}{\begin{center} Eigenstate \end{center}}  & \parbox[c]{7.5cm}{ \begin{center}   Energy \end{center}}  & \parbox[c]{2cm}{ \begin{center} Example \end{center}}  \\ \hline 
\hline 
8 & $\ket{2,-1,2,\Gamma_2}$  & $\frac{J}{4}-2 t+W$   & -2. \\ 
9 & $\ket{2,-1,2,\Gamma_{3,1}}$  & $\frac{J}{4}+t+W$   & 1. \\ 
10 & $\ket{2,-1,2,\Gamma_{3,2}}$  & $\frac{J}{4}+t+W$   & 1. \\ 
\hline
\end{tabular} \\[2ex]
\begin{tabular}[t]{|r|l|c|c|}
\multicolumn{4}{c}{\large \bf \boldmath Eigenkets and eigenvalues for ${\rm  N_e}$=2 and   ${\rm m_s}$= $0$. } \\ \hline
\parbox[c]{1cm}{No}  & \parbox[c]{2.5cm}{\begin{center} Eigenstate \end{center}}  & \parbox[c]{7.5cm}{ \begin{center}   Energy \end{center}}  & \parbox[c]{2cm}{ \begin{center} Example \end{center}}  \\ \hline 
\hline 
11 & $\ket{2,0,0,\Gamma_1}$  & $\frac{1}{8} \left(-3 J+8 t+4 U+4 W-\sqrt{A_5}\right)$   & 0.298438 \\ 
12 & $\ket{2,0,0,\Gamma_1}$  & $\frac{1}{8} \left(-3 J+8 t+4 U+4 W+\sqrt{A_5}\right)$   & 6.70156 \\ 
13 & $\ket{2,0,2,\Gamma_2}$  & $\frac{J}{4}-2 t+W$   & -2. \\ 
14 & $\ket{2,0,0,\Gamma_{3,1}}$  & $\frac{1}{8} \left(-3 J-4 t+4 U+4 W-\sqrt{A_2}\right)$   & -1.31662 \\ 
15 & $\ket{2,0,0,\Gamma_{3,1}}$  & $\frac{1}{8} \left(-3 J-4 t+4 U+4 W+\sqrt{A_2}\right)$   & 5.31662 \\ 
16 & $\ket{2,0,2,\Gamma_{3,1}}$  & $\frac{J}{4}+t+W$   & 1. \\ 
17 & $\ket{2,0,0,\Gamma_{3,2}}$  & $\frac{1}{8} \left(-3 J-4 t+4 U+4 W-\sqrt{A_2}\right)$   & -1.31662 \\ 
18 & $\ket{2,0,0,\Gamma_{3,2}}$  & $\frac{1}{8} \left(-3 J-4 t+4 U+4 W+\sqrt{A_2}\right)$   & 5.31662 \\ 
19 & $\ket{2,0,2,\Gamma_{3,2}}$  & $\frac{J}{4}+t+W$   & 1. \\ 
\hline
\end{tabular} \\[2ex]
\begin{tabular}[t]{|r|l|c|c|}
\multicolumn{4}{c}{\large \bf \boldmath Eigenkets and eigenvalues for ${\rm  N_e}$=2 and   ${\rm m_s}$= $1$. } \\ \hline
\parbox[c]{1cm}{No}  & \parbox[c]{2.5cm}{\begin{center} Eigenstate \end{center}}  & \parbox[c]{7.5cm}{ \begin{center}   Energy \end{center}}  & \parbox[c]{2cm}{ \begin{center} Example \end{center}}  \\ \hline 
\hline 
20 & $\ket{2,1,2,\Gamma_2}$  & $\frac{J}{4}-2 t+W$   & -2. \\ 
21 & $\ket{2,1,2,\Gamma_{3,1}}$  & $\frac{J}{4}+t+W$   & 1. \\ 
22 & $\ket{2,1,2,\Gamma_{3,2}}$  & $\frac{J}{4}+t+W$   & 1. \\ 
\hline
\end{tabular} \\[2ex]
\begin{tabular}[t]{|r|l|c|c|}
\multicolumn{4}{c}{\large \bf \boldmath Eigenkets and eigenvalues for ${\rm  N_e}$=3 and   ${\rm m_s}$= $- \frac{3}{2} $. } \\ \hline
\parbox[c]{1cm}{No}  & \parbox[c]{2.5cm}{\begin{center} Eigenstate \end{center}}  & \parbox[c]{7.5cm}{ \begin{center}   Energy \end{center}}  & \parbox[c]{2cm}{ \begin{center} Example \end{center}}  \\ \hline 
\hline 
23 & $\ket{3,- \frac{3}{2} , \frac{15}{4} ,\Gamma_2}$  & $\frac{3}{4} (J+4 W)$   & 0. \\ 
\hline
\end{tabular} \\[2ex]
\begin{tabular}[t]{|r|l|c|c|}
\multicolumn{4}{c}{\large \bf \boldmath Eigenkets and eigenvalues for ${\rm  N_e}$=3 and   ${\rm m_s}$= $- \frac{1}{2} $. } \\ \hline
\parbox[c]{1cm}{No}  & \parbox[c]{2.5cm}{\begin{center} Eigenstate \end{center}}  & \parbox[c]{7.5cm}{ \begin{center}   Energy \end{center}}  & \parbox[c]{2cm}{ \begin{center} Example \end{center}}  \\ \hline 
\hline 
24 & $\ket{3,- \frac{1}{2} , \frac{3}{4} ,\Gamma_1}$  & $U+2 W$   & 5. \\ 
25 & $\ket{3,- \frac{1}{2} , \frac{3}{4} ,\Gamma_2}$  & $U+2 W$   & 5. \\ 
26 & $\ket{3,- \frac{1}{2} , \frac{15}{4} ,\Gamma_2}$  & $\frac{3 J}{4}+3 W$   & 0. \\ 
27 & $\ket{3,- \frac{1}{2} , \frac{3}{4} ,\Gamma_{3,1}}$  & $A_6$   & -1.07504 \\ 
28 & $\ket{3,- \frac{1}{2} , \frac{3}{4} ,\Gamma_{3,1}}$  & $A_{10}$   & 7.19825 \\ 
29 & $\ket{3,- \frac{1}{2} , \frac{3}{4} ,\Gamma_{3,1}}$  & $A_8$   & 3.87678 \\ 
30 & $\ket{3,- \frac{1}{2} , \frac{3}{4} ,\Gamma_{3,2}}$  & $A_6$   & -1.07504 \\ 
31 & $\ket{3,- \frac{1}{2} , \frac{3}{4} ,\Gamma_{3,2}}$  & $A_{10}$   & 7.19825 \\ 
32 & $\ket{3,- \frac{1}{2} , \frac{3}{4} ,\Gamma_{3,2}}$  & $A_8$   & 3.87678 \\ 
\hline
\end{tabular} \\[2ex]
\begin{tabular}[t]{|r|l|c|c|}
\multicolumn{4}{c}{\large \bf \boldmath Eigenkets and eigenvalues for ${\rm  N_e}$=3 and   ${\rm m_s}$= $\frac{1}{2} $. } \\ \hline
\parbox[c]{1cm}{No}  & \parbox[c]{2.5cm}{\begin{center} Eigenstate \end{center}}  & \parbox[c]{7.5cm}{ \begin{center}   Energy \end{center}}  & \parbox[c]{2cm}{ \begin{center} Example \end{center}}  \\ \hline 
\hline 
33 & $\ket{3,\frac{1}{2} , \frac{3}{4} ,\Gamma_1}$  & $U+2 W$   & 5. \\ 
34 & $\ket{3,\frac{1}{2} , \frac{3}{4} ,\Gamma_2}$  & $U+2 W$   & 5. \\ 
35 & $\ket{3,\frac{1}{2} , \frac{15}{4} ,\Gamma_2}$  & $\frac{3 J}{4}+3 W$   & 0. \\ 
36 & $\ket{3,\frac{1}{2} , \frac{3}{4} ,\Gamma_{3,1}}$  & $A_6$   & -1.07504 \\ 
37 & $\ket{3,\frac{1}{2} , \frac{3}{4} ,\Gamma_{3,1}}$  & $A_{11}$   & 7.19825 \\ 
38 & $\ket{3,\frac{1}{2} , \frac{3}{4} ,\Gamma_{3,1}}$  & $A_9$   & 3.87678 \\ 
39 & $\ket{3,\frac{1}{2} , \frac{3}{4} ,\Gamma_{3,2}}$  & $A_6$   & -1.07504 \\ 
40 & $\ket{3,\frac{1}{2} , \frac{3}{4} ,\Gamma_{3,2}}$  & $A_{11}$   & 7.19825 \\ 
41 & $\ket{3,\frac{1}{2} , \frac{3}{4} ,\Gamma_{3,2}}$  & $A_9$   & 3.87678 \\ 
\hline
\end{tabular} \\[2ex]
\begin{tabular}[t]{|r|l|c|c|}
\multicolumn{4}{c}{\large \bf \boldmath Eigenkets and eigenvalues for ${\rm  N_e}$=3 and   ${\rm m_s}$= $\frac{3}{2} $. } \\ \hline
\parbox[c]{1cm}{No}  & \parbox[c]{2.5cm}{\begin{center} Eigenstate \end{center}}  & \parbox[c]{7.5cm}{ \begin{center}   Energy \end{center}}  & \parbox[c]{2cm}{ \begin{center} Example \end{center}}  \\ \hline 
\hline 
42 & $\ket{3,\frac{3}{2} , \frac{15}{4} ,\Gamma_2}$  & $\frac{3 J}{4}+3 W$   & 0. \\ 
\hline
\end{tabular} \\[2ex]
\begin{tabular}[t]{|r|l|c|c|}
\multicolumn{4}{c}{\large \bf \boldmath Eigenkets and eigenvalues for ${\rm  N_e}$=4 and   ${\rm m_s}$= $-1$. } \\ \hline
\parbox[c]{1cm}{No}  & \parbox[c]{2.5cm}{\begin{center} Eigenstate \end{center}}  & \parbox[c]{7.5cm}{ \begin{center}   Energy \end{center}}  & \parbox[c]{2cm}{ \begin{center} Example \end{center}}  \\ \hline 
\hline 
43 & $\ket{4,-1,2,\Gamma_2}$  & $\frac{J}{4}+2 t+U+5 W$   & 7. \\ 
44 & $\ket{4,-1,2,\Gamma_{3,1}}$  & $\frac{J}{4}-t+U+5 W$   & 4. \\ 
45 & $\ket{4,-1,2,\Gamma_{3,2}}$  & $\frac{J}{4}-t+U+5 W$   & 4. \\ 
\hline
\end{tabular} \\[2ex]
\begin{tabular}[t]{|r|l|c|c|}
\multicolumn{4}{c}{\large \bf \boldmath Eigenkets and eigenvalues for ${\rm  N_e}$=4 and   ${\rm m_s}$= $0$. } \\ \hline
\parbox[c]{1cm}{No}  & \parbox[c]{2.5cm}{\begin{center} Eigenstate \end{center}}  & \parbox[c]{7.5cm}{ \begin{center}   Energy \end{center}}  & \parbox[c]{2cm}{ \begin{center} Example \end{center}}  \\ \hline 
\hline 
46 & $\ket{4,0,0,\Gamma_1}$  & $\frac{1}{8} \left(-3 J-8 t+12 U+36 W-\sqrt{A_3}\right)$   & 2. \\ 
47 & $\ket{4,0,0,\Gamma_1}$  & $\frac{1}{8} \left(-3 J-8 t+12 U+36 W+\sqrt{A_3}\right)$   & 11. \\ 
48 & $\ket{4,0,2,\Gamma_2}$  & $\frac{J}{4}+2 t+U+5 W$   & 7. \\ 
49 & $\ket{4,0,0,\Gamma_{3,1}}$  & $\frac{1}{8} \left(-3 J+4 t+12 U+36 W-\sqrt{A_4}\right)$   & 5.55051 \\ 
50 & $\ket{4,0,0,\Gamma_{3,1}}$  & $\frac{1}{8} \left(-3 J+4 t+12 U+36 W+\sqrt{A_4}\right)$   & 10.4495 \\ 
51 & $\ket{4,0,2,\Gamma_{3,1}}$  & $\frac{J}{4}-t+U+5 W$   & 4. \\ 
52 & $\ket{4,0,0,\Gamma_{3,2}}$  & $\frac{1}{8} \left(-3 J+4 t+12 U+36 W-\sqrt{A_4}\right)$   & 5.55051 \\ 
53 & $\ket{4,0,0,\Gamma_{3,2}}$  & $\frac{1}{8} \left(-3 J+4 t+12 U+36 W+\sqrt{A_4}\right)$   & 10.4495 \\ 
54 & $\ket{4,0,2,\Gamma_{3,2}}$  & $\frac{J}{4}-t+U+5 W$   & 4. \\ 
\hline
\end{tabular} \\[2ex]
\begin{tabular}[t]{|r|l|c|c|}
\multicolumn{4}{c}{\large \bf \boldmath Eigenkets and eigenvalues for ${\rm  N_e}$=4 and   ${\rm m_s}$= $1$. } \\ \hline
\parbox[c]{1cm}{No}  & \parbox[c]{2.5cm}{\begin{center} Eigenstate \end{center}}  & \parbox[c]{7.5cm}{ \begin{center}   Energy \end{center}}  & \parbox[c]{2cm}{ \begin{center} Example \end{center}}  \\ \hline 
\hline 
55 & $\ket{4,1,2,\Gamma_2}$  & $\frac{J}{4}+2 t+U+5 W$   & 7. \\ 
56 & $\ket{4,1,2,\Gamma_{3,1}}$  & $\frac{J}{4}-t+U+5 W$   & 4. \\ 
57 & $\ket{4,1,2,\Gamma_{3,2}}$  & $\frac{J}{4}-t+U+5 W$   & 4. \\ 
\hline
\end{tabular} \\[2ex]
\begin{tabular}[t]{|r|l|c|c|}
\multicolumn{4}{c}{\large \bf \boldmath Eigenkets and eigenvalues for ${\rm  N_e}$=5 and   ${\rm m_s}$= $- \frac{1}{2} $. } \\ \hline
\parbox[c]{1cm}{No}  & \parbox[c]{2.5cm}{\begin{center} Eigenstate \end{center}}  & \parbox[c]{7.5cm}{ \begin{center}   Energy \end{center}}  & \parbox[c]{2cm}{ \begin{center} Example \end{center}}  \\ \hline 
\hline 
58 & $\ket{5,- \frac{1}{2} , \frac{3}{4} ,\Gamma_1}$  & $-2 t+2 U+8 W$   & 8. \\ 
59 & $\ket{5,- \frac{1}{2} , \frac{3}{4} ,\Gamma_{3,1}}$  & $t+2 U+8 W$   & 11. \\ 
60 & $\ket{5,- \frac{1}{2} , \frac{3}{4} ,\Gamma_{3,2}}$  & $t+2 U+8 W$   & 11. \\ 
\hline
\end{tabular} \\[2ex]
\begin{tabular}[t]{|r|l|c|c|}
\multicolumn{4}{c}{\large \bf \boldmath Eigenkets and eigenvalues for ${\rm  N_e}$=5 and   ${\rm m_s}$= $\frac{1}{2} $. } \\ \hline
\parbox[c]{1cm}{No}  & \parbox[c]{2.5cm}{\begin{center} Eigenstate \end{center}}  & \parbox[c]{7.5cm}{ \begin{center}   Energy \end{center}}  & \parbox[c]{2cm}{ \begin{center} Example \end{center}}  \\ \hline 
\hline 
61 & $\ket{5,\frac{1}{2} , \frac{3}{4} ,\Gamma_1}$  & $-2 t+2 U+8 W$   & 8. \\ 
62 & $\ket{5,\frac{1}{2} , \frac{3}{4} ,\Gamma_{3,1}}$  & $t+2 U+8 W$   & 11. \\ 
63 & $\ket{5,\frac{1}{2} , \frac{3}{4} ,\Gamma_{3,2}}$  & $t+2 U+8 W$   & 11. \\ 
\hline
\end{tabular} \\[2ex]
\begin{tabular}[t]{|r|l|c|c|}
\multicolumn{4}{c}{\large \bf \boldmath Eigenkets and eigenvalues for ${\rm  N_e}$=6 and   ${\rm m_s}$= $0$. } \\ \hline
\parbox[c]{1cm}{No}  & \parbox[c]{2.5cm}{\begin{center} Eigenstate \end{center}}  & \parbox[c]{7.5cm}{ \begin{center}   Energy \end{center}}  & \parbox[c]{2cm}{ \begin{center} Example \end{center}}  \\ \hline 
\hline 
64 & $\ket{6,0,0,\Gamma_1}$  & $3 (U+4 W)$   & 15. \\ 
\hline
\end{tabular} \\

\parindent0cm
\subsection*{ List of abbreviations }
\beq 
Y &=& U-W+\frac{3}{4}J 
\nonumber \\
A_1 &=& 16 \left(27 t^2+Y^2\right)
 \nonumber \\
A_2 &=& 16 \left(9 t^2+2 Y t+Y^2\right)
 \nonumber \\
A_3 &=& 16 \left(36 t^2+4 Y t+Y^2\right)
 \nonumber \\
A_4 &=& 16 \left(9 t^2-2 Y t+Y^2\right)
 \nonumber \\
A_5 &=& 16 \left(36 t^2-4 Y t+Y^2\right)
 \nonumber \\
A_6 &=& U+2 W-\frac{Y}{3}-\frac{1}{6} \cos \left(\theta _1\right) \sqrt{A_1}
 \nonumber \\
A_7 &=& 4 (-3 U-6 W+Y)+2 \cos \left(\theta _1\right) \sqrt{A_1}
 \nonumber \\
A_8 &=& \frac{1}{12} \left(4 (3 U+6 W-Y)+\left(\cos \left(\theta _1\right)-\sqrt{3} \sin \left(\theta _1\right)\right) \sqrt{A_1}\right)
 \nonumber \\
A_9 &=& \frac{1}{12} \left(4 (3 U+6 W-Y)+\left(\cos \left(\theta _1\right)-\sqrt{3} \sin \left(\theta _1\right)\right) \sqrt{A_1}\right)
 \nonumber \\
A_{10} &=& \frac{1}{12} \left(4 (3 U+6 W-Y)+\left(\cos \left(\theta _1\right)+\sqrt{3} \sin \left(\theta _1\right)\right) \sqrt{A_1}\right)
 \nonumber \\
A_{11} &=& \frac{1}{12} \left(4 (3 U+6 W-Y)+\left(\cos \left(\theta _1\right)+\sqrt{3} \sin \left(\theta _1\right)\right) \sqrt{A_1}\right)
 \nonumber \\
\eeq \newpage \beq
\theta _1 &=& \frac{1}{3} \cos ^{-1}\left(\frac{64 Y^3}{A_1^{3/2}}\right)
\nonumber \eeq 

\small
{\subsubsection{\boldmath Eigenvectors for ${\rm  N_e}=0$ and   ${\rm m_s}$=  $0$.}
\beq
\ket{\Psi_{1}} & = & \ket{0,0,0,\Gamma_1} \nonumber \\ 
&=& 1
 \left ( \ket{000} \right) \nonumber 
\eeq
{\subsubsection{\boldmath Eigenvectors for ${\rm  N_e}=1$ and   ${\rm m_s}$=  $- \frac{1}{2} $.}
\beq
\ket{\Psi_{2}} & = & \ket{1,- \frac{1}{2} , \frac{3}{4} ,\Gamma_1} \nonumber \\ 
&=& \frac{1}{\sqrt{3}}
 \left ( \ket{00d} + \ket{0d0} + \ket{d00} \right) \nonumber 
\eeq
\beq
\ket{\Psi_{3}} & = & \ket{1,- \frac{1}{2} , \frac{3}{4} ,\Gamma_{3,1}} \nonumber \\ 
&=& \frac{1}{\sqrt{2}}
 \left ( \ket{00d} - \ket{0d0} \right) \nonumber 
\eeq
\beq
\ket{\Psi_{4}} & = & \ket{1,- \frac{1}{2} , \frac{3}{4} ,\Gamma_{3,2}} \nonumber \\ 
& = & \quad 
C_{4,1} \left ( 
\ket{00d} + \ket{0d0} \right) 
 \nonumber \\
& & + 
C_{4,2} \left ( 
\ket{d00} \right) 
\nonumber \eeq
\beq 
C_{4,1} &=& 
-\frac{1}{\sqrt{6}} \nonumber \\
C_{4,2} &=& 
\sqrt{\frac{2}{3}} \nonumber \\
 N_{4} &=& \sqrt{2 C_{4,1}^2+C_{4,2}^2} \nonumber \eeq 
{\subsubsection{\boldmath Eigenvectors for ${\rm  N_e}=1$ and   ${\rm m_s}$=  $\frac{1}{2} $.}
\beq
\ket{\Psi_{5}} & = & \ket{1,\frac{1}{2} , \frac{3}{4} ,\Gamma_1} \nonumber \\ 
&=& \frac{1}{\sqrt{3}}
 \left ( \ket{00u} + \ket{0u0} + \ket{u00} \right) \nonumber 
\eeq
\beq
\ket{\Psi_{6}} & = & \ket{1,\frac{1}{2} , \frac{3}{4} ,\Gamma_{3,1}} \nonumber \\ 
&=& \frac{1}{\sqrt{2}}
 \left ( \ket{00u} - \ket{0u0} \right) \nonumber 
\eeq
\beq
\ket{\Psi_{7}} & = & \ket{1,\frac{1}{2} , \frac{3}{4} ,\Gamma_{3,2}} \nonumber \\ 
& = & \quad 
C_{7,1} \left ( 
\ket{00u} + \ket{0u0} \right) 
 \nonumber \\
& & + 
C_{7,2} \left ( 
\ket{u00} \right) 
\nonumber \eeq
\beq 
C_{7,1} &=& 
-\frac{1}{\sqrt{6}} \nonumber \\
C_{7,2} &=& 
\sqrt{\frac{2}{3}} \nonumber \\
 N_{7} &=& \sqrt{2 C_{7,1}^2+C_{7,2}^2} \nonumber \eeq 
{\subsubsection{\boldmath Eigenvectors for ${\rm  N_e}=2$ and   ${\rm m_s}$=  $-1$.}
\beq
\ket{\Psi_{8}} & = & \ket{2,-1,2,\Gamma_2} \nonumber \\ 
&=& \frac{1}{\sqrt{3}}
 \left ( \ket{0dd} - \ket{d0d} + \ket{dd0} \right) \nonumber 
\eeq
\beq
\ket{\Psi_{9}} & = & \ket{2,-1,2,\Gamma_{3,1}} \nonumber \\ 
& = & \quad 
C_{9,1} \left ( 
\ket{0dd} \right) 
 \nonumber \\
& & + 
C_{9,2} \left ( 
\ket{d0d} - \ket{dd0} \right) 
\nonumber \eeq
\beq 
C_{9,1} &=& 
-\sqrt{\frac{2}{3}} \nonumber \\
C_{9,2} &=& 
-\frac{1}{\sqrt{6}} \nonumber \\
 N_{9} &=& \sqrt{C_{9,1}^2+2 C_{9,2}^2} \nonumber \eeq 
\beq
\ket{\Psi_{10}} & = & \ket{2,-1,2,\Gamma_{3,2}} \nonumber \\ 
&=& \frac{1}{\sqrt{2}}
 \left ( \ket{d0d} + \ket{dd0} \right) \nonumber 
\eeq
{\subsubsection{\boldmath Eigenvectors for ${\rm  N_e}=2$ and   ${\rm m_s}$=  $0$.}
\beq
\ket{\Psi_{11}} & = & \ket{2,0,0,\Gamma_1} \nonumber \\ 
& = & \quad 
C_{11,1} \left ( 
\ket{002} + \ket{020} + \ket{200} \right) 
 \nonumber \\
& & + 
C_{11,2} \left ( 
\ket{0du} - \ket{0ud} + \ket{d0u} + \ket{du0} - \ket{u0d} - \ket{ud0} \right) 
\nonumber \eeq
\beq 
C_{11,1} &=& 
2 \sqrt{\frac{2}{3}} t \nonumber \\
C_{11,2} &=& 
\frac{1}{8 \sqrt{6}} \left (
3 J-8 t+4 U-4 W+\sqrt{A_5} \right ) \nonumber \\
 N_{11} &=& \sqrt{3 C_{11,1}^2+6 C_{11,2}^2} \nonumber \eeq 
\beq
\ket{\Psi_{12}} & = & \ket{2,0,0,\Gamma_1} \nonumber \\ 
& = & \quad 
C_{12,1} \left ( 
\ket{002} + \ket{020} + \ket{200} \right) 
 \nonumber \\
& & + 
C_{12,2} \left ( 
\ket{0du} - \ket{0ud} + \ket{d0u} + \ket{du0} - \ket{u0d} - \ket{ud0} \right) 
\nonumber \eeq
\beq 
C_{12,1} &=& 
2 \sqrt{\frac{2}{3}} t \nonumber \\
C_{12,2} &=& 
\frac{1}{8 \sqrt{6}} \left (
3 J-8 t+4 U-4 W-\sqrt{A_5} \right ) \nonumber \\
 N_{12} &=& \sqrt{3 C_{12,1}^2+6 C_{12,2}^2} \nonumber \eeq 
\beq
\ket{\Psi_{13}} & = & \ket{2,0,2,\Gamma_2} \nonumber \\ 
&=& \frac{1}{\sqrt{6}}
 \left ( \ket{0du} + \ket{0ud} - \ket{d0u} + \ket{du0} - \ket{u0d} + \ket{ud0} \right) \nonumber 
\eeq
\beq
\ket{\Psi_{14}} & = & \ket{2,0,0,\Gamma_{3,1}} \nonumber \\ 
& = & \quad 
C_{14,1} \left ( 
\ket{002} - \ket{020} \right) 
 \nonumber \\
& & + 
C_{14,2} \left ( 
\ket{d0u} - \ket{du0} - \ket{u0d} + \ket{ud0} \right) 
\nonumber \eeq
\beq 
C_{14,1} &=& 
t \nonumber \\
C_{14,2} &=& 
\frac{1}{16} \left (
3 J+4 t+4 U-4 W+\sqrt{A_2} \right ) \nonumber \\
 N_{14} &=& \sqrt{2 C_{14,1}^2+4 C_{14,2}^2} \nonumber \eeq 
\beq
\ket{\Psi_{15}} & = & \ket{2,0,0,\Gamma_{3,1}} \nonumber \\ 
& = & \quad 
C_{15,1} \left ( 
\ket{002} - \ket{020} \right) 
 \nonumber \\
& & + 
C_{15,2} \left ( 
\ket{d0u} - \ket{du0} - \ket{u0d} + \ket{ud0} \right) 
\nonumber \eeq
\beq 
C_{15,1} &=& 
t \nonumber \\
C_{15,2} &=& 
\frac{1}{16} \left (
3 J+4 t+4 U-4 W-\sqrt{A_2} \right ) \nonumber \\
 N_{15} &=& \sqrt{2 C_{15,1}^2+4 C_{15,2}^2} \nonumber \eeq 
\beq
\ket{\Psi_{16}} & = & \ket{2,0,2,\Gamma_{3,1}} \nonumber \\ 
& = & \quad 
C_{16,1} \left ( 
\ket{0du} + \ket{0ud} \right) 
 \nonumber \\
& & + 
C_{16,2} \left ( 
\ket{d0u} - \ket{du0} + \ket{u0d} - \ket{ud0} \right) 
\nonumber \eeq
\beq 
C_{16,1} &=& 
-\frac{1}{\sqrt{3}} \nonumber \\
C_{16,2} &=& 
-\frac{1}{2 \sqrt{3}} \nonumber \\
 N_{16} &=& \sqrt{2 C_{16,1}^2+4 C_{16,2}^2} \nonumber \eeq 
\beq
\ket{\Psi_{17}} & = & \ket{2,0,0,\Gamma_{3,2}} \nonumber \\ 
& = & \quad 
C_{17,1} \left ( 
\ket{002} + \ket{020} \right) 
 \nonumber \\
& & + 
C_{17,2} \left ( 
\ket{0du} - \ket{0ud} \right) 
 \nonumber \\
& & + 
C_{17,3} \left ( 
\ket{200} \right) 
 \nonumber \\
& & + 
C_{17,4} \left ( 
\ket{d0u} + \ket{du0} - \ket{u0d} - \ket{ud0} \right) 
\nonumber \eeq
\beq 
C_{17,1} &=& 
-\frac{1}{8 \sqrt{6}} \left (
3 J+4 t+4 U-4 W-\sqrt{A_2} \right ) \nonumber \\
C_{17,2} &=& 
\sqrt{\frac{2}{3}} t \nonumber \\
C_{17,3} &=& 
\frac{1}{4 \sqrt{6}} \left (
3 J+4 t+4 U-4 W-\sqrt{A_2} \right ) \nonumber \\
C_{17,4} &=& 
-\frac{t}{\sqrt{6}} \nonumber \\
 N_{17} &=& \sqrt{2 C_{17,1}^2+2 C_{17,2}^2+C_{17,3}^2+4 C_{17,4}^2} \nonumber \eeq 
\beq
\ket{\Psi_{18}} & = & \ket{2,0,0,\Gamma_{3,2}} \nonumber \\ 
& = & \quad 
C_{18,1} \left ( 
\ket{002} + \ket{020} \right) 
 \nonumber \\
& & + 
C_{18,2} \left ( 
\ket{0du} - \ket{0ud} \right) 
 \nonumber \\
& & + 
C_{18,3} \left ( 
\ket{200} \right) 
 \nonumber \\
& & + 
C_{18,4} \left ( 
\ket{d0u} + \ket{du0} - \ket{u0d} - \ket{ud0} \right) 
\nonumber \eeq
\beq 
C_{18,1} &=& 
-\frac{1}{8 \sqrt{6}} \left (
3 J+4 t+4 U-4 W+\sqrt{A_2} \right ) \nonumber \\
C_{18,2} &=& 
\sqrt{\frac{2}{3}} t \nonumber \\
C_{18,3} &=& 
\frac{1}{4 \sqrt{6}} \left (
3 J+4 t+4 U-4 W+\sqrt{A_2} \right ) \nonumber \\
C_{18,4} &=& 
-\frac{t}{\sqrt{6}} \nonumber \\
 N_{18} &=& \sqrt{2 C_{18,1}^2+2 C_{18,2}^2+C_{18,3}^2+4 C_{18,4}^2} \nonumber \eeq 
\beq
\ket{\Psi_{19}} & = & \ket{2,0,2,\Gamma_{3,2}} \nonumber \\ 
&=& \frac{1}{2}
 \left ( \ket{d0u} + \ket{du0} + \ket{u0d} + \ket{ud0} \right) \nonumber 
\eeq
{\subsubsection{\boldmath Eigenvectors for ${\rm  N_e}=2$ and   ${\rm m_s}$=  $1$.}
\beq
\ket{\Psi_{20}} & = & \ket{2,1,2,\Gamma_2} \nonumber \\ 
&=& \frac{1}{\sqrt{3}}
 \left ( \ket{0uu} - \ket{u0u} + \ket{uu0} \right) \nonumber 
\eeq
\beq
\ket{\Psi_{21}} & = & \ket{2,1,2,\Gamma_{3,1}} \nonumber \\ 
& = & \quad 
C_{21,1} \left ( 
\ket{0uu} \right) 
 \nonumber \\
& & + 
C_{21,2} \left ( 
\ket{u0u} - \ket{uu0} \right) 
\nonumber \eeq
\beq 
C_{21,1} &=& 
-\sqrt{\frac{2}{3}} \nonumber \\
C_{21,2} &=& 
-\frac{1}{\sqrt{6}} \nonumber \\
 N_{21} &=& \sqrt{C_{21,1}^2+2 C_{21,2}^2} \nonumber \eeq 
\beq
\ket{\Psi_{22}} & = & \ket{2,1,2,\Gamma_{3,2}} \nonumber \\ 
&=& \frac{1}{\sqrt{2}}
 \left ( \ket{u0u} + \ket{uu0} \right) \nonumber 
\eeq
{\subsubsection{\boldmath Eigenvectors for ${\rm  N_e}=3$ and   ${\rm m_s}$=  $- \frac{3}{2} $.}
\beq
\ket{\Psi_{23}} & = & \ket{3,- \frac{3}{2} , \frac{15}{4} ,\Gamma_2} \nonumber \\ 
&=& 1
 \left ( \ket{ddd} \right) \nonumber 
\eeq
{\subsubsection{\boldmath Eigenvectors for ${\rm  N_e}=3$ and   ${\rm m_s}$=  $- \frac{1}{2} $.}
\beq
\ket{\Psi_{24}} & = & \ket{3,- \frac{1}{2} , \frac{3}{4} ,\Gamma_1} \nonumber \\ 
&=& \frac{1}{\sqrt{6}}
 \left ( \ket{02d} + \ket{0d2} + \ket{20d} + \ket{2d0} + \ket{d02} + \ket{d20} \right) \nonumber 
\eeq
\beq
\ket{\Psi_{25}} & = & \ket{3,- \frac{1}{2} , \frac{3}{4} ,\Gamma_2} \nonumber \\ 
&=& \frac{1}{\sqrt{6}}
 \left ( \ket{02d} - \ket{0d2} - \ket{20d} + \ket{2d0} + \ket{d02} - \ket{d20} \right) \nonumber 
\eeq
\beq
\ket{\Psi_{26}} & = & \ket{3,- \frac{1}{2} , \frac{15}{4} ,\Gamma_2} \nonumber \\ 
&=& \frac{1}{\sqrt{3}}
 \left ( \ket{ddu} + \ket{dud} + \ket{udd} \right) \nonumber 
\eeq
\beq
\ket{\Psi_{27}} & = & \ket{3,- \frac{1}{2} , \frac{3}{4} ,\Gamma_{3,1}} \nonumber \\ 
& = & \quad 
C_{27,1} \left ( 
\ket{02d} - \ket{0d2} \right) 
 \nonumber \\
& & + 
C_{27,2} \left ( 
\ket{20d} - \ket{2d0} \right) 
 \nonumber \\
& & + 
C_{27,3} \left ( 
\ket{d02} - \ket{d20} \right) 
 \nonumber \\
& & + 
C_{27,4} \left ( 
\ket{ddu} + \ket{dud} \right) 
 \nonumber \\
& & + 
C_{27,5} \left ( 
\ket{udd} \right) 
\nonumber \eeq
\beq 
C_{27,1} &=& 
-\frac{1}{4 \sqrt{6}} \left (
-12 t^2+3 J t+4 U t-4 W t+2 \cos \left(\theta _1\right) \sqrt{A_1} t \right ) \nonumber \\
C_{27,2} &=& 
-\frac{1}{4 \sqrt{6}} \left (
12 t^2+3 J t+4 U t-4 W t+2 \cos \left(\theta _1\right) \sqrt{A_1} t \right ) \nonumber \\
C_{27,3} &=& 
-\sqrt{6} t^2 \nonumber \\
C_{27,4} &=& 
\frac{1}{6 \sqrt{6}} \left (
18 t^2+2 U^2-3 J U-6 J W+20 U W \right . \nonumber \\
&& \hspace{1cm} 
+
 \left . 32 W^2-4 \cos \left(\theta _1\right) \sqrt{A_1} W-6 A_6^2-2 U \cos \left(\theta _1\right) \sqrt{A_1} \right) \nonumber \\
C_{27,5} &=& 
-\frac{1}{3 \sqrt{6}} \left (
18 t^2+2 U^2-3 J U-6 J W+20 U W \right . \nonumber \\
&& \hspace{1cm} 
+
 \left . 32 W^2-4 \cos \left(\theta _1\right) \sqrt{A_1} W-6 A_6^2-2 U \cos \left(\theta _1\right) \sqrt{A_1} \right) \nonumber \\
 N_{27} &=& \sqrt{2 C_{27,1}^2+2 C_{27,2}^2+2 C_{27,3}^2+2 C_{27,4}^2+C_{27,5}^2} \nonumber \eeq 
\beq
\ket{\Psi_{28}} & = & \ket{3,- \frac{1}{2} , \frac{3}{4} ,\Gamma_{3,1}} \nonumber \\ 
& = & \quad 
C_{28,1} \left ( 
\ket{02d} - \ket{0d2} \right) 
 \nonumber \\
& & + 
C_{28,2} \left ( 
\ket{20d} - \ket{2d0} \right) 
 \nonumber \\
& & + 
C_{28,3} \left ( 
\ket{d02} - \ket{d20} \right) 
 \nonumber \\
& & + 
C_{28,4} \left ( 
\ket{ddu} + \ket{dud} \right) 
 \nonumber \\
& & + 
C_{28,5} \left ( 
\ket{udd} \right) 
\nonumber \eeq
\beq 
C_{28,1} &=& 
\frac{1}{24} \left (
12 \sqrt{6} t^2-3 \sqrt{6} J t-4 \sqrt{6} U t+4 \sqrt{6} W t+\sqrt{6} \cos \left(\theta _1\right) \sqrt{A_1} t+3 \sqrt{2} \sin \left(\theta _1\right) \sqrt{A_1} t \right ) \nonumber \\
C_{28,2} &=& 
-\frac{1}{12 \sqrt{2}} \left (
12 \sqrt{3} t^2+3 \sqrt{3} J t+4 \sqrt{3} U t-4 \sqrt{3} W t-\sqrt{3} \cos \left(\theta _1\right) \sqrt{A_1} t-3 \sin \left(\theta _1\right) \sqrt{A_1} t \right ) \nonumber \\
C_{28,3} &=& 
-\sqrt{6} t^2 \nonumber \\
C_{28,4} &=& 
\frac{1}{\sqrt{6}} \left (
3 t^2-U^2-4 W^2-4 U W \right . \nonumber \\
&& \hspace{1cm} 
+
 \left . -A_{10}^2+2 U A_{10}+4 W A_{10} \right) \nonumber \\
C_{28,5} &=& 
-\sqrt{\frac{2}{3}} \left (
3 t^2-U^2-4 W^2-4 U W \right . \nonumber \\
&& \hspace{1cm} 
+
 \left . -A_{10}^2+2 U A_{10}+4 W A_{10} \right) \nonumber \\
 N_{28} &=& \sqrt{2 C_{28,1}^2+2 C_{28,2}^2+2 C_{28,3}^2+2 C_{28,4}^2+C_{28,5}^2} \nonumber \eeq 
\beq
\ket{\Psi_{29}} & = & \ket{3,- \frac{1}{2} , \frac{3}{4} ,\Gamma_{3,1}} \nonumber \\ 
& = & \quad 
C_{29,1} \left ( 
\ket{02d} - \ket{0d2} \right) 
 \nonumber \\
& & + 
C_{29,2} \left ( 
\ket{20d} - \ket{2d0} \right) 
 \nonumber \\
& & + 
C_{29,3} \left ( 
\ket{d02} - \ket{d20} \right) 
 \nonumber \\
& & + 
C_{29,4} \left ( 
\ket{ddu} + \ket{dud} \right) 
 \nonumber \\
& & + 
C_{29,5} \left ( 
\ket{udd} \right) 
\nonumber \eeq
\beq 
C_{29,1} &=& 
-\frac{1}{12 \sqrt{2}} \left (
-12 \sqrt{3} t^2+3 \sqrt{3} J t+4 \sqrt{3} U t-4 \sqrt{3} W t-\sqrt{3} \cos \left(\theta _1\right) \sqrt{A_1} t+3 \sin \left(\theta _1\right) \sqrt{A_1} t \right ) \nonumber \\
C_{29,2} &=& 
-\frac{1}{12 \sqrt{2}} \left (
12 \sqrt{3} t^2+3 \sqrt{3} J t+4 \sqrt{3} U t-4 \sqrt{3} W t-\sqrt{3} \cos \left(\theta _1\right) \sqrt{A_1} t+3 \sin \left(\theta _1\right) \sqrt{A_1} t \right ) \nonumber \\
C_{29,3} &=& 
-\sqrt{6} t^2 \nonumber \\
C_{29,4} &=& 
\frac{1}{\sqrt{6}} \left (
3 t^2-U^2-4 W^2-4 U W \right . \nonumber \\
&& \hspace{1cm} 
+
 \left . -A_8^2+2 U A_8+4 W A_8 \right) \nonumber \\
C_{29,5} &=& 
-\sqrt{\frac{2}{3}} \left (
3 t^2-U^2-4 W^2-4 U W \right . \nonumber \\
&& \hspace{1cm} 
+
 \left . -A_8^2+2 U A_8+4 W A_8 \right) \nonumber \\
 N_{29} &=& \sqrt{2 C_{29,1}^2+2 C_{29,2}^2+2 C_{29,3}^2+2 C_{29,4}^2+C_{29,5}^2} \nonumber \eeq 
\beq
\ket{\Psi_{30}} & = & \ket{3,- \frac{1}{2} , \frac{3}{4} ,\Gamma_{3,2}} \nonumber \\ 
& = & \quad 
C_{30,1} \left ( 
\ket{02d} + \ket{0d2} \right) 
 \nonumber \\
& & + 
C_{30,2} \left ( 
\ket{20d} + \ket{2d0} \right) 
 \nonumber \\
& & + 
C_{30,3} \left ( 
\ket{d02} + \ket{d20} \right) 
 \nonumber \\
& & + 
C_{30,4} \left ( 
\ket{ddu} - \ket{dud} \right) 
\nonumber \eeq
\beq 
C_{30,1} &=& 
-\frac{1}{12} \left (
18 t^2-3 J t-4 U t+4 W t+\cos \left(\theta _1\right) \sqrt{A_1} t \right ) \nonumber \\
C_{30,2} &=& 
\frac{1}{96} \left (
9 J^2-48 t J+144 t^2+32 U^2-64 t U \right . \nonumber \\
&& \hspace{1cm} 
 \left . 272 W^2-72 J W+64 t W+128 U W+6 J \cos \left(\theta _1\right) \sqrt{A_1} \right. \nonumber \\
&& \hspace{1cm} 
+
 \left . -48 A_6^2+16 t \cos \left(\theta _1\right) \sqrt{A_1}-8 U \cos \left(\theta _1\right) \sqrt{A_1}-40 W \cos \left(\theta _1\right) \sqrt{A_1} \right) \nonumber \\
C_{30,3} &=& 
-\frac{1}{96} \left (
9 J^2-24 t J-72 W J+32 U^2-32 t U \right . \nonumber \\
&& \hspace{1cm} 
+
 \left . 272 W^2+32 t W+128 U W+6 J \cos \left(\theta _1\right) \sqrt{A_1}+8 t \cos \left(\theta _1\right) \sqrt{A_1} \right. \nonumber \\
&& \hspace{1cm} 
 \left . -48 A_6^2-8 U \cos \left(\theta _1\right) \sqrt{A_1}-40 W \cos \left(\theta _1\right) \sqrt{A_1} \right) \nonumber \\
C_{30,4} &=& 
\frac{1}{8} \left (
-12 t^2+3 J t+4 U t-4 W t+2 \cos \left(\theta _1\right) \sqrt{A_1} t \right ) \nonumber \\
 N_{30} &=& \sqrt{2} \sqrt{C_{30,1}^2+C_{30,2}^2+C_{30,3}^2+C_{30,4}^2} \nonumber \eeq 
\beq
\ket{\Psi_{31}} & = & \ket{3,- \frac{1}{2} , \frac{3}{4} ,\Gamma_{3,2}} \nonumber \\ 
& = & \quad 
C_{31,1} \left ( 
\ket{02d} + \ket{0d2} \right) 
 \nonumber \\
& & + 
C_{31,2} \left ( 
\ket{20d} + \ket{2d0} \right) 
 \nonumber \\
& & + 
C_{31,3} \left ( 
\ket{d02} + \ket{d20} \right) 
 \nonumber \\
& & + 
C_{31,4} \left ( 
\ket{ddu} - \ket{dud} \right) 
\nonumber \eeq
\beq 
C_{31,1} &=& 
\frac{1}{24} \left (
-36 t^2+6 J t+8 U t-8 W t+\cos \left(\theta _1\right) \sqrt{A_1} t+\sqrt{3} \sin \left(\theta _1\right) \sqrt{A_1} t \right ) \nonumber \\
C_{31,2} &=& 
-\frac{1}{48} \left (
-72 t^2+33 J t+8 U t-116 W t-18 J U-36 J W \right . \nonumber \\
&& \hspace{1cm} 
+
 \left . 144 W^2+72 U W+18 J A_{10}+t \cos \left(\theta _1\right) \sqrt{A_1}+\sqrt{3} t \sin \left(\theta _1\right) \sqrt{A_1} \right. \nonumber \\
&& \hspace{1cm} 
+
 \left . 24 A_{10}^2+36 t A_{10}-24 U A_{10}-120 W A_{10} \right) \nonumber \\
C_{31,3} &=& 
-\frac{1}{48} \left (
-21 J t+8 U t+100 W t+18 J U+36 J W \right . \nonumber \\
&& \hspace{1cm} 
+
 \left . -144 W^2-72 U W-18 J A_{10}+t \cos \left(\theta _1\right) \sqrt{A_1}+\sqrt{3} t \sin \left(\theta _1\right) \sqrt{A_1} \right. \nonumber \\
&& \hspace{1cm} 
 \left . -24 A_{10}^2-36 t A_{10}+24 U A_{10}+120 W A_{10} \right) \nonumber \\
C_{31,4} &=& 
-\frac{1}{8} \left (
12 t^2-3 J t-4 U t+4 W t+\cos \left(\theta _1\right) \sqrt{A_1} t+\sqrt{3} \sin \left(\theta _1\right) \sqrt{A_1} t \right ) \nonumber \\
 N_{31} &=& \sqrt{2} \sqrt{C_{31,1}^2+C_{31,2}^2+C_{31,3}^2+C_{31,4}^2} \nonumber \eeq 
\beq
\ket{\Psi_{32}} & = & \ket{3,- \frac{1}{2} , \frac{3}{4} ,\Gamma_{3,2}} \nonumber \\ 
& = & \quad 
C_{32,1} \left ( 
\ket{02d} + \ket{0d2} \right) 
 \nonumber \\
& & + 
C_{32,2} \left ( 
\ket{20d} + \ket{2d0} \right) 
 \nonumber \\
& & + 
C_{32,3} \left ( 
\ket{d02} + \ket{d20} \right) 
 \nonumber \\
& & + 
C_{32,4} \left ( 
\ket{ddu} - \ket{dud} \right) 
\nonumber \eeq
\beq 
C_{32,1} &=& 
-\frac{1}{24} \left (
36 t^2-6 J t-8 U t+8 W t-\cos \left(\theta _1\right) \sqrt{A_1} t+\sqrt{3} \sin \left(\theta _1\right) \sqrt{A_1} t \right ) \nonumber \\
C_{32,2} &=& 
\frac{1}{48} \left (
72 t^2-33 J t-8 U t+116 W t+18 J U+36 J W \right . \nonumber \\
&& \hspace{1cm} 
+
 \left . -144 W^2-72 U W-18 J A_8-t \cos \left(\theta _1\right) \sqrt{A_1}+\sqrt{3} t \sin \left(\theta _1\right) \sqrt{A_1} \right. \nonumber \\
&& \hspace{1cm} 
 \left . -24 A_8^2-36 t A_8+24 U A_8+120 W A_8 \right) \nonumber \\
C_{32,3} &=& 
\frac{1}{48} \left (
21 J t-8 U t-100 W t-18 J U-36 J W \right . \nonumber \\
&& \hspace{1cm} 
+
 \left . 144 W^2+72 U W+18 J A_8-t \cos \left(\theta _1\right) \sqrt{A_1}+\sqrt{3} t \sin \left(\theta _1\right) \sqrt{A_1} \right. \nonumber \\
&& \hspace{1cm} 
+
 \left . 24 A_8^2+36 t A_8-24 U A_8-120 W A_8 \right) \nonumber \\
C_{32,4} &=& 
\frac{1}{8} \left (
-12 t^2+3 J t+4 U t-4 W t-\cos \left(\theta _1\right) \sqrt{A_1} t+\sqrt{3} \sin \left(\theta _1\right) \sqrt{A_1} t \right ) \nonumber \\
 N_{32} &=& \sqrt{2} \sqrt{C_{32,1}^2+C_{32,2}^2+C_{32,3}^2+C_{32,4}^2} \nonumber \eeq 
{\subsubsection{\boldmath Eigenvectors for ${\rm  N_e}=3$ and   ${\rm m_s}$=  $\frac{1}{2} $.}
\beq
\ket{\Psi_{33}} & = & \ket{3,\frac{1}{2} , \frac{3}{4} ,\Gamma_1} \nonumber \\ 
&=& \frac{1}{\sqrt{6}}
 \left ( \ket{02u} + \ket{0u2} + \ket{20u} + \ket{2u0} + \ket{u02} + \ket{u20} \right) \nonumber 
\eeq
\beq
\ket{\Psi_{34}} & = & \ket{3,\frac{1}{2} , \frac{3}{4} ,\Gamma_2} \nonumber \\ 
&=& \frac{1}{\sqrt{6}}
 \left ( \ket{02u} - \ket{0u2} - \ket{20u} + \ket{2u0} + \ket{u02} - \ket{u20} \right) \nonumber 
\eeq
\beq
\ket{\Psi_{35}} & = & \ket{3,\frac{1}{2} , \frac{15}{4} ,\Gamma_2} \nonumber \\ 
&=& \frac{1}{\sqrt{3}}
 \left ( \ket{duu} + \ket{udu} + \ket{uud} \right) \nonumber 
\eeq
\beq
\ket{\Psi_{36}} & = & \ket{3,\frac{1}{2} , \frac{3}{4} ,\Gamma_{3,1}} \nonumber \\ 
& = & \quad 
C_{36,1} \left ( 
\ket{02u} - \ket{0u2} \right) 
 \nonumber \\
& & + 
C_{36,2} \left ( 
\ket{20u} - \ket{2u0} \right) 
 \nonumber \\
& & + 
C_{36,3} \left ( 
\ket{duu} \right) 
 \nonumber \\
& & + 
C_{36,4} \left ( 
\ket{u02} - \ket{u20} \right) 
 \nonumber \\
& & + 
C_{36,5} \left ( 
\ket{udu} + \ket{uud} \right) 
\nonumber \eeq
\beq 
C_{36,1} &=& 
-\frac{1}{12} \left (
18 t^2-3 J t-4 U t+4 W t+\cos \left(\theta _1\right) \sqrt{A_1} t \right ) \nonumber \\
C_{36,2} &=& 
\frac{1}{288} \left (
432 t^2+198 J t+48 U t+108 J U+216 J W \right . \nonumber \\
&& \hspace{1cm} 
 \left . -864 W^2-696 t W-432 U W+9 J A_7-12 t \cos \left(\theta _1\right) \sqrt{A_1} \right. \nonumber \\
&& \hspace{1cm} 
 \left . -A_7^2-18 t A_7-12 U A_7-60 W A_7 \right) \nonumber \\
C_{36,3} &=& 
-\frac{1}{12} \left (
36 t^2+3 J t+4 U t-4 W t+2 \cos \left(\theta _1\right) \sqrt{A_1} t \right ) \nonumber \\
C_{36,4} &=& 
\frac{1}{288} \left (
864 t^2+126 J t-48 U t+108 J U+216 J W \right . \nonumber \\
&& \hspace{1cm} 
 \left . -864 W^2-600 t W-432 U W+9 J A_7+12 t \cos \left(\theta _1\right) \sqrt{A_1} \right. \nonumber \\
&& \hspace{1cm} 
 \left . -A_7^2-18 t A_7-12 U A_7-60 W A_7 \right) \nonumber \\
C_{36,5} &=& 
\frac{1}{24} \left (
36 t^2+3 J t+4 U t-4 W t+2 \cos \left(\theta _1\right) \sqrt{A_1} t \right ) \nonumber \\
 N_{36} &=& \sqrt{2 C_{36,1}^2+2 C_{36,2}^2+C_{36,3}^2+2 C_{36,4}^2+2 C_{36,5}^2} \nonumber \eeq 
\beq
\ket{\Psi_{37}} & = & \ket{3,\frac{1}{2} , \frac{3}{4} ,\Gamma_{3,1}} \nonumber \\ 
& = & \quad 
C_{37,1} \left ( 
\ket{02u} - \ket{0u2} \right) 
 \nonumber \\
& & + 
C_{37,2} \left ( 
\ket{20u} - \ket{2u0} \right) 
 \nonumber \\
& & + 
C_{37,3} \left ( 
\ket{duu} \right) 
 \nonumber \\
& & + 
C_{37,4} \left ( 
\ket{u02} - \ket{u20} \right) 
 \nonumber \\
& & + 
C_{37,5} \left ( 
\ket{udu} + \ket{uud} \right) 
\nonumber \eeq
\beq 
C_{37,1} &=& 
\frac{1}{24} \left (
-36 t^2+6 J t+8 U t-8 W t+\cos \left(\theta _1\right) \sqrt{A_1} t+\sqrt{3} \sin \left(\theta _1\right) \sqrt{A_1} t \right ) \nonumber \\
C_{37,2} &=& 
\frac{1}{48} \left (
72 t^2+33 J t+8 U t-116 W t+18 J U+36 J W \right . \nonumber \\
&& \hspace{1cm} 
+
 \left . -144 W^2-72 U W-18 J A_{11}+t \cos \left(\theta _1\right) \sqrt{A_1}+\sqrt{3} t \sin \left(\theta _1\right) \sqrt{A_1} \right. \nonumber \\
&& \hspace{1cm} 
+
 \left . -24 A_{11}^2+36 t A_{11}+24 U A_{11}+120 W A_{11} \right) \nonumber \\
C_{37,3} &=& 
\frac{1}{12} \left (
-36 t^2-3 J t-4 U t+4 W t+\cos \left(\theta _1\right) \sqrt{A_1} t+\sqrt{3} \sin \left(\theta _1\right) \sqrt{A_1} t \right ) \nonumber \\
C_{37,4} &=& 
-\frac{1}{48} \left (
-144 t^2-21 J t+8 U t+100 W t-18 J U-36 J W \right . \nonumber \\
&& \hspace{1cm} 
+
 \left . 144 W^2+72 U W+18 J A_{11}+t \cos \left(\theta _1\right) \sqrt{A_1}+\sqrt{3} t \sin \left(\theta _1\right) \sqrt{A_1} \right. \nonumber \\
&& \hspace{1cm} 
 \left . 24 A_{11}^2-36 t A_{11}-24 U A_{11}-120 W A_{11} \right) \nonumber \\
C_{37,5} &=& 
-\frac{1}{24} \left (
-36 t^2-3 J t-4 U t+4 W t+\cos \left(\theta _1\right) \sqrt{A_1} t+\sqrt{3} \sin \left(\theta _1\right) \sqrt{A_1} t \right ) \nonumber \\
 N_{37} &=& \sqrt{2 C_{37,1}^2+2 C_{37,2}^2+C_{37,3}^2+2 C_{37,4}^2+2 C_{37,5}^2} \nonumber \eeq 
\beq
\ket{\Psi_{38}} & = & \ket{3,\frac{1}{2} , \frac{3}{4} ,\Gamma_{3,1}} \nonumber \\ 
& = & \quad 
C_{38,1} \left ( 
\ket{02u} - \ket{0u2} \right) 
 \nonumber \\
& & + 
C_{38,2} \left ( 
\ket{20u} - \ket{2u0} \right) 
 \nonumber \\
& & + 
C_{38,3} \left ( 
\ket{duu} \right) 
 \nonumber \\
& & + 
C_{38,4} \left ( 
\ket{u02} - \ket{u20} \right) 
 \nonumber \\
& & + 
C_{38,5} \left ( 
\ket{udu} + \ket{uud} \right) 
\nonumber \eeq
\beq 
C_{38,1} &=& 
-\frac{1}{24} \left (
36 t^2-6 J t-8 U t+8 W t-\cos \left(\theta _1\right) \sqrt{A_1} t+\sqrt{3} \sin \left(\theta _1\right) \sqrt{A_1} t \right ) \nonumber \\
C_{38,2} &=& 
-\frac{1}{48} \left (
-72 t^2-33 J t-8 U t+116 W t-18 J U-36 J W \right . \nonumber \\
&& \hspace{1cm} 
+
 \left . 144 W^2+72 U W+18 J A_9-t \cos \left(\theta _1\right) \sqrt{A_1}+\sqrt{3} t \sin \left(\theta _1\right) \sqrt{A_1} \right. \nonumber \\
&& \hspace{1cm} 
 \left . 24 A_9^2-36 t A_9-24 U A_9-120 W A_9 \right) \nonumber \\
C_{38,3} &=& 
-\frac{1}{12} \left (
36 t^2+3 J t+4 U t-4 W t-\cos \left(\theta _1\right) \sqrt{A_1} t+\sqrt{3} \sin \left(\theta _1\right) \sqrt{A_1} t \right ) \nonumber \\
C_{38,4} &=& 
\frac{1}{48} \left (
144 t^2+21 J t-8 U t-100 W t+18 J U+36 J W \right . \nonumber \\
&& \hspace{1cm} 
+
 \left . -144 W^2-72 U W-18 J A_9-t \cos \left(\theta _1\right) \sqrt{A_1}+\sqrt{3} t \sin \left(\theta _1\right) \sqrt{A_1} \right. \nonumber \\
&& \hspace{1cm} 
+
 \left . -24 A_9^2+36 t A_9+24 U A_9+120 W A_9 \right) \nonumber \\
C_{38,5} &=& 
\frac{1}{24} \left (
36 t^2+3 J t+4 U t-4 W t-\cos \left(\theta _1\right) \sqrt{A_1} t+\sqrt{3} \sin \left(\theta _1\right) \sqrt{A_1} t \right ) \nonumber \\
 N_{38} &=& \sqrt{2 C_{38,1}^2+2 C_{38,2}^2+C_{38,3}^2+2 C_{38,4}^2+2 C_{38,5}^2} \nonumber \eeq 
\beq
\ket{\Psi_{39}} & = & \ket{3,\frac{1}{2} , \frac{3}{4} ,\Gamma_{3,2}} \nonumber \\ 
& = & \quad 
C_{39,1} \left ( 
\ket{02u} + \ket{0u2} \right) 
 \nonumber \\
& & + 
C_{39,2} \left ( 
\ket{20u} + \ket{2u0} \right) 
 \nonumber \\
& & + 
C_{39,3} \left ( 
\ket{u02} + \ket{u20} \right) 
 \nonumber \\
& & + 
C_{39,4} \left ( 
\ket{udu} - \ket{uud} \right) 
\nonumber \eeq
\beq 
C_{39,1} &=& 
-\frac{1}{12} \left (
18 t^2-3 J t-4 U t+4 W t+\cos \left(\theta _1\right) \sqrt{A_1} t \right ) \nonumber \\
C_{39,2} &=& 
-\frac{1}{288} \left (
-432 t^2+198 J t+48 U t-108 J U-216 J W \right . \nonumber \\
&& \hspace{1cm} 
 \left . 864 W^2-696 t W+432 U W-9 J A_7-12 t \cos \left(\theta _1\right) \sqrt{A_1} \right. \nonumber \\
&& \hspace{1cm} 
 \left . A_7^2-18 t A_7+12 U A_7+60 W A_7 \right) \nonumber \\
C_{39,3} &=& 
\frac{1}{288} \left (
126 J t-48 U t-600 W t-108 J U-216 J W \right . \nonumber \\
&& \hspace{1cm} 
+
 \left . 864 W^2+432 U W-9 J A_7-18 t A_7+12 t \cos \left(\theta _1\right) \sqrt{A_1} \right. \nonumber \\
&& \hspace{1cm} 
+
 \left . A_7^2+12 U A_7+60 W A_7 \right) \nonumber \\
C_{39,4} &=& 
\frac{1}{8} \left (
-12 t^2+3 J t+4 U t-4 W t+2 \cos \left(\theta _1\right) \sqrt{A_1} t \right ) \nonumber \\
 N_{39} &=& \sqrt{2} \sqrt{C_{39,1}^2+C_{39,2}^2+C_{39,3}^2+C_{39,4}^2} \nonumber \eeq 
\beq
\ket{\Psi_{40}} & = & \ket{3,\frac{1}{2} , \frac{3}{4} ,\Gamma_{3,2}} \nonumber \\ 
& = & \quad 
C_{40,1} \left ( 
\ket{02u} + \ket{0u2} \right) 
 \nonumber \\
& & + 
C_{40,2} \left ( 
\ket{20u} + \ket{2u0} \right) 
 \nonumber \\
& & + 
C_{40,3} \left ( 
\ket{u02} + \ket{u20} \right) 
 \nonumber \\
& & + 
C_{40,4} \left ( 
\ket{udu} - \ket{uud} \right) 
\nonumber \eeq
\beq 
C_{40,1} &=& 
\frac{1}{24} \left (
-36 t^2+6 J t+8 U t-8 W t+\cos \left(\theta _1\right) \sqrt{A_1} t+\sqrt{3} \sin \left(\theta _1\right) \sqrt{A_1} t \right ) \nonumber \\
C_{40,2} &=& 
-\frac{1}{48} \left (
-72 t^2+33 J t+8 U t-116 W t-18 J U-36 J W \right . \nonumber \\
&& \hspace{1cm} 
+
 \left . 144 W^2+72 U W+18 J A_{11}+t \cos \left(\theta _1\right) \sqrt{A_1}+\sqrt{3} t \sin \left(\theta _1\right) \sqrt{A_1} \right. \nonumber \\
&& \hspace{1cm} 
+
 \left . 24 A_{11}^2+36 t A_{11}-24 U A_{11}-120 W A_{11} \right) \nonumber \\
C_{40,3} &=& 
-\frac{1}{48} \left (
-21 J t+8 U t+100 W t+18 J U+36 J W \right . \nonumber \\
&& \hspace{1cm} 
+
 \left . -144 W^2-72 U W-18 J A_{11}+t \cos \left(\theta _1\right) \sqrt{A_1}+\sqrt{3} t \sin \left(\theta _1\right) \sqrt{A_1} \right. \nonumber \\
&& \hspace{1cm} 
 \left . -24 A_{11}^2-36 t A_{11}+24 U A_{11}+120 W A_{11} \right) \nonumber \\
C_{40,4} &=& 
-\frac{1}{8} \left (
12 t^2-3 J t-4 U t+4 W t+\cos \left(\theta _1\right) \sqrt{A_1} t+\sqrt{3} \sin \left(\theta _1\right) \sqrt{A_1} t \right ) \nonumber \\
 N_{40} &=& \sqrt{2} \sqrt{C_{40,1}^2+C_{40,2}^2+C_{40,3}^2+C_{40,4}^2} \nonumber \eeq 
\beq
\ket{\Psi_{41}} & = & \ket{3,\frac{1}{2} , \frac{3}{4} ,\Gamma_{3,2}} \nonumber \\ 
& = & \quad 
C_{41,1} \left ( 
\ket{02u} + \ket{0u2} \right) 
 \nonumber \\
& & + 
C_{41,2} \left ( 
\ket{20u} + \ket{2u0} \right) 
 \nonumber \\
& & + 
C_{41,3} \left ( 
\ket{u02} + \ket{u20} \right) 
 \nonumber \\
& & + 
C_{41,4} \left ( 
\ket{udu} - \ket{uud} \right) 
\nonumber \eeq
\beq 
C_{41,1} &=& 
-\frac{1}{24} \left (
36 t^2-6 J t-8 U t+8 W t-\cos \left(\theta _1\right) \sqrt{A_1} t+\sqrt{3} \sin \left(\theta _1\right) \sqrt{A_1} t \right ) \nonumber \\
C_{41,2} &=& 
\frac{1}{48} \left (
72 t^2-33 J t-8 U t+116 W t+18 J U+36 J W \right . \nonumber \\
&& \hspace{1cm} 
+
 \left . -144 W^2-72 U W-18 J A_9-t \cos \left(\theta _1\right) \sqrt{A_1}+\sqrt{3} t \sin \left(\theta _1\right) \sqrt{A_1} \right. \nonumber \\
&& \hspace{1cm} 
 \left . -24 A_9^2-36 t A_9+24 U A_9+120 W A_9 \right) \nonumber \\
C_{41,3} &=& 
\frac{1}{48} \left (
21 J t-8 U t-100 W t-18 J U-36 J W \right . \nonumber \\
&& \hspace{1cm} 
+
 \left . 144 W^2+72 U W+18 J A_9-t \cos \left(\theta _1\right) \sqrt{A_1}+\sqrt{3} t \sin \left(\theta _1\right) \sqrt{A_1} \right. \nonumber \\
&& \hspace{1cm} 
+
 \left . 24 A_9^2+36 t A_9-24 U A_9-120 W A_9 \right) \nonumber \\
C_{41,4} &=& 
\frac{1}{8} \left (
-12 t^2+3 J t+4 U t-4 W t-\cos \left(\theta _1\right) \sqrt{A_1} t+\sqrt{3} \sin \left(\theta _1\right) \sqrt{A_1} t \right ) \nonumber \\
 N_{41} &=& \sqrt{2} \sqrt{C_{41,1}^2+C_{41,2}^2+C_{41,3}^2+C_{41,4}^2} \nonumber \eeq 
{\subsubsection{\boldmath Eigenvectors for ${\rm  N_e}=3$ and   ${\rm m_s}$=  $\frac{3}{2} $.}
\beq
\ket{\Psi_{42}} & = & \ket{3,\frac{3}{2} , \frac{15}{4} ,\Gamma_2} \nonumber \\ 
&=& 1
 \left ( \ket{uuu} \right) \nonumber 
\eeq
{\subsubsection{\boldmath Eigenvectors for ${\rm  N_e}=4$ and   ${\rm m_s}$=  $-1$.}
\beq
\ket{\Psi_{43}} & = & \ket{4,-1,2,\Gamma_2} \nonumber \\ 
&=& \frac{1}{\sqrt{3}}
 \left ( \ket{2dd} - \ket{d2d} + \ket{dd2} \right) \nonumber 
\eeq
\beq
\ket{\Psi_{44}} & = & \ket{4,-1,2,\Gamma_{3,1}} \nonumber \\ 
& = & \quad 
C_{44,1} \left ( 
\ket{2dd} \right) 
 \nonumber \\
& & + 
C_{44,2} \left ( 
\ket{d2d} - \ket{dd2} \right) 
\nonumber \eeq
\beq 
C_{44,1} &=& 
\sqrt{\frac{2}{3}} \nonumber \\
C_{44,2} &=& 
\frac{1}{\sqrt{6}} \nonumber \\
 N_{44} &=& \sqrt{C_{44,1}^2+2 C_{44,2}^2} \nonumber \eeq 
\beq
\ket{\Psi_{45}} & = & \ket{4,-1,2,\Gamma_{3,2}} \nonumber \\ 
&=& \frac{1}{\sqrt{2}}
 \left ( \ket{d2d} + \ket{dd2} \right) \nonumber 
\eeq
{\subsubsection{\boldmath Eigenvectors for ${\rm  N_e}=4$ and   ${\rm m_s}$=  $0$.}
\beq
\ket{\Psi_{46}} & = & \ket{4,0,0,\Gamma_1} \nonumber \\ 
& = & \quad 
C_{46,1} \left ( 
\ket{022} + \ket{202} + \ket{220} \right) 
 \nonumber \\
& & + 
C_{46,2} \left ( 
\ket{2du} - \ket{2ud} + \ket{d2u} + \ket{du2} - \ket{u2d} - \ket{ud2} \right) 
\nonumber \eeq
\beq 
C_{46,1} &=& 
2 \sqrt{\frac{2}{3}} t \nonumber \\
C_{46,2} &=& 
\frac{1}{8 \sqrt{6}} \left (
3 J+8 t+4 U-4 W+\sqrt{A_3} \right ) \nonumber \\
 N_{46} &=& \sqrt{3 C_{46,1}^2+6 C_{46,2}^2} \nonumber \eeq 
\beq
\ket{\Psi_{47}} & = & \ket{4,0,0,\Gamma_1} \nonumber \\ 
& = & \quad 
C_{47,1} \left ( 
\ket{022} + \ket{202} + \ket{220} \right) 
 \nonumber \\
& & + 
C_{47,2} \left ( 
\ket{2du} - \ket{2ud} + \ket{d2u} + \ket{du2} - \ket{u2d} - \ket{ud2} \right) 
\nonumber \eeq
\beq 
C_{47,1} &=& 
2 \sqrt{\frac{2}{3}} t \nonumber \\
C_{47,2} &=& 
\frac{1}{8 \sqrt{6}} \left (
3 J+8 t+4 U-4 W-\sqrt{A_3} \right ) \nonumber \\
 N_{47} &=& \sqrt{3 C_{47,1}^2+6 C_{47,2}^2} \nonumber \eeq 
\beq
\ket{\Psi_{48}} & = & \ket{4,0,2,\Gamma_2} \nonumber \\ 
&=& \frac{1}{\sqrt{6}}
 \left ( \ket{2du} + \ket{2ud} - \ket{d2u} + \ket{du2} - \ket{u2d} + \ket{ud2} \right) \nonumber 
\eeq
\beq
\ket{\Psi_{49}} & = & \ket{4,0,0,\Gamma_{3,1}} \nonumber \\ 
& = & \quad 
C_{49,1} \left ( 
\ket{202} - \ket{220} \right) 
 \nonumber \\
& & + 
C_{49,2} \left ( 
\ket{d2u} - \ket{du2} - \ket{u2d} + \ket{ud2} \right) 
\nonumber \eeq
\beq 
C_{49,1} &=& 
-\frac{1}{8 \sqrt{2}} \left (
3 J-4 t+4 U-4 W-\sqrt{A_4} \right ) \nonumber \\
C_{49,2} &=& 
-\frac{t}{\sqrt{2}} \nonumber \\
 N_{49} &=& \sqrt{2 C_{49,1}^2+4 C_{49,2}^2} \nonumber \eeq 
\beq
\ket{\Psi_{50}} & = & \ket{4,0,0,\Gamma_{3,1}} \nonumber \\ 
& = & \quad 
C_{50,1} \left ( 
\ket{202} - \ket{220} \right) 
 \nonumber \\
& & + 
C_{50,2} \left ( 
\ket{d2u} - \ket{du2} - \ket{u2d} + \ket{ud2} \right) 
\nonumber \eeq
\beq 
C_{50,1} &=& 
-\frac{1}{8 \sqrt{2}} \left (
3 J-4 t+4 U-4 W+\sqrt{A_4} \right ) \nonumber \\
C_{50,2} &=& 
-\frac{t}{\sqrt{2}} \nonumber \\
 N_{50} &=& \sqrt{2 C_{50,1}^2+4 C_{50,2}^2} \nonumber \eeq 
\beq
\ket{\Psi_{51}} & = & \ket{4,0,2,\Gamma_{3,1}} \nonumber \\ 
& = & \quad 
C_{51,1} \left ( 
\ket{2du} + \ket{2ud} \right) 
 \nonumber \\
& & + 
C_{51,2} \left ( 
\ket{d2u} - \ket{du2} + \ket{u2d} - \ket{ud2} \right) 
\nonumber \eeq
\beq 
C_{51,1} &=& 
\frac{1}{\sqrt{3}} \nonumber \\
C_{51,2} &=& 
\frac{1}{2 \sqrt{3}} \nonumber \\
 N_{51} &=& \sqrt{2 C_{51,1}^2+4 C_{51,2}^2} \nonumber \eeq 
\beq
\ket{\Psi_{52}} & = & \ket{4,0,0,\Gamma_{3,2}} \nonumber \\ 
& = & \quad 
C_{52,1} \left ( 
\ket{022} \right) 
 \nonumber \\
& & + 
C_{52,2} \left ( 
\ket{202} + \ket{220} \right) 
 \nonumber \\
& & + 
C_{52,3} \left ( 
\ket{2du} - \ket{2ud} \right) 
 \nonumber \\
& & + 
C_{52,4} \left ( 
\ket{d2u} + \ket{du2} - \ket{u2d} - \ket{ud2} \right) 
\nonumber \eeq
\beq 
C_{52,1} &=& 
\frac{2 t}{\sqrt{3}} \nonumber \\
C_{52,2} &=& 
-\frac{t}{\sqrt{3}} \nonumber \\
C_{52,3} &=& 
-\frac{1}{8 \sqrt{3}} \left (
3 J-4 t+4 U-4 W+\sqrt{A_4} \right ) \nonumber \\
C_{52,4} &=& 
\frac{1}{16 \sqrt{3}} \left (
3 J-4 t+4 U-4 W+\sqrt{A_4} \right ) \nonumber \\
 N_{52} &=& \sqrt{C_{52,1}^2+2 \left(C_{52,2}^2+C_{52,3}^2+2 C_{52,4}^2\right)} \nonumber \eeq 
\beq
\ket{\Psi_{53}} & = & \ket{4,0,0,\Gamma_{3,2}} \nonumber \\ 
& = & \quad 
C_{53,1} \left ( 
\ket{022} \right) 
 \nonumber \\
& & + 
C_{53,2} \left ( 
\ket{202} + \ket{220} \right) 
 \nonumber \\
& & + 
C_{53,3} \left ( 
\ket{2du} - \ket{2ud} \right) 
 \nonumber \\
& & + 
C_{53,4} \left ( 
\ket{d2u} + \ket{du2} - \ket{u2d} - \ket{ud2} \right) 
\nonumber \eeq
\beq 
C_{53,1} &=& 
\frac{2 t}{\sqrt{3}} \nonumber \\
C_{53,2} &=& 
-\frac{t}{\sqrt{3}} \nonumber \\
C_{53,3} &=& 
-\frac{1}{8 \sqrt{3}} \left (
3 J-4 t+4 U-4 W-\sqrt{A_4} \right ) \nonumber \\
C_{53,4} &=& 
\frac{1}{16 \sqrt{3}} \left (
3 J-4 t+4 U-4 W-\sqrt{A_4} \right ) \nonumber \\
 N_{53} &=& \sqrt{C_{53,1}^2+2 \left(C_{53,2}^2+C_{53,3}^2+2 C_{53,4}^2\right)} \nonumber \eeq 
\beq
\ket{\Psi_{54}} & = & \ket{4,0,2,\Gamma_{3,2}} \nonumber \\ 
&=& \frac{1}{2}
 \left ( \ket{d2u} + \ket{du2} + \ket{u2d} + \ket{ud2} \right) \nonumber 
\eeq
{\subsubsection{\boldmath Eigenvectors for ${\rm  N_e}=4$ and   ${\rm m_s}$=  $1$.}
\beq
\ket{\Psi_{55}} & = & \ket{4,1,2,\Gamma_2} \nonumber \\ 
&=& \frac{1}{\sqrt{3}}
 \left ( \ket{2uu} - \ket{u2u} + \ket{uu2} \right) \nonumber 
\eeq
\beq
\ket{\Psi_{56}} & = & \ket{4,1,2,\Gamma_{3,1}} \nonumber \\ 
& = & \quad 
C_{56,1} \left ( 
\ket{2uu} \right) 
 \nonumber \\
& & + 
C_{56,2} \left ( 
\ket{u2u} - \ket{uu2} \right) 
\nonumber \eeq
\beq 
C_{56,1} &=& 
\sqrt{\frac{2}{3}} \nonumber \\
C_{56,2} &=& 
\frac{1}{\sqrt{6}} \nonumber \\
 N_{56} &=& \sqrt{C_{56,1}^2+2 C_{56,2}^2} \nonumber \eeq 
\beq
\ket{\Psi_{57}} & = & \ket{4,1,2,\Gamma_{3,2}} \nonumber \\ 
&=& \frac{1}{\sqrt{2}}
 \left ( \ket{u2u} + \ket{uu2} \right) \nonumber 
\eeq
{\subsubsection{\boldmath Eigenvectors for ${\rm  N_e}=5$ and   ${\rm m_s}$=  $- \frac{1}{2} $.}
\beq
\ket{\Psi_{58}} & = & \ket{5,- \frac{1}{2} , \frac{3}{4} ,\Gamma_1} \nonumber \\ 
&=& \frac{1}{\sqrt{3}}
 \left ( \ket{22d} + \ket{2d2} + \ket{d22} \right) \nonumber 
\eeq
\beq
\ket{\Psi_{59}} & = & \ket{5,- \frac{1}{2} , \frac{3}{4} ,\Gamma_{3,1}} \nonumber \\ 
&=& \frac{1}{\sqrt{2}}
 \left ( \ket{22d} - \ket{2d2} \right) \nonumber 
\eeq
\beq
\ket{\Psi_{60}} & = & \ket{5,- \frac{1}{2} , \frac{3}{4} ,\Gamma_{3,2}} \nonumber \\ 
& = & \quad 
C_{60,1} \left ( 
\ket{22d} + \ket{2d2} \right) 
 \nonumber \\
& & + 
C_{60,2} \left ( 
\ket{d22} \right) 
\nonumber \eeq
\beq 
C_{60,1} &=& 
\frac{1}{\sqrt{6}} \nonumber \\
C_{60,2} &=& 
-\sqrt{\frac{2}{3}} \nonumber \\
 N_{60} &=& \sqrt{2 C_{60,1}^2+C_{60,2}^2} \nonumber \eeq 
{\subsubsection{\boldmath Eigenvectors for ${\rm  N_e}=5$ and   ${\rm m_s}$=  $\frac{1}{2} $.}
\beq
\ket{\Psi_{61}} & = & \ket{5,\frac{1}{2} , \frac{3}{4} ,\Gamma_1} \nonumber \\ 
&=& \frac{1}{\sqrt{3}}
 \left ( \ket{22u} + \ket{2u2} + \ket{u22} \right) \nonumber 
\eeq
\beq
\ket{\Psi_{62}} & = & \ket{5,\frac{1}{2} , \frac{3}{4} ,\Gamma_{3,1}} \nonumber \\ 
&=& \frac{1}{\sqrt{2}}
 \left ( \ket{22u} - \ket{2u2} \right) \nonumber 
\eeq
\beq
\ket{\Psi_{63}} & = & \ket{5,\frac{1}{2} , \frac{3}{4} ,\Gamma_{3,2}} \nonumber \\ 
& = & \quad 
C_{63,1} \left ( 
\ket{22u} + \ket{2u2} \right) 
 \nonumber \\
& & + 
C_{63,2} \left ( 
\ket{u22} \right) 
\nonumber \eeq
\beq 
C_{63,1} &=& 
\frac{1}{\sqrt{6}} \nonumber \\
C_{63,2} &=& 
-\sqrt{\frac{2}{3}} \nonumber \\
 N_{63} &=& \sqrt{2 C_{63,1}^2+C_{63,2}^2} \nonumber \eeq 
{\subsubsection{\boldmath Eigenvectors for ${\rm  N_e}=6$ and   ${\rm m_s}$=  $0$.}
\beq
\ket{\Psi_{64}} & = & \ket{6,0,0,\Gamma_1} \nonumber \\ 
&=& 1
 \left ( \ket{222} \right) \nonumber 
\eeq
}

\subsection{The tetrahedral cluster}
\parindent0cm
\begin{tabular}[t]{|r|l|c|c|}
\multicolumn{4}{c}{\bf \boldmath Eigenkets and eigenvalues for ${\rm  N_e}$=0 and   ${\rm m_s}$= $0$. } \\[1ex] \hline
\parbox[c]{1cm}{\hfill No}  & \parbox[c][2.5ex]{2.5cm}{\begin{center} Eigenstate \end{center}}  & \parbox[c][2.5ex]{7.5cm}{ \begin{center}   Energy \end{center}}  & \parbox[c][2.5ex]{2cm}{ \begin{center} Example \end{center}}  \\ \hline 
\hline 
1 & $\ket{0,0,0,\Gamma_1}$  &\parbox[c][5.0ex]{7cm}{ \begin{center} $0$ \end{center}}  & 0. \\ 
\hline
\end{tabular} \\[2ex]
\begin{tabular}[t]{|r|l|c|c|}
\multicolumn{4}{c}{\bf \boldmath Eigenkets and eigenvalues for ${\rm  N_e}$=1 and   ${\rm m_s}$= $- \frac{1}{2} $. } \\[1ex] \hline
\parbox[c]{1cm}{\hfill No}  & \parbox[c][2.5ex]{2.5cm}{\begin{center} Eigenstate \end{center}}  & \parbox[c][2.5ex]{7.5cm}{ \begin{center}   Energy \end{center}}  & \parbox[c][2.5ex]{2cm}{ \begin{center} Example \end{center}}  \\ \hline 
\hline 
2 & $\ket{1,- \frac{1}{2} , \frac{3}{4} ,\Gamma_1}$  &\parbox[c][5.0ex]{7cm}{ \begin{center} $3 t$ \end{center}}  & 3. \\ 
3 & $\ket{1,- \frac{1}{2} , \frac{3}{4} ,\Gamma_{4,1}}$  &\parbox[c][5.0ex]{7cm}{ \begin{center} $-t$ \end{center}}  & -1. \\ 
4 & $\ket{1,- \frac{1}{2} , \frac{3}{4} ,\Gamma_{4,2}}$  &\parbox[c][5.0ex]{7cm}{ \begin{center} $-t$ \end{center}}  & -1. \\ 
5 & $\ket{1,- \frac{1}{2} , \frac{3}{4} ,\Gamma_{4,3}}$  &\parbox[c][5.0ex]{7cm}{ \begin{center} $-t$ \end{center}}  & -1. \\ 
\hline
\end{tabular} \\[2ex]
\begin{tabular}[t]{|r|l|c|c|}
\multicolumn{4}{c}{\bf \boldmath Eigenkets and eigenvalues for ${\rm  N_e}$=1 and   ${\rm m_s}$= $\frac{1}{2} $. } \\[1ex] \hline
\parbox[c]{1cm}{\hfill No}  & \parbox[c][2.5ex]{2.5cm}{\begin{center} Eigenstate \end{center}}  & \parbox[c][2.5ex]{7.5cm}{ \begin{center}   Energy \end{center}}  & \parbox[c][2.5ex]{2cm}{ \begin{center} Example \end{center}}  \\ \hline 
\hline 
6 & $\ket{1,\frac{1}{2} , \frac{3}{4} ,\Gamma_1}$  &\parbox[c][5.0ex]{7cm}{ \begin{center} $3 t$ \end{center}}  & 3. \\ 
7 & $\ket{1,\frac{1}{2} , \frac{3}{4} ,\Gamma_{4,1}}$  &\parbox[c][5.0ex]{7cm}{ \begin{center} $-t$ \end{center}}  & -1. \\ 
8 & $\ket{1,\frac{1}{2} , \frac{3}{4} ,\Gamma_{4,2}}$  &\parbox[c][5.0ex]{7cm}{ \begin{center} $-t$ \end{center}}  & -1. \\ 
9 & $\ket{1,\frac{1}{2} , \frac{3}{4} ,\Gamma_{4,3}}$  &\parbox[c][5.0ex]{7cm}{ \begin{center} $-t$ \end{center}}  & -1. \\ 
\hline
\end{tabular} \\[2ex]
\begin{tabular}[t]{|r|l|c|c|}
\multicolumn{4}{c}{\bf \boldmath Eigenkets and eigenvalues for ${\rm  N_e}$=2 and   ${\rm m_s}$= $-1$. } \\[1ex] \hline
\parbox[c]{1cm}{\hfill No}  & \parbox[c][2.5ex]{2.5cm}{\begin{center} Eigenstate \end{center}}  & \parbox[c][2.5ex]{7.5cm}{ \begin{center}   Energy \end{center}}  & \parbox[c][2.5ex]{2cm}{ \begin{center} Example \end{center}}  \\ \hline 
\hline 
10 & $\ket{2,-1,2,\Gamma_{4,1}}$  &\parbox[c][5.0ex]{7cm}{ \begin{center} $\frac{J}{4}+2 t+W$ \end{center}}  & 2. \\ 
11 & $\ket{2,-1,2,\Gamma_{4,2}}$  &\parbox[c][5.0ex]{7cm}{ \begin{center} $\frac{J}{4}+2 t+W$ \end{center}}  & 2. \\ 
12 & $\ket{2,-1,2,\Gamma_{4,3}}$  &\parbox[c][5.0ex]{7cm}{ \begin{center} $\frac{J}{4}+2 t+W$ \end{center}}  & 2. \\ 
13 & $\ket{2,-1,2,\Gamma_{5,1}}$  &\parbox[c][5.0ex]{7cm}{ \begin{center} $\frac{J}{4}-2 t+W$ \end{center}}  & -2. \\ 
14 & $\ket{2,-1,2,\Gamma_{5,2}}$  &\parbox[c][5.0ex]{7cm}{ \begin{center} $\frac{J}{4}-2 t+W$ \end{center}}  & -2. \\ 
15 & $\ket{2,-1,2,\Gamma_{5,3}}$  &\parbox[c][5.0ex]{7cm}{ \begin{center} $\frac{J}{4}-2 t+W$ \end{center}}  & -2. \\ 
\hline
\end{tabular} \\[2ex]
\begin{tabular}[t]{|r|l|c|c|}
\multicolumn{4}{c}{\bf \boldmath Eigenkets and eigenvalues for ${\rm  N_e}$=2 and   ${\rm m_s}$= $0$. } \\[1ex] \hline
\parbox[c]{1cm}{\hfill No}  & \parbox[c][2.5ex]{2.5cm}{\begin{center} Eigenstate \end{center}}  & \parbox[c][2.5ex]{7.5cm}{ \begin{center}   Energy \end{center}}  & \parbox[c][2.5ex]{2cm}{ \begin{center} Example \end{center}}  \\ \hline 
\hline 
16 & $\ket{2,0,0,\Gamma_1}$  &\parbox[c][5.0ex]{7cm}{ \begin{center} $\frac{1}{8} \left(-3 J+16 t+4 U+4 W-\sqrt{A_{23}}\right)$ \end{center}}  & 1. \\ 
17 & $\ket{2,0,0,\Gamma_1}$  &\parbox[c][5.0ex]{7cm}{ \begin{center} $\frac{1}{8} \left(-3 J+16 t+4 U+4 W+\sqrt{A_{23}}\right)$ \end{center}}  & 8. \\ 
18 & $\ket{2,0,0,\Gamma_{3,1}}$  &\parbox[c][5.0ex]{7cm}{ \begin{center} $-\frac{3 J}{4}-2 t+W$ \end{center}}  & -2. \\ 
19 & $\ket{2,0,0,\Gamma_{3,2}}$  &\parbox[c][5.0ex]{7cm}{ \begin{center} $-\frac{3 J}{4}-2 t+W$ \end{center}}  & -2. \\ 
20 & $\ket{2,0,0,\Gamma_{4,1}}$  &\parbox[c][5.0ex]{7cm}{ \begin{center} $\frac{1}{8} \left(-3 J+4 U+4 W-\sqrt{A_2}\right)$ \end{center}}  & -0.701562 \\ 
21 & $\ket{2,0,0,\Gamma_{4,1}}$  &\parbox[c][5.0ex]{7cm}{ \begin{center} $\frac{1}{8} \left(-3 J+4 U+4 W+\sqrt{A_2}\right)$ \end{center}}  & 5.70156 \\ 
22 & $\ket{2,0,2,\Gamma_{4,1}}$  &\parbox[c][5.0ex]{7cm}{ \begin{center} $\frac{J}{4}+2 t+W$ \end{center}}  & 2. \\ 
23 & $\ket{2,0,0,\Gamma_{4,2}}$  &\parbox[c][5.0ex]{7cm}{ \begin{center} $\frac{1}{8} \left(-3 J+4 U+4 W-\sqrt{A_2}\right)$ \end{center}}  & -0.701562 \\ 
24 & $\ket{2,0,0,\Gamma_{4,2}}$  &\parbox[c][5.0ex]{7cm}{ \begin{center} $\frac{1}{8} \left(-3 J+4 U+4 W+\sqrt{A_2}\right)$ \end{center}}  & 5.70156 \\ 
25 & $\ket{2,0,2,\Gamma_{4,2}}$  &\parbox[c][5.0ex]{7cm}{ \begin{center} $\frac{J}{4}+2 t+W$ \end{center}}  & 2. \\ 
26 & $\ket{2,0,0,\Gamma_{4,3}}$  &\parbox[c][5.0ex]{7cm}{ \begin{center} $\frac{1}{8} \left(-3 J+4 U+4 W-\sqrt{A_2}\right)$ \end{center}}  & -0.701562 \\ 
27 & $\ket{2,0,0,\Gamma_{4,3}}$  &\parbox[c][5.0ex]{7cm}{ \begin{center} $\frac{1}{8} \left(-3 J+4 U+4 W+\sqrt{A_2}\right)$ \end{center}}  & 5.70156 \\ 
28 & $\ket{2,0,2,\Gamma_{4,3}}$  &\parbox[c][5.0ex]{7cm}{ \begin{center} $\frac{J}{4}+2 t+W$ \end{center}}  & 2. \\ 
29 & $\ket{2,0,2,\Gamma_{5,1}}$  &\parbox[c][5.0ex]{7cm}{ \begin{center} $\frac{J}{4}-2 t+W$ \end{center}}  & -2. \\ 
30 & $\ket{2,0,2,\Gamma_{5,2}}$  &\parbox[c][5.0ex]{7cm}{ \begin{center} $\frac{J}{4}-2 t+W$ \end{center}}  & -2. \\ 
31 & $\ket{2,0,2,\Gamma_{5,3}}$  &\parbox[c][5.0ex]{7cm}{ \begin{center} $\frac{J}{4}-2 t+W$ \end{center}}  & -2. \\ 
\hline
\end{tabular} \\[2ex]
\begin{tabular}[t]{|r|l|c|c|}
\multicolumn{4}{c}{\bf \boldmath Eigenkets and eigenvalues for ${\rm  N_e}$=2 and   ${\rm m_s}$= $1$. } \\[1ex] \hline
\parbox[c]{1cm}{\hfill No}  & \parbox[c][2.5ex]{2.5cm}{\begin{center} Eigenstate \end{center}}  & \parbox[c][2.5ex]{7.5cm}{ \begin{center}   Energy \end{center}}  & \parbox[c][2.5ex]{2cm}{ \begin{center} Example \end{center}}  \\ \hline 
\hline 
32 & $\ket{2,1,2,\Gamma_{4,1}}$  &\parbox[c][5.0ex]{7cm}{ \begin{center} $\frac{J}{4}+2 t+W$ \end{center}}  & 2. \\ 
33 & $\ket{2,1,2,\Gamma_{4,2}}$  &\parbox[c][5.0ex]{7cm}{ \begin{center} $\frac{J}{4}+2 t+W$ \end{center}}  & 2. \\ 
34 & $\ket{2,1,2,\Gamma_{4,3}}$  &\parbox[c][5.0ex]{7cm}{ \begin{center} $\frac{J}{4}+2 t+W$ \end{center}}  & 2. \\ 
35 & $\ket{2,1,2,\Gamma_{5,1}}$  &\parbox[c][5.0ex]{7cm}{ \begin{center} $\frac{J}{4}-2 t+W$ \end{center}}  & -2. \\ 
36 & $\ket{2,1,2,\Gamma_{5,2}}$  &\parbox[c][5.0ex]{7cm}{ \begin{center} $\frac{J}{4}-2 t+W$ \end{center}}  & -2. \\ 
37 & $\ket{2,1,2,\Gamma_{5,3}}$  &\parbox[c][5.0ex]{7cm}{ \begin{center} $\frac{J}{4}-2 t+W$ \end{center}}  & -2. \\ 
\hline
\end{tabular} \\[2ex]
\begin{tabular}[t]{|r|l|c|c|}
\multicolumn{4}{c}{\bf \boldmath Eigenkets and eigenvalues for ${\rm  N_e}$=3 and   ${\rm m_s}$= $- \frac{3}{2} $. } \\[1ex] \hline
\parbox[c]{1cm}{\hfill No}  & \parbox[c][2.5ex]{2.5cm}{\begin{center} Eigenstate \end{center}}  & \parbox[c][2.5ex]{7.5cm}{ \begin{center}   Energy \end{center}}  & \parbox[c][2.5ex]{2cm}{ \begin{center} Example \end{center}}  \\ \hline 
\hline 
38 & $\ket{3,- \frac{3}{2} , \frac{15}{4} ,\Gamma_2}$  &\parbox[c][5.0ex]{7cm}{ \begin{center} $\frac{3}{4} (J-4 t+4 W)$ \end{center}}  & -3. \\ 
39 & $\ket{3,- \frac{3}{2} , \frac{15}{4} ,\Gamma_{5,1}}$  &\parbox[c][5.0ex]{7cm}{ \begin{center} $\frac{3 J}{4}+t+3 W$ \end{center}}  & 1. \\ 
40 & $\ket{3,- \frac{3}{2} , \frac{15}{4} ,\Gamma_{5,2}}$  &\parbox[c][5.0ex]{7cm}{ \begin{center} $\frac{3 J}{4}+t+3 W$ \end{center}}  & 1. \\ 
41 & $\ket{3,- \frac{3}{2} , \frac{15}{4} ,\Gamma_{5,3}}$  &\parbox[c][5.0ex]{7cm}{ \begin{center} $\frac{3 J}{4}+t+3 W$ \end{center}}  & 1. \\ 
\hline
\end{tabular} \\[2ex]
\begin{tabular}[t]{|r|l|c|c|}
\multicolumn{4}{c}{\bf \boldmath Eigenkets and eigenvalues for ${\rm  N_e}$=3 and   ${\rm m_s}$= $- \frac{1}{2} $. } \\[1ex] \hline
\parbox[c]{1cm}{\hfill No}  & \parbox[c][2.5ex]{2.5cm}{\begin{center} Eigenstate \end{center}}  & \parbox[c][2.5ex]{7.5cm}{ \begin{center}   Energy \end{center}}  & \parbox[c][2.5ex]{2cm}{ \begin{center} Example \end{center}}  \\ \hline 
\hline 
42 & $\ket{3,- \frac{1}{2} , \frac{3}{4} ,\Gamma_1}$  &\parbox[c][5.0ex]{7cm}{ \begin{center} $t+U+2 W$ \end{center}}  & 6. \\ 
43 & $\ket{3,- \frac{1}{2} , \frac{15}{4} ,\Gamma_2}$  &\parbox[c][5.0ex]{7cm}{ \begin{center} $\frac{3}{4} (J-4 t+4 W)$ \end{center}}  & -3. \\ 
44 & $\ket{3,- \frac{1}{2} , \frac{3}{4} ,\Gamma_{3,1}}$  &\parbox[c][5.0ex]{7cm}{ \begin{center} $\frac{1}{8} \left(-3 J-8 t+4 U+20 W-\sqrt{A_8}\right)$ \end{center}}  & -0.791288 \\ 
45 & $\ket{3,- \frac{1}{2} , \frac{3}{4} ,\Gamma_{3,1}}$  &\parbox[c][5.0ex]{7cm}{ \begin{center} $\frac{1}{8} \left(-3 J-8 t+4 U+20 W+\sqrt{A_8}\right)$ \end{center}}  & 3.79129 \\ 
46 & $\ket{3,- \frac{1}{2} , \frac{3}{4} ,\Gamma_{3,2}}$  &\parbox[c][5.0ex]{7cm}{ \begin{center} $\frac{1}{8} \left(-3 J-8 t+4 U+20 W-\sqrt{A_8}\right)$ \end{center}}  & -0.791288 \\ 
47 & $\ket{3,- \frac{1}{2} , \frac{3}{4} ,\Gamma_{3,2}}$  &\parbox[c][5.0ex]{7cm}{ \begin{center} $\frac{1}{8} \left(-3 J-8 t+4 U+20 W+\sqrt{A_8}\right)$ \end{center}}  & 3.79129 \\ 
48 & $\ket{3,- \frac{1}{2} , \frac{3}{4} ,\Gamma_{4,1}}$  &\parbox[c][5.0ex]{7cm}{ \begin{center} $-\frac{J}{4}+t+\frac{2 U}{3}+\frac{7 W}{3}-\frac{1}{6} \cos \left(\theta _5\right) \sqrt{A_{10}}$ \end{center}}  & -0.248211 \\ 
49 & $\ket{3,- \frac{1}{2} , \frac{3}{4} ,\Gamma_{4,1}}$  &\parbox[c][5.0ex]{7cm}{ \begin{center} $A_{26}$ \end{center}}  & 8.51895 \\ 
50 & $\ket{3,- \frac{1}{2} , \frac{3}{4} ,\Gamma_{4,1}}$  &\parbox[c][5.0ex]{7cm}{ \begin{center} $A_{25}$ \end{center}}  & 4.72926 \\ 
51 & $\ket{3,- \frac{1}{2} , \frac{3}{4} ,\Gamma_{4,2}}$  &\parbox[c][5.0ex]{7cm}{ \begin{center} $-\frac{J}{4}+t+\frac{2 U}{3}+\frac{7 W}{3}-\frac{1}{6} \cos \left(\theta _5\right) \sqrt{A_{10}}$ \end{center}}  & -0.248211 \\ 
52 & $\ket{3,- \frac{1}{2} , \frac{3}{4} ,\Gamma_{4,2}}$  &\parbox[c][5.0ex]{7cm}{ \begin{center} $A_{26}$ \end{center}}  & 8.51895 \\ 
53 & $\ket{3,- \frac{1}{2} , \frac{3}{4} ,\Gamma_{4,2}}$  &\parbox[c][5.0ex]{7cm}{ \begin{center} $A_{25}$ \end{center}}  & 4.72926 \\ 
54 & $\ket{3,- \frac{1}{2} , \frac{3}{4} ,\Gamma_{4,3}}$  &\parbox[c][5.0ex]{7cm}{ \begin{center} $-\frac{J}{4}+t+\frac{2 U}{3}+\frac{7 W}{3}-\frac{1}{6} \cos \left(\theta _5\right) \sqrt{A_{10}}$ \end{center}}  & -0.248211 \\ 
55 & $\ket{3,- \frac{1}{2} , \frac{3}{4} ,\Gamma_{4,3}}$  &\parbox[c][5.0ex]{7cm}{ \begin{center} $A_{26}$ \end{center}}  & 8.51895 \\ 
56 & $\ket{3,- \frac{1}{2} , \frac{3}{4} ,\Gamma_{4,3}}$  &\parbox[c][5.0ex]{7cm}{ \begin{center} $A_{25}$ \end{center}}  & 4.72926 \\ 
57 & $\ket{3,- \frac{1}{2} , \frac{3}{4} ,\Gamma_{5,1}}$  &\parbox[c][5.0ex]{7cm}{ \begin{center} $\frac{1}{8} \left(-3 J-8 t+4 U+20 W-\sqrt{A_6}\right)$ \end{center}}  & -2.40512 \\ 
58 & $\ket{3,- \frac{1}{2} , \frac{3}{4} ,\Gamma_{5,1}}$  &\parbox[c][5.0ex]{7cm}{ \begin{center} $\frac{1}{8} \left(-3 J-8 t+4 U+20 W+\sqrt{A_6}\right)$ \end{center}}  & 5.40512 \\ 
59 & $\ket{3,- \frac{1}{2} , \frac{15}{4} ,\Gamma_{5,1}}$  &\parbox[c][5.0ex]{7cm}{ \begin{center} $\frac{3 J}{4}+t+3 W$ \end{center}}  & 1. \\ 
60 & $\ket{3,- \frac{1}{2} , \frac{3}{4} ,\Gamma_{5,2}}$  &\parbox[c][5.0ex]{7cm}{ \begin{center} $\frac{1}{8} \left(-3 J-8 t+4 U+20 W-\sqrt{A_6}\right)$ \end{center}}  & -2.40512 \\ 
61 & $\ket{3,- \frac{1}{2} , \frac{3}{4} ,\Gamma_{5,2}}$  &\parbox[c][5.0ex]{7cm}{ \begin{center} $\frac{1}{8} \left(-3 J-8 t+4 U+20 W+\sqrt{A_6}\right)$ \end{center}}  & 5.40512 \\ 
62 & $\ket{3,- \frac{1}{2} , \frac{15}{4} ,\Gamma_{5,2}}$  &\parbox[c][5.0ex]{7cm}{ \begin{center} $\frac{3 J}{4}+t+3 W$ \end{center}}  & 1. \\ 
63 & $\ket{3,- \frac{1}{2} , \frac{3}{4} ,\Gamma_{5,3}}$  &\parbox[c][5.0ex]{7cm}{ \begin{center} $\frac{1}{8} \left(-3 J-8 t+4 U+20 W-\sqrt{A_6}\right)$ \end{center}}  & -2.40512 \\ 
64 & $\ket{3,- \frac{1}{2} , \frac{3}{4} ,\Gamma_{5,3}}$  &\parbox[c][5.0ex]{7cm}{ \begin{center} $\frac{1}{8} \left(-3 J-8 t+4 U+20 W+\sqrt{A_6}\right)$ \end{center}}  & 5.40512 \\ 
65 & $\ket{3,- \frac{1}{2} , \frac{15}{4} ,\Gamma_{5,3}}$  &\parbox[c][5.0ex]{7cm}{ \begin{center} $\frac{3 J}{4}+t+3 W$ \end{center}}  & 1. \\ 
\hline
\end{tabular} \\[2ex]
\begin{tabular}[t]{|r|l|c|c|}
\multicolumn{4}{c}{\bf \boldmath Eigenkets and eigenvalues for ${\rm  N_e}$=3 and   ${\rm m_s}$= $\frac{1}{2} $. } \\[1ex] \hline
\parbox[c]{1cm}{\hfill No}  & \parbox[c][2.5ex]{2.5cm}{\begin{center} Eigenstate \end{center}}  & \parbox[c][2.5ex]{7.5cm}{ \begin{center}   Energy \end{center}}  & \parbox[c][2.5ex]{2cm}{ \begin{center} Example \end{center}}  \\ \hline 
\hline 
66 & $\ket{3,\frac{1}{2} , \frac{3}{4} ,\Gamma_1}$  &\parbox[c][5.0ex]{7cm}{ \begin{center} $t+U+2 W$ \end{center}}  & 6. \\ 
67 & $\ket{3,\frac{1}{2} , \frac{15}{4} ,\Gamma_2}$  &\parbox[c][5.0ex]{7cm}{ \begin{center} $\frac{3}{4} (J-4 t+4 W)$ \end{center}}  & -3. \\ 
68 & $\ket{3,\frac{1}{2} , \frac{3}{4} ,\Gamma_{3,1}}$  &\parbox[c][5.0ex]{7cm}{ \begin{center} $\frac{1}{8} \left(-3 J-8 t+4 U+20 W-\sqrt{A_8}\right)$ \end{center}}  & -0.791288 \\ 
69 & $\ket{3,\frac{1}{2} , \frac{3}{4} ,\Gamma_{3,1}}$  &\parbox[c][5.0ex]{7cm}{ \begin{center} $\frac{1}{8} \left(-3 J-8 t+4 U+20 W+\sqrt{A_8}\right)$ \end{center}}  & 3.79129 \\ 
70 & $\ket{3,\frac{1}{2} , \frac{3}{4} ,\Gamma_{3,2}}$  &\parbox[c][5.0ex]{7cm}{ \begin{center} $\frac{1}{8} \left(-3 J-8 t+4 U+20 W-\sqrt{A_8}\right)$ \end{center}}  & -0.791288 \\ 
71 & $\ket{3,\frac{1}{2} , \frac{3}{4} ,\Gamma_{3,2}}$  &\parbox[c][5.0ex]{7cm}{ \begin{center} $\frac{1}{8} \left(-3 J-8 t+4 U+20 W+\sqrt{A_8}\right)$ \end{center}}  & 3.79129 \\ 
72 & $\ket{3,\frac{1}{2} , \frac{3}{4} ,\Gamma_{4,1}}$  &\parbox[c][5.0ex]{7cm}{ \begin{center} $-\frac{J}{4}+t+\frac{2 U}{3}+\frac{7 W}{3}-\frac{1}{6} \cos \left(\theta _5\right) \sqrt{A_{10}}$ \end{center}}  & -0.248211 \\ 
73 & $\ket{3,\frac{1}{2} , \frac{3}{4} ,\Gamma_{4,1}}$  &\parbox[c][5.0ex]{7cm}{ \begin{center} $A_{26}$ \end{center}}  & 8.51895 \\ 
74 & $\ket{3,\frac{1}{2} , \frac{3}{4} ,\Gamma_{4,1}}$  &\parbox[c][5.0ex]{7cm}{ \begin{center} $A_{25}$ \end{center}}  & 4.72926 \\ 
75 & $\ket{3,\frac{1}{2} , \frac{3}{4} ,\Gamma_{4,2}}$  &\parbox[c][5.0ex]{7cm}{ \begin{center} $-\frac{J}{4}+t+\frac{2 U}{3}+\frac{7 W}{3}-\frac{1}{6} \cos \left(\theta _5\right) \sqrt{A_{10}}$ \end{center}}  & -0.248211 \\ 
76 & $\ket{3,\frac{1}{2} , \frac{3}{4} ,\Gamma_{4,2}}$  &\parbox[c][5.0ex]{7cm}{ \begin{center} $A_{26}$ \end{center}}  & 8.51895 \\ 
77 & $\ket{3,\frac{1}{2} , \frac{3}{4} ,\Gamma_{4,2}}$  &\parbox[c][5.0ex]{7cm}{ \begin{center} $A_{25}$ \end{center}}  & 4.72926 \\ 
78 & $\ket{3,\frac{1}{2} , \frac{3}{4} ,\Gamma_{4,3}}$  &\parbox[c][5.0ex]{7cm}{ \begin{center} $-\frac{J}{4}+t+\frac{2 U}{3}+\frac{7 W}{3}-\frac{1}{6} \cos \left(\theta _5\right) \sqrt{A_{10}}$ \end{center}}  & -0.248211 \\ 
79 & $\ket{3,\frac{1}{2} , \frac{3}{4} ,\Gamma_{4,3}}$  &\parbox[c][5.0ex]{7cm}{ \begin{center} $A_{26}$ \end{center}}  & 8.51895 \\ 
80 & $\ket{3,\frac{1}{2} , \frac{3}{4} ,\Gamma_{4,3}}$  &\parbox[c][5.0ex]{7cm}{ \begin{center} $A_{25}$ \end{center}}  & 4.72926 \\ 
81 & $\ket{3,\frac{1}{2} , \frac{3}{4} ,\Gamma_{5,1}}$  &\parbox[c][5.0ex]{7cm}{ \begin{center} $\frac{1}{8} \left(-3 J-8 t+4 U+20 W-\sqrt{A_6}\right)$ \end{center}}  & -2.40512 \\ 
82 & $\ket{3,\frac{1}{2} , \frac{3}{4} ,\Gamma_{5,1}}$  &\parbox[c][5.0ex]{7cm}{ \begin{center} $\frac{1}{8} \left(-3 J-8 t+4 U+20 W+\sqrt{A_6}\right)$ \end{center}}  & 5.40512 \\ 
83 & $\ket{3,\frac{1}{2} , \frac{15}{4} ,\Gamma_{5,1}}$  &\parbox[c][5.0ex]{7cm}{ \begin{center} $\frac{3 J}{4}+t+3 W$ \end{center}}  & 1. \\ 
84 & $\ket{3,\frac{1}{2} , \frac{3}{4} ,\Gamma_{5,2}}$  &\parbox[c][5.0ex]{7cm}{ \begin{center} $\frac{1}{8} \left(-3 J-8 t+4 U+20 W-\sqrt{A_6}\right)$ \end{center}}  & -2.40512 \\ 
85 & $\ket{3,\frac{1}{2} , \frac{3}{4} ,\Gamma_{5,2}}$  &\parbox[c][5.0ex]{7cm}{ \begin{center} $\frac{1}{8} \left(-3 J-8 t+4 U+20 W+\sqrt{A_6}\right)$ \end{center}}  & 5.40512 \\ 
86 & $\ket{3,\frac{1}{2} , \frac{15}{4} ,\Gamma_{5,2}}$  &\parbox[c][5.0ex]{7cm}{ \begin{center} $\frac{3 J}{4}+t+3 W$ \end{center}}  & 1. \\ 
87 & $\ket{3,\frac{1}{2} , \frac{3}{4} ,\Gamma_{5,3}}$  &\parbox[c][5.0ex]{7cm}{ \begin{center} $\frac{1}{8} \left(-3 J-8 t+4 U+20 W-\sqrt{A_6}\right)$ \end{center}}  & -2.40512 \\ 
88 & $\ket{3,\frac{1}{2} , \frac{3}{4} ,\Gamma_{5,3}}$  &\parbox[c][5.0ex]{7cm}{ \begin{center} $\frac{1}{8} \left(-3 J-8 t+4 U+20 W+\sqrt{A_6}\right)$ \end{center}}  & 5.40512 \\ 
89 & $\ket{3,\frac{1}{2} , \frac{15}{4} ,\Gamma_{5,3}}$  &\parbox[c][5.0ex]{7cm}{ \begin{center} $\frac{3 J}{4}+t+3 W$ \end{center}}  & 1. \\ 
\hline
\end{tabular} \\[2ex]
\begin{tabular}[t]{|r|l|c|c|}
\multicolumn{4}{c}{\bf \boldmath Eigenkets and eigenvalues for ${\rm  N_e}$=3 and   ${\rm m_s}$= $\frac{3}{2} $. } \\[1ex] \hline
\parbox[c]{1cm}{\hfill No}  & \parbox[c][2.5ex]{2.5cm}{\begin{center} Eigenstate \end{center}}  & \parbox[c][2.5ex]{7.5cm}{ \begin{center}   Energy \end{center}}  & \parbox[c][2.5ex]{2cm}{ \begin{center} Example \end{center}}  \\ \hline 
\hline 
90 & $\ket{3,\frac{3}{2} , \frac{15}{4} ,\Gamma_2}$  &\parbox[c][5.0ex]{7cm}{ \begin{center} $\frac{3}{4} (J-4 t+4 W)$ \end{center}}  & -3. \\ 
91 & $\ket{3,\frac{3}{2} , \frac{15}{4} ,\Gamma_{5,1}}$  &\parbox[c][5.0ex]{7cm}{ \begin{center} $\frac{3 J}{4}+t+3 W$ \end{center}}  & 1. \\ 
92 & $\ket{3,\frac{3}{2} , \frac{15}{4} ,\Gamma_{5,2}}$  &\parbox[c][5.0ex]{7cm}{ \begin{center} $\frac{3 J}{4}+t+3 W$ \end{center}}  & 1. \\ 
93 & $\ket{3,\frac{3}{2} , \frac{15}{4} ,\Gamma_{5,3}}$  &\parbox[c][5.0ex]{7cm}{ \begin{center} $\frac{3 J}{4}+t+3 W$ \end{center}}  & 1. \\ 
\hline
\end{tabular} \\[2ex]
\begin{tabular}[t]{|r|l|c|c|}
\multicolumn{4}{c}{\bf \boldmath Eigenkets and eigenvalues for ${\rm  N_e}$=4 and   ${\rm m_s}$= $-2$. } \\[1ex] \hline
\parbox[c]{1cm}{\hfill No}  & \parbox[c][2.5ex]{2.5cm}{\begin{center} Eigenstate \end{center}}  & \parbox[c][2.5ex]{7.5cm}{ \begin{center}   Energy \end{center}}  & \parbox[c][2.5ex]{2cm}{ \begin{center} Example \end{center}}  \\ \hline 
\hline 
94 & $\ket{4,-2,6,\Gamma_2}$  &\parbox[c][5.0ex]{7cm}{ \begin{center} $\frac{3}{2} (J+4 W)$ \end{center}}  & 0. \\ 
\hline
\end{tabular} \\[2ex]
\begin{tabular}[t]{|r|l|c|c|}
\multicolumn{4}{c}{\bf \boldmath Eigenkets and eigenvalues for ${\rm  N_e}$=4 and   ${\rm m_s}$= $-1$. } \\[1ex] \hline
\parbox[c]{1cm}{\hfill No}  & \parbox[c][2.5ex]{2.5cm}{\begin{center} Eigenstate \end{center}}  & \parbox[c][2.5ex]{7.5cm}{ \begin{center}   Energy \end{center}}  & \parbox[c][2.5ex]{2cm}{ \begin{center} Example \end{center}}  \\ \hline 
\hline 
95 & $\ket{4,-1,2,\Gamma_2}$  &\parbox[c][5.0ex]{7cm}{ \begin{center} $\frac{J}{4}+U+5 W$ \end{center}}  & 5. \\ 
96 & $\ket{4,-1,6,\Gamma_2}$  &\parbox[c][5.0ex]{7cm}{ \begin{center} $\frac{3}{2} (J+4 W)$ \end{center}}  & 0. \\ 
97 & $\ket{4,-1,2,\Gamma_{3,1}}$  &\parbox[c][5.0ex]{7cm}{ \begin{center} $\frac{J}{4}+U+5 W$ \end{center}}  & 5. \\ 
98 & $\ket{4,-1,2,\Gamma_{3,2}}$  &\parbox[c][5.0ex]{7cm}{ \begin{center} $\frac{J}{4}+U+5 W$ \end{center}}  & 5. \\ 
99 & $\ket{4,-1,2,\Gamma_{4,1}}$  &\parbox[c][5.0ex]{7cm}{ \begin{center} $\frac{J}{4}+U+5 W$ \end{center}}  & 5. \\ 
100 & $\ket{4,-1,2,\Gamma_{4,2}}$  &\parbox[c][5.0ex]{7cm}{ \begin{center} $\frac{J}{4}+U+5 W$ \end{center}}  & 5. \\ 
101 & $\ket{4,-1,2,\Gamma_{4,3}}$  &\parbox[c][5.0ex]{7cm}{ \begin{center} $\frac{J}{4}+U+5 W$ \end{center}}  & 5. \\ 
102 & $\ket{4,-1,2,\Gamma_{5,1}}$  &\parbox[c][5.0ex]{7cm}{ \begin{center} $\frac{2}{3} (U+8 W)-\frac{1}{6} \cos \left(\theta _4\right) \sqrt{A_3}$ \end{center}}  & -1.5136 \\ 
103 & $\ket{4,-1,2,\Gamma_{5,1}}$  &\parbox[c][5.0ex]{7cm}{ \begin{center} $A_{21}$ \end{center}}  & 8.34789 \\ 
104 & $\ket{4,-1,2,\Gamma_{5,1}}$  &\parbox[c][5.0ex]{7cm}{ \begin{center} $A_{20}$ \end{center}}  & 3.16571 \\ 
105 & $\ket{4,-1,2,\Gamma_{5,2}}$  &\parbox[c][5.0ex]{7cm}{ \begin{center} $\frac{2}{3} (U+8 W)-\frac{1}{6} \cos \left(\theta _4\right) \sqrt{A_3}$ \end{center}}  & -1.5136 \\ 
106 & $\ket{4,-1,2,\Gamma_{5,2}}$  &\parbox[c][5.0ex]{7cm}{ \begin{center} $A_{21}$ \end{center}}  & 8.34789 \\ 
107 & $\ket{4,-1,2,\Gamma_{5,2}}$  &\parbox[c][5.0ex]{7cm}{ \begin{center} $A_{20}$ \end{center}}  & 3.16571 \\ 
108 & $\ket{4,-1,2,\Gamma_{5,3}}$  &\parbox[c][5.0ex]{7cm}{ \begin{center} $\frac{2}{3} (U+8 W)-\frac{1}{6} \cos \left(\theta _4\right) \sqrt{A_3}$ \end{center}}  & -1.5136 \\ 
109 & $\ket{4,-1,2,\Gamma_{5,3}}$  &\parbox[c][5.0ex]{7cm}{ \begin{center} $A_{21}$ \end{center}}  & 8.34789 \\ 
110 & $\ket{4,-1,2,\Gamma_{5,3}}$  &\parbox[c][5.0ex]{7cm}{ \begin{center} $A_{20}$ \end{center}}  & 3.16571 \\ 
\hline
\end{tabular} \\[2ex]
\begin{tabular}[t]{|r|l|c|c|}
\multicolumn{4}{c}{\bf \boldmath Eigenkets and eigenvalues for ${\rm  N_e}$=4 and   ${\rm m_s}$= $0$. } \\[1ex] \hline
\parbox[c]{1cm}{\hfill No}  & \parbox[c][2.5ex]{2.5cm}{\begin{center} Eigenstate \end{center}}  & \parbox[c][2.5ex]{7.5cm}{ \begin{center}   Energy \end{center}}  & \parbox[c][2.5ex]{2cm}{ \begin{center} Example \end{center}}  \\ \hline 
\hline 
111 & $\ket{4,0,0,\Gamma_1}$  &\parbox[c][5.0ex]{7cm}{ \begin{center} $\frac{1}{8} \left(-3 J+12 U+36 W-\sqrt{A_4}\right)$ \end{center}}  & 2.78301 \\ 
112 & $\ket{4,0,0,\Gamma_1}$  &\parbox[c][5.0ex]{7cm}{ \begin{center} $\frac{1}{8} \left(-3 J+12 U+36 W+\sqrt{A_4}\right)$ \end{center}}  & 12.217 \\ 
113 & $\ket{4,0,2,\Gamma_2}$  &\parbox[c][5.0ex]{7cm}{ \begin{center} $\frac{J}{4}+U+5 W$ \end{center}}  & 5. \\ 
114 & $\ket{4,0,6,\Gamma_2}$  &\parbox[c][5.0ex]{7cm}{ \begin{center} $\frac{3}{2} (J+4 W)$ \end{center}}  & 0. \\ 
115 & $\ket{4,0,0,\Gamma_{3,1}}$  &\parbox[c][5.0ex]{7cm}{ \begin{center} $-\frac{3 J}{4}+U+5 W-\frac{\cos \left(\theta _2\right) \sqrt{A_2}}{2 \sqrt{3}}$ \end{center}}  & -1.84429 \\ 
116 & $\ket{4,0,0,\Gamma_{3,1}}$  &\parbox[c][5.0ex]{7cm}{ \begin{center} $A_{15}$ \end{center}}  & 10.8443 \\ 
117 & $\ket{4,0,0,\Gamma_{3,1}}$  &\parbox[c][5.0ex]{7cm}{ \begin{center} $A_{14}$ \end{center}}  & 6. \\ 
118 & $\ket{4,0,2,\Gamma_{3,1}}$  &\parbox[c][5.0ex]{7cm}{ \begin{center} $\frac{J}{4}+U+5 W$ \end{center}}  & 5. \\ 
119 & $\ket{4,0,0,\Gamma_{3,2}}$  &\parbox[c][5.0ex]{7cm}{ \begin{center} $-\frac{3 J}{4}+U+5 W-\frac{\cos \left(\theta _2\right) \sqrt{A_2}}{2 \sqrt{3}}$ \end{center}}  & -1.84429 \\ 
120 & $\ket{4,0,0,\Gamma_{3,2}}$  &\parbox[c][5.0ex]{7cm}{ \begin{center} $A_{15}$ \end{center}}  & 10.8443 \\ 
121 & $\ket{4,0,0,\Gamma_{3,2}}$  &\parbox[c][5.0ex]{7cm}{ \begin{center} $A_{14}$ \end{center}}  & 6. \\ 
122 & $\ket{4,0,2,\Gamma_{3,2}}$  &\parbox[c][5.0ex]{7cm}{ \begin{center} $\frac{J}{4}+U+5 W$ \end{center}}  & 5. \\ 
123 & $\ket{4,0,0,\Gamma_{4,1}}$  &\parbox[c][5.0ex]{7cm}{ \begin{center} $\frac{1}{6} (-3 J+8 U+28 W)-\frac{1}{6} \cos \left(\theta _3\right) \sqrt{A_3}$ \end{center}}  & 1.65211 \\ 
124 & $\ket{4,0,0,\Gamma_{4,1}}$  &\parbox[c][5.0ex]{7cm}{ \begin{center} $A_{19}$ \end{center}}  & 11.5136 \\ 
125 & $\ket{4,0,0,\Gamma_{4,1}}$  &\parbox[c][5.0ex]{7cm}{ \begin{center} $A_{18}$ \end{center}}  & 6.83429 \\ 
126 & $\ket{4,0,2,\Gamma_{4,1}}$  &\parbox[c][5.0ex]{7cm}{ \begin{center} $\frac{J}{4}+U+5 W$ \end{center}}  & 5. \\ 
127 & $\ket{4,0,0,\Gamma_{4,2}}$  &\parbox[c][5.0ex]{7cm}{ \begin{center} $\frac{1}{6} (-3 J+8 U+28 W)-\frac{1}{6} \cos \left(\theta _3\right) \sqrt{A_3}$ \end{center}}  & 1.65211 \\ 
128 & $\ket{4,0,0,\Gamma_{4,2}}$  &\parbox[c][5.0ex]{7cm}{ \begin{center} $A_{19}$ \end{center}}  & 11.5136 \\ 
129 & $\ket{4,0,0,\Gamma_{4,2}}$  &\parbox[c][5.0ex]{7cm}{ \begin{center} $A_{18}$ \end{center}}  & 6.83429 \\ 
130 & $\ket{4,0,2,\Gamma_{4,2}}$  &\parbox[c][5.0ex]{7cm}{ \begin{center} $\frac{J}{4}+U+5 W$ \end{center}}  & 5. \\ 
131 & $\ket{4,0,0,\Gamma_{4,3}}$  &\parbox[c][5.0ex]{7cm}{ \begin{center} $\frac{1}{6} (-3 J+8 U+28 W)-\frac{1}{6} \cos \left(\theta _3\right) \sqrt{A_3}$ \end{center}}  & 1.65211 \\ 
132 & $\ket{4,0,0,\Gamma_{4,3}}$  &\parbox[c][5.0ex]{7cm}{ \begin{center} $A_{19}$ \end{center}}  & 11.5136 \\ 
133 & $\ket{4,0,0,\Gamma_{4,3}}$  &\parbox[c][5.0ex]{7cm}{ \begin{center} $A_{18}$ \end{center}}  & 6.83429 \\ 
134 & $\ket{4,0,2,\Gamma_{4,3}}$  &\parbox[c][5.0ex]{7cm}{ \begin{center} $\frac{J}{4}+U+5 W$ \end{center}}  & 5. \\ 
135 & $\ket{4,0,0,\Gamma_{5,1}}$  &\parbox[c][5.0ex]{7cm}{ \begin{center} $-\frac{3 J}{4}+U+5 W$ \end{center}}  & 5. \\ 
136 & $\ket{4,0,2,\Gamma_{5,1}}$  &\parbox[c][5.0ex]{7cm}{ \begin{center} $\frac{2}{3} (U+8 W)-\frac{1}{6} \cos \left(\theta _4\right) \sqrt{A_3}$ \end{center}}  & -1.5136 \\ 
137 & $\ket{4,0,2,\Gamma_{5,1}}$  &\parbox[c][5.0ex]{7cm}{ \begin{center} $A_{21}$ \end{center}}  & 8.34789 \\ 
138 & $\ket{4,0,2,\Gamma_{5,1}}$  &\parbox[c][5.0ex]{7cm}{ \begin{center} $A_{20}$ \end{center}}  & 3.16571 \\ 
139 & $\ket{4,0,0,\Gamma_{5,2}}$  &\parbox[c][5.0ex]{7cm}{ \begin{center} $-\frac{3 J}{4}+U+5 W$ \end{center}}  & 5. \\ 
140 & $\ket{4,0,2,\Gamma_{5,2}}$  &\parbox[c][5.0ex]{7cm}{ \begin{center} $\frac{2}{3} (U+8 W)-\frac{1}{6} \cos \left(\theta _4\right) \sqrt{A_3}$ \end{center}}  & -1.5136 \\ 
141 & $\ket{4,0,2,\Gamma_{5,2}}$  &\parbox[c][5.0ex]{7cm}{ \begin{center} $A_{21}$ \end{center}}  & 8.34789 \\ 
142 & $\ket{4,0,2,\Gamma_{5,2}}$  &\parbox[c][5.0ex]{7cm}{ \begin{center} $A_{20}$ \end{center}}  & 3.16571 \\ 
143 & $\ket{4,0,0,\Gamma_{5,3}}$  &\parbox[c][5.0ex]{7cm}{ \begin{center} $-\frac{3 J}{4}+U+5 W$ \end{center}}  & 5. \\ 
144 & $\ket{4,0,2,\Gamma_{5,3}}$  &\parbox[c][5.0ex]{7cm}{ \begin{center} $\frac{2}{3} (U+8 W)-\frac{1}{6} \cos \left(\theta _4\right) \sqrt{A_3}$ \end{center}}  & -1.5136 \\ 
145 & $\ket{4,0,2,\Gamma_{5,3}}$  &\parbox[c][5.0ex]{7cm}{ \begin{center} $A_{21}$ \end{center}}  & 8.34789 \\ 
146 & $\ket{4,0,2,\Gamma_{5,3}}$  &\parbox[c][5.0ex]{7cm}{ \begin{center} $A_{20}$ \end{center}}  & 3.16571 \\ 
\hline
\end{tabular} \\[2ex]
\begin{tabular}[t]{|r|l|c|c|}
\multicolumn{4}{c}{\bf \boldmath Eigenkets and eigenvalues for ${\rm  N_e}$=4 and   ${\rm m_s}$= $1$. } \\[1ex] \hline
\parbox[c]{1cm}{\hfill No}  & \parbox[c][2.5ex]{2.5cm}{\begin{center} Eigenstate \end{center}}  & \parbox[c][2.5ex]{7.5cm}{ \begin{center}   Energy \end{center}}  & \parbox[c][2.5ex]{2cm}{ \begin{center} Example \end{center}}  \\ \hline 
\hline 
147 & $\ket{4,1,2,\Gamma_2}$  &\parbox[c][5.0ex]{7cm}{ \begin{center} $\frac{J}{4}+U+5 W$ \end{center}}  & 5. \\ 
148 & $\ket{4,1,6,\Gamma_2}$  &\parbox[c][5.0ex]{7cm}{ \begin{center} $\frac{3}{2} (J+4 W)$ \end{center}}  & 0. \\ 
149 & $\ket{4,1,2,\Gamma_{3,1}}$  &\parbox[c][5.0ex]{7cm}{ \begin{center} $\frac{J}{4}+U+5 W$ \end{center}}  & 5. \\ 
150 & $\ket{4,1,2,\Gamma_{3,2}}$  &\parbox[c][5.0ex]{7cm}{ \begin{center} $\frac{J}{4}+U+5 W$ \end{center}}  & 5. \\ 
151 & $\ket{4,1,2,\Gamma_{4,1}}$  &\parbox[c][5.0ex]{7cm}{ \begin{center} $\frac{J}{4}+U+5 W$ \end{center}}  & 5. \\ 
152 & $\ket{4,1,2,\Gamma_{4,2}}$  &\parbox[c][5.0ex]{7cm}{ \begin{center} $\frac{J}{4}+U+5 W$ \end{center}}  & 5. \\ 
153 & $\ket{4,1,2,\Gamma_{4,3}}$  &\parbox[c][5.0ex]{7cm}{ \begin{center} $\frac{J}{4}+U+5 W$ \end{center}}  & 5. \\ 
154 & $\ket{4,1,2,\Gamma_{5,1}}$  &\parbox[c][5.0ex]{7cm}{ \begin{center} $\frac{2}{3} (U+8 W)-\frac{1}{6} \cos \left(\theta _4\right) \sqrt{A_3}$ \end{center}}  & -1.5136 \\ 
155 & $\ket{4,1,2,\Gamma_{5,1}}$  &\parbox[c][5.0ex]{7cm}{ \begin{center} $A_{21}$ \end{center}}  & 8.34789 \\ 
156 & $\ket{4,1,2,\Gamma_{5,1}}$  &\parbox[c][5.0ex]{7cm}{ \begin{center} $A_{20}$ \end{center}}  & 3.16571 \\ 
157 & $\ket{4,1,2,\Gamma_{5,2}}$  &\parbox[c][5.0ex]{7cm}{ \begin{center} $\frac{2}{3} (U+8 W)-\frac{1}{6} \cos \left(\theta _4\right) \sqrt{A_3}$ \end{center}}  & -1.5136 \\ 
158 & $\ket{4,1,2,\Gamma_{5,2}}$  &\parbox[c][5.0ex]{7cm}{ \begin{center} $A_{21}$ \end{center}}  & 8.34789 \\ 
159 & $\ket{4,1,2,\Gamma_{5,2}}$  &\parbox[c][5.0ex]{7cm}{ \begin{center} $A_{20}$ \end{center}}  & 3.16571 \\ 
160 & $\ket{4,1,2,\Gamma_{5,3}}$  &\parbox[c][5.0ex]{7cm}{ \begin{center} $\frac{2}{3} (U+8 W)-\frac{1}{6} \cos \left(\theta _4\right) \sqrt{A_3}$ \end{center}}  & -1.5136 \\ 
161 & $\ket{4,1,2,\Gamma_{5,3}}$  &\parbox[c][5.0ex]{7cm}{ \begin{center} $A_{21}$ \end{center}}  & 8.34789 \\ 
162 & $\ket{4,1,2,\Gamma_{5,3}}$  &\parbox[c][5.0ex]{7cm}{ \begin{center} $A_{20}$ \end{center}}  & 3.16571 \\ 
\hline
\end{tabular} \\[2ex]
\begin{tabular}[t]{|r|l|c|c|}
\multicolumn{4}{c}{\bf \boldmath Eigenkets and eigenvalues for ${\rm  N_e}$=4 and   ${\rm m_s}$= $2$. } \\[1ex] \hline
\parbox[c]{1cm}{\hfill No}  & \parbox[c][2.5ex]{2.5cm}{\begin{center} Eigenstate \end{center}}  & \parbox[c][2.5ex]{7.5cm}{ \begin{center}   Energy \end{center}}  & \parbox[c][2.5ex]{2cm}{ \begin{center} Example \end{center}}  \\ \hline 
\hline 
163 & $\ket{4,2,6,\Gamma_2}$  &\parbox[c][5.0ex]{7cm}{ \begin{center} $\frac{3}{2} (J+4 W)$ \end{center}}  & 0. \\ 
\hline
\end{tabular} \\[2ex]
\begin{tabular}[t]{|r|l|c|c|}
\multicolumn{4}{c}{\bf \boldmath Eigenkets and eigenvalues for ${\rm  N_e}$=5 and   ${\rm m_s}$= $- \frac{3}{2} $. } \\[1ex] \hline
\parbox[c]{1cm}{\hfill No}  & \parbox[c][2.5ex]{2.5cm}{\begin{center} Eigenstate \end{center}}  & \parbox[c][2.5ex]{7.5cm}{ \begin{center}   Energy \end{center}}  & \parbox[c][2.5ex]{2cm}{ \begin{center} Example \end{center}}  \\ \hline 
\hline 
164 & $\ket{5,- \frac{3}{2} , \frac{15}{4} ,\Gamma_2}$  &\parbox[c][5.0ex]{7cm}{ \begin{center} $\frac{3 J}{4}+3 t+U+9 W$ \end{center}}  & 8. \\ 
165 & $\ket{5,- \frac{3}{2} , \frac{15}{4} ,\Gamma_{5,1}}$  &\parbox[c][5.0ex]{7cm}{ \begin{center} $\frac{3 J}{4}-t+U+9 W$ \end{center}}  & 4. \\ 
166 & $\ket{5,- \frac{3}{2} , \frac{15}{4} ,\Gamma_{5,2}}$  &\parbox[c][5.0ex]{7cm}{ \begin{center} $\frac{3 J}{4}-t+U+9 W$ \end{center}}  & 4. \\ 
167 & $\ket{5,- \frac{3}{2} , \frac{15}{4} ,\Gamma_{5,3}}$  &\parbox[c][5.0ex]{7cm}{ \begin{center} $\frac{3 J}{4}-t+U+9 W$ \end{center}}  & 4. \\ 
\hline
\end{tabular} \\[2ex]
\begin{tabular}[t]{|r|l|c|c|}
\multicolumn{4}{c}{\bf \boldmath Eigenkets and eigenvalues for ${\rm  N_e}$=5 and   ${\rm m_s}$= $- \frac{1}{2} $. } \\[1ex] \hline
\parbox[c]{1cm}{\hfill No}  & \parbox[c][2.5ex]{2.5cm}{\begin{center} Eigenstate \end{center}}  & \parbox[c][2.5ex]{7.5cm}{ \begin{center}   Energy \end{center}}  & \parbox[c][2.5ex]{2cm}{ \begin{center} Example \end{center}}  \\ \hline 
\hline 
168 & $\ket{5,- \frac{1}{2} , \frac{3}{4} ,\Gamma_1}$  &\parbox[c][5.0ex]{7cm}{ \begin{center} $-t+2 U+8 W$ \end{center}}  & 9. \\ 
169 & $\ket{5,- \frac{1}{2} , \frac{15}{4} ,\Gamma_2}$  &\parbox[c][5.0ex]{7cm}{ \begin{center} $\frac{3 J}{4}+3 t+U+9 W$ \end{center}}  & 8. \\ 
170 & $\ket{5,- \frac{1}{2} , \frac{3}{4} ,\Gamma_{3,1}}$  &\parbox[c][5.0ex]{7cm}{ \begin{center} $\frac{1}{8} \left(-3 J+8 t+12 U+68 W-\sqrt{A_6}\right)$ \end{center}}  & 4.59488 \\ 
171 & $\ket{5,- \frac{1}{2} , \frac{3}{4} ,\Gamma_{3,1}}$  &\parbox[c][5.0ex]{7cm}{ \begin{center} $\frac{1}{8} \left(-3 J+8 t+12 U+68 W+\sqrt{A_6}\right)$ \end{center}}  & 12.4051 \\ 
172 & $\ket{5,- \frac{1}{2} , \frac{3}{4} ,\Gamma_{3,2}}$  &\parbox[c][5.0ex]{7cm}{ \begin{center} $\frac{1}{8} \left(-3 J+8 t+12 U+68 W-\sqrt{A_6}\right)$ \end{center}}  & 4.59488 \\ 
173 & $\ket{5,- \frac{1}{2} , \frac{3}{4} ,\Gamma_{3,2}}$  &\parbox[c][5.0ex]{7cm}{ \begin{center} $\frac{1}{8} \left(-3 J+8 t+12 U+68 W+\sqrt{A_6}\right)$ \end{center}}  & 12.4051 \\ 
174 & $\ket{5,- \frac{1}{2} , \frac{3}{4} ,\Gamma_{4,1}}$  &\parbox[c][5.0ex]{7cm}{ \begin{center} $-\frac{J}{4}-t+\frac{5 U}{3}+\frac{25 W}{3}-\frac{1}{6} \cos \left(\theta _1\right) \sqrt{A_1}$ \end{center}}  & 1.54341 \\ 
175 & $\ket{5,- \frac{1}{2} , \frac{3}{4} ,\Gamma_{4,1}}$  &\parbox[c][5.0ex]{7cm}{ \begin{center} $A_{13}$ \end{center}}  & 12.2755 \\ 
176 & $\ket{5,- \frac{1}{2} , \frac{3}{4} ,\Gamma_{4,1}}$  &\parbox[c][5.0ex]{7cm}{ \begin{center} $A_{12}$ \end{center}}  & 8.18113 \\ 
177 & $\ket{5,- \frac{1}{2} , \frac{3}{4} ,\Gamma_{4,2}}$  &\parbox[c][5.0ex]{7cm}{ \begin{center} $-\frac{J}{4}-t+\frac{5 U}{3}+\frac{25 W}{3}-\frac{1}{6} \cos \left(\theta _1\right) \sqrt{A_1}$ \end{center}}  & 1.54341 \\ 
178 & $\ket{5,- \frac{1}{2} , \frac{3}{4} ,\Gamma_{4,2}}$  &\parbox[c][5.0ex]{7cm}{ \begin{center} $A_{13}$ \end{center}}  & 12.2755 \\ 
179 & $\ket{5,- \frac{1}{2} , \frac{3}{4} ,\Gamma_{4,2}}$  &\parbox[c][5.0ex]{7cm}{ \begin{center} $A_{12}$ \end{center}}  & 8.18113 \\ 
180 & $\ket{5,- \frac{1}{2} , \frac{3}{4} ,\Gamma_{4,3}}$  &\parbox[c][5.0ex]{7cm}{ \begin{center} $-\frac{J}{4}-t+\frac{5 U}{3}+\frac{25 W}{3}-\frac{1}{6} \cos \left(\theta _1\right) \sqrt{A_1}$ \end{center}}  & 1.54341 \\ 
181 & $\ket{5,- \frac{1}{2} , \frac{3}{4} ,\Gamma_{4,3}}$  &\parbox[c][5.0ex]{7cm}{ \begin{center} $A_{13}$ \end{center}}  & 12.2755 \\ 
182 & $\ket{5,- \frac{1}{2} , \frac{3}{4} ,\Gamma_{4,3}}$  &\parbox[c][5.0ex]{7cm}{ \begin{center} $A_{12}$ \end{center}}  & 8.18113 \\ 
183 & $\ket{5,- \frac{1}{2} , \frac{3}{4} ,\Gamma_{5,1}}$  &\parbox[c][5.0ex]{7cm}{ \begin{center} $\frac{1}{8} \left(-3 J+8 t+12 U+68 W-\sqrt{A_8}\right)$ \end{center}}  & 6.20871 \\ 
184 & $\ket{5,- \frac{1}{2} , \frac{3}{4} ,\Gamma_{5,1}}$  &\parbox[c][5.0ex]{7cm}{ \begin{center} $\frac{1}{8} \left(-3 J+8 t+12 U+68 W+\sqrt{A_8}\right)$ \end{center}}  & 10.7913 \\ 
185 & $\ket{5,- \frac{1}{2} , \frac{15}{4} ,\Gamma_{5,1}}$  &\parbox[c][5.0ex]{7cm}{ \begin{center} $\frac{3 J}{4}-t+U+9 W$ \end{center}}  & 4. \\ 
186 & $\ket{5,- \frac{1}{2} , \frac{3}{4} ,\Gamma_{5,2}}$  &\parbox[c][5.0ex]{7cm}{ \begin{center} $\frac{1}{8} \left(-3 J+8 t+12 U+68 W-\sqrt{A_8}\right)$ \end{center}}  & 6.20871 \\ 
187 & $\ket{5,- \frac{1}{2} , \frac{3}{4} ,\Gamma_{5,2}}$  &\parbox[c][5.0ex]{7cm}{ \begin{center} $\frac{1}{8} \left(-3 J+8 t+12 U+68 W+\sqrt{A_8}\right)$ \end{center}}  & 10.7913 \\ 
188 & $\ket{5,- \frac{1}{2} , \frac{15}{4} ,\Gamma_{5,2}}$  &\parbox[c][5.0ex]{7cm}{ \begin{center} $\frac{3 J}{4}-t+U+9 W$ \end{center}}  & 4. \\ 
189 & $\ket{5,- \frac{1}{2} , \frac{3}{4} ,\Gamma_{5,3}}$  &\parbox[c][5.0ex]{7cm}{ \begin{center} $\frac{1}{8} \left(-3 J+8 t+12 U+68 W-\sqrt{A_8}\right)$ \end{center}}  & 6.20871 \\ 
190 & $\ket{5,- \frac{1}{2} , \frac{3}{4} ,\Gamma_{5,3}}$  &\parbox[c][5.0ex]{7cm}{ \begin{center} $\frac{1}{8} \left(-3 J+8 t+12 U+68 W+\sqrt{A_8}\right)$ \end{center}}  & 10.7913 \\ 
191 & $\ket{5,- \frac{1}{2} , \frac{15}{4} ,\Gamma_{5,3}}$  &\parbox[c][5.0ex]{7cm}{ \begin{center} $\frac{3 J}{4}-t+U+9 W$ \end{center}}  & 4. \\ 
\hline
\end{tabular} \\[2ex]
\begin{tabular}[t]{|r|l|c|c|}
\multicolumn{4}{c}{\bf \boldmath Eigenkets and eigenvalues for ${\rm  N_e}$=5 and   ${\rm m_s}$= $\frac{1}{2} $. } \\[1ex] \hline
\parbox[c]{1cm}{\hfill No}  & \parbox[c][2.5ex]{2.5cm}{\begin{center} Eigenstate \end{center}}  & \parbox[c][2.5ex]{7.5cm}{ \begin{center}   Energy \end{center}}  & \parbox[c][2.5ex]{2cm}{ \begin{center} Example \end{center}}  \\ \hline 
\hline 
192 & $\ket{5,\frac{1}{2} , \frac{3}{4} ,\Gamma_1}$  &\parbox[c][5.0ex]{7cm}{ \begin{center} $-t+2 U+8 W$ \end{center}}  & 9. \\ 
193 & $\ket{5,\frac{1}{2} , \frac{15}{4} ,\Gamma_2}$  &\parbox[c][5.0ex]{7cm}{ \begin{center} $\frac{3 J}{4}+3 t+U+9 W$ \end{center}}  & 8. \\ 
194 & $\ket{5,\frac{1}{2} , \frac{3}{4} ,\Gamma_{3,1}}$  &\parbox[c][5.0ex]{7cm}{ \begin{center} $\frac{1}{8} \left(-3 J+8 t+12 U+68 W-\sqrt{A_6}\right)$ \end{center}}  & 4.59488 \\ 
195 & $\ket{5,\frac{1}{2} , \frac{3}{4} ,\Gamma_{3,1}}$  &\parbox[c][5.0ex]{7cm}{ \begin{center} $\frac{1}{8} \left(-3 J+8 t+12 U+68 W+\sqrt{A_6}\right)$ \end{center}}  & 12.4051 \\ 
196 & $\ket{5,\frac{1}{2} , \frac{3}{4} ,\Gamma_{3,2}}$  &\parbox[c][5.0ex]{7cm}{ \begin{center} $\frac{1}{8} \left(-3 J+8 t+12 U+68 W-\sqrt{A_6}\right)$ \end{center}}  & 4.59488 \\ 
197 & $\ket{5,\frac{1}{2} , \frac{3}{4} ,\Gamma_{3,2}}$  &\parbox[c][5.0ex]{7cm}{ \begin{center} $\frac{1}{8} \left(-3 J+8 t+12 U+68 W+\sqrt{A_6}\right)$ \end{center}}  & 12.4051 \\ 
198 & $\ket{5,\frac{1}{2} , \frac{3}{4} ,\Gamma_{4,1}}$  &\parbox[c][5.0ex]{7cm}{ \begin{center} $-\frac{J}{4}-t+\frac{5 U}{3}+\frac{25 W}{3}-\frac{1}{6} \cos \left(\theta _1\right) \sqrt{A_1}$ \end{center}}  & 1.54341 \\ 
199 & $\ket{5,\frac{1}{2} , \frac{3}{4} ,\Gamma_{4,1}}$  &\parbox[c][5.0ex]{7cm}{ \begin{center} $A_{13}$ \end{center}}  & 12.2755 \\ 
200 & $\ket{5,\frac{1}{2} , \frac{3}{4} ,\Gamma_{4,1}}$  &\parbox[c][5.0ex]{7cm}{ \begin{center} $A_{12}$ \end{center}}  & 8.18113 \\ 
201 & $\ket{5,\frac{1}{2} , \frac{3}{4} ,\Gamma_{4,2}}$  &\parbox[c][5.0ex]{7cm}{ \begin{center} $-\frac{J}{4}-t+\frac{5 U}{3}+\frac{25 W}{3}-\frac{1}{6} \cos \left(\theta _1\right) \sqrt{A_1}$ \end{center}}  & 1.54341 \\ 
202 & $\ket{5,\frac{1}{2} , \frac{3}{4} ,\Gamma_{4,2}}$  &\parbox[c][5.0ex]{7cm}{ \begin{center} $A_{13}$ \end{center}}  & 12.2755 \\ 
203 & $\ket{5,\frac{1}{2} , \frac{3}{4} ,\Gamma_{4,2}}$  &\parbox[c][5.0ex]{7cm}{ \begin{center} $A_{12}$ \end{center}}  & 8.18113 \\ 
204 & $\ket{5,\frac{1}{2} , \frac{3}{4} ,\Gamma_{4,3}}$  &\parbox[c][5.0ex]{7cm}{ \begin{center} $-\frac{J}{4}-t+\frac{5 U}{3}+\frac{25 W}{3}-\frac{1}{6} \cos \left(\theta _1\right) \sqrt{A_1}$ \end{center}}  & 1.54341 \\ 
205 & $\ket{5,\frac{1}{2} , \frac{3}{4} ,\Gamma_{4,3}}$  &\parbox[c][5.0ex]{7cm}{ \begin{center} $A_{13}$ \end{center}}  & 12.2755 \\ 
206 & $\ket{5,\frac{1}{2} , \frac{3}{4} ,\Gamma_{4,3}}$  &\parbox[c][5.0ex]{7cm}{ \begin{center} $A_{12}$ \end{center}}  & 8.18113 \\ 
207 & $\ket{5,\frac{1}{2} , \frac{3}{4} ,\Gamma_{5,1}}$  &\parbox[c][5.0ex]{7cm}{ \begin{center} $\frac{1}{8} \left(-3 J+8 t+12 U+68 W-\sqrt{A_8}\right)$ \end{center}}  & 6.20871 \\ 
208 & $\ket{5,\frac{1}{2} , \frac{3}{4} ,\Gamma_{5,1}}$  &\parbox[c][5.0ex]{7cm}{ \begin{center} $\frac{1}{8} \left(-3 J+8 t+12 U+68 W+\sqrt{A_8}\right)$ \end{center}}  & 10.7913 \\ 
209 & $\ket{5,\frac{1}{2} , \frac{15}{4} ,\Gamma_{5,1}}$  &\parbox[c][5.0ex]{7cm}{ \begin{center} $\frac{3 J}{4}-t+U+9 W$ \end{center}}  & 4. \\ 
210 & $\ket{5,\frac{1}{2} , \frac{3}{4} ,\Gamma_{5,2}}$  &\parbox[c][5.0ex]{7cm}{ \begin{center} $\frac{1}{8} \left(-3 J+8 t+12 U+68 W-\sqrt{A_8}\right)$ \end{center}}  & 6.20871 \\ 
211 & $\ket{5,\frac{1}{2} , \frac{3}{4} ,\Gamma_{5,2}}$  &\parbox[c][5.0ex]{7cm}{ \begin{center} $\frac{1}{8} \left(-3 J+8 t+12 U+68 W+\sqrt{A_8}\right)$ \end{center}}  & 10.7913 \\ 
212 & $\ket{5,\frac{1}{2} , \frac{15}{4} ,\Gamma_{5,2}}$  &\parbox[c][5.0ex]{7cm}{ \begin{center} $\frac{3 J}{4}-t+U+9 W$ \end{center}}  & 4. \\ 
213 & $\ket{5,\frac{1}{2} , \frac{3}{4} ,\Gamma_{5,3}}$  &\parbox[c][5.0ex]{7cm}{ \begin{center} $\frac{1}{8} \left(-3 J+8 t+12 U+68 W-\sqrt{A_8}\right)$ \end{center}}  & 6.20871 \\ 
214 & $\ket{5,\frac{1}{2} , \frac{3}{4} ,\Gamma_{5,3}}$  &\parbox[c][5.0ex]{7cm}{ \begin{center} $\frac{1}{8} \left(-3 J+8 t+12 U+68 W+\sqrt{A_8}\right)$ \end{center}}  & 10.7913 \\ 
215 & $\ket{5,\frac{1}{2} , \frac{15}{4} ,\Gamma_{5,3}}$  &\parbox[c][5.0ex]{7cm}{ \begin{center} $\frac{3 J}{4}-t+U+9 W$ \end{center}}  & 4. \\ 
\hline
\end{tabular} \\[2ex]
\begin{tabular}[t]{|r|l|c|c|}
\multicolumn{4}{c}{\bf \boldmath Eigenkets and eigenvalues for ${\rm  N_e}$=5 and   ${\rm m_s}$= $\frac{3}{2} $. } \\[1ex] \hline
\parbox[c]{1cm}{\hfill No}  & \parbox[c][2.5ex]{2.5cm}{\begin{center} Eigenstate \end{center}}  & \parbox[c][2.5ex]{7.5cm}{ \begin{center}   Energy \end{center}}  & \parbox[c][2.5ex]{2cm}{ \begin{center} Example \end{center}}  \\ \hline 
\hline 
216 & $\ket{5,\frac{3}{2} , \frac{15}{4} ,\Gamma_2}$  &\parbox[c][5.0ex]{7cm}{ \begin{center} $\frac{3 J}{4}+3 t+U+9 W$ \end{center}}  & 8. \\ 
217 & $\ket{5,\frac{3}{2} , \frac{15}{4} ,\Gamma_{5,1}}$  &\parbox[c][5.0ex]{7cm}{ \begin{center} $\frac{3 J}{4}-t+U+9 W$ \end{center}}  & 4. \\ 
218 & $\ket{5,\frac{3}{2} , \frac{15}{4} ,\Gamma_{5,2}}$  &\parbox[c][5.0ex]{7cm}{ \begin{center} $\frac{3 J}{4}-t+U+9 W$ \end{center}}  & 4. \\ 
219 & $\ket{5,\frac{3}{2} , \frac{15}{4} ,\Gamma_{5,3}}$  &\parbox[c][5.0ex]{7cm}{ \begin{center} $\frac{3 J}{4}-t+U+9 W$ \end{center}}  & 4. \\ 
\hline
\end{tabular} \\[2ex]
\begin{tabular}[t]{|r|l|c|c|}
\multicolumn{4}{c}{\bf \boldmath Eigenkets and eigenvalues for ${\rm  N_e}$=6 and   ${\rm m_s}$= $-1$. } \\[1ex] \hline
\parbox[c]{1cm}{\hfill No}  & \parbox[c][2.5ex]{2.5cm}{\begin{center} Eigenstate \end{center}}  & \parbox[c][2.5ex]{7.5cm}{ \begin{center}   Energy \end{center}}  & \parbox[c][2.5ex]{2cm}{ \begin{center} Example \end{center}}  \\ \hline 
\hline 
220 & $\ket{6,-1,2,\Gamma_{4,1}}$  &\parbox[c][5.0ex]{7cm}{ \begin{center} $\frac{1}{4} (J-8 t+8 U+52 W)$ \end{center}}  & 8. \\ 
221 & $\ket{6,-1,2,\Gamma_{4,2}}$  &\parbox[c][5.0ex]{7cm}{ \begin{center} $\frac{1}{4} (J-8 t+8 U+52 W)$ \end{center}}  & 8. \\ 
222 & $\ket{6,-1,2,\Gamma_{4,3}}$  &\parbox[c][5.0ex]{7cm}{ \begin{center} $\frac{1}{4} (J-8 t+8 U+52 W)$ \end{center}}  & 8. \\ 
223 & $\ket{6,-1,2,\Gamma_{5,1}}$  &\parbox[c][5.0ex]{7cm}{ \begin{center} $\frac{1}{4} (J+8 t+8 U+52 W)$ \end{center}}  & 12. \\ 
224 & $\ket{6,-1,2,\Gamma_{5,2}}$  &\parbox[c][5.0ex]{7cm}{ \begin{center} $\frac{1}{4} (J+8 t+8 U+52 W)$ \end{center}}  & 12. \\ 
225 & $\ket{6,-1,2,\Gamma_{5,3}}$  &\parbox[c][5.0ex]{7cm}{ \begin{center} $\frac{1}{4} (J+8 t+8 U+52 W)$ \end{center}}  & 12. \\ 
\hline
\end{tabular} \\[2ex]
\begin{tabular}[t]{|r|l|c|c|}
\multicolumn{4}{c}{\bf \boldmath Eigenkets and eigenvalues for ${\rm  N_e}$=6 and   ${\rm m_s}$= $0$. } \\[1ex] \hline
\parbox[c]{1cm}{\hfill No}  & \parbox[c][2.5ex]{2.5cm}{\begin{center} Eigenstate \end{center}}  & \parbox[c][2.5ex]{7.5cm}{ \begin{center}   Energy \end{center}}  & \parbox[c][2.5ex]{2cm}{ \begin{center} Example \end{center}}  \\ \hline 
\hline 
226 & $\ket{6,0,0,\Gamma_1}$  &\parbox[c][5.0ex]{7cm}{ \begin{center} $\frac{1}{8} \left(-3 J-16 t+20 U+100 W-\sqrt{A_7}\right)$ \end{center}}  & 4.82109 \\ 
227 & $\ket{6,0,0,\Gamma_1}$  &\parbox[c][5.0ex]{7cm}{ \begin{center} $\frac{1}{8} \left(-3 J-16 t+20 U+100 W+\sqrt{A_7}\right)$ \end{center}}  & 16.1789 \\ 
228 & $\ket{6,0,0,\Gamma_{3,1}}$  &\parbox[c][5.0ex]{7cm}{ \begin{center} $-\frac{3 J}{4}+2 t+2 U+13 W$ \end{center}}  & 12. \\ 
229 & $\ket{6,0,0,\Gamma_{3,2}}$  &\parbox[c][5.0ex]{7cm}{ \begin{center} $-\frac{3 J}{4}+2 t+2 U+13 W$ \end{center}}  & 12. \\ 
230 & $\ket{6,0,0,\Gamma_{4,1}}$  &\parbox[c][5.0ex]{7cm}{ \begin{center} $\frac{1}{8} \left(-3 J+20 U+100 W-\sqrt{A_2}\right)$ \end{center}}  & 9.29844 \\ 
231 & $\ket{6,0,0,\Gamma_{4,1}}$  &\parbox[c][5.0ex]{7cm}{ \begin{center} $\frac{1}{8} \left(-3 J+20 U+100 W+\sqrt{A_2}\right)$ \end{center}}  & 15.7016 \\ 
232 & $\ket{6,0,2,\Gamma_{4,1}}$  &\parbox[c][5.0ex]{7cm}{ \begin{center} $\frac{1}{4} (J-8 t+8 U+52 W)$ \end{center}}  & 8. \\ 
233 & $\ket{6,0,0,\Gamma_{4,2}}$  &\parbox[c][5.0ex]{7cm}{ \begin{center} $\frac{1}{8} \left(-3 J+20 U+100 W-\sqrt{A_2}\right)$ \end{center}}  & 9.29844 \\ 
234 & $\ket{6,0,0,\Gamma_{4,2}}$  &\parbox[c][5.0ex]{7cm}{ \begin{center} $\frac{1}{8} \left(-3 J+20 U+100 W+\sqrt{A_2}\right)$ \end{center}}  & 15.7016 \\ 
235 & $\ket{6,0,2,\Gamma_{4,2}}$  &\parbox[c][5.0ex]{7cm}{ \begin{center} $\frac{1}{4} (J-8 t+8 U+52 W)$ \end{center}}  & 8. \\ 
236 & $\ket{6,0,0,\Gamma_{4,3}}$  &\parbox[c][5.0ex]{7cm}{ \begin{center} $\frac{1}{8} \left(-3 J+20 U+100 W-\sqrt{A_2}\right)$ \end{center}}  & 9.29844 \\ 
237 & $\ket{6,0,0,\Gamma_{4,3}}$  &\parbox[c][5.0ex]{7cm}{ \begin{center} $\frac{1}{8} \left(-3 J+20 U+100 W+\sqrt{A_2}\right)$ \end{center}}  & 15.7016 \\ 
238 & $\ket{6,0,2,\Gamma_{4,3}}$  &\parbox[c][5.0ex]{7cm}{ \begin{center} $\frac{1}{4} (J-8 t+8 U+52 W)$ \end{center}}  & 8. \\ 
239 & $\ket{6,0,2,\Gamma_{5,1}}$  &\parbox[c][5.0ex]{7cm}{ \begin{center} $\frac{1}{4} (J+8 t+8 U+52 W)$ \end{center}}  & 12. \\ 
240 & $\ket{6,0,2,\Gamma_{5,2}}$  &\parbox[c][5.0ex]{7cm}{ \begin{center} $\frac{1}{4} (J+8 t+8 U+52 W)$ \end{center}}  & 12. \\ 
241 & $\ket{6,0,2,\Gamma_{5,3}}$  &\parbox[c][5.0ex]{7cm}{ \begin{center} $\frac{1}{4} (J+8 t+8 U+52 W)$ \end{center}}  & 12. \\ 
\hline
\end{tabular} \\[2ex]
\begin{tabular}[t]{|r|l|c|c|}
\multicolumn{4}{c}{\bf \boldmath Eigenkets and eigenvalues for ${\rm  N_e}$=6 and   ${\rm m_s}$= $1$. } \\[1ex] \hline
\parbox[c]{1cm}{\hfill No}  & \parbox[c][2.5ex]{2.5cm}{\begin{center} Eigenstate \end{center}}  & \parbox[c][2.5ex]{7.5cm}{ \begin{center}   Energy \end{center}}  & \parbox[c][2.5ex]{2cm}{ \begin{center} Example \end{center}}  \\ \hline 
\hline 
242 & $\ket{6,1,2,\Gamma_{4,1}}$  &\parbox[c][5.0ex]{7cm}{ \begin{center} $\frac{1}{4} (J-8 t+8 U+52 W)$ \end{center}}  & 8. \\ 
243 & $\ket{6,1,2,\Gamma_{4,2}}$  &\parbox[c][5.0ex]{7cm}{ \begin{center} $\frac{1}{4} (J-8 t+8 U+52 W)$ \end{center}}  & 8. \\ 
244 & $\ket{6,1,2,\Gamma_{4,3}}$  &\parbox[c][5.0ex]{7cm}{ \begin{center} $\frac{1}{4} (J-8 t+8 U+52 W)$ \end{center}}  & 8. \\ 
245 & $\ket{6,1,2,\Gamma_{5,1}}$  &\parbox[c][5.0ex]{7cm}{ \begin{center} $\frac{1}{4} (J+8 t+8 U+52 W)$ \end{center}}  & 12. \\ 
246 & $\ket{6,1,2,\Gamma_{5,2}}$  &\parbox[c][5.0ex]{7cm}{ \begin{center} $\frac{1}{4} (J+8 t+8 U+52 W)$ \end{center}}  & 12. \\ 
247 & $\ket{6,1,2,\Gamma_{5,3}}$  &\parbox[c][5.0ex]{7cm}{ \begin{center} $\frac{1}{4} (J+8 t+8 U+52 W)$ \end{center}}  & 12. \\ 
\hline
\end{tabular} \\[2ex]
\begin{tabular}[t]{|r|l|c|c|}
\multicolumn{4}{c}{\bf \boldmath Eigenkets and eigenvalues for ${\rm  N_e}$=7 and   ${\rm m_s}$= $- \frac{1}{2} $. } \\[1ex] \hline
\parbox[c]{1cm}{\hfill No}  & \parbox[c][2.5ex]{2.5cm}{\begin{center} Eigenstate \end{center}}  & \parbox[c][2.5ex]{7.5cm}{ \begin{center}   Energy \end{center}}  & \parbox[c][2.5ex]{2cm}{ \begin{center} Example \end{center}}  \\ \hline 
\hline 
248 & $\ket{7,- \frac{1}{2} , \frac{3}{4} ,\Gamma_1}$  &\parbox[c][5.0ex]{7cm}{ \begin{center} $3 (-t+U+6 W)$ \end{center}}  & 12. \\ 
249 & $\ket{7,- \frac{1}{2} , \frac{3}{4} ,\Gamma_{4,1}}$  &\parbox[c][5.0ex]{7cm}{ \begin{center} $t+3 (U+6 W)$ \end{center}}  & 16. \\ 
250 & $\ket{7,- \frac{1}{2} , \frac{3}{4} ,\Gamma_{4,2}}$  &\parbox[c][5.0ex]{7cm}{ \begin{center} $t+3 (U+6 W)$ \end{center}}  & 16. \\ 
251 & $\ket{7,- \frac{1}{2} , \frac{3}{4} ,\Gamma_{4,3}}$  &\parbox[c][5.0ex]{7cm}{ \begin{center} $t+3 (U+6 W)$ \end{center}}  & 16. \\ 
\hline
\end{tabular} \\[2ex]
\begin{tabular}[t]{|r|l|c|c|}
\multicolumn{4}{c}{\bf \boldmath Eigenkets and eigenvalues for ${\rm  N_e}$=7 and   ${\rm m_s}$= $\frac{1}{2} $. } \\[1ex] \hline
\parbox[c]{1cm}{\hfill No}  & \parbox[c][2.5ex]{2.5cm}{\begin{center} Eigenstate \end{center}}  & \parbox[c][2.5ex]{7.5cm}{ \begin{center}   Energy \end{center}}  & \parbox[c][2.5ex]{2cm}{ \begin{center} Example \end{center}}  \\ \hline 
\hline 
252 & $\ket{7,\frac{1}{2} , \frac{3}{4} ,\Gamma_1}$  &\parbox[c][5.0ex]{7cm}{ \begin{center} $3 (-t+U+6 W)$ \end{center}}  & 12. \\ 
253 & $\ket{7,\frac{1}{2} , \frac{3}{4} ,\Gamma_{4,1}}$  &\parbox[c][5.0ex]{7cm}{ \begin{center} $t+3 (U+6 W)$ \end{center}}  & 16. \\ 
254 & $\ket{7,\frac{1}{2} , \frac{3}{4} ,\Gamma_{4,2}}$  &\parbox[c][5.0ex]{7cm}{ \begin{center} $t+3 (U+6 W)$ \end{center}}  & 16. \\ 
255 & $\ket{7,\frac{1}{2} , \frac{3}{4} ,\Gamma_{4,3}}$  &\parbox[c][5.0ex]{7cm}{ \begin{center} $t+3 (U+6 W)$ \end{center}}  & 16. \\ 
\hline
\end{tabular} \\[2ex]
\begin{tabular}[t]{|r|l|c|c|}
\multicolumn{4}{c}{\bf \boldmath Eigenkets and eigenvalues for ${\rm  N_e}$=8 and   ${\rm m_s}$= $0$. } \\[1ex] \hline
\parbox[c]{1cm}{\hfill No}  & \parbox[c][2.5ex]{2.5cm}{\begin{center} Eigenstate \end{center}}  & \parbox[c][2.5ex]{7.5cm}{ \begin{center}   Energy \end{center}}  & \parbox[c][2.5ex]{2cm}{ \begin{center} Example \end{center}}  \\ \hline 
\hline 
256 & $\ket{8,0,0,\Gamma_1}$  &\parbox[c][5.0ex]{7cm}{ \begin{center} $4 (U+6 W)$ \end{center}}  & 20. \\ 
\hline
\end{tabular} \\

\parindent0cm
\subsection*{ List of abbreviations }
\beq 
Y &=& U-W+\frac{3}{4} J  \nonumber \\
A_1 &=& 16 \left(48 t^2+3 Y t+Y^2\right)
 \nonumber \\
A_2 &=& 16 \left(16 t^2+Y^2\right)
 \nonumber \\
A_3 &=& 16 \left(48 t^2+Y^2\right)
 \nonumber \\
A_4 &=& 16 \left(64 t^2+Y^2\right)
 \nonumber \\
A_5 &=& 48 \left(16 t^2+Y^2\right)
 \nonumber \\
A_6 &=& 16 \left(16 t^2+4 Y t+Y^2\right)
 \nonumber \\
A_7 &=& 16 \left(64 t^2+8 Y t+Y^2\right)
 \nonumber \\
A_8 &=& 16 \left(16 t^2-4 Y t+Y^2\right)
 \nonumber \\
A_9 &=& 3 J-4 (4 t+U+W)
 \nonumber \\
A_{10} &=& 16 \left(48 t^2-3 Y t+Y^2\right)
 \nonumber \\
A_{11} &=& 3 J+4 (3 t-5 (U+5 W))+2 \cos \left(\theta _1\right) \sqrt{A_1}
 \nonumber \\
A_{12} &=& \frac{1}{12} \left(-3 J-12 t+20 U+100 W+\left(\cos \left(\theta _1\right)-\sqrt{3} \sin \left(\theta _1\right)\right) \sqrt{A_1}\right)
 \nonumber \\
A_{13} &=& \frac{1}{12} \left(-3 J-12 t+20 U+100 W+\left(\cos \left(\theta _1\right)+\sqrt{3} \sin \left(\theta _1\right)\right) \sqrt{A_1}\right)
 \nonumber \\
A_{14} &=& -\frac{3 J}{4}+U+5 W+\frac{1}{12} \left(\sqrt{3} \cos \left(\theta _2\right)-3 \sin \left(\theta _2\right)\right) \sqrt{A_2}
 \nonumber \\
A_{15} &=& -\frac{3 J}{4}+U+5 W+\frac{1}{12} \left(\sqrt{3} \cos \left(\theta _2\right)+3 \sin \left(\theta _2\right)\right) \sqrt{A_2}
 \nonumber \\
A_{16} &=& 3 J-8 U-28 W+\cos \left(\theta _3\right) \sqrt{A_3}
 \nonumber \\
A_{17} &=& \cos \left(\theta _4\right) \sqrt{A_3}-4 (U+8 W)
 \nonumber \\
A_{18} &=& \frac{1}{12} \left(-6 J+16 U+56 W+\left(\cos \left(\theta _3\right)-\sqrt{3} \sin \left(\theta _3\right)\right) \sqrt{A_3}\right)
 \nonumber \\
A_{19} &=& \frac{1}{12} \left(-6 J+16 U+56 W+\left(\cos \left(\theta _3\right)+\sqrt{3} \sin \left(\theta _3\right)\right) \sqrt{A_3}\right)
 \nonumber \\
A_{20} &=& \frac{1}{12} \left(8 (U+8 W)+\left(\cos \left(\theta _4\right)-\sqrt{3} \sin \left(\theta _4\right)\right) \sqrt{A_3}\right)
 \nonumber \\
A_{21} &=& \frac{1}{12} \left(8 (U+8 W)+\left(\cos \left(\theta _4\right)+\sqrt{3} \sin \left(\theta _4\right)\right) \sqrt{A_3}\right)
 \nonumber \\
A_{22} &=& 9 J-12 (U+5 W)+2 \cos \left(\theta _2\right) \sqrt{A_5}
 \nonumber \\
A_{23} &=& 768 t^2-256 U t+A_9^2+48 J U-64 U W
 \nonumber \\
A_{24} &=& 3 J-4 (3 t+2 U+7 W)+2 \cos \left(\theta _5\right) \sqrt{A_{10}}
 \nonumber \\
A_{25} &=& \frac{1}{12} \left(-3 J+12 t+8 U+28 W+\left(\cos \left(\theta _5\right)-\sqrt{3} \sin \left(\theta _5\right)\right) \sqrt{A_{10}}\right)
 \nonumber \\
A_{26} &=& \frac{1}{12} \left(-3 J+12 t+8 U+28 W+\left(\cos \left(\theta _5\right)+\sqrt{3} \sin \left(\theta _5\right)\right) \sqrt{A_{10}}\right)
 \nonumber \\
\theta _1 &=& \frac{1}{3} \cos ^{-1}\left(\frac{32 Y \left(36 t^2+9 Y t+2 Y^2\right)}{A_1^{3/2}}\right)
 \nonumber \\
\theta _2 &=& \frac{1}{3} \cos ^{-1}\left(\frac{768 \sqrt{3} t^2 Y}{A_2^{3/2}}\right)
 \nonumber \\
\theta _3 &=& \frac{1}{3} \cos ^{-1}\left(-\frac{64 Y \left(Y^2-36 t^2\right)}{A_3^{3/2}}\right)
 \nonumber \\
\theta _4 &=& \frac{1}{3} \cos ^{-1}\left(\frac{64 Y \left(Y^2-36 t^2\right)}{A_3^{3/2}}\right)
 \nonumber \\
\theta _5 &=& \frac{1}{3} \cos ^{-1}\left(\frac{32 Y \left(36 t^2-9 Y t+2 Y^2\right)}{A_{10}^{3/2}}\right)
\nonumber \eeq 

{\small
{\subsubsection{\boldmath Eigenvectors for ${\rm  N_e}=0$ and   ${\rm m_s}$= $0$.}
\beq
\ket{\Psi_{1}} & = & \ket{0,0,0,\Gamma_1} \nonumber \\ 
&=& 1
 \left ( \ket{0000} \right) \nonumber 
\eeq
{\subsubsection{\boldmath Eigenvectors for ${\rm  N_e}=1$ and   ${\rm m_s}$= $- \frac{1}{2} $.}
\beq
\ket{\Psi_{2}} & = & \ket{1,- \frac{1}{2} , \frac{3}{4} ,\Gamma_1} \nonumber \\ 
&=& \frac{1}{2}
 \left ( \ket{000d} + \ket{00d0} + \ket{0d00} + \ket{d000} \right) \nonumber 
\eeq
\beq
\ket{\Psi_{3}} & = & \ket{1,- \frac{1}{2} , \frac{3}{4} ,\Gamma_{4,1}} \nonumber \\ 
&=& \frac{1}{2}
 \left ( \ket{000d} + \ket{00d0} - \ket{0d00} - \ket{d000} \right) \nonumber 
\eeq
\beq
\ket{\Psi_{4}} & = & \ket{1,- \frac{1}{2} , \frac{3}{4} ,\Gamma_{4,2}} \nonumber \\ 
&=& \frac{1}{2}
 \left ( \ket{000d} - \ket{00d0} - \ket{0d00} + \ket{d000} \right) \nonumber 
\eeq
\beq
\ket{\Psi_{5}} & = & \ket{1,- \frac{1}{2} , \frac{3}{4} ,\Gamma_{4,3}} \nonumber \\ 
&=& \frac{1}{2}
 \left ( \ket{000d} - \ket{00d0} + \ket{0d00} - \ket{d000} \right) \nonumber 
\eeq
{\subsubsection{\boldmath Eigenvectors for ${\rm  N_e}=1$ and   ${\rm m_s}$= $\frac{1}{2} $.}
\beq
\ket{\Psi_{6}} & = & \ket{1,\frac{1}{2} , \frac{3}{4} ,\Gamma_1} \nonumber \\ 
&=& \frac{1}{2}
 \left ( \ket{000u} + \ket{00u0} + \ket{0u00} + \ket{u000} \right) \nonumber 
\eeq
\beq
\ket{\Psi_{7}} & = & \ket{1,\frac{1}{2} , \frac{3}{4} ,\Gamma_{4,1}} \nonumber \\ 
&=& \frac{1}{2}
 \left ( \ket{000u} + \ket{00u0} - \ket{0u00} - \ket{u000} \right) \nonumber 
\eeq
\beq
\ket{\Psi_{8}} & = & \ket{1,\frac{1}{2} , \frac{3}{4} ,\Gamma_{4,2}} \nonumber \\ 
&=& \frac{1}{2}
 \left ( \ket{000u} - \ket{00u0} - \ket{0u00} + \ket{u000} \right) \nonumber 
\eeq
\beq
\ket{\Psi_{9}} & = & \ket{1,\frac{1}{2} , \frac{3}{4} ,\Gamma_{4,3}} \nonumber \\ 
&=& \frac{1}{2}
 \left ( \ket{000u} - \ket{00u0} + \ket{0u00} - \ket{u000} \right) \nonumber 
\eeq
{\subsubsection{\boldmath Eigenvectors for ${\rm  N_e}=2$ and   ${\rm m_s}$= $-1$.}
\beq
\ket{\Psi_{10}} & = & \ket{2,-1,2,\Gamma_{4,1}} \nonumber \\ 
&=& \frac{1}{2}
 \left ( \ket{0d0d} + \ket{0dd0} + \ket{d00d} + \ket{d0d0} \right) \nonumber 
\eeq
\beq
\ket{\Psi_{11}} & = & \ket{2,-1,2,\Gamma_{4,2}} \nonumber \\ 
&=& \frac{1}{2}
 \left ( \ket{00dd} + \ket{0d0d} - \ket{d0d0} - \ket{dd00} \right) \nonumber 
\eeq
\beq
\ket{\Psi_{12}} & = & \ket{2,-1,2,\Gamma_{4,3}} \nonumber \\ 
&=& \frac{1}{2}
 \left ( \ket{00dd} - \ket{0dd0} + \ket{d00d} + \ket{dd00} \right) \nonumber 
\eeq
\beq
\ket{\Psi_{13}} & = & \ket{2,-1,2,\Gamma_{5,1}} \nonumber \\ 
&=& \frac{1}{2}
 \left ( \ket{00dd} + \ket{0dd0} - \ket{d00d} + \ket{dd00} \right) \nonumber 
\eeq
\beq
\ket{\Psi_{14}} & = & \ket{2,-1,2,\Gamma_{5,2}} \nonumber \\ 
&=& \frac{1}{2}
 \left ( \ket{0d0d} - \ket{0dd0} - \ket{d00d} + \ket{d0d0} \right) \nonumber 
\eeq
\beq
\ket{\Psi_{15}} & = & \ket{2,-1,2,\Gamma_{5,3}} \nonumber \\ 
&=& \frac{1}{2}
 \left ( \ket{00dd} - \ket{0d0d} + \ket{d0d0} - \ket{dd00} \right) \nonumber 
\eeq
{\subsubsection{\boldmath Eigenvectors for ${\rm  N_e}=2$ and   ${\rm m_s}$= $0$.}
\beq
\ket{\Psi_{16}} & = & \ket{2,0,0,\Gamma_1} \nonumber \\ 
& = & \quad 
C_{16,1} \left ( 
\ket{0002} + \ket{0020} + \ket{0200} + \ket{2000} \right) 
 \nonumber \\
& & + 
C_{16,2} \left ( 
\ket{00du} - \ket{00ud} + \ket{0d0u} + \ket{0du0} - \ket{0u0d} - \ket{0ud0} \right . \nonumber \\
&& \hspace{3em} 
 + 
\left . \ket{d00u} + \ket{d0u0} + \ket{du00} - \ket{u00d} - \ket{u0d0} - \ket{ud00}\right ) 
\nonumber \eeq
\beq 
C_{16,1} &=& 
\sqrt{3} t \nonumber \\
C_{16,2} &=& 
\frac{1}{16 \sqrt{3}} \left (
3 J-16 t+4 U-4 W+\sqrt{A_{23}} \right ) \nonumber \\
 N_{16} &=& 2 \sqrt{C_{16,1}^2+3 C_{16,2}^2} \nonumber \eeq 
\beq
\ket{\Psi_{17}} & = & \ket{2,0,0,\Gamma_1} \nonumber \\ 
& = & \quad 
C_{17,1} \left ( 
\ket{0002} + \ket{0020} + \ket{0200} + \ket{2000} \right) 
 \nonumber \\
& & + 
C_{17,2} \left ( 
\ket{00du} - \ket{00ud} + \ket{0d0u} + \ket{0du0} - \ket{0u0d} - \ket{0ud0} \right . \nonumber \\
&& \hspace{3em} 
 + 
\left . \ket{d00u} + \ket{d0u0} + \ket{du00} - \ket{u00d} - \ket{u0d0} - \ket{ud00}\right ) 
\nonumber \eeq
\beq 
C_{17,1} &=& 
\sqrt{3} t \nonumber \\
C_{17,2} &=& 
\frac{1}{16 \sqrt{3}} \left (
3 J-16 t+4 U-4 W-\sqrt{A_{23}} \right ) \nonumber \\
 N_{17} &=& 2 \sqrt{C_{17,1}^2+3 C_{17,2}^2} \nonumber \eeq 
\beq
\ket{\Psi_{18}} & = & \ket{2,0,0,\Gamma_{3,1}} \nonumber \\ 
& = & \quad 
C_{18,1} \left ( 
\ket{00du} - \ket{00ud} + \ket{0du0} - \ket{0ud0} + \ket{d00u} + \ket{du00} - \ket{u00d} - \ket{ud00} \right) 
 \nonumber \\
& & + 
C_{18,2} \left ( 
\ket{0d0u} - \ket{0u0d} + \ket{d0u0} - \ket{u0d0} \right) 
\nonumber \eeq
\beq 
C_{18,1} &=& 
-\frac{1}{2 \sqrt{6}} \nonumber \\
C_{18,2} &=& 
\frac{1}{\sqrt{6}} \nonumber \\
 N_{18} &=& 2 \sqrt{2 C_{18,1}^2+C_{18,2}^2} \nonumber \eeq 
\beq
\ket{\Psi_{19}} & = & \ket{2,0,0,\Gamma_{3,2}} \nonumber \\ 
&=& \frac{1}{2 \sqrt{2}}
 \left ( \ket{00du} - \ket{00ud} - \ket{0du0} + \ket{0ud0} - \ket{d00u} + \ket{du00} + \ket{u00d} - \ket{ud00} \right) \nonumber 
\eeq
\beq
\ket{\Psi_{20}} & = & \ket{2,0,0,\Gamma_{4,1}} \nonumber \\ 
& = & \quad 
C_{20,1} \left ( 
\ket{0002} + \ket{0020} - \ket{0200} - \ket{2000} \right) 
 \nonumber \\
& & + 
C_{20,2} \left ( 
\ket{00du} - \ket{00ud} - \ket{du00} + \ket{ud00} \right) 
\nonumber \eeq
\beq 
C_{20,1} &=& 
-t \nonumber \\
C_{20,2} &=& 
-\frac{1}{16} \left (
3 J+4 U-4 W+\sqrt{A_2} \right ) \nonumber \\
 N_{20} &=& 2 \sqrt{C_{20,1}^2+C_{20,2}^2} \nonumber \eeq 
\beq
\ket{\Psi_{21}} & = & \ket{2,0,0,\Gamma_{4,1}} \nonumber \\ 
& = & \quad 
C_{21,1} \left ( 
\ket{0002} + \ket{0020} - \ket{0200} - \ket{2000} \right) 
 \nonumber \\
& & + 
C_{21,2} \left ( 
\ket{00du} - \ket{00ud} - \ket{du00} + \ket{ud00} \right) 
\nonumber \eeq
\beq 
C_{21,1} &=& 
-t \nonumber \\
C_{21,2} &=& 
-\frac{1}{16} \left (
3 J+4 U-4 W-\sqrt{A_2} \right ) \nonumber \\
 N_{21} &=& 2 \sqrt{C_{21,1}^2+C_{21,2}^2} \nonumber \eeq 
\beq
\ket{\Psi_{22}} & = & \ket{2,0,2,\Gamma_{4,1}} \nonumber \\ 
&=& \frac{1}{2 \sqrt{2}}
 \left ( \ket{0d0u} + \ket{0du0} + \ket{0u0d} + \ket{0ud0} + \ket{d00u} + \ket{d0u0} + \ket{u00d} + \ket{u0d0} \right) \nonumber 
\eeq
\beq
\ket{\Psi_{23}} & = & \ket{2,0,0,\Gamma_{4,2}} \nonumber \\ 
& = & \quad 
C_{23,1} \left ( 
\ket{0002} - \ket{0020} - \ket{0200} + \ket{2000} \right) 
 \nonumber \\
& & + 
C_{23,2} \left ( 
\ket{0du0} - \ket{0ud0} - \ket{d00u} + \ket{u00d} \right) 
\nonumber \eeq
\beq 
C_{23,1} &=& 
t \nonumber \\
C_{23,2} &=& 
-\frac{1}{16} \left (
3 J+4 U-4 W+\sqrt{A_2} \right ) \nonumber \\
 N_{23} &=& 2 \sqrt{C_{23,1}^2+C_{23,2}^2} \nonumber \eeq 
\beq
\ket{\Psi_{24}} & = & \ket{2,0,0,\Gamma_{4,2}} \nonumber \\ 
& = & \quad 
C_{24,1} \left ( 
\ket{0002} - \ket{0020} - \ket{0200} + \ket{2000} \right) 
 \nonumber \\
& & + 
C_{24,2} \left ( 
\ket{0du0} - \ket{0ud0} - \ket{d00u} + \ket{u00d} \right) 
\nonumber \eeq
\beq 
C_{24,1} &=& 
t \nonumber \\
C_{24,2} &=& 
-\frac{1}{16} \left (
3 J+4 U-4 W-\sqrt{A_2} \right ) \nonumber \\
 N_{24} &=& 2 \sqrt{C_{24,1}^2+C_{24,2}^2} \nonumber \eeq 
\beq
\ket{\Psi_{25}} & = & \ket{2,0,2,\Gamma_{4,2}} \nonumber \\ 
&=& \frac{1}{2 \sqrt{2}}
 \left ( \ket{00du} + \ket{00ud} + \ket{0d0u} + \ket{0u0d} - \ket{d0u0} - \ket{du00} - \ket{u0d0} - \ket{ud00} \right) \nonumber 
\eeq
\beq
\ket{\Psi_{26}} & = & \ket{2,0,0,\Gamma_{4,3}} \nonumber \\ 
& = & \quad 
C_{26,1} \left ( 
\ket{0002} - \ket{0020} + \ket{0200} - \ket{2000} \right) 
 \nonumber \\
& & + 
C_{26,2} \left ( 
\ket{0d0u} - \ket{0u0d} - \ket{d0u0} + \ket{u0d0} \right) 
\nonumber \eeq
\beq 
C_{26,1} &=& 
-t \nonumber \\
C_{26,2} &=& 
-\frac{1}{16} \left (
3 J+4 U-4 W+\sqrt{A_2} \right ) \nonumber \\
 N_{26} &=& 2 \sqrt{C_{26,1}^2+C_{26,2}^2} \nonumber \eeq 
\beq
\ket{\Psi_{27}} & = & \ket{2,0,0,\Gamma_{4,3}} \nonumber \\ 
& = & \quad 
C_{27,1} \left ( 
\ket{0002} - \ket{0020} + \ket{0200} - \ket{2000} \right) 
 \nonumber \\
& & + 
C_{27,2} \left ( 
\ket{0d0u} - \ket{0u0d} - \ket{d0u0} + \ket{u0d0} \right) 
\nonumber \eeq
\beq 
C_{27,1} &=& 
-t \nonumber \\
C_{27,2} &=& 
-\frac{1}{16} \left (
3 J+4 U-4 W-\sqrt{A_2} \right ) \nonumber \\
 N_{27} &=& 2 \sqrt{C_{27,1}^2+C_{27,2}^2} \nonumber \eeq 
\beq
\ket{\Psi_{28}} & = & \ket{2,0,2,\Gamma_{4,3}} \nonumber \\ 
&=& \frac{1}{2 \sqrt{2}}
 \left ( \ket{00du} + \ket{00ud} - \ket{0du0} - \ket{0ud0} + \ket{d00u} + \ket{du00} + \ket{u00d} + \ket{ud00} \right) \nonumber 
\eeq
\beq
\ket{\Psi_{29}} & = & \ket{2,0,2,\Gamma_{5,1}} \nonumber \\ 
&=& \frac{1}{2 \sqrt{2}}
 \left ( \ket{00du} + \ket{00ud} + \ket{0du0} + \ket{0ud0} - \ket{d00u} + \ket{du00} - \ket{u00d} + \ket{ud00} \right) \nonumber 
\eeq
\beq
\ket{\Psi_{30}} & = & \ket{2,0,2,\Gamma_{5,2}} \nonumber \\ 
&=& \frac{1}{2 \sqrt{2}}
 \left ( \ket{0d0u} - \ket{0du0} + \ket{0u0d} - \ket{0ud0} - \ket{d00u} + \ket{d0u0} - \ket{u00d} + \ket{u0d0} \right) \nonumber 
\eeq
\beq
\ket{\Psi_{31}} & = & \ket{2,0,2,\Gamma_{5,3}} \nonumber \\ 
&=& \frac{1}{2 \sqrt{2}}
 \left ( \ket{00du} + \ket{00ud} - \ket{0d0u} - \ket{0u0d} + \ket{d0u0} - \ket{du00} + \ket{u0d0} - \ket{ud00} \right) \nonumber 
\eeq
{\subsubsection{\boldmath Eigenvectors for ${\rm  N_e}=2$ and   ${\rm m_s}$= $1$.}
\beq
\ket{\Psi_{32}} & = & \ket{2,1,2,\Gamma_{4,1}} \nonumber \\ 
&=& \frac{1}{2}
 \left ( \ket{0u0u} + \ket{0uu0} + \ket{u00u} + \ket{u0u0} \right) \nonumber 
\eeq
\beq
\ket{\Psi_{33}} & = & \ket{2,1,2,\Gamma_{4,2}} \nonumber \\ 
&=& \frac{1}{2}
 \left ( \ket{00uu} + \ket{0u0u} - \ket{u0u0} - \ket{uu00} \right) \nonumber 
\eeq
\beq
\ket{\Psi_{34}} & = & \ket{2,1,2,\Gamma_{4,3}} \nonumber \\ 
&=& \frac{1}{2}
 \left ( \ket{00uu} - \ket{0uu0} + \ket{u00u} + \ket{uu00} \right) \nonumber 
\eeq
\beq
\ket{\Psi_{35}} & = & \ket{2,1,2,\Gamma_{5,1}} \nonumber \\ 
&=& \frac{1}{2}
 \left ( \ket{00uu} + \ket{0uu0} - \ket{u00u} + \ket{uu00} \right) \nonumber 
\eeq
\beq
\ket{\Psi_{36}} & = & \ket{2,1,2,\Gamma_{5,2}} \nonumber \\ 
&=& \frac{1}{2}
 \left ( \ket{0u0u} - \ket{0uu0} - \ket{u00u} + \ket{u0u0} \right) \nonumber 
\eeq
\beq
\ket{\Psi_{37}} & = & \ket{2,1,2,\Gamma_{5,3}} \nonumber \\ 
&=& \frac{1}{2}
 \left ( \ket{00uu} - \ket{0u0u} + \ket{u0u0} - \ket{uu00} \right) \nonumber 
\eeq
{\subsubsection{\boldmath Eigenvectors for ${\rm  N_e}=3$ and   ${\rm m_s}$= $- \frac{3}{2} $.}
\beq
\ket{\Psi_{38}} & = & \ket{3,- \frac{3}{2} , \frac{15}{4} ,\Gamma_2} \nonumber \\ 
&=& \frac{1}{2}
 \left ( \ket{0ddd} - \ket{d0dd} + \ket{dd0d} - \ket{ddd0} \right) \nonumber 
\eeq
\beq
\ket{\Psi_{39}} & = & \ket{3,- \frac{3}{2} , \frac{15}{4} ,\Gamma_{5,1}} \nonumber \\ 
&=& \frac{1}{2}
 \left ( \ket{0ddd} + \ket{d0dd} + \ket{dd0d} + \ket{ddd0} \right) \nonumber 
\eeq
\beq
\ket{\Psi_{40}} & = & \ket{3,- \frac{3}{2} , \frac{15}{4} ,\Gamma_{5,2}} \nonumber \\ 
&=& \frac{1}{2}
 \left ( \ket{0ddd} - \ket{d0dd} - \ket{dd0d} + \ket{ddd0} \right) \nonumber 
\eeq
\beq
\ket{\Psi_{41}} & = & \ket{3,- \frac{3}{2} , \frac{15}{4} ,\Gamma_{5,3}} \nonumber \\ 
&=& \frac{1}{2}
 \left ( \ket{0ddd} + \ket{d0dd} - \ket{dd0d} - \ket{ddd0} \right) \nonumber 
\eeq
{\subsubsection{\boldmath Eigenvectors for ${\rm  N_e}=3$ and   ${\rm m_s}$= $- \frac{1}{2} $.}
\beq
\ket{\Psi_{42}} & = & \ket{3,- \frac{1}{2} , \frac{3}{4} ,\Gamma_1} \nonumber \\ 
&=& \frac{1}{2 \sqrt{3}}
 \left ( \ket{002d} + \ket{00d2} + \ket{020d} + \ket{02d0} + \ket{0d02} + \ket{0d20} \right . \nonumber \\
&& \hspace{3em} 
 + 
\left . \ket{200d} + \ket{20d0} + \ket{2d00} + \ket{d002} + \ket{d020} + \ket{d200} \right ) 
\nonumber \eeq
\beq
\ket{\Psi_{43}} & = & \ket{3,- \frac{1}{2} , \frac{15}{4} ,\Gamma_2} \nonumber \\ 
&=& \frac{1}{2 \sqrt{3}}
 \left ( \ket{0ddu} + \ket{0dud} + \ket{0udd} - \ket{d0du} - \ket{d0ud} + \ket{dd0u} \right . \nonumber \\
&& \hspace{3em} 
\left . -\ket{ddu0} + \ket{du0d} - \ket{dud0} - \ket{u0dd} + \ket{ud0d} - \ket{udd0} \right ) 
\nonumber \eeq
\beq
\ket{\Psi_{44}} & = & \ket{3,- \frac{1}{2} , \frac{3}{4} ,\Gamma_{3,1}} \nonumber \\ 
& = & \quad 
C_{44,1} \left ( 
\ket{002d} + \ket{00d2} + \ket{02d0} + \ket{0d20} + \ket{200d} + \ket{2d00} + \ket{d002} + \ket{d200} \right) 
 \nonumber \\
& & + 
C_{44,2} \left ( 
\ket{020d} + \ket{0d02} + \ket{20d0} + \ket{d020} \right) 
 \nonumber \\
& & + 
C_{44,3} \left ( 
\ket{0ddu} - \ket{0udd} - \ket{d0ud} - \ket{dd0u} + \ket{ddu0} + \ket{du0d} + \ket{u0dd} - \ket{udd0} \right) 
\nonumber \eeq
\beq 
C_{44,1} &=& 
-\frac{t}{2 \sqrt{2}} \nonumber \\
C_{44,2} &=& 
\frac{t}{\sqrt{2}} \nonumber \\
C_{44,3} &=& 
\frac{1}{16 \sqrt{2}} \left (
3 J-8 t+4 U-4 W+\sqrt{A_8} \right ) \nonumber \\
 N_{44} &=& 2 \sqrt{2 C_{44,1}^2+C_{44,2}^2+2 C_{44,3}^2} \nonumber \eeq 
\beq
\ket{\Psi_{45}} & = & \ket{3,- \frac{1}{2} , \frac{3}{4} ,\Gamma_{3,1}} \nonumber \\ 
& = & \quad 
C_{45,1} \left ( 
\ket{002d} + \ket{00d2} + \ket{02d0} + \ket{0d20} + \ket{200d} + \ket{2d00} + \ket{d002} + \ket{d200} \right) 
 \nonumber \\
& & + 
C_{45,2} \left ( 
\ket{020d} + \ket{0d02} + \ket{20d0} + \ket{d020} \right) 
 \nonumber \\
& & + 
C_{45,3} \left ( 
\ket{0ddu} - \ket{0udd} - \ket{d0ud} - \ket{dd0u} + \ket{ddu0} + \ket{du0d} + \ket{u0dd} - \ket{udd0} \right) 
\nonumber \eeq
\beq 
C_{45,1} &=& 
-\frac{t}{2 \sqrt{2}} \nonumber \\
C_{45,2} &=& 
\frac{t}{\sqrt{2}} \nonumber \\
C_{45,3} &=& 
\frac{1}{16 \sqrt{2}} \left (
3 J-8 t+4 U-4 W-\sqrt{A_8} \right ) \nonumber \\
 N_{45} &=& 2 \sqrt{2 C_{45,1}^2+C_{45,2}^2+2 C_{45,3}^2} \nonumber \eeq 
\beq
\ket{\Psi_{46}} & = & \ket{3,- \frac{1}{2} , \frac{3}{4} ,\Gamma_{3,2}} \nonumber \\ 
& = & \quad 
C_{46,1} \left ( 
\ket{002d} + \ket{00d2} - \ket{02d0} - \ket{0d20} - \ket{200d} + \ket{2d00} - \ket{d002} + \ket{d200} \right) 
 \nonumber \\
& & + 
C_{46,2} \left ( 
\ket{0ddu} + \ket{0udd} - \ket{d0ud} + \ket{dd0u} - \ket{ddu0} + \ket{du0d} - \ket{u0dd} - \ket{udd0} \right) 
 \nonumber \\
& & + 
C_{46,3} \left ( 
\ket{0dud} - \ket{d0du} - \ket{dud0} + \ket{ud0d} \right) 
\nonumber \eeq
\beq 
C_{46,1} &=& 
-\frac{1}{2} \sqrt{\frac{3}{2}} t \nonumber \\
C_{46,2} &=& 
\frac{1}{16 \sqrt{6}} \left (
3 J-8 t+4 U-4 W+\sqrt{A_8} \right ) \nonumber \\
C_{46,3} &=& 
-\frac{1}{8 \sqrt{6}} \left (
3 J-8 t+4 U-4 W+\sqrt{A_8} \right ) \nonumber \\
 N_{46} &=& 2 \sqrt{2 C_{46,1}^2+2 C_{46,2}^2+C_{46,3}^2} \nonumber \eeq 
\beq
\ket{\Psi_{47}} & = & \ket{3,- \frac{1}{2} , \frac{3}{4} ,\Gamma_{3,2}} \nonumber \\ 
& = & \quad 
C_{47,1} \left ( 
\ket{002d} + \ket{00d2} - \ket{02d0} - \ket{0d20} - \ket{200d} + \ket{2d00} - \ket{d002} + \ket{d200} \right) 
 \nonumber \\
& & + 
C_{47,2} \left ( 
\ket{0ddu} + \ket{0udd} - \ket{d0ud} + \ket{dd0u} - \ket{ddu0} + \ket{du0d} - \ket{u0dd} - \ket{udd0} \right) 
 \nonumber \\
& & + 
C_{47,3} \left ( 
\ket{0dud} - \ket{d0du} - \ket{dud0} + \ket{ud0d} \right) 
\nonumber \eeq
\beq 
C_{47,1} &=& 
-\frac{1}{2} \sqrt{\frac{3}{2}} t \nonumber \\
C_{47,2} &=& 
\frac{1}{16 \sqrt{6}} \left (
3 J-8 t+4 U-4 W-\sqrt{A_8} \right ) \nonumber \\
C_{47,3} &=& 
-\frac{1}{8 \sqrt{6}} \left (
3 J-8 t+4 U-4 W-\sqrt{A_8} \right ) \nonumber \\
 N_{47} &=& 2 \sqrt{2 C_{47,1}^2+2 C_{47,2}^2+C_{47,3}^2} \nonumber \eeq 
\beq
\ket{\Psi_{48}} & = & \ket{3,- \frac{1}{2} , \frac{3}{4} ,\Gamma_{4,1}} \nonumber \\ 
& = & \quad 
C_{48,1} \left ( 
\ket{002d} + \ket{00d2} - \ket{2d00} - \ket{d200} \right) 
 \nonumber \\
& & + 
C_{48,2} \left ( 
\ket{020d} + \ket{02d0} - \ket{0d02} - \ket{0d20} + \ket{200d} + \ket{20d0} - \ket{d002} - \ket{d020} \right) 
 \nonumber \\
& & + 
C_{48,3} \left ( 
\ket{0ddu} - \ket{0dud} + \ket{d0du} - \ket{d0ud} - \ket{du0d} - \ket{dud0} + \ket{ud0d} + \ket{udd0} \right) 
\nonumber \eeq
\beq 
C_{48,1} &=& 
-\frac{1}{12 \sqrt{2}} \left (
48 t^2+3 J t+4 U t-4 W t+2 \cos \left(\theta _5\right) \sqrt{A_{10}} t \right ) \nonumber \\
C_{48,2} &=& 
-\frac{1}{8 \sqrt{2}} \left (
-16 t^2+3 J t+4 U t-4 W t+2 \cos \left(\theta _5\right) \sqrt{A_{10}} t \right ) \nonumber \\
C_{48,3} &=& 
\frac{1}{288 \sqrt{2}} \left (
-720 t^2+36 J t-240 U t-48 U^2+72 J U \right . \nonumber \\
&& \hspace{1cm} 
+
 \left . -768 W^2+144 J W-624 t W-480 U W+24 t \cos \left(\theta _5\right) \sqrt{A_{10}} \right. \nonumber \\
&& \hspace{1cm} 
+
 \left . A_{24}^2+48 U \cos \left(\theta _5\right) \sqrt{A_{10}}+96 W \cos \left(\theta _5\right) \sqrt{A_{10}} \right) \nonumber \\
 N_{48} &=& 2 \sqrt{C_{48,1}^2+2 \left(C_{48,2}^2+C_{48,3}^2\right)} \nonumber \eeq 
\beq
\ket{\Psi_{49}} & = & \ket{3,- \frac{1}{2} , \frac{3}{4} ,\Gamma_{4,1}} \nonumber \\ 
& = & \quad 
C_{49,1} \left ( 
\ket{002d} + \ket{00d2} - \ket{2d00} - \ket{d200} \right) 
 \nonumber \\
& & + 
C_{49,2} \left ( 
\ket{020d} + \ket{02d0} - \ket{0d02} - \ket{0d20} + \ket{200d} + \ket{20d0} - \ket{d002} - \ket{d020} \right) 
 \nonumber \\
& & + 
C_{49,3} \left ( 
\ket{0ddu} - \ket{0dud} + \ket{d0du} - \ket{d0ud} - \ket{du0d} - \ket{dud0} + \ket{ud0d} + \ket{udd0} \right) 
\nonumber \eeq
\beq 
C_{49,1} &=& 
\frac{1}{12 \sqrt{2}} \left (
-48 t^2-3 J t-4 U t+4 W t+\cos \left(\theta _5\right) \sqrt{A_{10}} t+\sqrt{3} \sin \left(\theta _5\right) \sqrt{A_{10}} t \right ) \nonumber \\
C_{49,2} &=& 
\frac{1}{2 \sqrt{2}} \left (
t^2-3 U t-6 W t+3 A_{26} t \right ) \nonumber \\
C_{49,3} &=& 
-\frac{1}{2 \sqrt{2}} \left (
4 t^2-U t-2 W t-U^2-4 W^2-4 U W \right . \nonumber \\
&& \hspace{1cm} 
+
 \left . -A_{26}^2+t A_{26}+2 U A_{26}+4 W A_{26} \right) \nonumber \\
 N_{49} &=& 2 \sqrt{C_{49,1}^2+2 \left(C_{49,2}^2+C_{49,3}^2\right)} \nonumber \eeq 
\beq
\ket{\Psi_{50}} & = & \ket{3,- \frac{1}{2} , \frac{3}{4} ,\Gamma_{4,1}} \nonumber \\ 
& = & \quad 
C_{50,1} \left ( 
\ket{002d} + \ket{00d2} - \ket{2d00} - \ket{d200} \right) 
 \nonumber \\
& & + 
C_{50,2} \left ( 
\ket{020d} + \ket{02d0} - \ket{0d02} - \ket{0d20} + \ket{200d} + \ket{20d0} - \ket{d002} - \ket{d020} \right) 
 \nonumber \\
& & + 
C_{50,3} \left ( 
\ket{0ddu} - \ket{0dud} + \ket{d0du} - \ket{d0ud} - \ket{du0d} - \ket{dud0} + \ket{ud0d} + \ket{udd0} \right) 
\nonumber \eeq
\beq 
C_{50,1} &=& 
-\frac{1}{12 \sqrt{2}} \left (
48 t^2+3 J t+4 U t-4 W t-\cos \left(\theta _5\right) \sqrt{A_{10}} t+\sqrt{3} \sin \left(\theta _5\right) \sqrt{A_{10}} t \right ) \nonumber \\
C_{50,2} &=& 
\frac{1}{2 \sqrt{2}} \left (
t^2-3 U t-6 W t+3 A_{25} t \right ) \nonumber \\
C_{50,3} &=& 
-\frac{1}{2 \sqrt{2}} \left (
4 t^2-U t-2 W t-U^2-4 W^2-4 U W \right . \nonumber \\
&& \hspace{1cm} 
+
 \left . -A_{25}^2+t A_{25}+2 U A_{25}+4 W A_{25} \right) \nonumber \\
 N_{50} &=& 2 \sqrt{C_{50,1}^2+2 \left(C_{50,2}^2+C_{50,3}^2\right)} \nonumber \eeq 
\beq
\ket{\Psi_{51}} & = & \ket{3,- \frac{1}{2} , \frac{3}{4} ,\Gamma_{4,2}} \nonumber \\ 
& = & \quad 
C_{51,1} \left ( 
\ket{002d} - \ket{00d2} + \ket{020d} - \ket{0d02} - \ket{20d0} - \ket{2d00} + \ket{d020} + \ket{d200} \right) 
 \nonumber \\
& & + 
C_{51,2} \left ( 
\ket{02d0} + \ket{0d20} - \ket{200d} - \ket{d002} \right) 
 \nonumber \\
& & + 
C_{51,3} \left ( 
\ket{0dud} - \ket{0udd} + \ket{d0du} + \ket{dd0u} + \ket{ddu0} - \ket{dud0} - \ket{u0dd} - \ket{ud0d} \right) 
\nonumber \eeq
\beq 
C_{51,1} &=& 
-\frac{1}{8 \sqrt{2}} \left (
-16 t^2+3 J t+4 U t-4 W t+2 \cos \left(\theta _5\right) \sqrt{A_{10}} t \right ) \nonumber \\
C_{51,2} &=& 
\frac{1}{12 \sqrt{2}} \left (
48 t^2+3 J t+4 U t-4 W t+2 \cos \left(\theta _5\right) \sqrt{A_{10}} t \right ) \nonumber \\
C_{51,3} &=& 
-\frac{1}{288 \sqrt{2}} \left (
-720 t^2+36 J t-240 U t-48 U^2+72 J U \right . \nonumber \\
&& \hspace{1cm} 
+
 \left . -768 W^2+144 J W-624 t W-480 U W+24 t \cos \left(\theta _5\right) \sqrt{A_{10}} \right. \nonumber \\
&& \hspace{1cm} 
+
 \left . A_{24}^2+48 U \cos \left(\theta _5\right) \sqrt{A_{10}}+96 W \cos \left(\theta _5\right) \sqrt{A_{10}} \right) \nonumber \\
 N_{51} &=& 2 \sqrt{2 C_{51,1}^2+C_{51,2}^2+2 C_{51,3}^2} \nonumber \eeq 
\beq
\ket{\Psi_{52}} & = & \ket{3,- \frac{1}{2} , \frac{3}{4} ,\Gamma_{4,2}} \nonumber \\ 
& = & \quad 
C_{52,1} \left ( 
\ket{002d} - \ket{00d2} + \ket{020d} - \ket{0d02} - \ket{20d0} - \ket{2d00} + \ket{d020} + \ket{d200} \right) 
 \nonumber \\
& & + 
C_{52,2} \left ( 
\ket{02d0} + \ket{0d20} - \ket{200d} - \ket{d002} \right) 
 \nonumber \\
& & + 
C_{52,3} \left ( 
\ket{0dud} - \ket{0udd} + \ket{d0du} + \ket{dd0u} + \ket{ddu0} - \ket{dud0} - \ket{u0dd} - \ket{ud0d} \right) 
\nonumber \eeq
\beq 
C_{52,1} &=& 
\frac{1}{2 \sqrt{2}} \left (
t^2-3 U t-6 W t+3 A_{26} t \right ) \nonumber \\
C_{52,2} &=& 
-\frac{1}{12 \sqrt{2}} \left (
-48 t^2-3 J t-4 U t+4 W t+\cos \left(\theta _5\right) \sqrt{A_{10}} t+\sqrt{3} \sin \left(\theta _5\right) \sqrt{A_{10}} t \right ) \nonumber \\
C_{52,3} &=& 
\frac{1}{2 \sqrt{2}} \left (
4 t^2-U t-2 W t-U^2-4 W^2-4 U W \right . \nonumber \\
&& \hspace{1cm} 
+
 \left . -A_{26}^2+t A_{26}+2 U A_{26}+4 W A_{26} \right) \nonumber \\
 N_{52} &=& 2 \sqrt{2 C_{52,1}^2+C_{52,2}^2+2 C_{52,3}^2} \nonumber \eeq 
\beq
\ket{\Psi_{53}} & = & \ket{3,- \frac{1}{2} , \frac{3}{4} ,\Gamma_{4,2}} \nonumber \\ 
& = & \quad 
C_{53,1} \left ( 
\ket{002d} - \ket{00d2} + \ket{020d} - \ket{0d02} - \ket{20d0} - \ket{2d00} + \ket{d020} + \ket{d200} \right) 
 \nonumber \\
& & + 
C_{53,2} \left ( 
\ket{02d0} + \ket{0d20} - \ket{200d} - \ket{d002} \right) 
 \nonumber \\
& & + 
C_{53,3} \left ( 
\ket{0dud} - \ket{0udd} + \ket{d0du} + \ket{dd0u} + \ket{ddu0} - \ket{dud0} - \ket{u0dd} - \ket{ud0d} \right) 
\nonumber \eeq
\beq 
C_{53,1} &=& 
\frac{1}{2 \sqrt{2}} \left (
t^2-3 U t-6 W t+3 A_{25} t \right ) \nonumber \\
C_{53,2} &=& 
\frac{1}{12 \sqrt{2}} \left (
48 t^2+3 J t+4 U t-4 W t-\cos \left(\theta _5\right) \sqrt{A_{10}} t+\sqrt{3} \sin \left(\theta _5\right) \sqrt{A_{10}} t \right ) \nonumber \\
C_{53,3} &=& 
\frac{1}{2 \sqrt{2}} \left (
4 t^2-U t-2 W t-U^2-4 W^2-4 U W \right . \nonumber \\
&& \hspace{1cm} 
+
 \left . -A_{25}^2+t A_{25}+2 U A_{25}+4 W A_{25} \right) \nonumber \\
 N_{53} &=& 2 \sqrt{2 C_{53,1}^2+C_{53,2}^2+2 C_{53,3}^2} \nonumber \eeq 
\beq
\ket{\Psi_{54}} & = & \ket{3,- \frac{1}{2} , \frac{3}{4} ,\Gamma_{4,3}} \nonumber \\ 
& = & \quad 
C_{54,1} \left ( 
\ket{002d} - \ket{00d2} - \ket{02d0} + \ket{0d20} + \ket{200d} + \ket{2d00} - \ket{d002} - \ket{d200} \right) 
 \nonumber \\
& & + 
C_{54,2} \left ( 
\ket{020d} + \ket{0d02} - \ket{20d0} - \ket{d020} \right) 
 \nonumber \\
& & + 
C_{54,3} \left ( 
\ket{0ddu} - \ket{0udd} + \ket{d0ud} - \ket{dd0u} - \ket{ddu0} + \ket{du0d} - \ket{u0dd} + \ket{udd0} \right) 
\nonumber \eeq
\beq 
C_{54,1} &=& 
\frac{1}{8 \sqrt{2}} \left (
-16 t^2+3 J t+4 U t-4 W t+2 \cos \left(\theta _5\right) \sqrt{A_{10}} t \right ) \nonumber \\
C_{54,2} &=& 
\frac{1}{12 \sqrt{2}} \left (
48 t^2+3 J t+4 U t-4 W t+2 \cos \left(\theta _5\right) \sqrt{A_{10}} t \right ) \nonumber \\
C_{54,3} &=& 
\frac{1}{288 \sqrt{2}} \left (
-720 t^2+36 J t-240 U t-48 U^2+72 J U \right . \nonumber \\
&& \hspace{1cm} 
+
 \left . -768 W^2+144 J W-624 t W-480 U W+24 t \cos \left(\theta _5\right) \sqrt{A_{10}} \right. \nonumber \\
&& \hspace{1cm} 
+
 \left . A_{24}^2+48 U \cos \left(\theta _5\right) \sqrt{A_{10}}+96 W \cos \left(\theta _5\right) \sqrt{A_{10}} \right) \nonumber \\
 N_{54} &=& 2 \sqrt{2 C_{54,1}^2+C_{54,2}^2+2 C_{54,3}^2} \nonumber \eeq 
\beq
\ket{\Psi_{55}} & = & \ket{3,- \frac{1}{2} , \frac{3}{4} ,\Gamma_{4,3}} \nonumber \\ 
& = & \quad 
C_{55,1} \left ( 
\ket{002d} - \ket{00d2} - \ket{02d0} + \ket{0d20} + \ket{200d} + \ket{2d00} - \ket{d002} - \ket{d200} \right) 
 \nonumber \\
& & + 
C_{55,2} \left ( 
\ket{020d} + \ket{0d02} - \ket{20d0} - \ket{d020} \right) 
 \nonumber \\
& & + 
C_{55,3} \left ( 
\ket{0ddu} - \ket{0udd} + \ket{d0ud} - \ket{dd0u} - \ket{ddu0} + \ket{du0d} - \ket{u0dd} + \ket{udd0} \right) 
\nonumber \eeq
\beq 
C_{55,1} &=& 
-\frac{1}{2 \sqrt{2}} \left (
t^2-3 U t-6 W t+3 A_{26} t \right ) \nonumber \\
C_{55,2} &=& 
-\frac{1}{12 \sqrt{2}} \left (
-48 t^2-3 J t-4 U t+4 W t+\cos \left(\theta _5\right) \sqrt{A_{10}} t+\sqrt{3} \sin \left(\theta _5\right) \sqrt{A_{10}} t \right ) \nonumber \\
C_{55,3} &=& 
-\frac{1}{2 \sqrt{2}} \left (
4 t^2-U t-2 W t-U^2-4 W^2-4 U W \right . \nonumber \\
&& \hspace{1cm} 
+
 \left . -A_{26}^2+t A_{26}+2 U A_{26}+4 W A_{26} \right) \nonumber \\
 N_{55} &=& 2 \sqrt{2 C_{55,1}^2+C_{55,2}^2+2 C_{55,3}^2} \nonumber \eeq 
\beq
\ket{\Psi_{56}} & = & \ket{3,- \frac{1}{2} , \frac{3}{4} ,\Gamma_{4,3}} \nonumber \\ 
& = & \quad 
C_{56,1} \left ( 
\ket{002d} - \ket{00d2} - \ket{02d0} + \ket{0d20} + \ket{200d} + \ket{2d00} - \ket{d002} - \ket{d200} \right) 
 \nonumber \\
& & + 
C_{56,2} \left ( 
\ket{020d} + \ket{0d02} - \ket{20d0} - \ket{d020} \right) 
 \nonumber \\
& & + 
C_{56,3} \left ( 
\ket{0ddu} - \ket{0udd} + \ket{d0ud} - \ket{dd0u} - \ket{ddu0} + \ket{du0d} - \ket{u0dd} + \ket{udd0} \right) 
\nonumber \eeq
\beq 
C_{56,1} &=& 
-\frac{1}{2 \sqrt{2}} \left (
t^2-3 U t-6 W t+3 A_{25} t \right ) \nonumber \\
C_{56,2} &=& 
\frac{1}{12 \sqrt{2}} \left (
48 t^2+3 J t+4 U t-4 W t-\cos \left(\theta _5\right) \sqrt{A_{10}} t+\sqrt{3} \sin \left(\theta _5\right) \sqrt{A_{10}} t \right ) \nonumber \\
C_{56,3} &=& 
-\frac{1}{2 \sqrt{2}} \left (
4 t^2-U t-2 W t-U^2-4 W^2-4 U W \right . \nonumber \\
&& \hspace{1cm} 
+
 \left . -A_{25}^2+t A_{25}+2 U A_{25}+4 W A_{25} \right) \nonumber \\
 N_{56} &=& 2 \sqrt{2 C_{56,1}^2+C_{56,2}^2+2 C_{56,3}^2} \nonumber \eeq 
\beq
\ket{\Psi_{57}} & = & \ket{3,- \frac{1}{2} , \frac{3}{4} ,\Gamma_{5,1}} \nonumber \\ 
& = & \quad 
C_{57,1} \left ( 
\ket{002d} - \ket{00d2} + \ket{02d0} - \ket{0d20} - \ket{200d} + \ket{2d00} + \ket{d002} - \ket{d200} \right) 
 \nonumber \\
& & + 
C_{57,2} \left ( 
\ket{0ddu} + \ket{0udd} + \ket{d0ud} + \ket{dd0u} + \ket{ddu0} + \ket{du0d} + \ket{u0dd} + \ket{udd0} \right) 
 \nonumber \\
& & + 
C_{57,3} \left ( 
\ket{0dud} + \ket{d0du} + \ket{dud0} + \ket{ud0d} \right) 
\nonumber \eeq
\beq 
C_{57,1} &=& 
-\frac{1}{2} \sqrt{\frac{3}{2}} t \nonumber \\
C_{57,2} &=& 
\frac{1}{16 \sqrt{6}} \left (
3 J+8 t+4 U-4 W+\sqrt{A_6} \right ) \nonumber \\
C_{57,3} &=& 
-\frac{1}{8 \sqrt{6}} \left (
3 J+8 t+4 U-4 W+\sqrt{A_6} \right ) \nonumber \\
 N_{57} &=& 2 \sqrt{2 C_{57,1}^2+2 C_{57,2}^2+C_{57,3}^2} \nonumber \eeq 
\beq
\ket{\Psi_{58}} & = & \ket{3,- \frac{1}{2} , \frac{3}{4} ,\Gamma_{5,1}} \nonumber \\ 
& = & \quad 
C_{58,1} \left ( 
\ket{002d} - \ket{00d2} + \ket{02d0} - \ket{0d20} - \ket{200d} + \ket{2d00} + \ket{d002} - \ket{d200} \right) 
 \nonumber \\
& & + 
C_{58,2} \left ( 
\ket{0ddu} + \ket{0udd} + \ket{d0ud} + \ket{dd0u} + \ket{ddu0} + \ket{du0d} + \ket{u0dd} + \ket{udd0} \right) 
 \nonumber \\
& & + 
C_{58,3} \left ( 
\ket{0dud} + \ket{d0du} + \ket{dud0} + \ket{ud0d} \right) 
\nonumber \eeq
\beq 
C_{58,1} &=& 
-\frac{1}{2} \sqrt{\frac{3}{2}} t \nonumber \\
C_{58,2} &=& 
\frac{1}{16 \sqrt{6}} \left (
3 J+8 t+4 U-4 W-\sqrt{A_6} \right ) \nonumber \\
C_{58,3} &=& 
-\frac{1}{8 \sqrt{6}} \left (
3 J+8 t+4 U-4 W-\sqrt{A_6} \right ) \nonumber \\
 N_{58} &=& 2 \sqrt{2 C_{58,1}^2+2 C_{58,2}^2+C_{58,3}^2} \nonumber \eeq 
\beq
\ket{\Psi_{59}} & = & \ket{3,- \frac{1}{2} , \frac{15}{4} ,\Gamma_{5,1}} \nonumber \\ 
&=& \frac{1}{2 \sqrt{3}}
 \left ( \ket{0ddu} + \ket{0dud} + \ket{0udd} + \ket{d0du} + \ket{d0ud} + \ket{dd0u} \right . \nonumber \\
&& \hspace{3em} 
 + 
\left . \ket{ddu0} + \ket{du0d} + \ket{dud0} + \ket{u0dd} + \ket{ud0d} + \ket{udd0} \right ) 
\nonumber \eeq
\beq
\ket{\Psi_{60}} & = & \ket{3,- \frac{1}{2} , \frac{3}{4} ,\Gamma_{5,2}} \nonumber \\ 
& = & \quad 
C_{60,1} \left ( 
\ket{020d} - \ket{02d0} - \ket{0d02} + \ket{0d20} - \ket{200d} + \ket{20d0} + \ket{d002} - \ket{d020} \right) 
 \nonumber \\
& & + 
C_{60,2} \left ( 
\ket{0ddu} + \ket{0dud} - \ket{d0du} - \ket{d0ud} - \ket{du0d} + \ket{dud0} - \ket{ud0d} + \ket{udd0} \right) 
 \nonumber \\
& & + 
C_{60,3} \left ( 
\ket{0udd} - \ket{dd0u} + \ket{ddu0} - \ket{u0dd} \right) 
\nonumber \eeq
\beq 
C_{60,1} &=& 
-\frac{1}{2} \sqrt{\frac{3}{2}} t \nonumber \\
C_{60,2} &=& 
-\frac{1}{16 \sqrt{6}} \left (
3 J+8 t+4 U-4 W+\sqrt{A_6} \right ) \nonumber \\
C_{60,3} &=& 
\frac{1}{8 \sqrt{6}} \left (
3 J+8 t+4 U-4 W+\sqrt{A_6} \right ) \nonumber \\
 N_{60} &=& 2 \sqrt{2 C_{60,1}^2+2 C_{60,2}^2+C_{60,3}^2} \nonumber \eeq 
\beq
\ket{\Psi_{61}} & = & \ket{3,- \frac{1}{2} , \frac{3}{4} ,\Gamma_{5,2}} \nonumber \\ 
& = & \quad 
C_{61,1} \left ( 
\ket{020d} - \ket{02d0} - \ket{0d02} + \ket{0d20} - \ket{200d} + \ket{20d0} + \ket{d002} - \ket{d020} \right) 
 \nonumber \\
& & + 
C_{61,2} \left ( 
\ket{0ddu} + \ket{0dud} - \ket{d0du} - \ket{d0ud} - \ket{du0d} + \ket{dud0} - \ket{ud0d} + \ket{udd0} \right) 
 \nonumber \\
& & + 
C_{61,3} \left ( 
\ket{0udd} - \ket{dd0u} + \ket{ddu0} - \ket{u0dd} \right) 
\nonumber \eeq
\beq 
C_{61,1} &=& 
-\frac{1}{2} \sqrt{\frac{3}{2}} t \nonumber \\
C_{61,2} &=& 
-\frac{1}{16 \sqrt{6}} \left (
3 J+8 t+4 U-4 W-\sqrt{A_6} \right ) \nonumber \\
C_{61,3} &=& 
\frac{1}{8 \sqrt{6}} \left (
3 J+8 t+4 U-4 W-\sqrt{A_6} \right ) \nonumber \\
 N_{61} &=& 2 \sqrt{2 C_{61,1}^2+2 C_{61,2}^2+C_{61,3}^2} \nonumber \eeq 
\beq
\ket{\Psi_{62}} & = & \ket{3,- \frac{1}{2} , \frac{15}{4} ,\Gamma_{5,2}} \nonumber \\ 
&=& \frac{1}{2 \sqrt{3}}
 \left ( \ket{0ddu} + \ket{0dud} + \ket{0udd} - \ket{d0du} - \ket{d0ud} - \ket{dd0u} \right . \nonumber \\
&& \hspace{3em} 
 + 
\left . \ket{ddu0} - \ket{du0d} + \ket{dud0} - \ket{u0dd} - \ket{ud0d} + \ket{udd0} \right ) 
\nonumber \eeq
\beq
\ket{\Psi_{63}} & = & \ket{3,- \frac{1}{2} , \frac{3}{4} ,\Gamma_{5,3}} \nonumber \\ 
& = & \quad 
C_{63,1} \left ( 
\ket{002d} - \ket{00d2} - \ket{020d} + \ket{0d02} + \ket{20d0} - \ket{2d00} - \ket{d020} + \ket{d200} \right) 
 \nonumber \\
& & + 
C_{63,2} \left ( 
\ket{0ddu} + \ket{d0ud} - \ket{du0d} - \ket{udd0} \right) 
 \nonumber \\
& & + 
C_{63,3} \left ( 
\ket{0dud} + \ket{0udd} + \ket{d0du} - \ket{dd0u} - \ket{ddu0} - \ket{dud0} + \ket{u0dd} - \ket{ud0d} \right) 
\nonumber \eeq
\beq 
C_{63,1} &=& 
-\frac{1}{2} \sqrt{\frac{3}{2}} t \nonumber \\
C_{63,2} &=& 
-\frac{1}{8 \sqrt{6}} \left (
3 J+8 t+4 U-4 W+\sqrt{A_6} \right ) \nonumber \\
C_{63,3} &=& 
\frac{1}{16 \sqrt{6}} \left (
3 J+8 t+4 U-4 W+\sqrt{A_6} \right ) \nonumber \\
 N_{63} &=& 2 \sqrt{2 C_{63,1}^2+C_{63,2}^2+2 C_{63,3}^2} \nonumber \eeq 
\beq
\ket{\Psi_{64}} & = & \ket{3,- \frac{1}{2} , \frac{3}{4} ,\Gamma_{5,3}} \nonumber \\ 
& = & \quad 
C_{64,1} \left ( 
\ket{002d} - \ket{00d2} - \ket{020d} + \ket{0d02} + \ket{20d0} - \ket{2d00} - \ket{d020} + \ket{d200} \right) 
 \nonumber \\
& & + 
C_{64,2} \left ( 
\ket{0ddu} + \ket{d0ud} - \ket{du0d} - \ket{udd0} \right) 
 \nonumber \\
& & + 
C_{64,3} \left ( 
\ket{0dud} + \ket{0udd} + \ket{d0du} - \ket{dd0u} - \ket{ddu0} - \ket{dud0} + \ket{u0dd} - \ket{ud0d} \right) 
\nonumber \eeq
\beq 
C_{64,1} &=& 
-\frac{1}{2} \sqrt{\frac{3}{2}} t \nonumber \\
C_{64,2} &=& 
-\frac{1}{8 \sqrt{6}} \left (
3 J+8 t+4 U-4 W-\sqrt{A_6} \right ) \nonumber \\
C_{64,3} &=& 
\frac{1}{16 \sqrt{6}} \left (
3 J+8 t+4 U-4 W-\sqrt{A_6} \right ) \nonumber \\
 N_{64} &=& 2 \sqrt{2 C_{64,1}^2+C_{64,2}^2+2 C_{64,3}^2} \nonumber \eeq 
\beq
\ket{\Psi_{65}} & = & \ket{3,- \frac{1}{2} , \frac{15}{4} ,\Gamma_{5,3}} \nonumber \\ 
&=& \frac{1}{2 \sqrt{3}}
 \left ( \ket{0ddu} + \ket{0dud} + \ket{0udd} + \ket{d0du} + \ket{d0ud} - \ket{dd0u} \right . \nonumber \\
&& \hspace{3em} 
\left . -\ket{ddu0} - \ket{du0d} - \ket{dud0} + \ket{u0dd} - \ket{ud0d} - \ket{udd0} \right ) 
\nonumber \eeq
{\subsubsection{\boldmath Eigenvectors for ${\rm  N_e}=3$ and   ${\rm m_s}$= $\frac{1}{2} $.}
\beq
\ket{\Psi_{66}} & = & \ket{3,\frac{1}{2} , \frac{3}{4} ,\Gamma_1} \nonumber \\ 
&=& \frac{1}{2 \sqrt{3}}
 \left ( \ket{002u} + \ket{00u2} + \ket{020u} + \ket{02u0} + \ket{0u02} + \ket{0u20} \right . \nonumber \\
&& \hspace{3em} 
 + 
\left . \ket{200u} + \ket{20u0} + \ket{2u00} + \ket{u002} + \ket{u020} + \ket{u200} \right ) 
\nonumber \eeq
\beq
\ket{\Psi_{67}} & = & \ket{3,\frac{1}{2} , \frac{15}{4} ,\Gamma_2} \nonumber \\ 
&=& \frac{1}{2 \sqrt{3}}
 \left ( \ket{0duu} + \ket{0udu} + \ket{0uud} - \ket{d0uu} + \ket{du0u} - \ket{duu0} \right . \nonumber \\
&& \hspace{3em} 
\left . -\ket{u0du} - \ket{u0ud} + \ket{ud0u} - \ket{udu0} + \ket{uu0d} - \ket{uud0} \right ) 
\nonumber \eeq
\beq
\ket{\Psi_{68}} & = & \ket{3,\frac{1}{2} , \frac{3}{4} ,\Gamma_{3,1}} \nonumber \\ 
& = & \quad 
C_{68,1} \left ( 
\ket{002u} + \ket{00u2} + \ket{02u0} + \ket{0u20} + \ket{200u} + \ket{2u00} + \ket{u002} + \ket{u200} \right) 
 \nonumber \\
& & + 
C_{68,2} \left ( 
\ket{020u} + \ket{0u02} + \ket{20u0} + \ket{u020} \right) 
 \nonumber \\
& & + 
C_{68,3} \left ( 
\ket{0duu} - \ket{0uud} - \ket{d0uu} + \ket{duu0} + \ket{u0du} - \ket{ud0u} + \ket{uu0d} - \ket{uud0} \right) 
\nonumber \eeq
\beq 
C_{68,1} &=& 
-\frac{t}{2 \sqrt{2}} \nonumber \\
C_{68,2} &=& 
\frac{t}{\sqrt{2}} \nonumber \\
C_{68,3} &=& 
\frac{1}{16 \sqrt{2}} \left (
3 J-8 t+4 U-4 W+\sqrt{A_8} \right ) \nonumber \\
 N_{68} &=& 2 \sqrt{2 C_{68,1}^2+C_{68,2}^2+2 C_{68,3}^2} \nonumber \eeq 
\beq
\ket{\Psi_{69}} & = & \ket{3,\frac{1}{2} , \frac{3}{4} ,\Gamma_{3,1}} \nonumber \\ 
& = & \quad 
C_{69,1} \left ( 
\ket{002u} + \ket{00u2} + \ket{02u0} + \ket{0u20} + \ket{200u} + \ket{2u00} + \ket{u002} + \ket{u200} \right) 
 \nonumber \\
& & + 
C_{69,2} \left ( 
\ket{020u} + \ket{0u02} + \ket{20u0} + \ket{u020} \right) 
 \nonumber \\
& & + 
C_{69,3} \left ( 
\ket{0duu} - \ket{0uud} - \ket{d0uu} + \ket{duu0} + \ket{u0du} - \ket{ud0u} + \ket{uu0d} - \ket{uud0} \right) 
\nonumber \eeq
\beq 
C_{69,1} &=& 
-\frac{t}{2 \sqrt{2}} \nonumber \\
C_{69,2} &=& 
\frac{t}{\sqrt{2}} \nonumber \\
C_{69,3} &=& 
\frac{1}{16 \sqrt{2}} \left (
3 J-8 t+4 U-4 W-\sqrt{A_8} \right ) \nonumber \\
 N_{69} &=& 2 \sqrt{2 C_{69,1}^2+C_{69,2}^2+2 C_{69,3}^2} \nonumber \eeq 
\beq
\ket{\Psi_{70}} & = & \ket{3,\frac{1}{2} , \frac{3}{4} ,\Gamma_{3,2}} \nonumber \\ 
& = & \quad 
C_{70,1} \left ( 
\ket{002u} + \ket{00u2} - \ket{02u0} - \ket{0u20} - \ket{200u} + \ket{2u00} - \ket{u002} + \ket{u200} \right) 
 \nonumber \\
& & + 
C_{70,2} \left ( 
\ket{0duu} + \ket{0uud} - \ket{d0uu} - \ket{duu0} - \ket{u0du} + \ket{ud0u} + \ket{uu0d} - \ket{uud0} \right) 
 \nonumber \\
& & + 
C_{70,3} \left ( 
\ket{0udu} + \ket{du0u} - \ket{u0ud} - \ket{udu0} \right) 
\nonumber \eeq
\beq 
C_{70,1} &=& 
\frac{1}{2} \sqrt{\frac{3}{2}} t \nonumber \\
C_{70,2} &=& 
\frac{1}{16 \sqrt{6}} \left (
3 J-8 t+4 U-4 W+\sqrt{A_8} \right ) \nonumber \\
C_{70,3} &=& 
-\frac{1}{8 \sqrt{6}} \left (
3 J-8 t+4 U-4 W+\sqrt{A_8} \right ) \nonumber \\
 N_{70} &=& 2 \sqrt{2 C_{70,1}^2+2 C_{70,2}^2+C_{70,3}^2} \nonumber \eeq 
\beq
\ket{\Psi_{71}} & = & \ket{3,\frac{1}{2} , \frac{3}{4} ,\Gamma_{3,2}} \nonumber \\ 
& = & \quad 
C_{71,1} \left ( 
\ket{002u} + \ket{00u2} - \ket{02u0} - \ket{0u20} - \ket{200u} + \ket{2u00} - \ket{u002} + \ket{u200} \right) 
 \nonumber \\
& & + 
C_{71,2} \left ( 
\ket{0duu} + \ket{0uud} - \ket{d0uu} - \ket{duu0} - \ket{u0du} + \ket{ud0u} + \ket{uu0d} - \ket{uud0} \right) 
 \nonumber \\
& & + 
C_{71,3} \left ( 
\ket{0udu} + \ket{du0u} - \ket{u0ud} - \ket{udu0} \right) 
\nonumber \eeq
\beq 
C_{71,1} &=& 
\frac{1}{2} \sqrt{\frac{3}{2}} t \nonumber \\
C_{71,2} &=& 
\frac{1}{16 \sqrt{6}} \left (
3 J-8 t+4 U-4 W-\sqrt{A_8} \right ) \nonumber \\
C_{71,3} &=& 
-\frac{1}{8 \sqrt{6}} \left (
3 J-8 t+4 U-4 W-\sqrt{A_8} \right ) \nonumber \\
 N_{71} &=& 2 \sqrt{2 C_{71,1}^2+2 C_{71,2}^2+C_{71,3}^2} \nonumber \eeq 
\beq
\ket{\Psi_{72}} & = & \ket{3,\frac{1}{2} , \frac{3}{4} ,\Gamma_{4,1}} \nonumber \\ 
& = & \quad 
C_{72,1} \left ( 
\ket{002u} + \ket{00u2} - \ket{2u00} - \ket{u200} \right) 
 \nonumber \\
& & + 
C_{72,2} \left ( 
\ket{020u} + \ket{02u0} - \ket{0u02} - \ket{0u20} + \ket{200u} + \ket{20u0} - \ket{u002} - \ket{u020} \right) 
 \nonumber \\
& & + 
C_{72,3} \left ( 
\ket{0udu} - \ket{0uud} - \ket{du0u} - \ket{duu0} + \ket{u0du} - \ket{u0ud} + \ket{ud0u} + \ket{udu0} \right) 
\nonumber \eeq
\beq 
C_{72,1} &=& 
-\frac{1}{12 \sqrt{2}} \left (
48 t^2+3 J t+4 U t-4 W t+2 \cos \left(\theta _5\right) \sqrt{A_{10}} t \right ) \nonumber \\
C_{72,2} &=& 
-\frac{1}{8 \sqrt{2}} \left (
-16 t^2+3 J t+4 U t-4 W t+2 \cos \left(\theta _5\right) \sqrt{A_{10}} t \right ) \nonumber \\
C_{72,3} &=& 
\frac{1}{288 \sqrt{2}} \left (
-720 t^2+36 J t-240 U t-48 U^2+72 J U \right . \nonumber \\
&& \hspace{1cm} 
+
 \left . -768 W^2+144 J W-624 t W-480 U W+24 t \cos \left(\theta _5\right) \sqrt{A_{10}} \right. \nonumber \\
&& \hspace{1cm} 
+
 \left . A_{24}^2+48 U \cos \left(\theta _5\right) \sqrt{A_{10}}+96 W \cos \left(\theta _5\right) \sqrt{A_{10}} \right) \nonumber \\
 N_{72} &=& 2 \sqrt{C_{72,1}^2+2 \left(C_{72,2}^2+C_{72,3}^2\right)} \nonumber \eeq 
\beq
\ket{\Psi_{73}} & = & \ket{3,\frac{1}{2} , \frac{3}{4} ,\Gamma_{4,1}} \nonumber \\ 
& = & \quad 
C_{73,1} \left ( 
\ket{002u} + \ket{00u2} - \ket{2u00} - \ket{u200} \right) 
 \nonumber \\
& & + 
C_{73,2} \left ( 
\ket{020u} + \ket{02u0} - \ket{0u02} - \ket{0u20} + \ket{200u} + \ket{20u0} - \ket{u002} - \ket{u020} \right) 
 \nonumber \\
& & + 
C_{73,3} \left ( 
\ket{0udu} - \ket{0uud} - \ket{du0u} - \ket{duu0} + \ket{u0du} - \ket{u0ud} + \ket{ud0u} + \ket{udu0} \right) 
\nonumber \eeq
\beq 
C_{73,1} &=& 
\frac{1}{12 \sqrt{2}} \left (
-48 t^2-3 J t-4 U t+4 W t+\cos \left(\theta _5\right) \sqrt{A_{10}} t+\sqrt{3} \sin \left(\theta _5\right) \sqrt{A_{10}} t \right ) \nonumber \\
C_{73,2} &=& 
\frac{1}{2 \sqrt{2}} \left (
t^2-3 U t-6 W t+3 A_{26} t \right ) \nonumber \\
C_{73,3} &=& 
-\frac{1}{2 \sqrt{2}} \left (
4 t^2-U t-2 W t-U^2-4 W^2-4 U W \right . \nonumber \\
&& \hspace{1cm} 
+
 \left . -A_{26}^2+t A_{26}+2 U A_{26}+4 W A_{26} \right) \nonumber \\
 N_{73} &=& 2 \sqrt{C_{73,1}^2+2 \left(C_{73,2}^2+C_{73,3}^2\right)} \nonumber \eeq 
\beq
\ket{\Psi_{74}} & = & \ket{3,\frac{1}{2} , \frac{3}{4} ,\Gamma_{4,1}} \nonumber \\ 
& = & \quad 
C_{74,1} \left ( 
\ket{002u} + \ket{00u2} - \ket{2u00} - \ket{u200} \right) 
 \nonumber \\
& & + 
C_{74,2} \left ( 
\ket{020u} + \ket{02u0} - \ket{0u02} - \ket{0u20} + \ket{200u} + \ket{20u0} - \ket{u002} - \ket{u020} \right) 
 \nonumber \\
& & + 
C_{74,3} \left ( 
\ket{0udu} - \ket{0uud} - \ket{du0u} - \ket{duu0} + \ket{u0du} - \ket{u0ud} + \ket{ud0u} + \ket{udu0} \right) 
\nonumber \eeq
\beq 
C_{74,1} &=& 
-\frac{1}{12 \sqrt{2}} \left (
48 t^2+3 J t+4 U t-4 W t-\cos \left(\theta _5\right) \sqrt{A_{10}} t+\sqrt{3} \sin \left(\theta _5\right) \sqrt{A_{10}} t \right ) \nonumber \\
C_{74,2} &=& 
\frac{1}{2 \sqrt{2}} \left (
t^2-3 U t-6 W t+3 A_{25} t \right ) \nonumber \\
C_{74,3} &=& 
-\frac{1}{2 \sqrt{2}} \left (
4 t^2-U t-2 W t-U^2-4 W^2-4 U W \right . \nonumber \\
&& \hspace{1cm} 
+
 \left . -A_{25}^2+t A_{25}+2 U A_{25}+4 W A_{25} \right) \nonumber \\
 N_{74} &=& 2 \sqrt{C_{74,1}^2+2 \left(C_{74,2}^2+C_{74,3}^2\right)} \nonumber \eeq 
\beq
\ket{\Psi_{75}} & = & \ket{3,\frac{1}{2} , \frac{3}{4} ,\Gamma_{4,2}} \nonumber \\ 
& = & \quad 
C_{75,1} \left ( 
\ket{002u} - \ket{00u2} + \ket{020u} - \ket{0u02} - \ket{20u0} - \ket{2u00} + \ket{u020} + \ket{u200} \right) 
 \nonumber \\
& & + 
C_{75,2} \left ( 
\ket{02u0} + \ket{0u20} - \ket{200u} - \ket{u002} \right) 
 \nonumber \\
& & + 
C_{75,3} \left ( 
\ket{0duu} - \ket{0udu} + \ket{d0uu} + \ket{du0u} - \ket{u0ud} + \ket{udu0} - \ket{uu0d} - \ket{uud0} \right) 
\nonumber \eeq
\beq 
C_{75,1} &=& 
-\frac{1}{8 \sqrt{2}} \left (
-16 t^2+3 J t+4 U t-4 W t+2 \cos \left(\theta _5\right) \sqrt{A_{10}} t \right ) \nonumber \\
C_{75,2} &=& 
\frac{1}{12 \sqrt{2}} \left (
48 t^2+3 J t+4 U t-4 W t+2 \cos \left(\theta _5\right) \sqrt{A_{10}} t \right ) \nonumber \\
C_{75,3} &=& 
-\frac{1}{288 \sqrt{2}} \left (
-720 t^2+36 J t-240 U t-48 U^2+72 J U \right . \nonumber \\
&& \hspace{1cm} 
+
 \left . -768 W^2+144 J W-624 t W-480 U W+24 t \cos \left(\theta _5\right) \sqrt{A_{10}} \right. \nonumber \\
&& \hspace{1cm} 
+
 \left . A_{24}^2+48 U \cos \left(\theta _5\right) \sqrt{A_{10}}+96 W \cos \left(\theta _5\right) \sqrt{A_{10}} \right) \nonumber \\
 N_{75} &=& 2 \sqrt{2 C_{75,1}^2+C_{75,2}^2+2 C_{75,3}^2} \nonumber \eeq 
\beq
\ket{\Psi_{76}} & = & \ket{3,\frac{1}{2} , \frac{3}{4} ,\Gamma_{4,2}} \nonumber \\ 
& = & \quad 
C_{76,1} \left ( 
\ket{002u} - \ket{00u2} + \ket{020u} - \ket{0u02} - \ket{20u0} - \ket{2u00} + \ket{u020} + \ket{u200} \right) 
 \nonumber \\
& & + 
C_{76,2} \left ( 
\ket{02u0} + \ket{0u20} - \ket{200u} - \ket{u002} \right) 
 \nonumber \\
& & + 
C_{76,3} \left ( 
\ket{0duu} - \ket{0udu} + \ket{d0uu} + \ket{du0u} - \ket{u0ud} + \ket{udu0} - \ket{uu0d} - \ket{uud0} \right) 
\nonumber \eeq
\beq 
C_{76,1} &=& 
\frac{1}{2 \sqrt{2}} \left (
t^2-3 U t-6 W t+3 A_{26} t \right ) \nonumber \\
C_{76,2} &=& 
-\frac{1}{12 \sqrt{2}} \left (
-48 t^2-3 J t-4 U t+4 W t+\cos \left(\theta _5\right) \sqrt{A_{10}} t+\sqrt{3} \sin \left(\theta _5\right) \sqrt{A_{10}} t \right ) \nonumber \\
C_{76,3} &=& 
\frac{1}{2 \sqrt{2}} \left (
4 t^2-U t-2 W t-U^2-4 W^2-4 U W \right . \nonumber \\
&& \hspace{1cm} 
+
 \left . -A_{26}^2+t A_{26}+2 U A_{26}+4 W A_{26} \right) \nonumber \\
 N_{76} &=& 2 \sqrt{2 C_{76,1}^2+C_{76,2}^2+2 C_{76,3}^2} \nonumber \eeq 
\beq
\ket{\Psi_{77}} & = & \ket{3,\frac{1}{2} , \frac{3}{4} ,\Gamma_{4,2}} \nonumber \\ 
& = & \quad 
C_{77,1} \left ( 
\ket{002u} - \ket{00u2} + \ket{020u} - \ket{0u02} - \ket{20u0} - \ket{2u00} + \ket{u020} + \ket{u200} \right) 
 \nonumber \\
& & + 
C_{77,2} \left ( 
\ket{02u0} + \ket{0u20} - \ket{200u} - \ket{u002} \right) 
 \nonumber \\
& & + 
C_{77,3} \left ( 
\ket{0duu} - \ket{0udu} + \ket{d0uu} + \ket{du0u} - \ket{u0ud} + \ket{udu0} - \ket{uu0d} - \ket{uud0} \right) 
\nonumber \eeq
\beq 
C_{77,1} &=& 
\frac{1}{2 \sqrt{2}} \left (
t^2-3 U t-6 W t+3 A_{25} t \right ) \nonumber \\
C_{77,2} &=& 
\frac{1}{12 \sqrt{2}} \left (
48 t^2+3 J t+4 U t-4 W t-\cos \left(\theta _5\right) \sqrt{A_{10}} t+\sqrt{3} \sin \left(\theta _5\right) \sqrt{A_{10}} t \right ) \nonumber \\
C_{77,3} &=& 
\frac{1}{2 \sqrt{2}} \left (
4 t^2-U t-2 W t-U^2-4 W^2-4 U W \right . \nonumber \\
&& \hspace{1cm} 
+
 \left . -A_{25}^2+t A_{25}+2 U A_{25}+4 W A_{25} \right) \nonumber \\
 N_{77} &=& 2 \sqrt{2 C_{77,1}^2+C_{77,2}^2+2 C_{77,3}^2} \nonumber \eeq 
\beq
\ket{\Psi_{78}} & = & \ket{3,\frac{1}{2} , \frac{3}{4} ,\Gamma_{4,3}} \nonumber \\ 
& = & \quad 
C_{78,1} \left ( 
\ket{002u} - \ket{00u2} - \ket{02u0} + \ket{0u20} + \ket{200u} + \ket{2u00} - \ket{u002} - \ket{u200} \right) 
 \nonumber \\
& & + 
C_{78,2} \left ( 
\ket{020u} + \ket{0u02} - \ket{20u0} - \ket{u020} \right) 
 \nonumber \\
& & + 
C_{78,3} \left ( 
\ket{0duu} - \ket{0uud} + \ket{d0uu} - \ket{duu0} - \ket{u0du} - \ket{ud0u} + \ket{uu0d} + \ket{uud0} \right) 
\nonumber \eeq
\beq 
C_{78,1} &=& 
\frac{1}{8 \sqrt{2}} \left (
-16 t^2+3 J t+4 U t-4 W t+2 \cos \left(\theta _5\right) \sqrt{A_{10}} t \right ) \nonumber \\
C_{78,2} &=& 
\frac{1}{12 \sqrt{2}} \left (
48 t^2+3 J t+4 U t-4 W t+2 \cos \left(\theta _5\right) \sqrt{A_{10}} t \right ) \nonumber \\
C_{78,3} &=& 
\frac{1}{288 \sqrt{2}} \left (
-720 t^2+36 J t-240 U t-48 U^2+72 J U \right . \nonumber \\
&& \hspace{1cm} 
+
 \left . -768 W^2+144 J W-624 t W-480 U W+24 t \cos \left(\theta _5\right) \sqrt{A_{10}} \right. \nonumber \\
&& \hspace{1cm} 
+
 \left . A_{24}^2+48 U \cos \left(\theta _5\right) \sqrt{A_{10}}+96 W \cos \left(\theta _5\right) \sqrt{A_{10}} \right) \nonumber \\
 N_{78} &=& 2 \sqrt{2 C_{78,1}^2+C_{78,2}^2+2 C_{78,3}^2} \nonumber \eeq 
\beq
\ket{\Psi_{79}} & = & \ket{3,\frac{1}{2} , \frac{3}{4} ,\Gamma_{4,3}} \nonumber \\ 
& = & \quad 
C_{79,1} \left ( 
\ket{002u} - \ket{00u2} - \ket{02u0} + \ket{0u20} + \ket{200u} + \ket{2u00} - \ket{u002} - \ket{u200} \right) 
 \nonumber \\
& & + 
C_{79,2} \left ( 
\ket{020u} + \ket{0u02} - \ket{20u0} - \ket{u020} \right) 
 \nonumber \\
& & + 
C_{79,3} \left ( 
\ket{0duu} - \ket{0uud} + \ket{d0uu} - \ket{duu0} - \ket{u0du} - \ket{ud0u} + \ket{uu0d} + \ket{uud0} \right) 
\nonumber \eeq
\beq 
C_{79,1} &=& 
-\frac{1}{2 \sqrt{2}} \left (
t^2-3 U t-6 W t+3 A_{26} t \right ) \nonumber \\
C_{79,2} &=& 
-\frac{1}{12 \sqrt{2}} \left (
-48 t^2-3 J t-4 U t+4 W t+\cos \left(\theta _5\right) \sqrt{A_{10}} t+\sqrt{3} \sin \left(\theta _5\right) \sqrt{A_{10}} t \right ) \nonumber \\
C_{79,3} &=& 
-\frac{1}{2 \sqrt{2}} \left (
4 t^2-U t-2 W t-U^2-4 W^2-4 U W \right . \nonumber \\
&& \hspace{1cm} 
+
 \left . -A_{26}^2+t A_{26}+2 U A_{26}+4 W A_{26} \right) \nonumber \\
 N_{79} &=& 2 \sqrt{2 C_{79,1}^2+C_{79,2}^2+2 C_{79,3}^2} \nonumber \eeq 
\beq
\ket{\Psi_{80}} & = & \ket{3,\frac{1}{2} , \frac{3}{4} ,\Gamma_{4,3}} \nonumber \\ 
& = & \quad 
C_{80,1} \left ( 
\ket{002u} - \ket{00u2} - \ket{02u0} + \ket{0u20} + \ket{200u} + \ket{2u00} - \ket{u002} - \ket{u200} \right) 
 \nonumber \\
& & + 
C_{80,2} \left ( 
\ket{020u} + \ket{0u02} - \ket{20u0} - \ket{u020} \right) 
 \nonumber \\
& & + 
C_{80,3} \left ( 
\ket{0duu} - \ket{0uud} + \ket{d0uu} - \ket{duu0} - \ket{u0du} - \ket{ud0u} + \ket{uu0d} + \ket{uud0} \right) 
\nonumber \eeq
\beq 
C_{80,1} &=& 
-\frac{1}{2 \sqrt{2}} \left (
t^2-3 U t-6 W t+3 A_{25} t \right ) \nonumber \\
C_{80,2} &=& 
\frac{1}{12 \sqrt{2}} \left (
48 t^2+3 J t+4 U t-4 W t-\cos \left(\theta _5\right) \sqrt{A_{10}} t+\sqrt{3} \sin \left(\theta _5\right) \sqrt{A_{10}} t \right ) \nonumber \\
C_{80,3} &=& 
-\frac{1}{2 \sqrt{2}} \left (
4 t^2-U t-2 W t-U^2-4 W^2-4 U W \right . \nonumber \\
&& \hspace{1cm} 
+
 \left . -A_{25}^2+t A_{25}+2 U A_{25}+4 W A_{25} \right) \nonumber \\
 N_{80} &=& 2 \sqrt{2 C_{80,1}^2+C_{80,2}^2+2 C_{80,3}^2} \nonumber \eeq 
\beq
\ket{\Psi_{81}} & = & \ket{3,\frac{1}{2} , \frac{3}{4} ,\Gamma_{5,1}} \nonumber \\ 
& = & \quad 
C_{81,1} \left ( 
\ket{002u} - \ket{00u2} + \ket{02u0} - \ket{0u20} - \ket{200u} + \ket{2u00} + \ket{u002} - \ket{u200} \right) 
 \nonumber \\
& & + 
C_{81,2} \left ( 
\ket{0duu} + \ket{0uud} + \ket{d0uu} + \ket{duu0} + \ket{u0du} + \ket{ud0u} + \ket{uu0d} + \ket{uud0} \right) 
 \nonumber \\
& & + 
C_{81,3} \left ( 
\ket{0udu} + \ket{du0u} + \ket{u0ud} + \ket{udu0} \right) 
\nonumber \eeq
\beq 
C_{81,1} &=& 
\frac{1}{2} \sqrt{\frac{3}{2}} t \nonumber \\
C_{81,2} &=& 
\frac{1}{16 \sqrt{6}} \left (
3 J+8 t+4 U-4 W+\sqrt{A_6} \right ) \nonumber \\
C_{81,3} &=& 
-\frac{1}{8 \sqrt{6}} \left (
3 J+8 t+4 U-4 W+\sqrt{A_6} \right ) \nonumber \\
 N_{81} &=& 2 \sqrt{2 C_{81,1}^2+2 C_{81,2}^2+C_{81,3}^2} \nonumber \eeq 
\beq
\ket{\Psi_{82}} & = & \ket{3,\frac{1}{2} , \frac{3}{4} ,\Gamma_{5,1}} \nonumber \\ 
& = & \quad 
C_{82,1} \left ( 
\ket{002u} - \ket{00u2} + \ket{02u0} - \ket{0u20} - \ket{200u} + \ket{2u00} + \ket{u002} - \ket{u200} \right) 
 \nonumber \\
& & + 
C_{82,2} \left ( 
\ket{0duu} + \ket{0uud} + \ket{d0uu} + \ket{duu0} + \ket{u0du} + \ket{ud0u} + \ket{uu0d} + \ket{uud0} \right) 
 \nonumber \\
& & + 
C_{82,3} \left ( 
\ket{0udu} + \ket{du0u} + \ket{u0ud} + \ket{udu0} \right) 
\nonumber \eeq
\beq 
C_{82,1} &=& 
\frac{1}{2} \sqrt{\frac{3}{2}} t \nonumber \\
C_{82,2} &=& 
\frac{1}{16 \sqrt{6}} \left (
3 J+8 t+4 U-4 W-\sqrt{A_6} \right ) \nonumber \\
C_{82,3} &=& 
-\frac{1}{8 \sqrt{6}} \left (
3 J+8 t+4 U-4 W-\sqrt{A_6} \right ) \nonumber \\
 N_{82} &=& 2 \sqrt{2 C_{82,1}^2+2 C_{82,2}^2+C_{82,3}^2} \nonumber \eeq 
\beq
\ket{\Psi_{83}} & = & \ket{3,\frac{1}{2} , \frac{15}{4} ,\Gamma_{5,1}} \nonumber \\ 
&=& \frac{1}{2 \sqrt{3}}
 \left ( \ket{0duu} + \ket{0udu} + \ket{0uud} + \ket{d0uu} + \ket{du0u} + \ket{duu0} \right . \nonumber \\
&& \hspace{3em} 
 + 
\left . \ket{u0du} + \ket{u0ud} + \ket{ud0u} + \ket{udu0} + \ket{uu0d} + \ket{uud0} \right ) 
\nonumber \eeq
\beq
\ket{\Psi_{84}} & = & \ket{3,\frac{1}{2} , \frac{3}{4} ,\Gamma_{5,2}} \nonumber \\ 
& = & \quad 
C_{84,1} \left ( 
\ket{020u} - \ket{02u0} - \ket{0u02} + \ket{0u20} - \ket{200u} + \ket{20u0} + \ket{u002} - \ket{u020} \right) 
 \nonumber \\
& & + 
C_{84,2} \left ( 
\ket{0duu} - \ket{d0uu} - \ket{uu0d} + \ket{uud0} \right) 
 \nonumber \\
& & + 
C_{84,3} \left ( 
\ket{0udu} + \ket{0uud} - \ket{du0u} + \ket{duu0} - \ket{u0du} - \ket{u0ud} - \ket{ud0u} + \ket{udu0} \right) 
\nonumber \eeq
\beq 
C_{84,1} &=& 
\frac{1}{2} \sqrt{\frac{3}{2}} t \nonumber \\
C_{84,2} &=& 
\frac{1}{8 \sqrt{6}} \left (
3 J+8 t+4 U-4 W+\sqrt{A_6} \right ) \nonumber \\
C_{84,3} &=& 
-\frac{1}{16 \sqrt{6}} \left (
3 J+8 t+4 U-4 W+\sqrt{A_6} \right ) \nonumber \\
 N_{84} &=& 2 \sqrt{2 C_{84,1}^2+C_{84,2}^2+2 C_{84,3}^2} \nonumber \eeq 
\beq
\ket{\Psi_{85}} & = & \ket{3,\frac{1}{2} , \frac{3}{4} ,\Gamma_{5,2}} \nonumber \\ 
& = & \quad 
C_{85,1} \left ( 
\ket{020u} - \ket{02u0} - \ket{0u02} + \ket{0u20} - \ket{200u} + \ket{20u0} + \ket{u002} - \ket{u020} \right) 
 \nonumber \\
& & + 
C_{85,2} \left ( 
\ket{0duu} - \ket{d0uu} - \ket{uu0d} + \ket{uud0} \right) 
 \nonumber \\
& & + 
C_{85,3} \left ( 
\ket{0udu} + \ket{0uud} - \ket{du0u} + \ket{duu0} - \ket{u0du} - \ket{u0ud} - \ket{ud0u} + \ket{udu0} \right) 
\nonumber \eeq
\beq 
C_{85,1} &=& 
\frac{1}{2} \sqrt{\frac{3}{2}} t \nonumber \\
C_{85,2} &=& 
\frac{1}{8 \sqrt{6}} \left (
3 J+8 t+4 U-4 W-\sqrt{A_6} \right ) \nonumber \\
C_{85,3} &=& 
-\frac{1}{16 \sqrt{6}} \left (
3 J+8 t+4 U-4 W-\sqrt{A_6} \right ) \nonumber \\
 N_{85} &=& 2 \sqrt{2 C_{85,1}^2+C_{85,2}^2+2 C_{85,3}^2} \nonumber \eeq 
\beq
\ket{\Psi_{86}} & = & \ket{3,\frac{1}{2} , \frac{15}{4} ,\Gamma_{5,2}} \nonumber \\ 
&=& \frac{1}{2 \sqrt{3}}
 \left ( \ket{0duu} + \ket{0udu} + \ket{0uud} - \ket{d0uu} - \ket{du0u} + \ket{duu0} \right . \nonumber \\
&& \hspace{3em} 
\left . -\ket{u0du} - \ket{u0ud} - \ket{ud0u} + \ket{udu0} - \ket{uu0d} + \ket{uud0} \right ) 
\nonumber \eeq
\beq
\ket{\Psi_{87}} & = & \ket{3,\frac{1}{2} , \frac{3}{4} ,\Gamma_{5,3}} \nonumber \\ 
& = & \quad 
C_{87,1} \left ( 
\ket{002u} - \ket{00u2} - \ket{020u} + \ket{0u02} + \ket{20u0} - \ket{2u00} - \ket{u020} + \ket{u200} \right) 
 \nonumber \\
& & + 
C_{87,2} \left ( 
\ket{0duu} + \ket{0udu} + \ket{d0uu} - \ket{du0u} + \ket{u0ud} - \ket{udu0} - \ket{uu0d} - \ket{uud0} \right) 
 \nonumber \\
& & + 
C_{87,3} \left ( 
\ket{0uud} - \ket{duu0} + \ket{u0du} - \ket{ud0u} \right) 
\nonumber \eeq
\beq 
C_{87,1} &=& 
-\frac{1}{2} \sqrt{\frac{3}{2}} t \nonumber \\
C_{87,2} &=& 
-\frac{1}{16 \sqrt{6}} \left (
3 J+8 t+4 U-4 W+\sqrt{A_6} \right ) \nonumber \\
C_{87,3} &=& 
\frac{1}{8 \sqrt{6}} \left (
3 J+8 t+4 U-4 W+\sqrt{A_6} \right ) \nonumber \\
 N_{87} &=& 2 \sqrt{2 C_{87,1}^2+2 C_{87,2}^2+C_{87,3}^2} \nonumber \eeq 
\beq
\ket{\Psi_{88}} & = & \ket{3,\frac{1}{2} , \frac{3}{4} ,\Gamma_{5,3}} \nonumber \\ 
& = & \quad 
C_{88,1} \left ( 
\ket{002u} - \ket{00u2} - \ket{020u} + \ket{0u02} + \ket{20u0} - \ket{2u00} - \ket{u020} + \ket{u200} \right) 
 \nonumber \\
& & + 
C_{88,2} \left ( 
\ket{0duu} + \ket{0udu} + \ket{d0uu} - \ket{du0u} + \ket{u0ud} - \ket{udu0} - \ket{uu0d} - \ket{uud0} \right) 
 \nonumber \\
& & + 
C_{88,3} \left ( 
\ket{0uud} - \ket{duu0} + \ket{u0du} - \ket{ud0u} \right) 
\nonumber \eeq
\beq 
C_{88,1} &=& 
-\frac{1}{2} \sqrt{\frac{3}{2}} t \nonumber \\
C_{88,2} &=& 
-\frac{1}{16 \sqrt{6}} \left (
3 J+8 t+4 U-4 W-\sqrt{A_6} \right ) \nonumber \\
C_{88,3} &=& 
\frac{1}{8 \sqrt{6}} \left (
3 J+8 t+4 U-4 W-\sqrt{A_6} \right ) \nonumber \\
 N_{88} &=& 2 \sqrt{2 C_{88,1}^2+2 C_{88,2}^2+C_{88,3}^2} \nonumber \eeq 
\beq
\ket{\Psi_{89}} & = & \ket{3,\frac{1}{2} , \frac{15}{4} ,\Gamma_{5,3}} \nonumber \\ 
&=& \frac{1}{2 \sqrt{3}}
 \left ( \ket{0duu} + \ket{0udu} + \ket{0uud} + \ket{d0uu} - \ket{du0u} - \ket{duu0} \right . \nonumber \\
&& \hspace{3em} 
 + 
\left . \ket{u0du} + \ket{u0ud} - \ket{ud0u} - \ket{udu0} - \ket{uu0d} - \ket{uud0} \right ) 
\nonumber \eeq
{\subsubsection{\boldmath Eigenvectors for ${\rm  N_e}=3$ and   ${\rm m_s}$= $\frac{3}{2} $.}
\beq
\ket{\Psi_{90}} & = & \ket{3,\frac{3}{2} , \frac{15}{4} ,\Gamma_2} \nonumber \\ 
&=& \frac{1}{2}
 \left ( \ket{0uuu} - \ket{u0uu} + \ket{uu0u} - \ket{uuu0} \right) \nonumber 
\eeq
\beq
\ket{\Psi_{91}} & = & \ket{3,\frac{3}{2} , \frac{15}{4} ,\Gamma_{5,1}} \nonumber \\ 
&=& \frac{1}{2}
 \left ( \ket{0uuu} + \ket{u0uu} + \ket{uu0u} + \ket{uuu0} \right) \nonumber 
\eeq
\beq
\ket{\Psi_{92}} & = & \ket{3,\frac{3}{2} , \frac{15}{4} ,\Gamma_{5,2}} \nonumber \\ 
&=& \frac{1}{2}
 \left ( \ket{0uuu} - \ket{u0uu} - \ket{uu0u} + \ket{uuu0} \right) \nonumber 
\eeq
\beq
\ket{\Psi_{93}} & = & \ket{3,\frac{3}{2} , \frac{15}{4} ,\Gamma_{5,3}} \nonumber \\ 
&=& \frac{1}{2}
 \left ( \ket{0uuu} + \ket{u0uu} - \ket{uu0u} - \ket{uuu0} \right) \nonumber 
\eeq
{\subsubsection{\boldmath Eigenvectors for ${\rm  N_e}=4$ and   ${\rm m_s}$= $-2$.}
\beq
\ket{\Psi_{94}} & = & \ket{4,-2,6,\Gamma_2} \nonumber \\ 
&=& 1
 \left ( \ket{dddd} \right) \nonumber 
\eeq
{\subsubsection{\boldmath Eigenvectors for ${\rm  N_e}=4$ and   ${\rm m_s}$= $-1$.}
\beq
\ket{\Psi_{95}} & = & \ket{4,-1,2,\Gamma_2} \nonumber \\ 
&=& \frac{1}{2 \sqrt{3}}
 \left ( \ket{02dd} - \ket{0d2d} + \ket{0dd2} - \ket{20dd} + \ket{2d0d} - \ket{2dd0} \right . \nonumber \\
&& \hspace{3em} 
 + 
\left . \ket{d02d} - \ket{d0d2} - \ket{d20d} + \ket{d2d0} + \ket{dd02} - \ket{dd20} \right ) 
\nonumber \eeq
\beq
\ket{\Psi_{96}} & = & \ket{4,-1,6,\Gamma_2} \nonumber \\ 
&=& \frac{1}{2}
 \left ( \ket{dddu} + \ket{ddud} + \ket{dudd} + \ket{uddd} \right) \nonumber 
\eeq
\beq
\ket{\Psi_{97}} & = & \ket{4,-1,2,\Gamma_{3,1}} \nonumber \\ 
&=& \frac{1}{2 \sqrt{2}}
 \left ( \ket{02dd} - \ket{0dd2} - \ket{20dd} + \ket{2dd0} - \ket{d02d} + \ket{d20d} + \ket{dd02} - \ket{dd20} \right) \nonumber 
\eeq
\beq
\ket{\Psi_{98}} & = & \ket{4,-1,2,\Gamma_{3,2}} \nonumber \\ 
& = & \quad 
C_{98,1} \left ( 
\ket{02dd} + \ket{0dd2} - \ket{20dd} - \ket{2dd0} + \ket{d02d} - \ket{d20d} + \ket{dd02} - \ket{dd20} \right) 
 \nonumber \\
& & + 
C_{98,2} \left ( 
\ket{0d2d} - \ket{2d0d} + \ket{d0d2} - \ket{d2d0} \right) 
\nonumber \eeq
\beq 
C_{98,1} &=& 
-\frac{1}{2 \sqrt{6}} \nonumber \\
C_{98,2} &=& 
-\frac{1}{\sqrt{6}} \nonumber \\
 N_{98} &=& 2 \sqrt{2 C_{98,1}^2+C_{98,2}^2} \nonumber \eeq 
\beq
\ket{\Psi_{99}} & = & \ket{4,-1,2,\Gamma_{4,1}} \nonumber \\ 
&=& \frac{1}{2 \sqrt{2}}
 \left ( \ket{0d2d} + \ket{0dd2} + \ket{2d0d} + \ket{2dd0} + \ket{d02d} + \ket{d0d2} + \ket{d20d} + \ket{d2d0} \right) \nonumber 
\eeq
\beq
\ket{\Psi_{100}} & = & \ket{4,-1,2,\Gamma_{4,2}} \nonumber \\ 
&=& \frac{1}{2 \sqrt{2}}
 \left ( \ket{02dd} + \ket{0d2d} + \ket{20dd} + \ket{2d0d} - \ket{d0d2} - \ket{d2d0} - \ket{dd02} - \ket{dd20} \right) \nonumber 
\eeq
\beq
\ket{\Psi_{101}} & = & \ket{4,-1,2,\Gamma_{4,3}} \nonumber \\ 
&=& \frac{1}{2 \sqrt{2}}
 \left ( \ket{02dd} - \ket{0dd2} + \ket{20dd} - \ket{2dd0} + \ket{d02d} + \ket{d20d} + \ket{dd02} + \ket{dd20} \right) \nonumber 
\eeq
\beq
\ket{\Psi_{102}} & = & \ket{4,-1,2,\Gamma_{5,1}} \nonumber \\ 
& = & \quad 
C_{102,1} \left ( 
\ket{02dd} + \ket{0dd2} + \ket{20dd} + \ket{2dd0} - \ket{d02d} - \ket{d20d} + \ket{dd02} + \ket{dd20} \right) 
 \nonumber \\
& & + 
C_{102,2} \left ( 
\ket{0d2d} - \ket{2d0d} - \ket{d0d2} + \ket{d2d0} \right) 
 \nonumber \\
& & + 
C_{102,3} \left ( 
\ket{dddu} - \ket{ddud} + \ket{dudd} - \ket{uddd} \right) 
\nonumber \eeq
\beq 
C_{102,1} &=& 
-\frac{1}{12} \left (
3 J t+4 U t-4 W t+2 \cos \left(\theta _4\right) \sqrt{A_3} t \right ) \nonumber \\
C_{102,2} &=& 
-4 t^2 \nonumber \\
C_{102,3} &=& 
-\frac{1}{288} \left (
9 J^2+24 U J-24 W J-1152 t^2-48 U^2-1056 U W \right . \nonumber \\
&& \hspace{1cm} 
 \left . -4080 W^2+240 \cos \left(\theta _4\right) \sqrt{A_3} W+4 A_{17}^2+12 J \cos \left(\theta _4\right) \sqrt{A_3}+48 U \cos \left(\theta _4\right) \sqrt{A_3} \right) \nonumber \\
 N_{102} &=& 2 \sqrt{2 C_{102,1}^2+C_{102,2}^2+C_{102,3}^2} \nonumber \eeq 
\beq
\ket{\Psi_{103}} & = & \ket{4,-1,2,\Gamma_{5,1}} \nonumber \\ 
& = & \quad 
C_{103,1} \left ( 
\ket{02dd} + \ket{0dd2} + \ket{20dd} + \ket{2dd0} - \ket{d02d} - \ket{d20d} + \ket{dd02} + \ket{dd20} \right) 
 \nonumber \\
& & + 
C_{103,2} \left ( 
\ket{0d2d} - \ket{2d0d} - \ket{d0d2} + \ket{d2d0} \right) 
 \nonumber \\
& & + 
C_{103,3} \left ( 
\ket{dddu} - \ket{ddud} + \ket{dudd} - \ket{uddd} \right) 
\nonumber \eeq
\beq 
C_{103,1} &=& 
\frac{1}{12} \left (
-3 J t-4 U t+4 W t+\cos \left(\theta _4\right) \sqrt{A_3} t+\sqrt{3} \sin \left(\theta _4\right) \sqrt{A_3} t \right ) \nonumber \\
C_{103,2} &=& 
-4 t^2 \nonumber \\
C_{103,3} &=& 
-\frac{1}{32} \left (
J^2+8 U J+40 W J-128 t^2+16 U^2+160 U W \right . \nonumber \\
&& \hspace{1cm} 
 \left . 400 W^2-160 A_{21} W+16 A_{21}^2-8 J A_{21}-32 U A_{21} \right) \nonumber \\
 N_{103} &=& 2 \sqrt{2 C_{103,1}^2+C_{103,2}^2+C_{103,3}^2} \nonumber \eeq 
\beq
\ket{\Psi_{104}} & = & \ket{4,-1,2,\Gamma_{5,1}} \nonumber \\ 
& = & \quad 
C_{104,1} \left ( 
\ket{02dd} + \ket{0dd2} + \ket{20dd} + \ket{2dd0} - \ket{d02d} - \ket{d20d} + \ket{dd02} + \ket{dd20} \right) 
 \nonumber \\
& & + 
C_{104,2} \left ( 
\ket{0d2d} - \ket{2d0d} - \ket{d0d2} + \ket{d2d0} \right) 
 \nonumber \\
& & + 
C_{104,3} \left ( 
\ket{dddu} - \ket{ddud} + \ket{dudd} - \ket{uddd} \right) 
\nonumber \eeq
\beq 
C_{104,1} &=& 
-\frac{1}{12} \left (
3 J t+4 U t-4 W t-\cos \left(\theta _4\right) \sqrt{A_3} t+\sqrt{3} \sin \left(\theta _4\right) \sqrt{A_3} t \right ) \nonumber \\
C_{104,2} &=& 
-4 t^2 \nonumber \\
C_{104,3} &=& 
-\frac{1}{32} \left (
J^2+8 U J+40 W J-128 t^2+16 U^2+160 U W \right . \nonumber \\
&& \hspace{1cm} 
 \left . 400 W^2-160 A_{20} W+16 A_{20}^2-8 J A_{20}-32 U A_{20} \right) \nonumber \\
 N_{104} &=& 2 \sqrt{2 C_{104,1}^2+C_{104,2}^2+C_{104,3}^2} \nonumber \eeq 
\beq
\ket{\Psi_{105}} & = & \ket{4,-1,2,\Gamma_{5,2}} \nonumber \\ 
& = & \quad 
C_{105,1} \left ( 
\ket{02dd} - \ket{20dd} - \ket{dd02} + \ket{dd20} \right) 
 \nonumber \\
& & + 
C_{105,2} \left ( 
\ket{0d2d} - \ket{0dd2} + \ket{2d0d} - \ket{2dd0} - \ket{d02d} + \ket{d0d2} - \ket{d20d} + \ket{d2d0} \right) 
 \nonumber \\
& & + 
C_{105,3} \left ( 
\ket{dddu} + \ket{ddud} - \ket{dudd} - \ket{uddd} \right) 
\nonumber \eeq
\beq 
C_{105,1} &=& 
4 t^2 \nonumber \\
C_{105,2} &=& 
\frac{1}{12} \left (
3 J t+4 U t-4 W t+2 \cos \left(\theta _4\right) \sqrt{A_3} t \right ) \nonumber \\
C_{105,3} &=& 
-\frac{1}{288} \left (
9 J^2+24 U J-24 W J-1152 t^2-48 U^2-1056 U W \right . \nonumber \\
&& \hspace{1cm} 
 \left . -4080 W^2+240 \cos \left(\theta _4\right) \sqrt{A_3} W+4 A_{17}^2+12 J \cos \left(\theta _4\right) \sqrt{A_3}+48 U \cos \left(\theta _4\right) \sqrt{A_3} \right) \nonumber \\
 N_{105} &=& 2 \sqrt{C_{105,1}^2+2 C_{105,2}^2+C_{105,3}^2} \nonumber \eeq 
\beq
\ket{\Psi_{106}} & = & \ket{4,-1,2,\Gamma_{5,2}} \nonumber \\ 
& = & \quad 
C_{106,1} \left ( 
\ket{02dd} - \ket{20dd} - \ket{dd02} + \ket{dd20} \right) 
 \nonumber \\
& & + 
C_{106,2} \left ( 
\ket{0d2d} - \ket{0dd2} + \ket{2d0d} - \ket{2dd0} - \ket{d02d} + \ket{d0d2} - \ket{d20d} + \ket{d2d0} \right) 
 \nonumber \\
& & + 
C_{106,3} \left ( 
\ket{dddu} + \ket{ddud} - \ket{dudd} - \ket{uddd} \right) 
\nonumber \eeq
\beq 
C_{106,1} &=& 
4 t^2 \nonumber \\
C_{106,2} &=& 
-\frac{1}{12} \left (
-3 J t-4 U t+4 W t+\cos \left(\theta _4\right) \sqrt{A_3} t+\sqrt{3} \sin \left(\theta _4\right) \sqrt{A_3} t \right ) \nonumber \\
C_{106,3} &=& 
-\frac{1}{32} \left (
J^2+8 U J+40 W J-128 t^2+16 U^2+160 U W \right . \nonumber \\
&& \hspace{1cm} 
 \left . 400 W^2-160 A_{21} W+16 A_{21}^2-8 J A_{21}-32 U A_{21} \right) \nonumber \\
 N_{106} &=& 2 \sqrt{C_{106,1}^2+2 C_{106,2}^2+C_{106,3}^2} \nonumber \eeq 
\beq
\ket{\Psi_{107}} & = & \ket{4,-1,2,\Gamma_{5,2}} \nonumber \\ 
& = & \quad 
C_{107,1} \left ( 
\ket{02dd} - \ket{20dd} - \ket{dd02} + \ket{dd20} \right) 
 \nonumber \\
& & + 
C_{107,2} \left ( 
\ket{0d2d} - \ket{0dd2} + \ket{2d0d} - \ket{2dd0} - \ket{d02d} + \ket{d0d2} - \ket{d20d} + \ket{d2d0} \right) 
 \nonumber \\
& & + 
C_{107,3} \left ( 
\ket{dddu} + \ket{ddud} - \ket{dudd} - \ket{uddd} \right) 
\nonumber \eeq
\beq 
C_{107,1} &=& 
4 t^2 \nonumber \\
C_{107,2} &=& 
\frac{1}{12} \left (
3 J t+4 U t-4 W t-\cos \left(\theta _4\right) \sqrt{A_3} t+\sqrt{3} \sin \left(\theta _4\right) \sqrt{A_3} t \right ) \nonumber \\
C_{107,3} &=& 
-\frac{1}{32} \left (
J^2+8 U J+40 W J-128 t^2+16 U^2+160 U W \right . \nonumber \\
&& \hspace{1cm} 
 \left . 400 W^2-160 A_{20} W+16 A_{20}^2-8 J A_{20}-32 U A_{20} \right) \nonumber \\
 N_{107} &=& 2 \sqrt{C_{107,1}^2+2 C_{107,2}^2+C_{107,3}^2} \nonumber \eeq 
\beq
\ket{\Psi_{108}} & = & \ket{4,-1,2,\Gamma_{5,3}} \nonumber \\ 
& = & \quad 
C_{108,1} \left ( 
\ket{02dd} - \ket{0d2d} + \ket{20dd} - \ket{2d0d} + \ket{d0d2} + \ket{d2d0} - \ket{dd02} - \ket{dd20} \right) 
 \nonumber \\
& & + 
C_{108,2} \left ( 
\ket{0dd2} - \ket{2dd0} - \ket{d02d} + \ket{d20d} \right) 
 \nonumber \\
& & + 
C_{108,3} \left ( 
\ket{dddu} - \ket{ddud} - \ket{dudd} + \ket{uddd} \right) 
\nonumber \eeq
\beq 
C_{108,1} &=& 
-\frac{1}{12} \left (
3 J t+4 U t-4 W t+2 \cos \left(\theta _4\right) \sqrt{A_3} t \right ) \nonumber \\
C_{108,2} &=& 
4 t^2 \nonumber \\
C_{108,3} &=& 
\frac{1}{288} \left (
9 J^2+24 U J-24 W J-1152 t^2-48 U^2-1056 U W \right . \nonumber \\
&& \hspace{1cm} 
 \left . -4080 W^2+240 \cos \left(\theta _4\right) \sqrt{A_3} W+4 A_{17}^2+12 J \cos \left(\theta _4\right) \sqrt{A_3}+48 U \cos \left(\theta _4\right) \sqrt{A_3} \right) \nonumber \\
 N_{108} &=& 2 \sqrt{2 C_{108,1}^2+C_{108,2}^2+C_{108,3}^2} \nonumber \eeq 
\beq
\ket{\Psi_{109}} & = & \ket{4,-1,2,\Gamma_{5,3}} \nonumber \\ 
& = & \quad 
C_{109,1} \left ( 
\ket{02dd} - \ket{0d2d} + \ket{20dd} - \ket{2d0d} + \ket{d0d2} + \ket{d2d0} - \ket{dd02} - \ket{dd20} \right) 
 \nonumber \\
& & + 
C_{109,2} \left ( 
\ket{0dd2} - \ket{2dd0} - \ket{d02d} + \ket{d20d} \right) 
 \nonumber \\
& & + 
C_{109,3} \left ( 
\ket{dddu} - \ket{ddud} - \ket{dudd} + \ket{uddd} \right) 
\nonumber \eeq
\beq 
C_{109,1} &=& 
\frac{1}{12} \left (
-3 J t-4 U t+4 W t+\cos \left(\theta _4\right) \sqrt{A_3} t+\sqrt{3} \sin \left(\theta _4\right) \sqrt{A_3} t \right ) \nonumber \\
C_{109,2} &=& 
4 t^2 \nonumber \\
C_{109,3} &=& 
\frac{1}{32} \left (
J^2+8 U J+40 W J-128 t^2+16 U^2+160 U W \right . \nonumber \\
&& \hspace{1cm} 
 \left . 400 W^2-160 A_{21} W+16 A_{21}^2-8 J A_{21}-32 U A_{21} \right) \nonumber \\
 N_{109} &=& 2 \sqrt{2 C_{109,1}^2+C_{109,2}^2+C_{109,3}^2} \nonumber \eeq 
\beq
\ket{\Psi_{110}} & = & \ket{4,-1,2,\Gamma_{5,3}} \nonumber \\ 
& = & \quad 
C_{110,1} \left ( 
\ket{02dd} - \ket{0d2d} + \ket{20dd} - \ket{2d0d} + \ket{d0d2} + \ket{d2d0} - \ket{dd02} - \ket{dd20} \right) 
 \nonumber \\
& & + 
C_{110,2} \left ( 
\ket{0dd2} - \ket{2dd0} - \ket{d02d} + \ket{d20d} \right) 
 \nonumber \\
& & + 
C_{110,3} \left ( 
\ket{dddu} - \ket{ddud} - \ket{dudd} + \ket{uddd} \right) 
\nonumber \eeq
\beq 
C_{110,1} &=& 
-\frac{1}{12} \left (
3 J t+4 U t-4 W t-\cos \left(\theta _4\right) \sqrt{A_3} t+\sqrt{3} \sin \left(\theta _4\right) \sqrt{A_3} t \right ) \nonumber \\
C_{110,2} &=& 
4 t^2 \nonumber \\
C_{110,3} &=& 
\frac{1}{32} \left (
J^2+8 U J+40 W J-128 t^2+16 U^2+160 U W \right . \nonumber \\
&& \hspace{1cm} 
 \left . 400 W^2-160 A_{20} W+16 A_{20}^2-8 J A_{20}-32 U A_{20} \right) \nonumber \\
 N_{110} &=& 2 \sqrt{2 C_{110,1}^2+C_{110,2}^2+C_{110,3}^2} \nonumber \eeq 
{\subsubsection{\boldmath Eigenvectors for ${\rm  N_e}=4$ and   ${\rm m_s}$= $0$.}
\beq
\ket{\Psi_{111}} & = & \ket{4,0,0,\Gamma_1} \nonumber \\ 
& = & \quad 
C_{111,1} \left ( 
\ket{0022} + \ket{0202} + \ket{0220} + \ket{2002} + \ket{2020} + \ket{2200} \right) 
 \nonumber \\
& & + 
C_{111,2} \left ( 
\ket{02du} - \ket{02ud} + \ket{0d2u} + \ket{0du2} - \ket{0u2d} - \ket{0ud2} + \ket{20du} - \ket{20ud} \right . \nonumber \\
&& \hspace{3em} 
 + 
\left . \ket{2d0u} + \ket{2du0} - \ket{2u0d} - \ket{2ud0} + \ket{d02u} + \ket{d0u2} + \ket{d20u} + \ket{d2u0} \right . \nonumber \\
&& \hspace{3em} 
 + 
\left . \ket{du02} + \ket{du20} - \ket{u02d} - \ket{u0d2} - \ket{u20d} - \ket{u2d0} - \ket{ud02} - \ket{ud20} \right ) \nonumber \\
\nonumber \eeq
\beq 
C_{111,1} &=& 
2 \sqrt{\frac{2}{3}} t \nonumber \\
C_{111,2} &=& 
\frac{1}{16 \sqrt{6}} \left (
3 J+4 U-4 W+\sqrt{A_4} \right ) \nonumber \\
 N_{111} &=& \sqrt{6 C_{111,1}^2+24 C_{111,2}^2} \nonumber \eeq 
\beq
\ket{\Psi_{112}} & = & \ket{4,0,0,\Gamma_1} \nonumber \\ 
& = & \quad 
C_{112,1} \left ( 
\ket{0022} + \ket{0202} + \ket{0220} + \ket{2002} + \ket{2020} + \ket{2200} \right) 
 \nonumber \\
& & + 
C_{112,2} \left ( 
\ket{02du} - \ket{02ud} + \ket{0d2u} + \ket{0du2} - \ket{0u2d} - \ket{0ud2} + \ket{20du} - \ket{20ud} \right . \nonumber \\
&& \hspace{3em} 
 + 
\left . \ket{2d0u} + \ket{2du0} - \ket{2u0d} - \ket{2ud0} + \ket{d02u} + \ket{d0u2} + \ket{d20u} + \ket{d2u0} \right . \nonumber \\
&& \hspace{3em} 
 + 
\left . \ket{du02} + \ket{du20} - \ket{u02d} - \ket{u0d2} - \ket{u20d} - \ket{u2d0} - \ket{ud02} - \ket{ud20} \right ) \nonumber \\
\nonumber \eeq
\beq 
C_{112,1} &=& 
2 \sqrt{\frac{2}{3}} t \nonumber \\
C_{112,2} &=& 
\frac{1}{16 \sqrt{6}} \left (
3 J+4 U-4 W-\sqrt{A_4} \right ) \nonumber \\
 N_{112} &=& \sqrt{6 C_{112,1}^2+24 C_{112,2}^2} \nonumber \eeq 
\beq
\ket{\Psi_{113}} & = & \ket{4,0,2,\Gamma_2} \nonumber \\ 
&=& \frac{1}{2 \sqrt{6}}
\ket{02du} + \ket{02ud} - \ket{0d2u} + \ket{0du2} - \ket{0u2d} + \ket{0ud2} - \ket{20du} - \ket{20ud}  \left (  \right . \nonumber \\
&& \hspace{3em} 
 + 
\left . \ket{2d0u} - \ket{2du0} + \ket{2u0d} - \ket{2ud0} + \ket{d02u} - \ket{d0u2} - \ket{d20u} + \ket{d2u0}\right . \nonumber \\
&& \hspace{3em} 
 + 
\left . \ket{du02} - \ket{du20} + \ket{u02d} - \ket{u0d2} - \ket{u20d} + \ket{u2d0} + \ket{ud02} - \ket{ud20}\right ) \nonumber 
\eeq
\beq
\ket{\Psi_{114}} & = & \ket{4,0,6,\Gamma_2} \nonumber \\ 
&=& \frac{1}{\sqrt{6}}
 \left ( \ket{dduu} + \ket{dudu} + \ket{duud} + \ket{uddu} + \ket{udud} + \ket{uudd} \right) \nonumber 
\eeq
\beq
\ket{\Psi_{115}} & = & \ket{4,0,0,\Gamma_{3,1}} \nonumber \\ 
& = & \quad 
C_{115,1} \left ( 
\ket{0022} + \ket{0220} + \ket{2002} + \ket{2200} \right) 
 \nonumber \\
& & + 
C_{115,2} \left ( 
\ket{0202} + \ket{2020} \right) 
 \nonumber \\
& & + 
C_{115,3} \left ( 
\ket{02du} - \ket{02ud} + \ket{0du2} - \ket{0ud2} + \ket{20du} - \ket{20ud} + \ket{2du0} - \ket{2ud0} \right . \nonumber \\
&& \hspace{3em} 
 + 
\left . \ket{d02u} + \ket{d20u} + \ket{du02} + \ket{du20} - \ket{u02d} - \ket{u20d} - \ket{ud02} - \ket{ud20}\right ) 
\nonumber \\
& & + 
C_{115,4} \left ( 
\ket{0d2u} - \ket{0u2d} + \ket{2d0u} - \ket{2u0d} + \ket{d0u2} + \ket{d2u0} - \ket{u0d2} - \ket{u2d0} \right) 
 \nonumber \\
& & + 
C_{115,5} \left ( 
\ket{dduu} - \ket{duud} - \ket{uddu} + \ket{uudd} \right) 
\nonumber \eeq
\beq 
C_{115,1} &=& 
-2 t^2 \nonumber \\
C_{115,2} &=& 
4 t^2 \nonumber \\
C_{115,3} &=& 
\frac{1}{24} \left (
9 J t+12 U t-12 W t+2 \cos \left(\theta _2\right) \sqrt{A_5} t \right ) \nonumber \\
C_{115,4} &=& 
-\frac{1}{12} \left (
9 J t+12 U t-12 W t+2 \cos \left(\theta _2\right) \sqrt{A_5} t \right ) \nonumber \\
C_{115,5} &=& 
-\frac{1}{288} \left (
576 t^2-288 U^2-2880 W^2+216 J U+432 J W-2016 U W \right . \nonumber \\
&& \hspace{1cm} 
+
 \left . -A_{22}^2+9 J A_{22}-36 U A_{22}-108 W A_{22} \right) \nonumber \\
 N_{115} &=& \sqrt{4 C_{115,1}^2+2 C_{115,2}^2+16 C_{115,3}^2+8 C_{115,4}^2+4 C_{115,5}^2} \nonumber \eeq 
\beq
\ket{\Psi_{116}} & = & \ket{4,0,0,\Gamma_{3,1}} \nonumber \\ 
& = & \quad 
C_{116,1} \left ( 
\ket{0022} + \ket{0220} + \ket{2002} + \ket{2200} \right) 
 \nonumber \\
& & + 
C_{116,2} \left ( 
\ket{0202} + \ket{2020} \right) 
 \nonumber \\
& & + 
C_{116,3} \left ( 
\ket{02du} - \ket{02ud} + \ket{0du2} - \ket{0ud2} + \ket{20du} - \ket{20ud} + \ket{2du0} - \ket{2ud0} \right . \nonumber \\
&& \hspace{3em} 
 + 
\left . \ket{d02u} + \ket{d20u} + \ket{du02} + \ket{du20} - \ket{u02d} - \ket{u20d} - \ket{ud02} - \ket{ud20}\right ) 
\nonumber \\
& & + 
C_{116,4} \left ( 
\ket{0d2u} - \ket{0u2d} + \ket{2d0u} - \ket{2u0d} + \ket{d0u2} + \ket{d2u0} - \ket{u0d2} - \ket{u2d0} \right) 
 \nonumber \\
& & + 
C_{116,5} \left ( 
\ket{dduu} - \ket{duud} - \ket{uddu} + \ket{uudd} \right) 
\nonumber \eeq
\beq 
C_{116,1} &=& 
-2 t^2 \nonumber \\
C_{116,2} &=& 
4 t^2 \nonumber \\
C_{116,3} &=& 
-\frac{1}{24} \left (
-9 J t-12 U t+12 W t+\sqrt{3} \cos \left(\theta _2\right) \sqrt{A_2} t+3 \sin \left(\theta _2\right) \sqrt{A_2} t \right ) \nonumber \\
C_{116,4} &=& 
\frac{1}{12} \left (
-9 J t-12 U t+12 W t+\sqrt{3} \cos \left(\theta _2\right) \sqrt{A_2} t+3 \sin \left(\theta _2\right) \sqrt{A_2} t \right ) \nonumber \\
C_{116,5} &=& 
-\frac{1}{8} \left (
16 t^2-8 U^2-80 W^2+6 J U+12 J W-56 U W \right . \nonumber \\
&& \hspace{1cm} 
 \left . -4 A_{15}^2-3 J A_{15}+12 U A_{15}+36 W A_{15} \right) \nonumber \\
 N_{116} &=& \sqrt{4 C_{116,1}^2+2 C_{116,2}^2+16 C_{116,3}^2+8 C_{116,4}^2+4 C_{116,5}^2} \nonumber \eeq 
\beq
\ket{\Psi_{117}} & = & \ket{4,0,0,\Gamma_{3,1}} \nonumber \\ 
& = & \quad 
C_{117,1} \left ( 
\ket{0022} + \ket{0220} + \ket{2002} + \ket{2200} \right) 
 \nonumber \\
& & + 
C_{117,2} \left ( 
\ket{0202} + \ket{2020} \right) 
 \nonumber \\
& & + 
C_{117,3} \left ( 
\ket{02du} - \ket{02ud} + \ket{0du2} - \ket{0ud2} + \ket{20du} - \ket{20ud} + \ket{2du0} - \ket{2ud0} \right . \nonumber \\
&& \hspace{3em} 
 + 
\left . \ket{d02u} + \ket{d20u} + \ket{du02} + \ket{du20} - \ket{u02d} - \ket{u20d} - \ket{ud02} - \ket{ud20}\right ) 
\nonumber \\
& & + 
C_{117,4} \left ( 
\ket{0d2u} - \ket{0u2d} + \ket{2d0u} - \ket{2u0d} + \ket{d0u2} + \ket{d2u0} - \ket{u0d2} - \ket{u2d0} \right) 
 \nonumber \\
& & + 
C_{117,5} \left ( 
\ket{dduu} - \ket{duud} - \ket{uddu} + \ket{uudd} \right) 
\nonumber \eeq
\beq 
C_{117,1} &=& 
-2 t^2 \nonumber \\
C_{117,2} &=& 
4 t^2 \nonumber \\
C_{117,3} &=& 
-\frac{1}{24} \left (
-9 J t-12 U t+12 W t+\sqrt{3} \cos \left(\theta _2\right) \sqrt{A_2} t-3 \sin \left(\theta _2\right) \sqrt{A_2} t \right ) \nonumber \\
C_{117,4} &=& 
\frac{1}{12} \left (
-9 J t-12 U t+12 W t+\sqrt{3} \cos \left(\theta _2\right) \sqrt{A_2} t-3 \sin \left(\theta _2\right) \sqrt{A_2} t \right ) \nonumber \\
C_{117,5} &=& 
-\frac{1}{8} \left (
16 t^2-8 U^2-80 W^2+6 J U+12 J W-56 U W \right . \nonumber \\
&& \hspace{1cm} 
 \left . -4 A_{14}^2-3 J A_{14}+12 U A_{14}+36 W A_{14} \right) \nonumber \\
 N_{117} &=& \sqrt{4 C_{117,1}^2+2 C_{117,2}^2+16 C_{117,3}^2+8 C_{117,4}^2+4 C_{117,5}^2} \nonumber \eeq 
\beq
\ket{\Psi_{118}} & = & \ket{4,0,2,\Gamma_{3,1}} \nonumber \\ 
&=& \frac{1}{4}
 \left ( \ket{02du} + \ket{02ud} - \ket{0du2} - \ket{0ud2} - \ket{20du} - \ket{20ud} + \ket{2du0} + \ket{2ud0} \right . \nonumber \\
&& \hspace{3em} 
\left . -\ket{d02u} + \ket{d20u} + \ket{du02} - \ket{du20} - \ket{u02d} + \ket{u20d} + \ket{ud02} - \ket{ud20} \right ) 
\nonumber \eeq
\beq
\ket{\Psi_{119}} & = & \ket{4,0,0,\Gamma_{3,2}} \nonumber \\ 
& = & \quad 
C_{119,1} \left ( 
\ket{0022} - \ket{0220} - \ket{2002} + \ket{2200} \right) 
 \nonumber \\
& & + 
C_{119,2} \left ( 
\ket{02du} - \ket{02ud} - \ket{0du2} + \ket{0ud2} + \ket{20du} - \ket{20ud} - \ket{2du0} + \ket{2ud0} \right . \nonumber \\
&& \hspace{3em} 
\left . -\ket{d02u} - \ket{d20u} + \ket{du02} + \ket{du20} + \ket{u02d} + \ket{u20d} - \ket{ud02} - \ket{ud20}\right ) 
\nonumber \\
& & + 
C_{119,3} \left ( 
\ket{dduu} + \ket{duud} + \ket{uddu} + \ket{uudd} \right) 
 \nonumber \\
& & + 
C_{119,4} \left ( 
\ket{dudu} + \ket{udud} \right) 
\nonumber \eeq
\beq 
C_{119,1} &=& 
-\frac{1}{288} \left (
81 J^2-216 U J-1080 W J+1728 t^2+144 U^2+1440 U W \right . \nonumber \\
&& \hspace{1cm} 
+
 \left . 3600 W^2-264 \cos \left(\theta _2\right) \sqrt{A_5} W-A_{22}^2+54 J \cos \left(\theta _2\right) \sqrt{A_5}-24 U \cos \left(\theta _2\right) \sqrt{A_5} \right) \nonumber \\
C_{119,2} &=& 
\frac{1}{24} \left (
9 J t+12 U t-12 W t-2 \cos \left(\theta _2\right) \sqrt{A_5} t \right ) \nonumber \\
C_{119,3} &=& 
2 t^2 \nonumber \\
C_{119,4} &=& 
-4 t^2 \nonumber \\
 N_{119} &=& \sqrt{4 C_{119,1}^2+16 C_{119,2}^2+4 C_{119,3}^2+2 C_{119,4}^2} \nonumber \eeq 
\beq
\ket{\Psi_{120}} & = & \ket{4,0,0,\Gamma_{3,2}} \nonumber \\ 
& = & \quad 
C_{120,1} \left ( 
\ket{0022} - \ket{0220} - \ket{2002} + \ket{2200} \right) 
 \nonumber \\
& & + 
C_{120,2} \left ( 
\ket{02du} - \ket{02ud} - \ket{0du2} + \ket{0ud2} + \ket{20du} - \ket{20ud} - \ket{2du0} + \ket{2ud0} \right . \nonumber \\
&& \hspace{3em} 
\left . -\ket{d02u} - \ket{d20u} + \ket{du02} + \ket{du20} + \ket{u02d} + \ket{u20d} - \ket{ud02} - \ket{ud20}\right ) 
\nonumber \\
& & + 
C_{120,3} \left ( 
\ket{dduu} + \ket{duud} + \ket{uddu} + \ket{uudd} \right) 
 \nonumber \\
& & + 
C_{120,4} \left ( 
\ket{dudu} + \ket{udud} \right) 
\nonumber \eeq
\beq 
C_{120,1} &=& 
\frac{1}{16} \left (
9 J^2-12 U J-96 W J-96 t^2+240 W^2+48 U W \right . \nonumber \\
&& \hspace{1cm} 
+
 \left . 8 A_{15}^2+18 J A_{15}-8 U A_{15}-88 W A_{15} \right) \nonumber \\
C_{120,2} &=& 
\frac{1}{24} \left (
9 J t+12 U t-12 W t+\sqrt{3} \cos \left(\theta _2\right) \sqrt{A_2} t+3 \sin \left(\theta _2\right) \sqrt{A_2} t \right ) \nonumber \\
C_{120,3} &=& 
2 t^2 \nonumber \\
C_{120,4} &=& 
-4 t^2 \nonumber \\
 N_{120} &=& \sqrt{4 C_{120,1}^2+16 C_{120,2}^2+4 C_{120,3}^2+2 C_{120,4}^2} \nonumber \eeq 
\beq
\ket{\Psi_{121}} & = & \ket{4,0,0,\Gamma_{3,2}} \nonumber \\ 
& = & \quad 
C_{121,1} \left ( 
\ket{0022} - \ket{0220} - \ket{2002} + \ket{2200} \right) 
 \nonumber \\
& & + 
C_{121,2} \left ( 
\ket{02du} - \ket{02ud} - \ket{0du2} + \ket{0ud2} + \ket{20du} - \ket{20ud} - \ket{2du0} + \ket{2ud0} \right . \nonumber \\
&& \hspace{3em} 
\left . -\ket{d02u} - \ket{d20u} + \ket{du02} + \ket{du20} + \ket{u02d} + \ket{u20d} - \ket{ud02} - \ket{ud20}\right ) 
\nonumber \\
& & + 
C_{121,3} \left ( 
\ket{dduu} + \ket{duud} + \ket{uddu} + \ket{uudd} \right) 
 \nonumber \\
& & + 
C_{121,4} \left ( 
\ket{dudu} + \ket{udud} \right) 
\nonumber \eeq
\beq 
C_{121,1} &=& 
\frac{1}{16} \left (
9 J^2-12 U J-96 W J-96 t^2+240 W^2+48 U W \right . \nonumber \\
&& \hspace{1cm} 
+
 \left . 8 A_{14}^2+18 J A_{14}-8 U A_{14}-88 W A_{14} \right) \nonumber \\
C_{121,2} &=& 
\frac{1}{24} \left (
9 J t+12 U t-12 W t+\sqrt{3} \cos \left(\theta _2\right) \sqrt{A_2} t-3 \sin \left(\theta _2\right) \sqrt{A_2} t \right ) \nonumber \\
C_{121,3} &=& 
2 t^2 \nonumber \\
C_{121,4} &=& 
-4 t^2 \nonumber \\
 N_{121} &=& \sqrt{4 C_{121,1}^2+16 C_{121,2}^2+4 C_{121,3}^2+2 C_{121,4}^2} \nonumber \eeq 
\beq
\ket{\Psi_{122}} & = & \ket{4,0,2,\Gamma_{3,2}} \nonumber \\ 
& = & \quad 
C_{122,1} \left ( 
\ket{02du} + \ket{02ud} + \ket{0du2} + \ket{0ud2} - \ket{20du} - \ket{20ud} - \ket{2du0} - \ket{2ud0} \right . \nonumber \\
&& \hspace{3em} 
 + 
\left . \ket{d02u} - \ket{d20u} + \ket{du02} - \ket{du20} + \ket{u02d} - \ket{u20d} + \ket{ud02} - \ket{ud20}\right ) 
\nonumber \\
& & + 
C_{122,2} \left ( 
\ket{0d2u} + \ket{0u2d} - \ket{2d0u} - \ket{2u0d} + \ket{d0u2} - \ket{d2u0} + \ket{u0d2} - \ket{u2d0} \right) 
\nonumber \eeq
\beq 
C_{122,1} &=& 
-\frac{1}{4 \sqrt{3}} \nonumber \\
C_{122,2} &=& 
-\frac{1}{2 \sqrt{3}} \nonumber \\
 N_{122} &=& \sqrt{16 C_{122,1}^2+8 C_{122,2}^2} \nonumber \eeq 
\beq
\ket{\Psi_{123}} & = & \ket{4,0,0,\Gamma_{4,1}} \nonumber \\ 
& = & \quad 
C_{123,1} \left ( 
\ket{0022} - \ket{2200} \right) 
 \nonumber \\
& & + 
C_{123,2} \left ( 
\ket{02du} - \ket{02ud} + \ket{20du} - \ket{20ud} - \ket{du02} - \ket{du20} + \ket{ud02} + \ket{ud20} \right) 
 \nonumber \\
& & + 
C_{123,3} \left ( 
\ket{0d2u} + \ket{0du2} - \ket{0u2d} - \ket{0ud2} - \ket{2d0u} - \ket{2du0} + \ket{2u0d} + \ket{2ud0} \right . \nonumber \\
&& \hspace{3em} 
 + 
\left . \ket{d02u} + \ket{d0u2} - \ket{d20u} - \ket{d2u0} - \ket{u02d} - \ket{u0d2} + \ket{u20d} + \ket{u2d0}\right ) 
\nonumber \eeq
\beq 
C_{123,1} &=& 
-4 \sqrt{2} t^2 \nonumber \\
C_{123,2} &=& 
\frac{1}{144 \sqrt{2}} \left (
576 t^2-144 U^2-1440 W^2+108 J U+216 J W-1008 U W \right . \nonumber \\
&& \hspace{1cm} 
+
 \left . -2 A_{16}^2+9 J A_{16}-36 U A_{16}-108 W A_{16} \right) \nonumber \\
C_{123,3} &=& 
-\frac{1}{6 \sqrt{2}} \left (
3 J t+4 U t-4 W t+\cos \left(\theta _3\right) \sqrt{A_3} t \right ) \nonumber \\
 N_{123} &=& \sqrt{2 C_{123,1}^2+8 C_{123,2}^2+16 C_{123,3}^2} \nonumber \eeq 
\beq
\ket{\Psi_{124}} & = & \ket{4,0,0,\Gamma_{4,1}} \nonumber \\ 
& = & \quad 
C_{124,1} \left ( 
\ket{0022} - \ket{2200} \right) 
 \nonumber \\
& & + 
C_{124,2} \left ( 
\ket{02du} - \ket{02ud} + \ket{20du} - \ket{20ud} - \ket{du02} - \ket{du20} + \ket{ud02} + \ket{ud20} \right) 
 \nonumber \\
& & + 
C_{124,3} \left ( 
\ket{0d2u} + \ket{0du2} - \ket{0u2d} - \ket{0ud2} - \ket{2d0u} - \ket{2du0} + \ket{2u0d} + \ket{2ud0} \right . \nonumber \\
&& \hspace{3em} 
 + 
\left . \ket{d02u} + \ket{d0u2} - \ket{d20u} - \ket{d2u0} - \ket{u02d} - \ket{u0d2} + \ket{u20d} + \ket{u2d0}\right ) 
\nonumber \eeq
\beq 
C_{124,1} &=& 
-4 \sqrt{2} t^2 \nonumber \\
C_{124,2} &=& 
\frac{1}{8 \sqrt{2}} \left (
32 t^2-8 U^2-80 W^2+6 J U+12 J W-56 U W \right . \nonumber \\
&& \hspace{1cm} 
 \left . -4 A_{19}^2-3 J A_{19}+12 U A_{19}+36 W A_{19} \right) \nonumber \\
C_{124,3} &=& 
\frac{1}{12 \sqrt{2}} \left (
-6 J t-8 U t+8 W t+\cos \left(\theta _3\right) \sqrt{A_3} t+\sqrt{3} \sin \left(\theta _3\right) \sqrt{A_3} t \right ) \nonumber \\
 N_{124} &=& \sqrt{2 C_{124,1}^2+8 C_{124,2}^2+16 C_{124,3}^2} \nonumber \eeq 
\beq
\ket{\Psi_{125}} & = & \ket{4,0,0,\Gamma_{4,1}} \nonumber \\ 
& = & \quad 
C_{125,1} \left ( 
\ket{0022} - \ket{2200} \right) 
 \nonumber \\
& & + 
C_{125,2} \left ( 
\ket{02du} - \ket{02ud} + \ket{20du} - \ket{20ud} - \ket{du02} - \ket{du20} + \ket{ud02} + \ket{ud20} \right) 
 \nonumber \\
& & + 
C_{125,3} \left ( 
\ket{0d2u} + \ket{0du2} - \ket{0u2d} - \ket{0ud2} - \ket{2d0u} - \ket{2du0} + \ket{2u0d} + \ket{2ud0} \right . \nonumber \\
&& \hspace{3em} 
 + 
\left . \ket{d02u} + \ket{d0u2} - \ket{d20u} - \ket{d2u0} - \ket{u02d} - \ket{u0d2} + \ket{u20d} + \ket{u2d0}\right ) 
\nonumber \eeq
\beq 
C_{125,1} &=& 
-4 \sqrt{2} t^2 \nonumber \\
C_{125,2} &=& 
\frac{1}{8 \sqrt{2}} \left (
32 t^2-8 U^2-80 W^2+6 J U+12 J W-56 U W \right . \nonumber \\
&& \hspace{1cm} 
 \left . -4 A_{18}^2-3 J A_{18}+12 U A_{18}+36 W A_{18} \right) \nonumber \\
C_{125,3} &=& 
-\frac{1}{12 \sqrt{2}} \left (
6 J t+8 U t-8 W t-\cos \left(\theta _3\right) \sqrt{A_3} t+\sqrt{3} \sin \left(\theta _3\right) \sqrt{A_3} t \right ) \nonumber \\
 N_{125} &=& \sqrt{2 C_{125,1}^2+8 C_{125,2}^2+16 C_{125,3}^2} \nonumber \eeq 
\beq
\ket{\Psi_{126}} & = & \ket{4,0,2,\Gamma_{4,1}} \nonumber \\ 
&=& \frac{1}{4}
 \left ( \ket{0d2u} + \ket{0du2} + \ket{0u2d} + \ket{0ud2} + \ket{2d0u} + \ket{2du0} + \ket{2u0d} + \ket{2ud0} \right . \nonumber \\
&& \hspace{3em} 
 + 
\left . \ket{d02u} + \ket{d0u2} + \ket{d20u} + \ket{d2u0} + \ket{u02d} + \ket{u0d2} + \ket{u20d} + \ket{u2d0} \right ) 
\nonumber \eeq
\beq
\ket{\Psi_{127}} & = & \ket{4,0,0,\Gamma_{4,2}} \nonumber \\ 
& = & \quad 
C_{127,1} \left ( 
\ket{0220} - \ket{2002} \right) 
 \nonumber \\
& & + 
C_{127,2} \left ( 
\ket{02du} - \ket{02ud} + \ket{0d2u} - \ket{0u2d} - \ket{20du} + \ket{20ud} - \ket{2d0u} + \ket{2u0d} \right . \nonumber \\
&& \hspace{3em} 
\left . -\ket{d0u2} + \ket{d2u0} - \ket{du02} + \ket{du20} + \ket{u0d2} - \ket{u2d0} + \ket{ud02} - \ket{ud20}\right ) 
\nonumber \\
& & + 
C_{127,3} \left ( 
\ket{0du2} - \ket{0ud2} + \ket{2du0} - \ket{2ud0} - \ket{d02u} - \ket{d20u} + \ket{u02d} + \ket{u20d} \right) 
\nonumber \eeq
\beq 
C_{127,1} &=& 
-\frac{1}{144 \sqrt{2}} \left (
81 J^2-216 U J-1080 W J-1152 t^2+144 U^2+1440 U W \right . \nonumber \\
&& \hspace{1cm} 
 \left . 3600 W^2+240 A_{16} W+4 A_{16}^2-36 J A_{16}+48 U A_{16} \right) \nonumber \\
C_{127,2} &=& 
\frac{1}{12 \sqrt{2}} \left (
3 J t+4 U t-4 W t-2 \cos \left(\theta _3\right) \sqrt{A_3} t \right ) \nonumber \\
C_{127,3} &=& 
-2 \sqrt{2} t^2 \nonumber \\
 N_{127} &=& \sqrt{2 C_{127,1}^2+16 C_{127,2}^2+8 C_{127,3}^2} \nonumber \eeq 
\beq
\ket{\Psi_{128}} & = & \ket{4,0,0,\Gamma_{4,2}} \nonumber \\ 
& = & \quad 
C_{128,1} \left ( 
\ket{0220} - \ket{2002} \right) 
 \nonumber \\
& & + 
C_{128,2} \left ( 
\ket{02du} - \ket{02ud} + \ket{0d2u} - \ket{0u2d} - \ket{20du} + \ket{20ud} - \ket{2d0u} + \ket{2u0d} \right . \nonumber \\
&& \hspace{3em} 
\left . -\ket{d0u2} + \ket{d2u0} - \ket{du02} + \ket{du20} + \ket{u0d2} - \ket{u2d0} + \ket{ud02} - \ket{ud20}\right ) 
\nonumber \\
& & + 
C_{128,3} \left ( 
\ket{0du2} - \ket{0ud2} + \ket{2du0} - \ket{2ud0} - \ket{d02u} - \ket{d20u} + \ket{u02d} + \ket{u20d} \right) 
\nonumber \eeq
\beq 
C_{128,1} &=& 
-\frac{1}{16 \sqrt{2}} \left (
9 J^2-24 U J-120 W J-128 t^2+16 U^2+160 U W \right . \nonumber \\
&& \hspace{1cm} 
+
 \left . 400 W^2-160 A_{19} W+16 A_{19}^2+24 J A_{19}-32 U A_{19} \right) \nonumber \\
C_{128,2} &=& 
\frac{1}{12 \sqrt{2}} \left (
3 J t+4 U t-4 W t+\cos \left(\theta _3\right) \sqrt{A_3} t+\sqrt{3} \sin \left(\theta _3\right) \sqrt{A_3} t \right ) \nonumber \\
C_{128,3} &=& 
-2 \sqrt{2} t^2 \nonumber \\
 N_{128} &=& \sqrt{2 C_{128,1}^2+16 C_{128,2}^2+8 C_{128,3}^2} \nonumber \eeq 
\beq
\ket{\Psi_{129}} & = & \ket{4,0,0,\Gamma_{4,2}} \nonumber \\ 
& = & \quad 
C_{129,1} \left ( 
\ket{0220} - \ket{2002} \right) 
 \nonumber \\
& & + 
C_{129,2} \left ( 
\ket{02du} - \ket{02ud} + \ket{0d2u} - \ket{0u2d} - \ket{20du} + \ket{20ud} - \ket{2d0u} + \ket{2u0d} \right . \nonumber \\
&& \hspace{3em} 
\left . -\ket{d0u2} + \ket{d2u0} - \ket{du02} + \ket{du20} + \ket{u0d2} - \ket{u2d0} + \ket{ud02} - \ket{ud20}\right ) 
\nonumber \\
& & + 
C_{129,3} \left ( 
\ket{0du2} - \ket{0ud2} + \ket{2du0} - \ket{2ud0} - \ket{d02u} - \ket{d20u} + \ket{u02d} + \ket{u20d} \right) 
\nonumber \eeq
\beq 
C_{129,1} &=& 
-\frac{1}{16 \sqrt{2}} \left (
9 J^2-24 U J-120 W J-128 t^2+16 U^2+160 U W \right . \nonumber \\
&& \hspace{1cm} 
+
 \left . 400 W^2-160 A_{18} W+16 A_{18}^2+24 J A_{18}-32 U A_{18} \right) \nonumber \\
C_{129,2} &=& 
-\frac{1}{12 \sqrt{2}} \left (
-3 J t-4 U t+4 W t-\cos \left(\theta _3\right) \sqrt{A_3} t+\sqrt{3} \sin \left(\theta _3\right) \sqrt{A_3} t \right ) \nonumber \\
C_{129,3} &=& 
-2 \sqrt{2} t^2 \nonumber \\
 N_{129} &=& \sqrt{2 C_{129,1}^2+16 C_{129,2}^2+8 C_{129,3}^2} \nonumber \eeq 
\beq
\ket{\Psi_{130}} & = & \ket{4,0,2,\Gamma_{4,2}} \nonumber \\ 
&=& \frac{1}{4}
 \left ( \ket{02du} + \ket{02ud} + \ket{0d2u} + \ket{0u2d} + \ket{20du} + \ket{20ud} + \ket{2d0u} + \ket{2u0d} \right . \nonumber \\
&& \hspace{3em} 
\left . -\ket{d0u2} - \ket{d2u0} - \ket{du02} - \ket{du20} - \ket{u0d2} - \ket{u2d0} - \ket{ud02} - \ket{ud20} \right ) 
\nonumber \eeq
\beq
\ket{\Psi_{131}} & = & \ket{4,0,0,\Gamma_{4,3}} \nonumber \\ 
& = & \quad 
C_{131,1} \left ( 
\ket{0202} - \ket{2020} \right) 
 \nonumber \\
& & + 
C_{131,2} \left ( 
\ket{02du} - \ket{02ud} + \ket{0du2} - \ket{0ud2} - \ket{20du} + \ket{20ud} - \ket{2du0} + \ket{2ud0} \right . \nonumber \\
&& \hspace{3em} 
\left . -\ket{d02u} + \ket{d20u} + \ket{du02} - \ket{du20} + \ket{u02d} - \ket{u20d} - \ket{ud02} + \ket{ud20}\right ) 
\nonumber \\
& & + 
C_{131,3} \left ( 
\ket{0d2u} - \ket{0u2d} + \ket{2d0u} - \ket{2u0d} - \ket{d0u2} - \ket{d2u0} + \ket{u0d2} + \ket{u2d0} \right) 
\nonumber \eeq
\beq 
C_{131,1} &=& 
-\frac{1}{144 \sqrt{2}} \left (
81 J^2-216 U J-1080 W J-1152 t^2+144 U^2+1440 U W \right . \nonumber \\
&& \hspace{1cm} 
 \left . 3600 W^2+240 A_{16} W+4 A_{16}^2-36 J A_{16}+48 U A_{16} \right) \nonumber \\
C_{131,2} &=& 
\frac{1}{12 \sqrt{2}} \left (
3 J t+4 U t-4 W t-2 \cos \left(\theta _3\right) \sqrt{A_3} t \right ) \nonumber \\
C_{131,3} &=& 
-2 \sqrt{2} t^2 \nonumber \\
 N_{131} &=& \sqrt{2 C_{131,1}^2+16 C_{131,2}^2+8 C_{131,3}^2} \nonumber \eeq 
\beq
\ket{\Psi_{132}} & = & \ket{4,0,0,\Gamma_{4,3}} \nonumber \\ 
& = & \quad 
C_{132,1} \left ( 
\ket{0202} - \ket{2020} \right) 
 \nonumber \\
& & + 
C_{132,2} \left ( 
\ket{02du} - \ket{02ud} + \ket{0du2} - \ket{0ud2} - \ket{20du} + \ket{20ud} - \ket{2du0} + \ket{2ud0} \right . \nonumber \\
&& \hspace{3em} 
\left . -\ket{d02u} + \ket{d20u} + \ket{du02} - \ket{du20} + \ket{u02d} - \ket{u20d} - \ket{ud02} + \ket{ud20}\right ) 
\nonumber \\
& & + 
C_{132,3} \left ( 
\ket{0d2u} - \ket{0u2d} + \ket{2d0u} - \ket{2u0d} - \ket{d0u2} - \ket{d2u0} + \ket{u0d2} + \ket{u2d0} \right) 
\nonumber \eeq
\beq 
C_{132,1} &=& 
-\frac{1}{16 \sqrt{2}} \left (
9 J^2-24 U J-120 W J-128 t^2+16 U^2+160 U W \right . \nonumber \\
&& \hspace{1cm} 
+
 \left . 400 W^2-160 A_{19} W+16 A_{19}^2+24 J A_{19}-32 U A_{19} \right) \nonumber \\
C_{132,2} &=& 
\frac{1}{12 \sqrt{2}} \left (
3 J t+4 U t-4 W t+\cos \left(\theta _3\right) \sqrt{A_3} t+\sqrt{3} \sin \left(\theta _3\right) \sqrt{A_3} t \right ) \nonumber \\
C_{132,3} &=& 
-2 \sqrt{2} t^2 \nonumber \\
 N_{132} &=& \sqrt{2 C_{132,1}^2+16 C_{132,2}^2+8 C_{132,3}^2} \nonumber \eeq 
\beq
\ket{\Psi_{133}} & = & \ket{4,0,0,\Gamma_{4,3}} \nonumber \\ 
& = & \quad 
C_{133,1} \left ( 
\ket{0202} - \ket{2020} \right) 
 \nonumber \\
& & + 
C_{133,2} \left ( 
\ket{02du} - \ket{02ud} + \ket{0du2} - \ket{0ud2} - \ket{20du} + \ket{20ud} - \ket{2du0} + \ket{2ud0} \right . \nonumber \\
&& \hspace{3em} 
\left . -\ket{d02u} + \ket{d20u} + \ket{du02} - \ket{du20} + \ket{u02d} - \ket{u20d} - \ket{ud02} + \ket{ud20}\right ) 
\nonumber \\
& & + 
C_{133,3} \left ( 
\ket{0d2u} - \ket{0u2d} + \ket{2d0u} - \ket{2u0d} - \ket{d0u2} - \ket{d2u0} + \ket{u0d2} + \ket{u2d0} \right) 
\nonumber \eeq
\beq 
C_{133,1} &=& 
-\frac{1}{16 \sqrt{2}} \left (
9 J^2-24 U J-120 W J-128 t^2+16 U^2+160 U W \right . \nonumber \\
&& \hspace{1cm} 
+
 \left . 400 W^2-160 A_{18} W+16 A_{18}^2+24 J A_{18}-32 U A_{18} \right) \nonumber \\
C_{133,2} &=& 
-\frac{1}{12 \sqrt{2}} \left (
-3 J t-4 U t+4 W t-\cos \left(\theta _3\right) \sqrt{A_3} t+\sqrt{3} \sin \left(\theta _3\right) \sqrt{A_3} t \right ) \nonumber \\
C_{133,3} &=& 
-2 \sqrt{2} t^2 \nonumber \\
 N_{133} &=& \sqrt{2 C_{133,1}^2+16 C_{133,2}^2+8 C_{133,3}^2} \nonumber \eeq 
\beq
\ket{\Psi_{134}} & = & \ket{4,0,2,\Gamma_{4,3}} \nonumber \\ 
&=& \frac{1}{4}
 \left ( \ket{02du} + \ket{02ud} - \ket{0du2} - \ket{0ud2} + \ket{20du} + \ket{20ud} - \ket{2du0} - \ket{2ud0} \right . \nonumber \\
&& \hspace{3em} 
 + 
\left . \ket{d02u} + \ket{d20u} + \ket{du02} + \ket{du20} + \ket{u02d} + \ket{u20d} + \ket{ud02} + \ket{ud20} \right ) 
\nonumber \eeq
\beq
\ket{\Psi_{135}} & = & \ket{4,0,0,\Gamma_{5,1}} \nonumber \\ 
&=& \frac{1}{4}
 \left ( \ket{02du} - \ket{02ud} - \ket{0du2} + \ket{0ud2} - \ket{20du} + \ket{20ud} + \ket{2du0} - \ket{2ud0} \right . \nonumber \\
&& \hspace{3em} 
 + 
\left . \ket{d02u} - \ket{d20u} + \ket{du02} - \ket{du20} - \ket{u02d} + \ket{u20d} - \ket{ud02} + \ket{ud20} \right ) 
\nonumber \eeq
\beq
\ket{\Psi_{136}} & = & \ket{4,0,2,\Gamma_{5,1}} \nonumber \\ 
& = & \quad 
C_{136,1} \left ( 
\ket{02du} + \ket{02ud} + \ket{0du2} + \ket{0ud2} + \ket{20du} + \ket{20ud} + \ket{2du0} + \ket{2ud0} \right . \nonumber \\
&& \hspace{3em} 
\left . -\ket{d02u} - \ket{d20u} + \ket{du02} + \ket{du20} - \ket{u02d} - \ket{u20d} + \ket{ud02} + \ket{ud20}\right ) 
\nonumber \\
& & + 
C_{136,2} \left ( 
\ket{0d2u} + \ket{0u2d} - \ket{2d0u} - \ket{2u0d} - \ket{d0u2} + \ket{d2u0} - \ket{u0d2} + \ket{u2d0} \right) 
 \nonumber \\
& & + 
C_{136,3} \left ( 
\ket{dudu} - \ket{udud} \right) 
\nonumber \eeq
\beq 
C_{136,1} &=& 
-\frac{1}{12 \sqrt{2}} \left (
3 J t+4 U t-4 W t+2 \cos \left(\theta _4\right) \sqrt{A_3} t \right ) \nonumber \\
C_{136,2} &=& 
-2 \sqrt{2} t^2 \nonumber \\
C_{136,3} &=& 
-\frac{1}{144 \sqrt{2}} \left (
9 J^2+24 U J-24 W J-1152 t^2-48 U^2-1056 U W \right . \nonumber \\
&& \hspace{1cm} 
 \left . -4080 W^2+240 \cos \left(\theta _4\right) \sqrt{A_3} W+4 A_{17}^2+12 J \cos \left(\theta _4\right) \sqrt{A_3}+48 U \cos \left(\theta _4\right) \sqrt{A_3} \right) \nonumber \\
 N_{136} &=& \sqrt{16 C_{136,1}^2+8 C_{136,2}^2+2 C_{136,3}^2} \nonumber \eeq 
\beq
\ket{\Psi_{137}} & = & \ket{4,0,2,\Gamma_{5,1}} \nonumber \\ 
& = & \quad 
C_{137,1} \left ( 
\ket{02du} + \ket{02ud} + \ket{0du2} + \ket{0ud2} + \ket{20du} + \ket{20ud} + \ket{2du0} + \ket{2ud0} \right . \nonumber \\
&& \hspace{3em} 
\left . -\ket{d02u} - \ket{d20u} + \ket{du02} + \ket{du20} - \ket{u02d} - \ket{u20d} + \ket{ud02} + \ket{ud20}\right ) 
\nonumber \\
& & + 
C_{137,2} \left ( 
\ket{0d2u} + \ket{0u2d} - \ket{2d0u} - \ket{2u0d} - \ket{d0u2} + \ket{d2u0} - \ket{u0d2} + \ket{u2d0} \right) 
 \nonumber \\
& & + 
C_{137,3} \left ( 
\ket{dudu} - \ket{udud} \right) 
\nonumber \eeq
\beq 
C_{137,1} &=& 
\frac{1}{12 \sqrt{2}} \left (
-3 J t-4 U t+4 W t+\cos \left(\theta _4\right) \sqrt{A_3} t+\sqrt{3} \sin \left(\theta _4\right) \sqrt{A_3} t \right ) \nonumber \\
C_{137,2} &=& 
-2 \sqrt{2} t^2 \nonumber \\
C_{137,3} &=& 
-\frac{1}{16 \sqrt{2}} \left (
J^2+8 U J+40 W J-128 t^2+16 U^2+160 U W \right . \nonumber \\
&& \hspace{1cm} 
 \left . 400 W^2-160 A_{21} W+16 A_{21}^2-8 J A_{21}-32 U A_{21} \right) \nonumber \\
 N_{137} &=& \sqrt{16 C_{137,1}^2+8 C_{137,2}^2+2 C_{137,3}^2} \nonumber \eeq 
\beq
\ket{\Psi_{138}} & = & \ket{4,0,2,\Gamma_{5,1}} \nonumber \\ 
& = & \quad 
C_{138,1} \left ( 
\ket{02du} + \ket{02ud} + \ket{0du2} + \ket{0ud2} + \ket{20du} + \ket{20ud} + \ket{2du0} + \ket{2ud0} \right . \nonumber \\
&& \hspace{3em} 
\left . -\ket{d02u} - \ket{d20u} + \ket{du02} + \ket{du20} - \ket{u02d} - \ket{u20d} + \ket{ud02} + \ket{ud20}\right ) 
\nonumber \\
& & + 
C_{138,2} \left ( 
\ket{0d2u} + \ket{0u2d} - \ket{2d0u} - \ket{2u0d} - \ket{d0u2} + \ket{d2u0} - \ket{u0d2} + \ket{u2d0} \right) 
 \nonumber \\
& & + 
C_{138,3} \left ( 
\ket{dudu} - \ket{udud} \right) 
\nonumber \eeq
\beq 
C_{138,1} &=& 
-\frac{1}{12 \sqrt{2}} \left (
3 J t+4 U t-4 W t-\cos \left(\theta _4\right) \sqrt{A_3} t+\sqrt{3} \sin \left(\theta _4\right) \sqrt{A_3} t \right ) \nonumber \\
C_{138,2} &=& 
-2 \sqrt{2} t^2 \nonumber \\
C_{138,3} &=& 
-\frac{1}{16 \sqrt{2}} \left (
J^2+8 U J+40 W J-128 t^2+16 U^2+160 U W \right . \nonumber \\
&& \hspace{1cm} 
 \left . 400 W^2-160 A_{20} W+16 A_{20}^2-8 J A_{20}-32 U A_{20} \right) \nonumber \\
 N_{138} &=& \sqrt{16 C_{138,1}^2+8 C_{138,2}^2+2 C_{138,3}^2} \nonumber \eeq 
\beq
\ket{\Psi_{139}} & = & \ket{4,0,0,\Gamma_{5,2}} \nonumber \\ 
&=& \frac{1}{4}
 \left ( \ket{0d2u} - \ket{0du2} - \ket{0u2d} + \ket{0ud2} - \ket{2d0u} + \ket{2du0} + \ket{2u0d} - \ket{2ud0} \right . \nonumber \\
&& \hspace{3em} 
\left . -\ket{d02u} + \ket{d0u2} + \ket{d20u} - \ket{d2u0} + \ket{u02d} - \ket{u0d2} - \ket{u20d} + \ket{u2d0} \right ) 
\nonumber \eeq
\beq
\ket{\Psi_{140}} & = & \ket{4,0,2,\Gamma_{5,2}} \nonumber \\ 
& = & \quad 
C_{140,1} \left ( 
\ket{02du} + \ket{02ud} - \ket{20du} - \ket{20ud} - \ket{du02} + \ket{du20} - \ket{ud02} + \ket{ud20} \right) 
 \nonumber \\
& & + 
C_{140,2} \left ( 
\ket{0d2u} - \ket{0du2} + \ket{0u2d} - \ket{0ud2} + \ket{2d0u} - \ket{2du0} + \ket{2u0d} - \ket{2ud0} \right . \nonumber \\
&& \hspace{3em} 
\left . -\ket{d02u} + \ket{d0u2} - \ket{d20u} + \ket{d2u0} - \ket{u02d} + \ket{u0d2} - \ket{u20d} + \ket{u2d0}\right ) 
\nonumber \\
& & + 
C_{140,3} \left ( 
\ket{dduu} - \ket{uudd} \right) 
\nonumber \eeq
\beq 
C_{140,1} &=& 
2 \sqrt{2} t^2 \nonumber \\
C_{140,2} &=& 
\frac{1}{12 \sqrt{2}} \left (
3 J t+4 U t-4 W t+2 \cos \left(\theta _4\right) \sqrt{A_3} t \right ) \nonumber \\
C_{140,3} &=& 
-\frac{1}{144 \sqrt{2}} \left (
9 J^2+24 U J-24 W J-1152 t^2-48 U^2-1056 U W \right . \nonumber \\
&& \hspace{1cm} 
 \left . -4080 W^2+240 \cos \left(\theta _4\right) \sqrt{A_3} W+4 A_{17}^2+12 J \cos \left(\theta _4\right) \sqrt{A_3}+48 U \cos \left(\theta _4\right) \sqrt{A_3} \right) \nonumber \\
 N_{140} &=& \sqrt{8 C_{140,1}^2+16 C_{140,2}^2+2 C_{140,3}^2} \nonumber \eeq 
\beq
\ket{\Psi_{141}} & = & \ket{4,0,2,\Gamma_{5,2}} \nonumber \\ 
& = & \quad 
C_{141,1} \left ( 
\ket{02du} + \ket{02ud} - \ket{20du} - \ket{20ud} - \ket{du02} + \ket{du20} - \ket{ud02} + \ket{ud20} \right) 
 \nonumber \\
& & + 
C_{141,2} \left ( 
\ket{0d2u} - \ket{0du2} + \ket{0u2d} - \ket{0ud2} + \ket{2d0u} - \ket{2du0} + \ket{2u0d} - \ket{2ud0} \right . \nonumber \\
&& \hspace{3em} 
\left . -\ket{d02u} + \ket{d0u2} - \ket{d20u} + \ket{d2u0} - \ket{u02d} + \ket{u0d2} - \ket{u20d} + \ket{u2d0}\right ) 
\nonumber \\
& & + 
C_{141,3} \left ( 
\ket{dduu} - \ket{uudd} \right) 
\nonumber \eeq
\beq 
C_{141,1} &=& 
2 \sqrt{2} t^2 \nonumber \\
C_{141,2} &=& 
-\frac{1}{12 \sqrt{2}} \left (
-3 J t-4 U t+4 W t+\cos \left(\theta _4\right) \sqrt{A_3} t+\sqrt{3} \sin \left(\theta _4\right) \sqrt{A_3} t \right ) \nonumber \\
C_{141,3} &=& 
-\frac{1}{16 \sqrt{2}} \left (
J^2+8 U J+40 W J-128 t^2+16 U^2+160 U W \right . \nonumber \\
&& \hspace{1cm} 
 \left . 400 W^2-160 A_{21} W+16 A_{21}^2-8 J A_{21}-32 U A_{21} \right) \nonumber \\
 N_{141} &=& \sqrt{8 C_{141,1}^2+16 C_{141,2}^2+2 C_{141,3}^2} \nonumber \eeq 
\beq
\ket{\Psi_{142}} & = & \ket{4,0,2,\Gamma_{5,2}} \nonumber \\ 
& = & \quad 
C_{142,1} \left ( 
\ket{02du} + \ket{02ud} - \ket{20du} - \ket{20ud} - \ket{du02} + \ket{du20} - \ket{ud02} + \ket{ud20} \right) 
 \nonumber \\
& & + 
C_{142,2} \left ( 
\ket{0d2u} - \ket{0du2} + \ket{0u2d} - \ket{0ud2} + \ket{2d0u} - \ket{2du0} + \ket{2u0d} - \ket{2ud0} \right . \nonumber \\
&& \hspace{3em} 
\left . -\ket{d02u} + \ket{d0u2} - \ket{d20u} + \ket{d2u0} - \ket{u02d} + \ket{u0d2} - \ket{u20d} + \ket{u2d0}\right ) 
\nonumber \\
& & + 
C_{142,3} \left ( 
\ket{dduu} - \ket{uudd} \right) 
\nonumber \eeq
\beq 
C_{142,1} &=& 
2 \sqrt{2} t^2 \nonumber \\
C_{142,2} &=& 
\frac{1}{12 \sqrt{2}} \left (
3 J t+4 U t-4 W t-\cos \left(\theta _4\right) \sqrt{A_3} t+\sqrt{3} \sin \left(\theta _4\right) \sqrt{A_3} t \right ) \nonumber \\
C_{142,3} &=& 
-\frac{1}{16 \sqrt{2}} \left (
J^2+8 U J+40 W J-128 t^2+16 U^2+160 U W \right . \nonumber \\
&& \hspace{1cm} 
 \left . 400 W^2-160 A_{20} W+16 A_{20}^2-8 J A_{20}-32 U A_{20} \right) \nonumber \\
 N_{142} &=& \sqrt{8 C_{142,1}^2+16 C_{142,2}^2+2 C_{142,3}^2} \nonumber \eeq 
\beq
\ket{\Psi_{143}} & = & \ket{4,0,0,\Gamma_{5,3}} \nonumber \\ 
&=& \frac{1}{4}
 \left ( \ket{02du} - \ket{02ud} - \ket{0d2u} + \ket{0u2d} - \ket{20du} + \ket{20ud} + \ket{2d0u} - \ket{2u0d} \right . \nonumber \\
&& \hspace{3em} 
 + 
\left . \ket{d0u2} - \ket{d2u0} - \ket{du02} + \ket{du20} - \ket{u0d2} + \ket{u2d0} + \ket{ud02} - \ket{ud20} \right ) 
\nonumber \eeq
\beq
\ket{\Psi_{144}} & = & \ket{4,0,2,\Gamma_{5,3}} \nonumber \\ 
& = & \quad 
C_{144,1} \left ( 
\ket{02du} + \ket{02ud} - \ket{0d2u} - \ket{0u2d} + \ket{20du} + \ket{20ud} - \ket{2d0u} - \ket{2u0d} \right . \nonumber \\
&& \hspace{3em} 
 + 
\left . \ket{d0u2} + \ket{d2u0} - \ket{du02} - \ket{du20} + \ket{u0d2} + \ket{u2d0} - \ket{ud02} - \ket{ud20}\right ) 
\nonumber \\
& & + 
C_{144,2} \left ( 
\ket{0du2} + \ket{0ud2} - \ket{2du0} - \ket{2ud0} - \ket{d02u} + \ket{d20u} - \ket{u02d} + \ket{u20d} \right) 
 \nonumber \\
& & + 
C_{144,3} \left ( 
\ket{duud} - \ket{uddu} \right) 
\nonumber \eeq
\beq 
C_{144,1} &=& 
-\frac{1}{12 \sqrt{2}} \left (
3 J t+4 U t-4 W t+2 \cos \left(\theta _4\right) \sqrt{A_3} t \right ) \nonumber \\
C_{144,2} &=& 
2 \sqrt{2} t^2 \nonumber \\
C_{144,3} &=& 
-\frac{1}{144 \sqrt{2}} \left (
9 J^2+24 U J-24 W J-1152 t^2-48 U^2-1056 U W \right . \nonumber \\
&& \hspace{1cm} 
 \left . -4080 W^2+240 \cos \left(\theta _4\right) \sqrt{A_3} W+4 A_{17}^2+12 J \cos \left(\theta _4\right) \sqrt{A_3}+48 U \cos \left(\theta _4\right) \sqrt{A_3} \right) \nonumber \\
 N_{144} &=& \sqrt{16 C_{144,1}^2+8 C_{144,2}^2+2 C_{144,3}^2} \nonumber \eeq 
\beq
\ket{\Psi_{145}} & = & \ket{4,0,2,\Gamma_{5,3}} \nonumber \\ 
& = & \quad 
C_{145,1} \left ( 
\ket{02du} + \ket{02ud} - \ket{0d2u} - \ket{0u2d} + \ket{20du} + \ket{20ud} - \ket{2d0u} - \ket{2u0d} \right . \nonumber \\
&& \hspace{3em} 
 + 
\left . \ket{d0u2} + \ket{d2u0} - \ket{du02} - \ket{du20} + \ket{u0d2} + \ket{u2d0} - \ket{ud02} - \ket{ud20}\right ) 
\nonumber \\
& & + 
C_{145,2} \left ( 
\ket{0du2} + \ket{0ud2} - \ket{2du0} - \ket{2ud0} - \ket{d02u} + \ket{d20u} - \ket{u02d} + \ket{u20d} \right) 
 \nonumber \\
& & + 
C_{145,3} \left ( 
\ket{duud} - \ket{uddu} \right) 
\nonumber \eeq
\beq 
C_{145,1} &=& 
\frac{1}{12 \sqrt{2}} \left (
-3 J t-4 U t+4 W t+\cos \left(\theta _4\right) \sqrt{A_3} t+\sqrt{3} \sin \left(\theta _4\right) \sqrt{A_3} t \right ) \nonumber \\
C_{145,2} &=& 
2 \sqrt{2} t^2 \nonumber \\
C_{145,3} &=& 
-\frac{1}{16 \sqrt{2}} \left (
J^2+8 U J+40 W J-128 t^2+16 U^2+160 U W \right . \nonumber \\
&& \hspace{1cm} 
 \left . 400 W^2-160 A_{21} W+16 A_{21}^2-8 J A_{21}-32 U A_{21} \right) \nonumber \\
 N_{145} &=& \sqrt{16 C_{145,1}^2+8 C_{145,2}^2+2 C_{145,3}^2} \nonumber \eeq 
\beq
\ket{\Psi_{146}} & = & \ket{4,0,2,\Gamma_{5,3}} \nonumber \\ 
& = & \quad 
C_{146,1} \left ( 
\ket{02du} + \ket{02ud} - \ket{0d2u} - \ket{0u2d} + \ket{20du} + \ket{20ud} - \ket{2d0u} - \ket{2u0d} \right . \nonumber \\
&& \hspace{3em} 
 + 
\left . \ket{d0u2} + \ket{d2u0} - \ket{du02} - \ket{du20} + \ket{u0d2} + \ket{u2d0} - \ket{ud02} - \ket{ud20}\right ) 
\nonumber \\
& & + 
C_{146,2} \left ( 
\ket{0du2} + \ket{0ud2} - \ket{2du0} - \ket{2ud0} - \ket{d02u} + \ket{d20u} - \ket{u02d} + \ket{u20d} \right) 
 \nonumber \\
& & + 
C_{146,3} \left ( 
\ket{duud} - \ket{uddu} \right) 
\nonumber \eeq
\beq 
C_{146,1} &=& 
-\frac{1}{12 \sqrt{2}} \left (
3 J t+4 U t-4 W t-\cos \left(\theta _4\right) \sqrt{A_3} t+\sqrt{3} \sin \left(\theta _4\right) \sqrt{A_3} t \right ) \nonumber \\
C_{146,2} &=& 
2 \sqrt{2} t^2 \nonumber \\
C_{146,3} &=& 
-\frac{1}{16 \sqrt{2}} \left (
J^2+8 U J+40 W J-128 t^2+16 U^2+160 U W \right . \nonumber \\
&& \hspace{1cm} 
 \left . 400 W^2-160 A_{20} W+16 A_{20}^2-8 J A_{20}-32 U A_{20} \right) \nonumber \\
 N_{146} &=& \sqrt{16 C_{146,1}^2+8 C_{146,2}^2+2 C_{146,3}^2} \nonumber \eeq 
{\subsubsection{\boldmath Eigenvectors for ${\rm  N_e}=4$ and   ${\rm m_s}$= $1$.}
\beq
\ket{\Psi_{147}} & = & \ket{4,1,2,\Gamma_2} \nonumber \\ 
&=& \frac{1}{2 \sqrt{3}}
 \left ( \ket{02uu} - \ket{0u2u} + \ket{0uu2} - \ket{20uu} + \ket{2u0u} - \ket{2uu0} \right . \nonumber \\
&& \hspace{3em} 
 + 
\left . \ket{u02u} - \ket{u0u2} - \ket{u20u} + \ket{u2u0} + \ket{uu02} - \ket{uu20} \right ) 
\nonumber \eeq
\beq
\ket{\Psi_{148}} & = & \ket{4,1,6,\Gamma_2} \nonumber \\ 
&=& \frac{1}{2}
 \left ( \ket{duuu} + \ket{uduu} + \ket{uudu} + \ket{uuud} \right) \nonumber 
\eeq
\beq
\ket{\Psi_{149}} & = & \ket{4,1,2,\Gamma_{3,1}} \nonumber \\ 
&=& \frac{1}{2 \sqrt{2}}
 \left ( \ket{02uu} - \ket{0uu2} - \ket{20uu} + \ket{2uu0} - \ket{u02u} + \ket{u20u} + \ket{uu02} - \ket{uu20} \right) \nonumber 
\eeq
\beq
\ket{\Psi_{150}} & = & \ket{4,1,2,\Gamma_{3,2}} \nonumber \\ 
& = & \quad 
C_{150,1} \left ( 
\ket{02uu} + \ket{0uu2} - \ket{20uu} - \ket{2uu0} + \ket{u02u} - \ket{u20u} + \ket{uu02} - \ket{uu20} \right) 
 \nonumber \\
& & + 
C_{150,2} \left ( 
\ket{0u2u} - \ket{2u0u} + \ket{u0u2} - \ket{u2u0} \right) 
\nonumber \eeq
\beq 
C_{150,1} &=& 
-\frac{1}{2 \sqrt{6}} \nonumber \\
C_{150,2} &=& 
-\frac{1}{\sqrt{6}} \nonumber \\
 N_{150} &=& 2 \sqrt{2 C_{150,1}^2+C_{150,2}^2} \nonumber \eeq 
\beq
\ket{\Psi_{151}} & = & \ket{4,1,2,\Gamma_{4,1}} \nonumber \\ 
&=& \frac{1}{2 \sqrt{2}}
 \left ( \ket{0u2u} + \ket{0uu2} + \ket{2u0u} + \ket{2uu0} + \ket{u02u} + \ket{u0u2} + \ket{u20u} + \ket{u2u0} \right) \nonumber 
\eeq
\beq
\ket{\Psi_{152}} & = & \ket{4,1,2,\Gamma_{4,2}} \nonumber \\ 
&=& \frac{1}{2 \sqrt{2}}
 \left ( \ket{02uu} + \ket{0u2u} + \ket{20uu} + \ket{2u0u} - \ket{u0u2} - \ket{u2u0} - \ket{uu02} - \ket{uu20} \right) \nonumber 
\eeq
\beq
\ket{\Psi_{153}} & = & \ket{4,1,2,\Gamma_{4,3}} \nonumber \\ 
&=& \frac{1}{2 \sqrt{2}}
 \left ( \ket{02uu} - \ket{0uu2} + \ket{20uu} - \ket{2uu0} + \ket{u02u} + \ket{u20u} + \ket{uu02} + \ket{uu20} \right) \nonumber 
\eeq
\beq
\ket{\Psi_{154}} & = & \ket{4,1,2,\Gamma_{5,1}} \nonumber \\ 
& = & \quad 
C_{154,1} \left ( 
\ket{02uu} + \ket{0uu2} + \ket{20uu} + \ket{2uu0} - \ket{u02u} - \ket{u20u} + \ket{uu02} + \ket{uu20} \right) 
 \nonumber \\
& & + 
C_{154,2} \left ( 
\ket{0u2u} - \ket{2u0u} - \ket{u0u2} + \ket{u2u0} \right) 
 \nonumber \\
& & + 
C_{154,3} \left ( 
\ket{duuu} - \ket{uduu} + \ket{uudu} - \ket{uuud} \right) 
\nonumber \eeq
\beq 
C_{154,1} &=& 
-\frac{1}{12} \left (
3 J t+4 U t-4 W t+2 \cos \left(\theta _4\right) \sqrt{A_3} t \right ) \nonumber \\
C_{154,2} &=& 
-4 t^2 \nonumber \\
C_{154,3} &=& 
-\frac{1}{288} \left (
9 J^2+24 U J-24 W J-1152 t^2-48 U^2-1056 U W \right . \nonumber \\
&& \hspace{1cm} 
 \left . -4080 W^2+240 \cos \left(\theta _4\right) \sqrt{A_3} W+4 A_{17}^2+12 J \cos \left(\theta _4\right) \sqrt{A_3}+48 U \cos \left(\theta _4\right) \sqrt{A_3} \right) \nonumber \\
 N_{154} &=& 2 \sqrt{2 C_{154,1}^2+C_{154,2}^2+C_{154,3}^2} \nonumber \eeq 
\beq
\ket{\Psi_{155}} & = & \ket{4,1,2,\Gamma_{5,1}} \nonumber \\ 
& = & \quad 
C_{155,1} \left ( 
\ket{02uu} + \ket{0uu2} + \ket{20uu} + \ket{2uu0} - \ket{u02u} - \ket{u20u} + \ket{uu02} + \ket{uu20} \right) 
 \nonumber \\
& & + 
C_{155,2} \left ( 
\ket{0u2u} - \ket{2u0u} - \ket{u0u2} + \ket{u2u0} \right) 
 \nonumber \\
& & + 
C_{155,3} \left ( 
\ket{duuu} - \ket{uduu} + \ket{uudu} - \ket{uuud} \right) 
\nonumber \eeq
\beq 
C_{155,1} &=& 
\frac{1}{12} \left (
-3 J t-4 U t+4 W t+\cos \left(\theta _4\right) \sqrt{A_3} t+\sqrt{3} \sin \left(\theta _4\right) \sqrt{A_3} t \right ) \nonumber \\
C_{155,2} &=& 
-4 t^2 \nonumber \\
C_{155,3} &=& 
-\frac{1}{32} \left (
J^2+8 U J+40 W J-128 t^2+16 U^2+160 U W \right . \nonumber \\
&& \hspace{1cm} 
 \left . 400 W^2-160 A_{21} W+16 A_{21}^2-8 J A_{21}-32 U A_{21} \right) \nonumber \\
 N_{155} &=& 2 \sqrt{2 C_{155,1}^2+C_{155,2}^2+C_{155,3}^2} \nonumber \eeq 
\beq
\ket{\Psi_{156}} & = & \ket{4,1,2,\Gamma_{5,1}} \nonumber \\ 
& = & \quad 
C_{156,1} \left ( 
\ket{02uu} + \ket{0uu2} + \ket{20uu} + \ket{2uu0} - \ket{u02u} - \ket{u20u} + \ket{uu02} + \ket{uu20} \right) 
 \nonumber \\
& & + 
C_{156,2} \left ( 
\ket{0u2u} - \ket{2u0u} - \ket{u0u2} + \ket{u2u0} \right) 
 \nonumber \\
& & + 
C_{156,3} \left ( 
\ket{duuu} - \ket{uduu} + \ket{uudu} - \ket{uuud} \right) 
\nonumber \eeq
\beq 
C_{156,1} &=& 
-\frac{1}{12} \left (
3 J t+4 U t-4 W t-\cos \left(\theta _4\right) \sqrt{A_3} t+\sqrt{3} \sin \left(\theta _4\right) \sqrt{A_3} t \right ) \nonumber \\
C_{156,2} &=& 
-4 t^2 \nonumber \\
C_{156,3} &=& 
-\frac{1}{32} \left (
J^2+8 U J+40 W J-128 t^2+16 U^2+160 U W \right . \nonumber \\
&& \hspace{1cm} 
 \left . 400 W^2-160 A_{20} W+16 A_{20}^2-8 J A_{20}-32 U A_{20} \right) \nonumber \\
 N_{156} &=& 2 \sqrt{2 C_{156,1}^2+C_{156,2}^2+C_{156,3}^2} \nonumber \eeq 
\beq
\ket{\Psi_{157}} & = & \ket{4,1,2,\Gamma_{5,2}} \nonumber \\ 
& = & \quad 
C_{157,1} \left ( 
\ket{02uu} - \ket{20uu} - \ket{uu02} + \ket{uu20} \right) 
 \nonumber \\
& & + 
C_{157,2} \left ( 
\ket{0u2u} - \ket{0uu2} + \ket{2u0u} - \ket{2uu0} - \ket{u02u} + \ket{u0u2} - \ket{u20u} + \ket{u2u0} \right) 
 \nonumber \\
& & + 
C_{157,3} \left ( 
\ket{duuu} + \ket{uduu} - \ket{uudu} - \ket{uuud} \right) 
\nonumber \eeq
\beq 
C_{157,1} &=& 
4 t^2 \nonumber \\
C_{157,2} &=& 
\frac{1}{12} \left (
3 J t+4 U t-4 W t+2 \cos \left(\theta _4\right) \sqrt{A_3} t \right ) \nonumber \\
C_{157,3} &=& 
-\frac{1}{288} \left (
9 J^2+24 U J-24 W J-1152 t^2-48 U^2-1056 U W \right . \nonumber \\
&& \hspace{1cm} 
 \left . -4080 W^2+240 \cos \left(\theta _4\right) \sqrt{A_3} W+4 A_{17}^2+12 J \cos \left(\theta _4\right) \sqrt{A_3}+48 U \cos \left(\theta _4\right) \sqrt{A_3} \right) \nonumber \\
 N_{157} &=& 2 \sqrt{C_{157,1}^2+2 C_{157,2}^2+C_{157,3}^2} \nonumber \eeq 
\beq
\ket{\Psi_{158}} & = & \ket{4,1,2,\Gamma_{5,2}} \nonumber \\ 
& = & \quad 
C_{158,1} \left ( 
\ket{02uu} - \ket{20uu} - \ket{uu02} + \ket{uu20} \right) 
 \nonumber \\
& & + 
C_{158,2} \left ( 
\ket{0u2u} - \ket{0uu2} + \ket{2u0u} - \ket{2uu0} - \ket{u02u} + \ket{u0u2} - \ket{u20u} + \ket{u2u0} \right) 
 \nonumber \\
& & + 
C_{158,3} \left ( 
\ket{duuu} + \ket{uduu} - \ket{uudu} - \ket{uuud} \right) 
\nonumber \eeq
\beq 
C_{158,1} &=& 
4 t^2 \nonumber \\
C_{158,2} &=& 
-\frac{1}{12} \left (
-3 J t-4 U t+4 W t+\cos \left(\theta _4\right) \sqrt{A_3} t+\sqrt{3} \sin \left(\theta _4\right) \sqrt{A_3} t \right ) \nonumber \\
C_{158,3} &=& 
-\frac{1}{32} \left (
J^2+8 U J+40 W J-128 t^2+16 U^2+160 U W \right . \nonumber \\
&& \hspace{1cm} 
 \left . 400 W^2-160 A_{21} W+16 A_{21}^2-8 J A_{21}-32 U A_{21} \right) \nonumber \\
 N_{158} &=& 2 \sqrt{C_{158,1}^2+2 C_{158,2}^2+C_{158,3}^2} \nonumber \eeq 
\beq
\ket{\Psi_{159}} & = & \ket{4,1,2,\Gamma_{5,2}} \nonumber \\ 
& = & \quad 
C_{159,1} \left ( 
\ket{02uu} - \ket{20uu} - \ket{uu02} + \ket{uu20} \right) 
 \nonumber \\
& & + 
C_{159,2} \left ( 
\ket{0u2u} - \ket{0uu2} + \ket{2u0u} - \ket{2uu0} - \ket{u02u} + \ket{u0u2} - \ket{u20u} + \ket{u2u0} \right) 
 \nonumber \\
& & + 
C_{159,3} \left ( 
\ket{duuu} + \ket{uduu} - \ket{uudu} - \ket{uuud} \right) 
\nonumber \eeq
\beq 
C_{159,1} &=& 
4 t^2 \nonumber \\
C_{159,2} &=& 
\frac{1}{12} \left (
3 J t+4 U t-4 W t-\cos \left(\theta _4\right) \sqrt{A_3} t+\sqrt{3} \sin \left(\theta _4\right) \sqrt{A_3} t \right ) \nonumber \\
C_{159,3} &=& 
-\frac{1}{32} \left (
J^2+8 U J+40 W J-128 t^2+16 U^2+160 U W \right . \nonumber \\
&& \hspace{1cm} 
 \left . 400 W^2-160 A_{20} W+16 A_{20}^2-8 J A_{20}-32 U A_{20} \right) \nonumber \\
 N_{159} &=& 2 \sqrt{C_{159,1}^2+2 C_{159,2}^2+C_{159,3}^2} \nonumber \eeq 
\beq
\ket{\Psi_{160}} & = & \ket{4,1,2,\Gamma_{5,3}} \nonumber \\ 
& = & \quad 
C_{160,1} \left ( 
\ket{02uu} - \ket{0u2u} + \ket{20uu} - \ket{2u0u} + \ket{u0u2} + \ket{u2u0} - \ket{uu02} - \ket{uu20} \right) 
 \nonumber \\
& & + 
C_{160,2} \left ( 
\ket{0uu2} - \ket{2uu0} - \ket{u02u} + \ket{u20u} \right) 
 \nonumber \\
& & + 
C_{160,3} \left ( 
\ket{duuu} - \ket{uduu} - \ket{uudu} + \ket{uuud} \right) 
\nonumber \eeq
\beq 
C_{160,1} &=& 
\frac{1}{12} \left (
3 J t+4 U t-4 W t+2 \cos \left(\theta _4\right) \sqrt{A_3} t \right ) \nonumber \\
C_{160,2} &=& 
-4 t^2 \nonumber \\
C_{160,3} &=& 
\frac{1}{288} \left (
9 J^2+24 U J-24 W J-1152 t^2-48 U^2-1056 U W \right . \nonumber \\
&& \hspace{1cm} 
 \left . -4080 W^2+240 \cos \left(\theta _4\right) \sqrt{A_3} W+4 A_{17}^2+12 J \cos \left(\theta _4\right) \sqrt{A_3}+48 U \cos \left(\theta _4\right) \sqrt{A_3} \right) \nonumber \\
 N_{160} &=& 2 \sqrt{2 C_{160,1}^2+C_{160,2}^2+C_{160,3}^2} \nonumber \eeq 
\beq
\ket{\Psi_{161}} & = & \ket{4,1,2,\Gamma_{5,3}} \nonumber \\ 
& = & \quad 
C_{161,1} \left ( 
\ket{02uu} - \ket{0u2u} + \ket{20uu} - \ket{2u0u} + \ket{u0u2} + \ket{u2u0} - \ket{uu02} - \ket{uu20} \right) 
 \nonumber \\
& & + 
C_{161,2} \left ( 
\ket{0uu2} - \ket{2uu0} - \ket{u02u} + \ket{u20u} \right) 
 \nonumber \\
& & + 
C_{161,3} \left ( 
\ket{duuu} - \ket{uduu} - \ket{uudu} + \ket{uuud} \right) 
\nonumber \eeq
\beq 
C_{161,1} &=& 
-\frac{1}{12} \left (
-3 J t-4 U t+4 W t+\cos \left(\theta _4\right) \sqrt{A_3} t+\sqrt{3} \sin \left(\theta _4\right) \sqrt{A_3} t \right ) \nonumber \\
C_{161,2} &=& 
-4 t^2 \nonumber \\
C_{161,3} &=& 
\frac{1}{32} \left (
J^2+8 U J+40 W J-128 t^2+16 U^2+160 U W \right . \nonumber \\
&& \hspace{1cm} 
 \left . 400 W^2-160 A_{21} W+16 A_{21}^2-8 J A_{21}-32 U A_{21} \right) \nonumber \\
 N_{161} &=& 2 \sqrt{2 C_{161,1}^2+C_{161,2}^2+C_{161,3}^2} \nonumber \eeq 
\beq
\ket{\Psi_{162}} & = & \ket{4,1,2,\Gamma_{5,3}} \nonumber \\ 
& = & \quad 
C_{162,1} \left ( 
\ket{02uu} - \ket{0u2u} + \ket{20uu} - \ket{2u0u} + \ket{u0u2} + \ket{u2u0} - \ket{uu02} - \ket{uu20} \right) 
 \nonumber \\
& & + 
C_{162,2} \left ( 
\ket{0uu2} - \ket{2uu0} - \ket{u02u} + \ket{u20u} \right) 
 \nonumber \\
& & + 
C_{162,3} \left ( 
\ket{duuu} - \ket{uduu} - \ket{uudu} + \ket{uuud} \right) 
\nonumber \eeq
\beq 
C_{162,1} &=& 
\frac{1}{12} \left (
3 J t+4 U t-4 W t-\cos \left(\theta _4\right) \sqrt{A_3} t+\sqrt{3} \sin \left(\theta _4\right) \sqrt{A_3} t \right ) \nonumber \\
C_{162,2} &=& 
-4 t^2 \nonumber \\
C_{162,3} &=& 
\frac{1}{32} \left (
J^2+8 U J+40 W J-128 t^2+16 U^2+160 U W \right . \nonumber \\
&& \hspace{1cm} 
 \left . 400 W^2-160 A_{20} W+16 A_{20}^2-8 J A_{20}-32 U A_{20} \right) \nonumber \\
 N_{162} &=& 2 \sqrt{2 C_{162,1}^2+C_{162,2}^2+C_{162,3}^2} \nonumber \eeq 
{\subsubsection{\boldmath Eigenvectors for ${\rm  N_e}=4$ and   ${\rm m_s}$= $2$.}
\beq
\ket{\Psi_{163}} & = & \ket{4,2,6,\Gamma_2} \nonumber \\ 
&=& 1
 \left ( \ket{uuuu} \right) \nonumber 
\eeq
{\subsubsection{\boldmath Eigenvectors for ${\rm  N_e}=5$ and   ${\rm m_s}$= $- \frac{3}{2} $.}
\beq
\ket{\Psi_{164}} & = & \ket{5,- \frac{3}{2} , \frac{15}{4} ,\Gamma_2} \nonumber \\ 
&=& \frac{1}{2}
 \left ( \ket{2ddd} - \ket{d2dd} + \ket{dd2d} - \ket{ddd2} \right) \nonumber 
\eeq
\beq
\ket{\Psi_{165}} & = & \ket{5,- \frac{3}{2} , \frac{15}{4} ,\Gamma_{5,1}} \nonumber \\ 
&=& \frac{1}{2}
 \left ( \ket{2ddd} + \ket{d2dd} + \ket{dd2d} + \ket{ddd2} \right) \nonumber 
\eeq
\beq
\ket{\Psi_{166}} & = & \ket{5,- \frac{3}{2} , \frac{15}{4} ,\Gamma_{5,2}} \nonumber \\ 
&=& \frac{1}{2}
 \left ( \ket{2ddd} - \ket{d2dd} - \ket{dd2d} + \ket{ddd2} \right) \nonumber 
\eeq
\beq
\ket{\Psi_{167}} & = & \ket{5,- \frac{3}{2} , \frac{15}{4} ,\Gamma_{5,3}} \nonumber \\ 
&=& \frac{1}{2}
 \left ( \ket{2ddd} + \ket{d2dd} - \ket{dd2d} - \ket{ddd2} \right) \nonumber 
\eeq
{\subsubsection{\boldmath Eigenvectors for ${\rm  N_e}=5$ and   ${\rm m_s}$= $- \frac{1}{2} $.}
\beq
\ket{\Psi_{168}} & = & \ket{5,- \frac{1}{2} , \frac{3}{4} ,\Gamma_1} \nonumber \\ 
&=& \frac{1}{2 \sqrt{3}}
 \left ( \ket{022d} + \ket{02d2} + \ket{0d22} + \ket{202d} + \ket{20d2} + \ket{220d} \right . \nonumber \\
&& \hspace{3em} 
 + 
\left . \ket{22d0} + \ket{2d02} + \ket{2d20} + \ket{d022} + \ket{d202} + \ket{d220} \right ) 
\nonumber \eeq
\beq
\ket{\Psi_{169}} & = & \ket{5,- \frac{1}{2} , \frac{15}{4} ,\Gamma_2} \nonumber \\ 
&=& \frac{1}{2 \sqrt{3}}
 \left ( \ket{2ddu} + \ket{2dud} + \ket{2udd} - \ket{d2du} - \ket{d2ud} + \ket{dd2u} \right . \nonumber \\
&& \hspace{3em} 
\left . -\ket{ddu2} + \ket{du2d} - \ket{dud2} - \ket{u2dd} + \ket{ud2d} - \ket{udd2} \right ) 
\nonumber \eeq
\beq
\ket{\Psi_{170}} & = & \ket{5,- \frac{1}{2} , \frac{3}{4} ,\Gamma_{3,1}} \nonumber \\ 
& = & \quad 
C_{170,1} \left ( 
\ket{022d} + \ket{0d22} + \ket{20d2} + \ket{220d} + \ket{22d0} + \ket{2d02} + \ket{d022} + \ket{d220} \right) 
 \nonumber \\
& & + 
C_{170,2} \left ( 
\ket{02d2} + \ket{202d} + \ket{2d20} + \ket{d202} \right) 
 \nonumber \\
& & + 
C_{170,3} \left ( 
\ket{2ddu} - \ket{2udd} - \ket{d2ud} - \ket{dd2u} + \ket{ddu2} + \ket{du2d} + \ket{u2dd} - \ket{udd2} \right) 
\nonumber \eeq
\beq 
C_{170,1} &=& 
-\frac{t}{2 \sqrt{2}} \nonumber \\
C_{170,2} &=& 
\frac{t}{\sqrt{2}} \nonumber \\
C_{170,3} &=& 
\frac{1}{16 \sqrt{2}} \left (
3 J+8 t+4 U-4 W+\sqrt{A_6} \right ) \nonumber \\
 N_{170} &=& 2 \sqrt{2 C_{170,1}^2+C_{170,2}^2+2 C_{170,3}^2} \nonumber \eeq 
\beq
\ket{\Psi_{171}} & = & \ket{5,- \frac{1}{2} , \frac{3}{4} ,\Gamma_{3,1}} \nonumber \\ 
& = & \quad 
C_{171,1} \left ( 
\ket{022d} + \ket{0d22} + \ket{20d2} + \ket{220d} + \ket{22d0} + \ket{2d02} + \ket{d022} + \ket{d220} \right) 
 \nonumber \\
& & + 
C_{171,2} \left ( 
\ket{02d2} + \ket{202d} + \ket{2d20} + \ket{d202} \right) 
 \nonumber \\
& & + 
C_{171,3} \left ( 
\ket{2ddu} - \ket{2udd} - \ket{d2ud} - \ket{dd2u} + \ket{ddu2} + \ket{du2d} + \ket{u2dd} - \ket{udd2} \right) 
\nonumber \eeq
\beq 
C_{171,1} &=& 
-\frac{t}{2 \sqrt{2}} \nonumber \\
C_{171,2} &=& 
\frac{t}{\sqrt{2}} \nonumber \\
C_{171,3} &=& 
\frac{1}{16 \sqrt{2}} \left (
3 J+8 t+4 U-4 W-\sqrt{A_6} \right ) \nonumber \\
 N_{171} &=& 2 \sqrt{2 C_{171,1}^2+C_{171,2}^2+2 C_{171,3}^2} \nonumber \eeq 
\beq
\ket{\Psi_{172}} & = & \ket{5,- \frac{1}{2} , \frac{3}{4} ,\Gamma_{3,2}} \nonumber \\ 
& = & \quad 
C_{172,1} \left ( 
\ket{022d} - \ket{0d22} + \ket{20d2} - \ket{220d} - \ket{22d0} + \ket{2d02} - \ket{d022} + \ket{d220} \right) 
 \nonumber \\
& & + 
C_{172,2} \left ( 
\ket{2ddu} + \ket{2udd} - \ket{d2ud} + \ket{dd2u} - \ket{ddu2} + \ket{du2d} - \ket{u2dd} - \ket{udd2} \right) 
 \nonumber \\
& & + 
C_{172,3} \left ( 
\ket{2dud} - \ket{d2du} - \ket{dud2} + \ket{ud2d} \right) 
\nonumber \eeq
\beq 
C_{172,1} &=& 
-\frac{1}{2} \sqrt{\frac{3}{2}} t \nonumber \\
C_{172,2} &=& 
-\frac{1}{16 \sqrt{6}} \left (
3 J+8 t+4 U-4 W+\sqrt{A_6} \right ) \nonumber \\
C_{172,3} &=& 
\frac{1}{8 \sqrt{6}} \left (
3 J+8 t+4 U-4 W+\sqrt{A_6} \right ) \nonumber \\
 N_{172} &=& 2 \sqrt{2 C_{172,1}^2+2 C_{172,2}^2+C_{172,3}^2} \nonumber \eeq 
\beq
\ket{\Psi_{173}} & = & \ket{5,- \frac{1}{2} , \frac{3}{4} ,\Gamma_{3,2}} \nonumber \\ 
& = & \quad 
C_{173,1} \left ( 
\ket{022d} - \ket{0d22} + \ket{20d2} - \ket{220d} - \ket{22d0} + \ket{2d02} - \ket{d022} + \ket{d220} \right) 
 \nonumber \\
& & + 
C_{173,2} \left ( 
\ket{2ddu} + \ket{2udd} - \ket{d2ud} + \ket{dd2u} - \ket{ddu2} + \ket{du2d} - \ket{u2dd} - \ket{udd2} \right) 
 \nonumber \\
& & + 
C_{173,3} \left ( 
\ket{2dud} - \ket{d2du} - \ket{dud2} + \ket{ud2d} \right) 
\nonumber \eeq
\beq 
C_{173,1} &=& 
-\frac{1}{2} \sqrt{\frac{3}{2}} t \nonumber \\
C_{173,2} &=& 
-\frac{1}{16 \sqrt{6}} \left (
3 J+8 t+4 U-4 W-\sqrt{A_6} \right ) \nonumber \\
C_{173,3} &=& 
\frac{1}{8 \sqrt{6}} \left (
3 J+8 t+4 U-4 W-\sqrt{A_6} \right ) \nonumber \\
 N_{173} &=& 2 \sqrt{2 C_{173,1}^2+2 C_{173,2}^2+C_{173,3}^2} \nonumber \eeq 
\beq
\ket{\Psi_{174}} & = & \ket{5,- \frac{1}{2} , \frac{3}{4} ,\Gamma_{4,1}} \nonumber \\ 
& = & \quad 
C_{174,1} \left ( 
\ket{022d} + \ket{02d2} + \ket{202d} + \ket{20d2} - \ket{2d02} - \ket{2d20} - \ket{d202} - \ket{d220} \right) 
 \nonumber \\
& & + 
C_{174,2} \left ( 
\ket{0d22} - \ket{220d} - \ket{22d0} + \ket{d022} \right) 
 \nonumber \\
& & + 
C_{174,3} \left ( 
\ket{2ddu} - \ket{2dud} + \ket{d2du} - \ket{d2ud} - \ket{du2d} - \ket{dud2} + \ket{ud2d} + \ket{udd2} \right) 
\nonumber \eeq
\beq 
C_{174,1} &=& 
\frac{1}{8 \sqrt{2}} \left (
16 t^2+3 J t+4 U t-4 W t+2 \cos \left(\theta _1\right) \sqrt{A_1} t \right ) \nonumber \\
C_{174,2} &=& 
-\frac{1}{12 \sqrt{2}} \left (
-48 t^2+3 J t+4 U t-4 W t+2 \cos \left(\theta _1\right) \sqrt{A_1} t \right ) \nonumber \\
C_{174,3} &=& 
\frac{1}{288 \sqrt{2}} \left (
576 t^2+288 U t+1152 W t-576 U^2-9216 W^2-4608 U W \right . \nonumber \\
&& \hspace{1cm} 
+
 \left . -A_{11}^2+12 t A_{11}-48 U A_{11}-192 W A_{11} \right) \nonumber \\
 N_{174} &=& 2 \sqrt{2 C_{174,1}^2+C_{174,2}^2+2 C_{174,3}^2} \nonumber \eeq 
\beq
\ket{\Psi_{175}} & = & \ket{5,- \frac{1}{2} , \frac{3}{4} ,\Gamma_{4,1}} \nonumber \\ 
& = & \quad 
C_{175,1} \left ( 
\ket{022d} + \ket{02d2} + \ket{202d} + \ket{20d2} - \ket{2d02} - \ket{2d20} - \ket{d202} - \ket{d220} \right) 
 \nonumber \\
& & + 
C_{175,2} \left ( 
\ket{0d22} - \ket{220d} - \ket{22d0} + \ket{d022} \right) 
 \nonumber \\
& & + 
C_{175,3} \left ( 
\ket{2ddu} - \ket{2dud} + \ket{d2du} - \ket{d2ud} - \ket{du2d} - \ket{dud2} + \ket{ud2d} + \ket{udd2} \right) 
\nonumber \eeq
\beq 
C_{175,1} &=& 
\frac{1}{2 \sqrt{2}} \left (
t^2+6 U t+24 W t-3 A_{13} t \right ) \nonumber \\
C_{175,2} &=& 
\frac{1}{12 \sqrt{2}} \left (
48 t^2-3 J t-4 U t+4 W t+\cos \left(\theta _1\right) \sqrt{A_1} t+\sqrt{3} \sin \left(\theta _1\right) \sqrt{A_1} t \right ) \nonumber \\
C_{175,3} &=& 
\frac{1}{2 \sqrt{2}} \left (
4 t^2+2 U t+8 W t-4 U^2-64 W^2-32 U W \right . \nonumber \\
&& \hspace{1cm} 
 \left . -A_{13}^2-t A_{13}+4 U A_{13}+16 W A_{13} \right) \nonumber \\
 N_{175} &=& 2 \sqrt{2 C_{175,1}^2+C_{175,2}^2+2 C_{175,3}^2} \nonumber \eeq 
\beq
\ket{\Psi_{176}} & = & \ket{5,- \frac{1}{2} , \frac{3}{4} ,\Gamma_{4,1}} \nonumber \\ 
& = & \quad 
C_{176,1} \left ( 
\ket{022d} + \ket{02d2} + \ket{202d} + \ket{20d2} - \ket{2d02} - \ket{2d20} - \ket{d202} - \ket{d220} \right) 
 \nonumber \\
& & + 
C_{176,2} \left ( 
\ket{0d22} - \ket{220d} - \ket{22d0} + \ket{d022} \right) 
 \nonumber \\
& & + 
C_{176,3} \left ( 
\ket{2ddu} - \ket{2dud} + \ket{d2du} - \ket{d2ud} - \ket{du2d} - \ket{dud2} + \ket{ud2d} + \ket{udd2} \right) 
\nonumber \eeq
\beq 
C_{176,1} &=& 
\frac{1}{2 \sqrt{2}} \left (
t^2+6 U t+24 W t-3 A_{12} t \right ) \nonumber \\
C_{176,2} &=& 
-\frac{1}{12 \sqrt{2}} \left (
-48 t^2+3 J t+4 U t-4 W t-\cos \left(\theta _1\right) \sqrt{A_1} t+\sqrt{3} \sin \left(\theta _1\right) \sqrt{A_1} t \right ) \nonumber \\
C_{176,3} &=& 
\frac{1}{2 \sqrt{2}} \left (
4 t^2+2 U t+8 W t-4 U^2-64 W^2-32 U W \right . \nonumber \\
&& \hspace{1cm} 
 \left . -A_{12}^2-t A_{12}+4 U A_{12}+16 W A_{12} \right) \nonumber \\
 N_{176} &=& 2 \sqrt{2 C_{176,1}^2+C_{176,2}^2+2 C_{176,3}^2} \nonumber \eeq 
\beq
\ket{\Psi_{177}} & = & \ket{5,- \frac{1}{2} , \frac{3}{4} ,\Gamma_{4,2}} \nonumber \\ 
& = & \quad 
C_{177,1} \left ( 
\ket{022d} - \ket{20d2} - \ket{2d02} + \ket{d220} \right) 
 \nonumber \\
& & + 
C_{177,2} \left ( 
\ket{02d2} + \ket{0d22} - \ket{202d} - \ket{220d} + \ket{22d0} + \ket{2d20} - \ket{d022} - \ket{d202} \right) 
 \nonumber \\
& & + 
C_{177,3} \left ( 
\ket{2dud} - \ket{2udd} + \ket{d2du} + \ket{dd2u} + \ket{ddu2} - \ket{dud2} - \ket{u2dd} - \ket{ud2d} \right) 
\nonumber \eeq
\beq 
C_{177,1} &=& 
\frac{1}{288} \left (
27 J^2-108 t J+36 U J+720 t^2-288 U^2+864 t U \right . \nonumber \\
&& \hspace{1cm} 
 \left . -10368 W^2-36 J W+4896 t W-3744 U W+18 J \cos \left(\theta _1\right) \sqrt{A_1} \right. \nonumber \\
&& \hspace{1cm} 
+
 \left . -A_{11}^2+48 t A_{11}-36 U A_{11}-204 W A_{11} \right) \nonumber \\
C_{177,2} &=& 
\frac{1}{12} \left (
24 t^2+3 J t+4 U t-4 W t-\cos \left(\theta _1\right) \sqrt{A_1} t \right ) \nonumber \\
C_{177,3} &=& 
-\frac{1}{24} \left (
-48 t^2+3 J t+4 U t-4 W t+2 \cos \left(\theta _1\right) \sqrt{A_1} t \right ) \nonumber \\
 N_{177} &=& 2 \sqrt{C_{177,1}^2+2 \left(C_{177,2}^2+C_{177,3}^2\right)} \nonumber \eeq 
\beq
\ket{\Psi_{178}} & = & \ket{5,- \frac{1}{2} , \frac{3}{4} ,\Gamma_{4,2}} \nonumber \\ 
& = & \quad 
C_{178,1} \left ( 
\ket{022d} - \ket{20d2} - \ket{2d02} + \ket{d220} \right) 
 \nonumber \\
& & + 
C_{178,2} \left ( 
\ket{02d2} + \ket{0d22} - \ket{202d} - \ket{220d} + \ket{22d0} + \ket{2d20} - \ket{d022} - \ket{d202} \right) 
 \nonumber \\
& & + 
C_{178,3} \left ( 
\ket{2dud} - \ket{2udd} + \ket{d2du} + \ket{dd2u} + \ket{ddu2} - \ket{dud2} - \ket{u2dd} - \ket{ud2d} \right) 
\nonumber \eeq
\beq 
C_{178,1} &=& 
-\frac{1}{32} \left (
-3 J^2+12 t J-4 U J+4 W J-80 t^2+32 U^2-96 t U \right . \nonumber \\
&& \hspace{1cm} 
+
 \left . 1152 W^2-544 t W+416 U W+J \cos \left(\theta _1\right) \sqrt{A_1}+\sqrt{3} J \sin \left(\theta _1\right) \sqrt{A_1} \right. \nonumber \\
&& \hspace{1cm} 
+
 \left . 16 A_{13}^2+64 t A_{13}-48 U A_{13}-272 W A_{13} \right) \nonumber \\
C_{178,2} &=& 
\frac{1}{24} \left (
48 t^2+6 J t+8 U t-8 W t+\cos \left(\theta _1\right) \sqrt{A_1} t+\sqrt{3} \sin \left(\theta _1\right) \sqrt{A_1} t \right ) \nonumber \\
C_{178,3} &=& 
\frac{1}{24} \left (
48 t^2-3 J t-4 U t+4 W t+\cos \left(\theta _1\right) \sqrt{A_1} t+\sqrt{3} \sin \left(\theta _1\right) \sqrt{A_1} t \right ) \nonumber \\
 N_{178} &=& 2 \sqrt{C_{178,1}^2+2 \left(C_{178,2}^2+C_{178,3}^2\right)} \nonumber \eeq 
\beq
\ket{\Psi_{179}} & = & \ket{5,- \frac{1}{2} , \frac{3}{4} ,\Gamma_{4,2}} \nonumber \\ 
& = & \quad 
C_{179,1} \left ( 
\ket{022d} - \ket{20d2} - \ket{2d02} + \ket{d220} \right) 
 \nonumber \\
& & + 
C_{179,2} \left ( 
\ket{02d2} + \ket{0d22} - \ket{202d} - \ket{220d} + \ket{22d0} + \ket{2d20} - \ket{d022} - \ket{d202} \right) 
 \nonumber \\
& & + 
C_{179,3} \left ( 
\ket{2dud} - \ket{2udd} + \ket{d2du} + \ket{dd2u} + \ket{ddu2} - \ket{dud2} - \ket{u2dd} - \ket{ud2d} \right) 
\nonumber \eeq
\beq 
C_{179,1} &=& 
\frac{1}{32} \left (
3 J^2-12 t J+4 U J-4 W J+80 t^2-32 U^2+96 t U \right . \nonumber \\
&& \hspace{1cm} 
+
 \left . -1152 W^2+544 t W-416 U W-J \cos \left(\theta _1\right) \sqrt{A_1}+\sqrt{3} J \sin \left(\theta _1\right) \sqrt{A_1} \right. \nonumber \\
&& \hspace{1cm} 
 \left . -16 A_{12}^2-64 t A_{12}+48 U A_{12}+272 W A_{12} \right) \nonumber \\
C_{179,2} &=& 
\frac{1}{24} \left (
48 t^2+6 J t+8 U t-8 W t+\cos \left(\theta _1\right) \sqrt{A_1} t-\sqrt{3} \sin \left(\theta _1\right) \sqrt{A_1} t \right ) \nonumber \\
C_{179,3} &=& 
-\frac{1}{24} \left (
-48 t^2+3 J t+4 U t-4 W t-\cos \left(\theta _1\right) \sqrt{A_1} t+\sqrt{3} \sin \left(\theta _1\right) \sqrt{A_1} t \right ) \nonumber \\
 N_{179} &=& 2 \sqrt{C_{179,1}^2+2 \left(C_{179,2}^2+C_{179,3}^2\right)} \nonumber \eeq 
\beq
\ket{\Psi_{180}} & = & \ket{5,- \frac{1}{2} , \frac{3}{4} ,\Gamma_{4,3}} \nonumber \\ 
& = & \quad 
C_{180,1} \left ( 
\ket{022d} + \ket{0d22} - \ket{20d2} + \ket{220d} - \ket{22d0} + \ket{2d02} - \ket{d022} - \ket{d220} \right) 
 \nonumber \\
& & + 
C_{180,2} \left ( 
\ket{02d2} - \ket{202d} - \ket{2d20} + \ket{d202} \right) 
 \nonumber \\
& & + 
C_{180,3} \left ( 
\ket{2ddu} - \ket{2udd} + \ket{d2ud} - \ket{dd2u} - \ket{ddu2} + \ket{du2d} - \ket{u2dd} + \ket{udd2} \right) 
\nonumber \eeq
\beq 
C_{180,1} &=& 
\frac{1}{12} \left (
24 t^2+3 J t+4 U t-4 W t-\cos \left(\theta _1\right) \sqrt{A_1} t \right ) \nonumber \\
C_{180,2} &=& 
\frac{1}{288} \left (
27 J^2-108 t J+36 U J+720 t^2-288 U^2+864 t U \right . \nonumber \\
&& \hspace{1cm} 
 \left . -10368 W^2-36 J W+4896 t W-3744 U W+18 J \cos \left(\theta _1\right) \sqrt{A_1} \right. \nonumber \\
&& \hspace{1cm} 
+
 \left . -A_{11}^2+48 t A_{11}-36 U A_{11}-204 W A_{11} \right) \nonumber \\
C_{180,3} &=& 
\frac{1}{24} \left (
-48 t^2+3 J t+4 U t-4 W t+2 \cos \left(\theta _1\right) \sqrt{A_1} t \right ) \nonumber \\
 N_{180} &=& 2 \sqrt{2 C_{180,1}^2+C_{180,2}^2+2 C_{180,3}^2} \nonumber \eeq 
\beq
\ket{\Psi_{181}} & = & \ket{5,- \frac{1}{2} , \frac{3}{4} ,\Gamma_{4,3}} \nonumber \\ 
& = & \quad 
C_{181,1} \left ( 
\ket{022d} + \ket{0d22} - \ket{20d2} + \ket{220d} - \ket{22d0} + \ket{2d02} - \ket{d022} - \ket{d220} \right) 
 \nonumber \\
& & + 
C_{181,2} \left ( 
\ket{02d2} - \ket{202d} - \ket{2d20} + \ket{d202} \right) 
 \nonumber \\
& & + 
C_{181,3} \left ( 
\ket{2ddu} - \ket{2udd} + \ket{d2ud} - \ket{dd2u} - \ket{ddu2} + \ket{du2d} - \ket{u2dd} + \ket{udd2} \right) 
\nonumber \eeq
\beq 
C_{181,1} &=& 
\frac{1}{8} \left (
20 t^2+3 J t-4 U t-36 W t+4 A_{13} t \right ) \nonumber \\
C_{181,2} &=& 
-\frac{1}{32} \left (
-3 J^2+12 t J-4 U J+4 W J-80 t^2+32 U^2-96 t U \right . \nonumber \\
&& \hspace{1cm} 
+
 \left . 1152 W^2-544 t W+416 U W+J \cos \left(\theta _1\right) \sqrt{A_1}+\sqrt{3} J \sin \left(\theta _1\right) \sqrt{A_1} \right. \nonumber \\
&& \hspace{1cm} 
+
 \left . 16 A_{13}^2+64 t A_{13}-48 U A_{13}-272 W A_{13} \right) \nonumber \\
C_{181,3} &=& 
-\frac{1}{24} \left (
48 t^2-3 J t-4 U t+4 W t+\cos \left(\theta _1\right) \sqrt{A_1} t+\sqrt{3} \sin \left(\theta _1\right) \sqrt{A_1} t \right ) \nonumber \\
 N_{181} &=& 2 \sqrt{2 C_{181,1}^2+C_{181,2}^2+2 C_{181,3}^2} \nonumber \eeq 
\beq
\ket{\Psi_{182}} & = & \ket{5,- \frac{1}{2} , \frac{3}{4} ,\Gamma_{4,3}} \nonumber \\ 
& = & \quad 
C_{182,1} \left ( 
\ket{022d} + \ket{0d22} - \ket{20d2} + \ket{220d} - \ket{22d0} + \ket{2d02} - \ket{d022} - \ket{d220} \right) 
 \nonumber \\
& & + 
C_{182,2} \left ( 
\ket{02d2} - \ket{202d} - \ket{2d20} + \ket{d202} \right) 
 \nonumber \\
& & + 
C_{182,3} \left ( 
\ket{2ddu} - \ket{2udd} + \ket{d2ud} - \ket{dd2u} - \ket{ddu2} + \ket{du2d} - \ket{u2dd} + \ket{udd2} \right) 
\nonumber \eeq
\beq 
C_{182,1} &=& 
\frac{1}{8} \left (
20 t^2+3 J t-4 U t-36 W t+4 A_{12} t \right ) \nonumber \\
C_{182,2} &=& 
\frac{1}{32} \left (
3 J^2-12 t J+4 U J-4 W J+80 t^2-32 U^2+96 t U \right . \nonumber \\
&& \hspace{1cm} 
+
 \left . -1152 W^2+544 t W-416 U W-J \cos \left(\theta _1\right) \sqrt{A_1}+\sqrt{3} J \sin \left(\theta _1\right) \sqrt{A_1} \right. \nonumber \\
&& \hspace{1cm} 
 \left . -16 A_{12}^2-64 t A_{12}+48 U A_{12}+272 W A_{12} \right) \nonumber \\
C_{182,3} &=& 
\frac{1}{24} \left (
-48 t^2+3 J t+4 U t-4 W t-\cos \left(\theta _1\right) \sqrt{A_1} t+\sqrt{3} \sin \left(\theta _1\right) \sqrt{A_1} t \right ) \nonumber \\
 N_{182} &=& 2 \sqrt{2 C_{182,1}^2+C_{182,2}^2+2 C_{182,3}^2} \nonumber \eeq 
\beq
\ket{\Psi_{183}} & = & \ket{5,- \frac{1}{2} , \frac{3}{4} ,\Gamma_{5,1}} \nonumber \\ 
& = & \quad 
C_{183,1} \left ( 
\ket{022d} - \ket{0d22} - \ket{20d2} - \ket{220d} + \ket{22d0} + \ket{2d02} + \ket{d022} - \ket{d220} \right) 
 \nonumber \\
& & + 
C_{183,2} \left ( 
\ket{2ddu} + \ket{2udd} + \ket{d2ud} + \ket{dd2u} + \ket{ddu2} + \ket{du2d} + \ket{u2dd} + \ket{udd2} \right) 
 \nonumber \\
& & + 
C_{183,3} \left ( 
\ket{2dud} + \ket{d2du} + \ket{dud2} + \ket{ud2d} \right) 
\nonumber \eeq
\beq 
C_{183,1} &=& 
\frac{1}{2} \sqrt{\frac{3}{2}} t \nonumber \\
C_{183,2} &=& 
\frac{1}{16 \sqrt{6}} \left (
3 J-8 t+4 U-4 W+\sqrt{A_8} \right ) \nonumber \\
C_{183,3} &=& 
-\frac{1}{8 \sqrt{6}} \left (
3 J-8 t+4 U-4 W+\sqrt{A_8} \right ) \nonumber \\
 N_{183} &=& 2 \sqrt{2 C_{183,1}^2+2 C_{183,2}^2+C_{183,3}^2} \nonumber \eeq 
\beq
\ket{\Psi_{184}} & = & \ket{5,- \frac{1}{2} , \frac{3}{4} ,\Gamma_{5,1}} \nonumber \\ 
& = & \quad 
C_{184,1} \left ( 
\ket{022d} - \ket{0d22} - \ket{20d2} - \ket{220d} + \ket{22d0} + \ket{2d02} + \ket{d022} - \ket{d220} \right) 
 \nonumber \\
& & + 
C_{184,2} \left ( 
\ket{2ddu} + \ket{2udd} + \ket{d2ud} + \ket{dd2u} + \ket{ddu2} + \ket{du2d} + \ket{u2dd} + \ket{udd2} \right) 
 \nonumber \\
& & + 
C_{184,3} \left ( 
\ket{2dud} + \ket{d2du} + \ket{dud2} + \ket{ud2d} \right) 
\nonumber \eeq
\beq 
C_{184,1} &=& 
\frac{1}{2} \sqrt{\frac{3}{2}} t \nonumber \\
C_{184,2} &=& 
\frac{1}{16 \sqrt{6}} \left (
3 J-8 t+4 U-4 W-\sqrt{A_8} \right ) \nonumber \\
C_{184,3} &=& 
-\frac{1}{8 \sqrt{6}} \left (
3 J-8 t+4 U-4 W-\sqrt{A_8} \right ) \nonumber \\
 N_{184} &=& 2 \sqrt{2 C_{184,1}^2+2 C_{184,2}^2+C_{184,3}^2} \nonumber \eeq 
\beq
\ket{\Psi_{185}} & = & \ket{5,- \frac{1}{2} , \frac{15}{4} ,\Gamma_{5,1}} \nonumber \\ 
&=& \frac{1}{2 \sqrt{3}}
 \left ( \ket{2ddu} + \ket{2dud} + \ket{2udd} + \ket{d2du} + \ket{d2ud} + \ket{dd2u} \right . \nonumber \\
&& \hspace{3em} 
 + 
\left . \ket{ddu2} + \ket{du2d} + \ket{dud2} + \ket{u2dd} + \ket{ud2d} + \ket{udd2} \right ) 
\nonumber \eeq
\beq
\ket{\Psi_{186}} & = & \ket{5,- \frac{1}{2} , \frac{3}{4} ,\Gamma_{5,2}} \nonumber \\ 
& = & \quad 
C_{186,1} \left ( 
\ket{022d} - \ket{02d2} - \ket{202d} + \ket{20d2} - \ket{2d02} + \ket{2d20} + \ket{d202} - \ket{d220} \right) 
 \nonumber \\
& & + 
C_{186,2} \left ( 
\ket{2ddu} + \ket{2dud} - \ket{d2du} - \ket{d2ud} - \ket{du2d} + \ket{dud2} - \ket{ud2d} + \ket{udd2} \right) 
 \nonumber \\
& & + 
C_{186,3} \left ( 
\ket{2udd} - \ket{dd2u} + \ket{ddu2} - \ket{u2dd} \right) 
\nonumber \eeq
\beq 
C_{186,1} &=& 
\frac{1}{2} \sqrt{\frac{3}{2}} t \nonumber \\
C_{186,2} &=& 
-\frac{1}{16 \sqrt{6}} \left (
3 J-8 t+4 U-4 W+\sqrt{A_8} \right ) \nonumber \\
C_{186,3} &=& 
\frac{1}{8 \sqrt{6}} \left (
3 J-8 t+4 U-4 W+\sqrt{A_8} \right ) \nonumber \\
 N_{186} &=& 2 \sqrt{2 C_{186,1}^2+2 C_{186,2}^2+C_{186,3}^2} \nonumber \eeq 
\beq
\ket{\Psi_{187}} & = & \ket{5,- \frac{1}{2} , \frac{3}{4} ,\Gamma_{5,2}} \nonumber \\ 
& = & \quad 
C_{187,1} \left ( 
\ket{022d} - \ket{02d2} - \ket{202d} + \ket{20d2} - \ket{2d02} + \ket{2d20} + \ket{d202} - \ket{d220} \right) 
 \nonumber \\
& & + 
C_{187,2} \left ( 
\ket{2ddu} + \ket{2dud} - \ket{d2du} - \ket{d2ud} - \ket{du2d} + \ket{dud2} - \ket{ud2d} + \ket{udd2} \right) 
 \nonumber \\
& & + 
C_{187,3} \left ( 
\ket{2udd} - \ket{dd2u} + \ket{ddu2} - \ket{u2dd} \right) 
\nonumber \eeq
\beq 
C_{187,1} &=& 
\frac{1}{2} \sqrt{\frac{3}{2}} t \nonumber \\
C_{187,2} &=& 
-\frac{1}{16 \sqrt{6}} \left (
3 J-8 t+4 U-4 W-\sqrt{A_8} \right ) \nonumber \\
C_{187,3} &=& 
\frac{1}{8 \sqrt{6}} \left (
3 J-8 t+4 U-4 W-\sqrt{A_8} \right ) \nonumber \\
 N_{187} &=& 2 \sqrt{2 C_{187,1}^2+2 C_{187,2}^2+C_{187,3}^2} \nonumber \eeq 
\beq
\ket{\Psi_{188}} & = & \ket{5,- \frac{1}{2} , \frac{15}{4} ,\Gamma_{5,2}} \nonumber \\ 
&=& \frac{1}{2 \sqrt{3}}
 \left ( \ket{2ddu} + \ket{2dud} + \ket{2udd} - \ket{d2du} - \ket{d2ud} - \ket{dd2u} \right . \nonumber \\
&& \hspace{3em} 
 + 
\left . \ket{ddu2} - \ket{du2d} + \ket{dud2} - \ket{u2dd} - \ket{ud2d} + \ket{udd2} \right ) 
\nonumber \eeq
\beq
\ket{\Psi_{189}} & = & \ket{5,- \frac{1}{2} , \frac{3}{4} ,\Gamma_{5,3}} \nonumber \\ 
& = & \quad 
C_{189,1} \left ( 
\ket{02d2} - \ket{0d22} - \ket{202d} + \ket{220d} - \ket{22d0} + \ket{2d20} + \ket{d022} - \ket{d202} \right) 
 \nonumber \\
& & + 
C_{189,2} \left ( 
\ket{2ddu} + \ket{d2ud} - \ket{du2d} - \ket{udd2} \right) 
 \nonumber \\
& & + 
C_{189,3} \left ( 
\ket{2dud} + \ket{2udd} + \ket{d2du} - \ket{dd2u} - \ket{ddu2} - \ket{dud2} + \ket{u2dd} - \ket{ud2d} \right) 
\nonumber \eeq
\beq 
C_{189,1} &=& 
-\frac{1}{2} \sqrt{\frac{3}{2}} t \nonumber \\
C_{189,2} &=& 
-\frac{1}{8 \sqrt{6}} \left (
3 J-8 t+4 U-4 W+\sqrt{A_8} \right ) \nonumber \\
C_{189,3} &=& 
\frac{1}{16 \sqrt{6}} \left (
3 J-8 t+4 U-4 W+\sqrt{A_8} \right ) \nonumber \\
 N_{189} &=& 2 \sqrt{2 C_{189,1}^2+C_{189,2}^2+2 C_{189,3}^2} \nonumber \eeq 
\beq
\ket{\Psi_{190}} & = & \ket{5,- \frac{1}{2} , \frac{3}{4} ,\Gamma_{5,3}} \nonumber \\ 
& = & \quad 
C_{190,1} \left ( 
\ket{02d2} - \ket{0d22} - \ket{202d} + \ket{220d} - \ket{22d0} + \ket{2d20} + \ket{d022} - \ket{d202} \right) 
 \nonumber \\
& & + 
C_{190,2} \left ( 
\ket{2ddu} + \ket{d2ud} - \ket{du2d} - \ket{udd2} \right) 
 \nonumber \\
& & + 
C_{190,3} \left ( 
\ket{2dud} + \ket{2udd} + \ket{d2du} - \ket{dd2u} - \ket{ddu2} - \ket{dud2} + \ket{u2dd} - \ket{ud2d} \right) 
\nonumber \eeq
\beq 
C_{190,1} &=& 
-\frac{1}{2} \sqrt{\frac{3}{2}} t \nonumber \\
C_{190,2} &=& 
-\frac{1}{8 \sqrt{6}} \left (
3 J-8 t+4 U-4 W-\sqrt{A_8} \right ) \nonumber \\
C_{190,3} &=& 
\frac{1}{16 \sqrt{6}} \left (
3 J-8 t+4 U-4 W-\sqrt{A_8} \right ) \nonumber \\
 N_{190} &=& 2 \sqrt{2 C_{190,1}^2+C_{190,2}^2+2 C_{190,3}^2} \nonumber \eeq 
\beq
\ket{\Psi_{191}} & = & \ket{5,- \frac{1}{2} , \frac{15}{4} ,\Gamma_{5,3}} \nonumber \\ 
&=& \frac{1}{2 \sqrt{3}}
 \left ( \ket{2ddu} + \ket{2dud} + \ket{2udd} + \ket{d2du} + \ket{d2ud} - \ket{dd2u} \right . \nonumber \\
&& \hspace{3em} 
\left . -\ket{ddu2} - \ket{du2d} - \ket{dud2} + \ket{u2dd} - \ket{ud2d} - \ket{udd2} \right ) 
\nonumber \eeq
{\subsubsection{\boldmath Eigenvectors for ${\rm  N_e}=5$ and   ${\rm m_s}$= $\frac{1}{2} $.}
\beq
\ket{\Psi_{192}} & = & \ket{5,\frac{1}{2} , \frac{3}{4} ,\Gamma_1} \nonumber \\ 
&=& \frac{1}{2 \sqrt{3}}
 \left ( \ket{022u} + \ket{02u2} + \ket{0u22} + \ket{202u} + \ket{20u2} + \ket{220u} \right . \nonumber \\
&& \hspace{3em} 
 + 
\left . \ket{22u0} + \ket{2u02} + \ket{2u20} + \ket{u022} + \ket{u202} + \ket{u220} \right ) 
\nonumber \eeq
\beq
\ket{\Psi_{193}} & = & \ket{5,\frac{1}{2} , \frac{15}{4} ,\Gamma_2} \nonumber \\ 
&=& \frac{1}{2 \sqrt{3}}
 \left ( \ket{2duu} + \ket{2udu} + \ket{2uud} - \ket{d2uu} + \ket{du2u} - \ket{duu2} \right . \nonumber \\
&& \hspace{3em} 
\left . -\ket{u2du} - \ket{u2ud} + \ket{ud2u} - \ket{udu2} + \ket{uu2d} - \ket{uud2} \right ) 
\nonumber \eeq
\beq
\ket{\Psi_{194}} & = & \ket{5,\frac{1}{2} , \frac{3}{4} ,\Gamma_{3,1}} \nonumber \\ 
& = & \quad 
C_{194,1} \left ( 
\ket{022u} + \ket{0u22} + \ket{20u2} + \ket{220u} + \ket{22u0} + \ket{2u02} + \ket{u022} + \ket{u220} \right) 
 \nonumber \\
& & + 
C_{194,2} \left ( 
\ket{02u2} + \ket{202u} + \ket{2u20} + \ket{u202} \right) 
 \nonumber \\
& & + 
C_{194,3} \left ( 
\ket{2duu} - \ket{2uud} - \ket{d2uu} + \ket{duu2} + \ket{u2du} - \ket{ud2u} + \ket{uu2d} - \ket{uud2} \right) 
\nonumber \eeq
\beq 
C_{194,1} &=& 
-\frac{t}{2 \sqrt{2}} \nonumber \\
C_{194,2} &=& 
\frac{t}{\sqrt{2}} \nonumber \\
C_{194,3} &=& 
\frac{1}{16 \sqrt{2}} \left (
3 J+8 t+4 U-4 W+\sqrt{A_6} \right ) \nonumber \\
 N_{194} &=& 2 \sqrt{2 C_{194,1}^2+C_{194,2}^2+2 C_{194,3}^2} \nonumber \eeq 
\beq
\ket{\Psi_{195}} & = & \ket{5,\frac{1}{2} , \frac{3}{4} ,\Gamma_{3,1}} \nonumber \\ 
& = & \quad 
C_{195,1} \left ( 
\ket{022u} + \ket{0u22} + \ket{20u2} + \ket{220u} + \ket{22u0} + \ket{2u02} + \ket{u022} + \ket{u220} \right) 
 \nonumber \\
& & + 
C_{195,2} \left ( 
\ket{02u2} + \ket{202u} + \ket{2u20} + \ket{u202} \right) 
 \nonumber \\
& & + 
C_{195,3} \left ( 
\ket{2duu} - \ket{2uud} - \ket{d2uu} + \ket{duu2} + \ket{u2du} - \ket{ud2u} + \ket{uu2d} - \ket{uud2} \right) 
\nonumber \eeq
\beq 
C_{195,1} &=& 
-\frac{t}{2 \sqrt{2}} \nonumber \\
C_{195,2} &=& 
\frac{t}{\sqrt{2}} \nonumber \\
C_{195,3} &=& 
\frac{1}{16 \sqrt{2}} \left (
3 J+8 t+4 U-4 W-\sqrt{A_6} \right ) \nonumber \\
 N_{195} &=& 2 \sqrt{2 C_{195,1}^2+C_{195,2}^2+2 C_{195,3}^2} \nonumber \eeq 
\beq
\ket{\Psi_{196}} & = & \ket{5,\frac{1}{2} , \frac{3}{4} ,\Gamma_{3,2}} \nonumber \\ 
& = & \quad 
C_{196,1} \left ( 
\ket{022u} - \ket{0u22} + \ket{20u2} - \ket{220u} - \ket{22u0} + \ket{2u02} - \ket{u022} + \ket{u220} \right) 
 \nonumber \\
& & + 
C_{196,2} \left ( 
\ket{2duu} + \ket{2uud} - \ket{d2uu} - \ket{duu2} - \ket{u2du} + \ket{ud2u} + \ket{uu2d} - \ket{uud2} \right) 
 \nonumber \\
& & + 
C_{196,3} \left ( 
\ket{2udu} + \ket{du2u} - \ket{u2ud} - \ket{udu2} \right) 
\nonumber \eeq
\beq 
C_{196,1} &=& 
\frac{1}{2} \sqrt{\frac{3}{2}} t \nonumber \\
C_{196,2} &=& 
-\frac{1}{16 \sqrt{6}} \left (
3 J+8 t+4 U-4 W+\sqrt{A_6} \right ) \nonumber \\
C_{196,3} &=& 
\frac{1}{8 \sqrt{6}} \left (
3 J+8 t+4 U-4 W+\sqrt{A_6} \right ) \nonumber \\
 N_{196} &=& 2 \sqrt{2 C_{196,1}^2+2 C_{196,2}^2+C_{196,3}^2} \nonumber \eeq 
\beq
\ket{\Psi_{197}} & = & \ket{5,\frac{1}{2} , \frac{3}{4} ,\Gamma_{3,2}} \nonumber \\ 
& = & \quad 
C_{197,1} \left ( 
\ket{022u} - \ket{0u22} + \ket{20u2} - \ket{220u} - \ket{22u0} + \ket{2u02} - \ket{u022} + \ket{u220} \right) 
 \nonumber \\
& & + 
C_{197,2} \left ( 
\ket{2duu} + \ket{2uud} - \ket{d2uu} - \ket{duu2} - \ket{u2du} + \ket{ud2u} + \ket{uu2d} - \ket{uud2} \right) 
 \nonumber \\
& & + 
C_{197,3} \left ( 
\ket{2udu} + \ket{du2u} - \ket{u2ud} - \ket{udu2} \right) 
\nonumber \eeq
\beq 
C_{197,1} &=& 
\frac{1}{2} \sqrt{\frac{3}{2}} t \nonumber \\
C_{197,2} &=& 
-\frac{1}{16 \sqrt{6}} \left (
3 J+8 t+4 U-4 W-\sqrt{A_6} \right ) \nonumber \\
C_{197,3} &=& 
\frac{1}{8 \sqrt{6}} \left (
3 J+8 t+4 U-4 W-\sqrt{A_6} \right ) \nonumber \\
 N_{197} &=& 2 \sqrt{2 C_{197,1}^2+2 C_{197,2}^2+C_{197,3}^2} \nonumber \eeq 
\beq
\ket{\Psi_{198}} & = & \ket{5,\frac{1}{2} , \frac{3}{4} ,\Gamma_{4,1}} \nonumber \\ 
& = & \quad 
C_{198,1} \left ( 
\ket{022u} + \ket{02u2} + \ket{202u} + \ket{20u2} - \ket{2u02} - \ket{2u20} - \ket{u202} - \ket{u220} \right) 
 \nonumber \\
& & + 
C_{198,2} \left ( 
\ket{0u22} - \ket{220u} - \ket{22u0} + \ket{u022} \right) 
 \nonumber \\
& & + 
C_{198,3} \left ( 
\ket{2udu} - \ket{2uud} - \ket{du2u} - \ket{duu2} + \ket{u2du} - \ket{u2ud} + \ket{ud2u} + \ket{udu2} \right) 
\nonumber \eeq
\beq 
C_{198,1} &=& 
\frac{1}{8 \sqrt{2}} \left (
16 t^2+3 J t+4 U t-4 W t+2 \cos \left(\theta _1\right) \sqrt{A_1} t \right ) \nonumber \\
C_{198,2} &=& 
-\frac{1}{12 \sqrt{2}} \left (
-48 t^2+3 J t+4 U t-4 W t+2 \cos \left(\theta _1\right) \sqrt{A_1} t \right ) \nonumber \\
C_{198,3} &=& 
\frac{1}{288 \sqrt{2}} \left (
576 t^2+288 U t+1152 W t-576 U^2-9216 W^2-4608 U W \right . \nonumber \\
&& \hspace{1cm} 
+
 \left . -A_{11}^2+12 t A_{11}-48 U A_{11}-192 W A_{11} \right) \nonumber \\
 N_{198} &=& 2 \sqrt{2 C_{198,1}^2+C_{198,2}^2+2 C_{198,3}^2} \nonumber \eeq 
\beq
\ket{\Psi_{199}} & = & \ket{5,\frac{1}{2} , \frac{3}{4} ,\Gamma_{4,1}} \nonumber \\ 
& = & \quad 
C_{199,1} \left ( 
\ket{022u} + \ket{02u2} + \ket{202u} + \ket{20u2} - \ket{2u02} - \ket{2u20} - \ket{u202} - \ket{u220} \right) 
 \nonumber \\
& & + 
C_{199,2} \left ( 
\ket{0u22} - \ket{220u} - \ket{22u0} + \ket{u022} \right) 
 \nonumber \\
& & + 
C_{199,3} \left ( 
\ket{2udu} - \ket{2uud} - \ket{du2u} - \ket{duu2} + \ket{u2du} - \ket{u2ud} + \ket{ud2u} + \ket{udu2} \right) 
\nonumber \eeq
\beq 
C_{199,1} &=& 
\frac{1}{2 \sqrt{2}} \left (
t^2+6 U t+24 W t-3 A_{13} t \right ) \nonumber \\
C_{199,2} &=& 
\frac{1}{12 \sqrt{2}} \left (
48 t^2-3 J t-4 U t+4 W t+\cos \left(\theta _1\right) \sqrt{A_1} t+\sqrt{3} \sin \left(\theta _1\right) \sqrt{A_1} t \right ) \nonumber \\
C_{199,3} &=& 
\frac{1}{2 \sqrt{2}} \left (
4 t^2+2 U t+8 W t-4 U^2-64 W^2-32 U W \right . \nonumber \\
&& \hspace{1cm} 
 \left . -A_{13}^2-t A_{13}+4 U A_{13}+16 W A_{13} \right) \nonumber \\
 N_{199} &=& 2 \sqrt{2 C_{199,1}^2+C_{199,2}^2+2 C_{199,3}^2} \nonumber \eeq 
\beq
\ket{\Psi_{200}} & = & \ket{5,\frac{1}{2} , \frac{3}{4} ,\Gamma_{4,1}} \nonumber \\ 
& = & \quad 
C_{200,1} \left ( 
\ket{022u} + \ket{02u2} + \ket{202u} + \ket{20u2} - \ket{2u02} - \ket{2u20} - \ket{u202} - \ket{u220} \right) 
 \nonumber \\
& & + 
C_{200,2} \left ( 
\ket{0u22} - \ket{220u} - \ket{22u0} + \ket{u022} \right) 
 \nonumber \\
& & + 
C_{200,3} \left ( 
\ket{2udu} - \ket{2uud} - \ket{du2u} - \ket{duu2} + \ket{u2du} - \ket{u2ud} + \ket{ud2u} + \ket{udu2} \right) 
\nonumber \eeq
\beq 
C_{200,1} &=& 
\frac{1}{2 \sqrt{2}} \left (
t^2+6 U t+24 W t-3 A_{12} t \right ) \nonumber \\
C_{200,2} &=& 
-\frac{1}{12 \sqrt{2}} \left (
-48 t^2+3 J t+4 U t-4 W t-\cos \left(\theta _1\right) \sqrt{A_1} t+\sqrt{3} \sin \left(\theta _1\right) \sqrt{A_1} t \right ) \nonumber \\
C_{200,3} &=& 
\frac{1}{2 \sqrt{2}} \left (
4 t^2+2 U t+8 W t-4 U^2-64 W^2-32 U W \right . \nonumber \\
&& \hspace{1cm} 
 \left . -A_{12}^2-t A_{12}+4 U A_{12}+16 W A_{12} \right) \nonumber \\
 N_{200} &=& 2 \sqrt{2 C_{200,1}^2+C_{200,2}^2+2 C_{200,3}^2} \nonumber \eeq 
\beq
\ket{\Psi_{201}} & = & \ket{5,\frac{1}{2} , \frac{3}{4} ,\Gamma_{4,2}} \nonumber \\ 
& = & \quad 
C_{201,1} \left ( 
\ket{022u} - \ket{20u2} - \ket{2u02} + \ket{u220} \right) 
 \nonumber \\
& & + 
C_{201,2} \left ( 
\ket{02u2} + \ket{0u22} - \ket{202u} - \ket{220u} + \ket{22u0} + \ket{2u20} - \ket{u022} - \ket{u202} \right) 
 \nonumber \\
& & + 
C_{201,3} \left ( 
\ket{2duu} - \ket{2udu} + \ket{d2uu} + \ket{du2u} - \ket{u2ud} + \ket{udu2} - \ket{uu2d} - \ket{uud2} \right) 
\nonumber \eeq
\beq 
C_{201,1} &=& 
\frac{1}{288} \left (
27 J^2-108 t J+36 U J+720 t^2-288 U^2+864 t U \right . \nonumber \\
&& \hspace{1cm} 
 \left . -10368 W^2-36 J W+4896 t W-3744 U W+18 J \cos \left(\theta _1\right) \sqrt{A_1} \right. \nonumber \\
&& \hspace{1cm} 
+
 \left . -A_{11}^2+48 t A_{11}-36 U A_{11}-204 W A_{11} \right) \nonumber \\
C_{201,2} &=& 
\frac{1}{12} \left (
24 t^2+3 J t+4 U t-4 W t-\cos \left(\theta _1\right) \sqrt{A_1} t \right ) \nonumber \\
C_{201,3} &=& 
-\frac{1}{24} \left (
-48 t^2+3 J t+4 U t-4 W t+2 \cos \left(\theta _1\right) \sqrt{A_1} t \right ) \nonumber \\
 N_{201} &=& 2 \sqrt{C_{201,1}^2+2 \left(C_{201,2}^2+C_{201,3}^2\right)} \nonumber \eeq 
\beq
\ket{\Psi_{202}} & = & \ket{5,\frac{1}{2} , \frac{3}{4} ,\Gamma_{4,2}} \nonumber \\ 
& = & \quad 
C_{202,1} \left ( 
\ket{022u} - \ket{20u2} - \ket{2u02} + \ket{u220} \right) 
 \nonumber \\
& & + 
C_{202,2} \left ( 
\ket{02u2} + \ket{0u22} - \ket{202u} - \ket{220u} + \ket{22u0} + \ket{2u20} - \ket{u022} - \ket{u202} \right) 
 \nonumber \\
& & + 
C_{202,3} \left ( 
\ket{2duu} - \ket{2udu} + \ket{d2uu} + \ket{du2u} - \ket{u2ud} + \ket{udu2} - \ket{uu2d} - \ket{uud2} \right) 
\nonumber \eeq
\beq 
C_{202,1} &=& 
-\frac{1}{32} \left (
-3 J^2+12 t J-4 U J+4 W J-80 t^2+32 U^2-96 t U \right . \nonumber \\
&& \hspace{1cm} 
+
 \left . 1152 W^2-544 t W+416 U W+J \cos \left(\theta _1\right) \sqrt{A_1}+\sqrt{3} J \sin \left(\theta _1\right) \sqrt{A_1} \right. \nonumber \\
&& \hspace{1cm} 
+
 \left . 16 A_{13}^2+64 t A_{13}-48 U A_{13}-272 W A_{13} \right) \nonumber \\
C_{202,2} &=& 
\frac{1}{24} \left (
48 t^2+6 J t+8 U t-8 W t+\cos \left(\theta _1\right) \sqrt{A_1} t+\sqrt{3} \sin \left(\theta _1\right) \sqrt{A_1} t \right ) \nonumber \\
C_{202,3} &=& 
\frac{1}{24} \left (
48 t^2-3 J t-4 U t+4 W t+\cos \left(\theta _1\right) \sqrt{A_1} t+\sqrt{3} \sin \left(\theta _1\right) \sqrt{A_1} t \right ) \nonumber \\
 N_{202} &=& 2 \sqrt{C_{202,1}^2+2 \left(C_{202,2}^2+C_{202,3}^2\right)} \nonumber \eeq 
\beq
\ket{\Psi_{203}} & = & \ket{5,\frac{1}{2} , \frac{3}{4} ,\Gamma_{4,2}} \nonumber \\ 
& = & \quad 
C_{203,1} \left ( 
\ket{022u} - \ket{20u2} - \ket{2u02} + \ket{u220} \right) 
 \nonumber \\
& & + 
C_{203,2} \left ( 
\ket{02u2} + \ket{0u22} - \ket{202u} - \ket{220u} + \ket{22u0} + \ket{2u20} - \ket{u022} - \ket{u202} \right) 
 \nonumber \\
& & + 
C_{203,3} \left ( 
\ket{2duu} - \ket{2udu} + \ket{d2uu} + \ket{du2u} - \ket{u2ud} + \ket{udu2} - \ket{uu2d} - \ket{uud2} \right) 
\nonumber \eeq
\beq 
C_{203,1} &=& 
\frac{1}{32} \left (
3 J^2-12 t J+4 U J-4 W J+80 t^2-32 U^2+96 t U \right . \nonumber \\
&& \hspace{1cm} 
+
 \left . -1152 W^2+544 t W-416 U W-J \cos \left(\theta _1\right) \sqrt{A_1}+\sqrt{3} J \sin \left(\theta _1\right) \sqrt{A_1} \right. \nonumber \\
&& \hspace{1cm} 
 \left . -16 A_{12}^2-64 t A_{12}+48 U A_{12}+272 W A_{12} \right) \nonumber \\
C_{203,2} &=& 
\frac{1}{24} \left (
48 t^2+6 J t+8 U t-8 W t+\cos \left(\theta _1\right) \sqrt{A_1} t-\sqrt{3} \sin \left(\theta _1\right) \sqrt{A_1} t \right ) \nonumber \\
C_{203,3} &=& 
-\frac{1}{24} \left (
-48 t^2+3 J t+4 U t-4 W t-\cos \left(\theta _1\right) \sqrt{A_1} t+\sqrt{3} \sin \left(\theta _1\right) \sqrt{A_1} t \right ) \nonumber \\
 N_{203} &=& 2 \sqrt{C_{203,1}^2+2 \left(C_{203,2}^2+C_{203,3}^2\right)} \nonumber \eeq 
\beq
\ket{\Psi_{204}} & = & \ket{5,\frac{1}{2} , \frac{3}{4} ,\Gamma_{4,3}} \nonumber \\ 
& = & \quad 
C_{204,1} \left ( 
\ket{022u} + \ket{0u22} - \ket{20u2} + \ket{220u} - \ket{22u0} + \ket{2u02} - \ket{u022} - \ket{u220} \right) 
 \nonumber \\
& & + 
C_{204,2} \left ( 
\ket{02u2} - \ket{202u} - \ket{2u20} + \ket{u202} \right) 
 \nonumber \\
& & + 
C_{204,3} \left ( 
\ket{2duu} - \ket{2uud} + \ket{d2uu} - \ket{duu2} - \ket{u2du} - \ket{ud2u} + \ket{uu2d} + \ket{uud2} \right) 
\nonumber \eeq
\beq 
C_{204,1} &=& 
-\frac{1}{8 \sqrt{2}} \left (
16 t^2+3 J t+4 U t-4 W t+2 \cos \left(\theta _1\right) \sqrt{A_1} t \right ) \nonumber \\
C_{204,2} &=& 
\frac{1}{12 \sqrt{2}} \left (
-48 t^2+3 J t+4 U t-4 W t+2 \cos \left(\theta _1\right) \sqrt{A_1} t \right ) \nonumber \\
C_{204,3} &=& 
\frac{1}{288 \sqrt{2}} \left (
576 t^2+288 U t+1152 W t-576 U^2-9216 W^2-4608 U W \right . \nonumber \\
&& \hspace{1cm} 
+
 \left . -A_{11}^2+12 t A_{11}-48 U A_{11}-192 W A_{11} \right) \nonumber \\
 N_{204} &=& 2 \sqrt{2 C_{204,1}^2+C_{204,2}^2+2 C_{204,3}^2} \nonumber \eeq 
\beq
\ket{\Psi_{205}} & = & \ket{5,\frac{1}{2} , \frac{3}{4} ,\Gamma_{4,3}} \nonumber \\ 
& = & \quad 
C_{205,1} \left ( 
\ket{022u} + \ket{0u22} - \ket{20u2} + \ket{220u} - \ket{22u0} + \ket{2u02} - \ket{u022} - \ket{u220} \right) 
 \nonumber \\
& & + 
C_{205,2} \left ( 
\ket{02u2} - \ket{202u} - \ket{2u20} + \ket{u202} \right) 
 \nonumber \\
& & + 
C_{205,3} \left ( 
\ket{2duu} - \ket{2uud} + \ket{d2uu} - \ket{duu2} - \ket{u2du} - \ket{ud2u} + \ket{uu2d} + \ket{uud2} \right) 
\nonumber \eeq
\beq 
C_{205,1} &=& 
-\frac{1}{2 \sqrt{2}} \left (
t^2+6 U t+24 W t-3 A_{13} t \right ) \nonumber \\
C_{205,2} &=& 
-\frac{1}{12 \sqrt{2}} \left (
48 t^2-3 J t-4 U t+4 W t+\cos \left(\theta _1\right) \sqrt{A_1} t+\sqrt{3} \sin \left(\theta _1\right) \sqrt{A_1} t \right ) \nonumber \\
C_{205,3} &=& 
\frac{1}{2 \sqrt{2}} \left (
4 t^2+2 U t+8 W t-4 U^2-64 W^2-32 U W \right . \nonumber \\
&& \hspace{1cm} 
 \left . -A_{13}^2-t A_{13}+4 U A_{13}+16 W A_{13} \right) \nonumber \\
 N_{205} &=& 2 \sqrt{2 C_{205,1}^2+C_{205,2}^2+2 C_{205,3}^2} \nonumber \eeq 
\beq
\ket{\Psi_{206}} & = & \ket{5,\frac{1}{2} , \frac{3}{4} ,\Gamma_{4,3}} \nonumber \\ 
& = & \quad 
C_{206,1} \left ( 
\ket{022u} + \ket{0u22} - \ket{20u2} + \ket{220u} - \ket{22u0} + \ket{2u02} - \ket{u022} - \ket{u220} \right) 
 \nonumber \\
& & + 
C_{206,2} \left ( 
\ket{02u2} - \ket{202u} - \ket{2u20} + \ket{u202} \right) 
 \nonumber \\
& & + 
C_{206,3} \left ( 
\ket{2duu} - \ket{2uud} + \ket{d2uu} - \ket{duu2} - \ket{u2du} - \ket{ud2u} + \ket{uu2d} + \ket{uud2} \right) 
\nonumber \eeq
\beq 
C_{206,1} &=& 
-\frac{1}{2 \sqrt{2}} \left (
t^2+6 U t+24 W t-3 A_{12} t \right ) \nonumber \\
C_{206,2} &=& 
\frac{1}{12 \sqrt{2}} \left (
-48 t^2+3 J t+4 U t-4 W t-\cos \left(\theta _1\right) \sqrt{A_1} t+\sqrt{3} \sin \left(\theta _1\right) \sqrt{A_1} t \right ) \nonumber \\
C_{206,3} &=& 
\frac{1}{2 \sqrt{2}} \left (
4 t^2+2 U t+8 W t-4 U^2-64 W^2-32 U W \right . \nonumber \\
&& \hspace{1cm} 
 \left . -A_{12}^2-t A_{12}+4 U A_{12}+16 W A_{12} \right) \nonumber \\
 N_{206} &=& 2 \sqrt{2 C_{206,1}^2+C_{206,2}^2+2 C_{206,3}^2} \nonumber \eeq 
\beq
\ket{\Psi_{207}} & = & \ket{5,\frac{1}{2} , \frac{3}{4} ,\Gamma_{5,1}} \nonumber \\ 
& = & \quad 
C_{207,1} \left ( 
\ket{022u} - \ket{0u22} - \ket{20u2} - \ket{220u} + \ket{22u0} + \ket{2u02} + \ket{u022} - \ket{u220} \right) 
 \nonumber \\
& & + 
C_{207,2} \left ( 
\ket{2duu} + \ket{2uud} + \ket{d2uu} + \ket{duu2} + \ket{u2du} + \ket{ud2u} + \ket{uu2d} + \ket{uud2} \right) 
 \nonumber \\
& & + 
C_{207,3} \left ( 
\ket{2udu} + \ket{du2u} + \ket{u2ud} + \ket{udu2} \right) 
\nonumber \eeq
\beq 
C_{207,1} &=& 
-\frac{1}{2} \sqrt{\frac{3}{2}} t \nonumber \\
C_{207,2} &=& 
\frac{1}{16 \sqrt{6}} \left (
3 J-8 t+4 U-4 W+\sqrt{A_8} \right ) \nonumber \\
C_{207,3} &=& 
-\frac{1}{8 \sqrt{6}} \left (
3 J-8 t+4 U-4 W+\sqrt{A_8} \right ) \nonumber \\
 N_{207} &=& 2 \sqrt{2 C_{207,1}^2+2 C_{207,2}^2+C_{207,3}^2} \nonumber \eeq 
\beq
\ket{\Psi_{208}} & = & \ket{5,\frac{1}{2} , \frac{3}{4} ,\Gamma_{5,1}} \nonumber \\ 
& = & \quad 
C_{208,1} \left ( 
\ket{022u} - \ket{0u22} - \ket{20u2} - \ket{220u} + \ket{22u0} + \ket{2u02} + \ket{u022} - \ket{u220} \right) 
 \nonumber \\
& & + 
C_{208,2} \left ( 
\ket{2duu} + \ket{2uud} + \ket{d2uu} + \ket{duu2} + \ket{u2du} + \ket{ud2u} + \ket{uu2d} + \ket{uud2} \right) 
 \nonumber \\
& & + 
C_{208,3} \left ( 
\ket{2udu} + \ket{du2u} + \ket{u2ud} + \ket{udu2} \right) 
\nonumber \eeq
\beq 
C_{208,1} &=& 
-\frac{1}{2} \sqrt{\frac{3}{2}} t \nonumber \\
C_{208,2} &=& 
\frac{1}{16 \sqrt{6}} \left (
3 J-8 t+4 U-4 W-\sqrt{A_8} \right ) \nonumber \\
C_{208,3} &=& 
-\frac{1}{8 \sqrt{6}} \left (
3 J-8 t+4 U-4 W-\sqrt{A_8} \right ) \nonumber \\
 N_{208} &=& 2 \sqrt{2 C_{208,1}^2+2 C_{208,2}^2+C_{208,3}^2} \nonumber \eeq 
\beq
\ket{\Psi_{209}} & = & \ket{5,\frac{1}{2} , \frac{15}{4} ,\Gamma_{5,1}} \nonumber \\ 
&=& \frac{1}{2 \sqrt{3}}
 \left ( \ket{2duu} + \ket{2udu} + \ket{2uud} + \ket{d2uu} + \ket{du2u} + \ket{duu2} \right . \nonumber \\
&& \hspace{3em} 
 + 
\left . \ket{u2du} + \ket{u2ud} + \ket{ud2u} + \ket{udu2} + \ket{uu2d} + \ket{uud2} \right ) 
\nonumber \eeq
\beq
\ket{\Psi_{210}} & = & \ket{5,\frac{1}{2} , \frac{3}{4} ,\Gamma_{5,2}} \nonumber \\ 
& = & \quad 
C_{210,1} \left ( 
\ket{022u} - \ket{02u2} - \ket{202u} + \ket{20u2} - \ket{2u02} + \ket{2u20} + \ket{u202} - \ket{u220} \right) 
 \nonumber \\
& & + 
C_{210,2} \left ( 
\ket{2duu} - \ket{d2uu} - \ket{uu2d} + \ket{uud2} \right) 
 \nonumber \\
& & + 
C_{210,3} \left ( 
\ket{2udu} + \ket{2uud} - \ket{du2u} + \ket{duu2} - \ket{u2du} - \ket{u2ud} - \ket{ud2u} + \ket{udu2} \right) 
\nonumber \eeq
\beq 
C_{210,1} &=& 
\frac{1}{2} \sqrt{\frac{3}{2}} t \nonumber \\
C_{210,2} &=& 
-\frac{1}{8 \sqrt{6}} \left (
3 J-8 t+4 U-4 W+\sqrt{A_8} \right ) \nonumber \\
C_{210,3} &=& 
\frac{1}{16 \sqrt{6}} \left (
3 J-8 t+4 U-4 W+\sqrt{A_8} \right ) \nonumber \\
 N_{210} &=& 2 \sqrt{2 C_{210,1}^2+C_{210,2}^2+2 C_{210,3}^2} \nonumber \eeq 
\beq
\ket{\Psi_{211}} & = & \ket{5,\frac{1}{2} , \frac{3}{4} ,\Gamma_{5,2}} \nonumber \\ 
& = & \quad 
C_{211,1} \left ( 
\ket{022u} - \ket{02u2} - \ket{202u} + \ket{20u2} - \ket{2u02} + \ket{2u20} + \ket{u202} - \ket{u220} \right) 
 \nonumber \\
& & + 
C_{211,2} \left ( 
\ket{2duu} - \ket{d2uu} - \ket{uu2d} + \ket{uud2} \right) 
 \nonumber \\
& & + 
C_{211,3} \left ( 
\ket{2udu} + \ket{2uud} - \ket{du2u} + \ket{duu2} - \ket{u2du} - \ket{u2ud} - \ket{ud2u} + \ket{udu2} \right) 
\nonumber \eeq
\beq 
C_{211,1} &=& 
\frac{1}{2} \sqrt{\frac{3}{2}} t \nonumber \\
C_{211,2} &=& 
-\frac{1}{8 \sqrt{6}} \left (
3 J-8 t+4 U-4 W-\sqrt{A_8} \right ) \nonumber \\
C_{211,3} &=& 
\frac{1}{16 \sqrt{6}} \left (
3 J-8 t+4 U-4 W-\sqrt{A_8} \right ) \nonumber \\
 N_{211} &=& 2 \sqrt{2 C_{211,1}^2+C_{211,2}^2+2 C_{211,3}^2} \nonumber \eeq 
\beq
\ket{\Psi_{212}} & = & \ket{5,\frac{1}{2} , \frac{15}{4} ,\Gamma_{5,2}} \nonumber \\ 
&=& \frac{1}{2 \sqrt{3}}
 \left ( \ket{2duu} + \ket{2udu} + \ket{2uud} - \ket{d2uu} - \ket{du2u} + \ket{duu2} \right . \nonumber \\
&& \hspace{3em} 
\left . -\ket{u2du} - \ket{u2ud} - \ket{ud2u} + \ket{udu2} - \ket{uu2d} + \ket{uud2} \right ) 
\nonumber \eeq
\beq
\ket{\Psi_{213}} & = & \ket{5,\frac{1}{2} , \frac{3}{4} ,\Gamma_{5,3}} \nonumber \\ 
& = & \quad 
C_{213,1} \left ( 
\ket{02u2} - \ket{0u22} - \ket{202u} + \ket{220u} - \ket{22u0} + \ket{2u20} + \ket{u022} - \ket{u202} \right) 
 \nonumber \\
& & + 
C_{213,2} \left ( 
\ket{2duu} + \ket{2udu} + \ket{d2uu} - \ket{du2u} + \ket{u2ud} - \ket{udu2} - \ket{uu2d} - \ket{uud2} \right) 
 \nonumber \\
& & + 
C_{213,3} \left ( 
\ket{2uud} - \ket{duu2} + \ket{u2du} - \ket{ud2u} \right) 
\nonumber \eeq
\beq 
C_{213,1} &=& 
-\frac{1}{2} \sqrt{\frac{3}{2}} t \nonumber \\
C_{213,2} &=& 
-\frac{1}{16 \sqrt{6}} \left (
3 J-8 t+4 U-4 W+\sqrt{A_8} \right ) \nonumber \\
C_{213,3} &=& 
\frac{1}{8 \sqrt{6}} \left (
3 J-8 t+4 U-4 W+\sqrt{A_8} \right ) \nonumber \\
 N_{213} &=& 2 \sqrt{2 C_{213,1}^2+2 C_{213,2}^2+C_{213,3}^2} \nonumber \eeq 
\beq
\ket{\Psi_{214}} & = & \ket{5,\frac{1}{2} , \frac{3}{4} ,\Gamma_{5,3}} \nonumber \\ 
& = & \quad 
C_{214,1} \left ( 
\ket{02u2} - \ket{0u22} - \ket{202u} + \ket{220u} - \ket{22u0} + \ket{2u20} + \ket{u022} - \ket{u202} \right) 
 \nonumber \\
& & + 
C_{214,2} \left ( 
\ket{2duu} + \ket{2udu} + \ket{d2uu} - \ket{du2u} + \ket{u2ud} - \ket{udu2} - \ket{uu2d} - \ket{uud2} \right) 
 \nonumber \\
& & + 
C_{214,3} \left ( 
\ket{2uud} - \ket{duu2} + \ket{u2du} - \ket{ud2u} \right) 
\nonumber \eeq
\beq 
C_{214,1} &=& 
-\frac{1}{2} \sqrt{\frac{3}{2}} t \nonumber \\
C_{214,2} &=& 
-\frac{1}{16 \sqrt{6}} \left (
3 J-8 t+4 U-4 W-\sqrt{A_8} \right ) \nonumber \\
C_{214,3} &=& 
\frac{1}{8 \sqrt{6}} \left (
3 J-8 t+4 U-4 W-\sqrt{A_8} \right ) \nonumber \\
 N_{214} &=& 2 \sqrt{2 C_{214,1}^2+2 C_{214,2}^2+C_{214,3}^2} \nonumber \eeq 
\beq
\ket{\Psi_{215}} & = & \ket{5,\frac{1}{2} , \frac{15}{4} ,\Gamma_{5,3}} \nonumber \\ 
&=& \frac{1}{2 \sqrt{3}}
 \left ( \ket{2duu} + \ket{2udu} + \ket{2uud} + \ket{d2uu} - \ket{du2u} - \ket{duu2} \right . \nonumber \\
&& \hspace{3em} 
 + 
\left . \ket{u2du} + \ket{u2ud} - \ket{ud2u} - \ket{udu2} - \ket{uu2d} - \ket{uud2} \right ) 
\nonumber \eeq
{\subsubsection{\boldmath Eigenvectors for ${\rm  N_e}=5$ and   ${\rm m_s}$= $\frac{3}{2} $.}
\beq
\ket{\Psi_{216}} & = & \ket{5,\frac{3}{2} , \frac{15}{4} ,\Gamma_2} \nonumber \\ 
&=& \frac{1}{2}
 \left ( \ket{2uuu} - \ket{u2uu} + \ket{uu2u} - \ket{uuu2} \right) \nonumber 
\eeq
\beq
\ket{\Psi_{217}} & = & \ket{5,\frac{3}{2} , \frac{15}{4} ,\Gamma_{5,1}} \nonumber \\ 
&=& \frac{1}{2}
 \left ( \ket{2uuu} + \ket{u2uu} + \ket{uu2u} + \ket{uuu2} \right) \nonumber 
\eeq
\beq
\ket{\Psi_{218}} & = & \ket{5,\frac{3}{2} , \frac{15}{4} ,\Gamma_{5,2}} \nonumber \\ 
&=& \frac{1}{2}
 \left ( \ket{2uuu} - \ket{u2uu} - \ket{uu2u} + \ket{uuu2} \right) \nonumber 
\eeq
\beq
\ket{\Psi_{219}} & = & \ket{5,\frac{3}{2} , \frac{15}{4} ,\Gamma_{5,3}} \nonumber \\ 
&=& \frac{1}{2}
 \left ( \ket{2uuu} + \ket{u2uu} - \ket{uu2u} - \ket{uuu2} \right) \nonumber 
\eeq
{\subsubsection{\boldmath Eigenvectors for ${\rm  N_e}=6$ and   ${\rm m_s}$= $-1$.}
\beq
\ket{\Psi_{220}} & = & \ket{6,-1,2,\Gamma_{4,1}} \nonumber \\ 
&=& \frac{1}{2}
 \left ( \ket{2d2d} + \ket{2dd2} + \ket{d22d} + \ket{d2d2} \right) \nonumber 
\eeq
\beq
\ket{\Psi_{221}} & = & \ket{6,-1,2,\Gamma_{4,2}} \nonumber \\ 
&=& \frac{1}{2}
 \left ( \ket{22dd} + \ket{2d2d} - \ket{d2d2} - \ket{dd22} \right) \nonumber 
\eeq
\beq
\ket{\Psi_{222}} & = & \ket{6,-1,2,\Gamma_{4,3}} \nonumber \\ 
&=& \frac{1}{2}
 \left ( \ket{22dd} - \ket{2dd2} + \ket{d22d} + \ket{dd22} \right) \nonumber 
\eeq
\beq
\ket{\Psi_{223}} & = & \ket{6,-1,2,\Gamma_{5,1}} \nonumber \\ 
&=& \frac{1}{2}
 \left ( \ket{22dd} + \ket{2dd2} - \ket{d22d} + \ket{dd22} \right) \nonumber 
\eeq
\beq
\ket{\Psi_{224}} & = & \ket{6,-1,2,\Gamma_{5,2}} \nonumber \\ 
&=& \frac{1}{2}
 \left ( \ket{2d2d} - \ket{2dd2} - \ket{d22d} + \ket{d2d2} \right) \nonumber 
\eeq
\beq
\ket{\Psi_{225}} & = & \ket{6,-1,2,\Gamma_{5,3}} \nonumber \\ 
&=& \frac{1}{2}
 \left ( \ket{22dd} - \ket{2d2d} + \ket{d2d2} - \ket{dd22} \right) \nonumber 
\eeq
{\subsubsection{\boldmath Eigenvectors for ${\rm  N_e}=6$ and   ${\rm m_s}$= $0$.}
\beq
\ket{\Psi_{226}} & = & \ket{6,0,0,\Gamma_1} \nonumber \\ 
& = & \quad 
C_{226,1} \left ( 
\ket{0222} + \ket{2022} + \ket{2202} + \ket{2220} \right) 
 \nonumber \\
& & + 
C_{226,2} \left ( 
\ket{22du} - \ket{22ud} + \ket{2d2u} + \ket{2du2} - \ket{2u2d} - \ket{2ud2} \right . \nonumber \\
&& \hspace{3em} 
 + 
\left . \ket{d22u} + \ket{d2u2} + \ket{du22} - \ket{u22d} - \ket{u2d2} - \ket{ud22}\right ) 
\nonumber \eeq
\beq 
C_{226,1} &=& 
\sqrt{3} t \nonumber \\
C_{226,2} &=& 
\frac{1}{16 \sqrt{3}} \left (
3 J+16 t+4 U-4 W+\sqrt{A_7} \right ) \nonumber \\
 N_{226} &=& 2 \sqrt{C_{226,1}^2+3 C_{226,2}^2} \nonumber \eeq 
\beq
\ket{\Psi_{227}} & = & \ket{6,0,0,\Gamma_1} \nonumber \\ 
& = & \quad 
C_{227,1} \left ( 
\ket{0222} + \ket{2022} + \ket{2202} + \ket{2220} \right) 
 \nonumber \\
& & + 
C_{227,2} \left ( 
\ket{22du} - \ket{22ud} + \ket{2d2u} + \ket{2du2} - \ket{2u2d} - \ket{2ud2} \right . \nonumber \\
&& \hspace{3em} 
 + 
\left . \ket{d22u} + \ket{d2u2} + \ket{du22} - \ket{u22d} - \ket{u2d2} - \ket{ud22}\right ) 
\nonumber \eeq
\beq 
C_{227,1} &=& 
\sqrt{3} t \nonumber \\
C_{227,2} &=& 
\frac{1}{16 \sqrt{3}} \left (
3 J+16 t+4 U-4 W-\sqrt{A_7} \right ) \nonumber \\
 N_{227} &=& 2 \sqrt{C_{227,1}^2+3 C_{227,2}^2} \nonumber \eeq 
\beq
\ket{\Psi_{228}} & = & \ket{6,0,0,\Gamma_{3,1}} \nonumber \\ 
& = & \quad 
C_{228,1} \left ( 
\ket{22du} - \ket{22ud} + \ket{2du2} - \ket{2ud2} + \ket{d22u} + \ket{du22} - \ket{u22d} - \ket{ud22} \right) 
 \nonumber \\
& & + 
C_{228,2} \left ( 
\ket{2d2u} - \ket{2u2d} + \ket{d2u2} - \ket{u2d2} \right) 
\nonumber \eeq
\beq 
C_{228,1} &=& 
-\frac{1}{2 \sqrt{6}} \nonumber \\
C_{228,2} &=& 
\frac{1}{\sqrt{6}} \nonumber \\
 N_{228} &=& 2 \sqrt{2 C_{228,1}^2+C_{228,2}^2} \nonumber \eeq 
\beq
\ket{\Psi_{229}} & = & \ket{6,0,0,\Gamma_{3,2}} \nonumber \\ 
&=& \frac{1}{2 \sqrt{2}}
 \left ( \ket{22du} - \ket{22ud} - \ket{2du2} + \ket{2ud2} - \ket{d22u} + \ket{du22} + \ket{u22d} - \ket{ud22} \right) \nonumber 
\eeq
\beq
\ket{\Psi_{230}} & = & \ket{6,0,0,\Gamma_{4,1}} \nonumber \\ 
& = & \quad 
C_{230,1} \left ( 
\ket{0222} + \ket{2022} - \ket{2202} - \ket{2220} \right) 
 \nonumber \\
& & + 
C_{230,2} \left ( 
\ket{22du} - \ket{22ud} - \ket{du22} + \ket{ud22} \right) 
\nonumber \eeq
\beq 
C_{230,1} &=& 
-t \nonumber \\
C_{230,2} &=& 
\frac{1}{16} \left (
3 J+4 U-4 W+\sqrt{A_2} \right ) \nonumber \\
 N_{230} &=& 2 \sqrt{C_{230,1}^2+C_{230,2}^2} \nonumber \eeq 
\beq
\ket{\Psi_{231}} & = & \ket{6,0,0,\Gamma_{4,1}} \nonumber \\ 
& = & \quad 
C_{231,1} \left ( 
\ket{0222} + \ket{2022} - \ket{2202} - \ket{2220} \right) 
 \nonumber \\
& & + 
C_{231,2} \left ( 
\ket{22du} - \ket{22ud} - \ket{du22} + \ket{ud22} \right) 
\nonumber \eeq
\beq 
C_{231,1} &=& 
-t \nonumber \\
C_{231,2} &=& 
\frac{1}{16} \left (
3 J+4 U-4 W-\sqrt{A_2} \right ) \nonumber \\
 N_{231} &=& 2 \sqrt{C_{231,1}^2+C_{231,2}^2} \nonumber \eeq 
\beq
\ket{\Psi_{232}} & = & \ket{6,0,2,\Gamma_{4,1}} \nonumber \\ 
&=& \frac{1}{2 \sqrt{2}}
 \left ( \ket{2d2u} + \ket{2du2} + \ket{2u2d} + \ket{2ud2} + \ket{d22u} + \ket{d2u2} + \ket{u22d} + \ket{u2d2} \right) \nonumber 
\eeq
\beq
\ket{\Psi_{233}} & = & \ket{6,0,0,\Gamma_{4,2}} \nonumber \\ 
& = & \quad 
C_{233,1} \left ( 
\ket{0222} - \ket{2022} - \ket{2202} + \ket{2220} \right) 
 \nonumber \\
& & + 
C_{233,2} \left ( 
\ket{2du2} - \ket{2ud2} - \ket{d22u} + \ket{u22d} \right) 
\nonumber \eeq
\beq 
C_{233,1} &=& 
-t \nonumber \\
C_{233,2} &=& 
\frac{1}{16} \left (
3 J+4 U-4 W+\sqrt{A_2} \right ) \nonumber \\
 N_{233} &=& 2 \sqrt{C_{233,1}^2+C_{233,2}^2} \nonumber \eeq 
\beq
\ket{\Psi_{234}} & = & \ket{6,0,0,\Gamma_{4,2}} \nonumber \\ 
& = & \quad 
C_{234,1} \left ( 
\ket{0222} - \ket{2022} - \ket{2202} + \ket{2220} \right) 
 \nonumber \\
& & + 
C_{234,2} \left ( 
\ket{2du2} - \ket{2ud2} - \ket{d22u} + \ket{u22d} \right) 
\nonumber \eeq
\beq 
C_{234,1} &=& 
-t \nonumber \\
C_{234,2} &=& 
\frac{1}{16} \left (
3 J+4 U-4 W-\sqrt{A_2} \right ) \nonumber \\
 N_{234} &=& 2 \sqrt{C_{234,1}^2+C_{234,2}^2} \nonumber \eeq 
\beq
\ket{\Psi_{235}} & = & \ket{6,0,2,\Gamma_{4,2}} \nonumber \\ 
&=& \frac{1}{2 \sqrt{2}}
 \left ( \ket{22du} + \ket{22ud} + \ket{2d2u} + \ket{2u2d} - \ket{d2u2} - \ket{du22} - \ket{u2d2} - \ket{ud22} \right) \nonumber 
\eeq
\beq
\ket{\Psi_{236}} & = & \ket{6,0,0,\Gamma_{4,3}} \nonumber \\ 
& = & \quad 
C_{236,1} \left ( 
\ket{0222} - \ket{2022} + \ket{2202} - \ket{2220} \right) 
 \nonumber \\
& & + 
C_{236,2} \left ( 
\ket{2d2u} - \ket{2u2d} - \ket{d2u2} + \ket{u2d2} \right) 
\nonumber \eeq
\beq 
C_{236,1} &=& 
-t \nonumber \\
C_{236,2} &=& 
\frac{1}{16} \left (
3 J+4 U-4 W+\sqrt{A_2} \right ) \nonumber \\
 N_{236} &=& 2 \sqrt{C_{236,1}^2+C_{236,2}^2} \nonumber \eeq 
\beq
\ket{\Psi_{237}} & = & \ket{6,0,0,\Gamma_{4,3}} \nonumber \\ 
& = & \quad 
C_{237,1} \left ( 
\ket{0222} - \ket{2022} + \ket{2202} - \ket{2220} \right) 
 \nonumber \\
& & + 
C_{237,2} \left ( 
\ket{2d2u} - \ket{2u2d} - \ket{d2u2} + \ket{u2d2} \right) 
\nonumber \eeq
\beq 
C_{237,1} &=& 
-t \nonumber \\
C_{237,2} &=& 
\frac{1}{16} \left (
3 J+4 U-4 W-\sqrt{A_2} \right ) \nonumber \\
 N_{237} &=& 2 \sqrt{C_{237,1}^2+C_{237,2}^2} \nonumber \eeq 
\beq
\ket{\Psi_{238}} & = & \ket{6,0,2,\Gamma_{4,3}} \nonumber \\ 
&=& \frac{1}{2 \sqrt{2}}
 \left ( \ket{22du} + \ket{22ud} - \ket{2du2} - \ket{2ud2} + \ket{d22u} + \ket{du22} + \ket{u22d} + \ket{ud22} \right) \nonumber 
\eeq
\beq
\ket{\Psi_{239}} & = & \ket{6,0,2,\Gamma_{5,1}} \nonumber \\ 
&=& \frac{1}{2 \sqrt{2}}
 \left ( \ket{22du} + \ket{22ud} + \ket{2du2} + \ket{2ud2} - \ket{d22u} + \ket{du22} - \ket{u22d} + \ket{ud22} \right) \nonumber 
\eeq
\beq
\ket{\Psi_{240}} & = & \ket{6,0,2,\Gamma_{5,2}} \nonumber \\ 
&=& \frac{1}{2 \sqrt{2}}
 \left ( \ket{2d2u} - \ket{2du2} + \ket{2u2d} - \ket{2ud2} - \ket{d22u} + \ket{d2u2} - \ket{u22d} + \ket{u2d2} \right) \nonumber 
\eeq
\beq
\ket{\Psi_{241}} & = & \ket{6,0,2,\Gamma_{5,3}} \nonumber \\ 
&=& \frac{1}{2 \sqrt{2}}
 \left ( \ket{22du} + \ket{22ud} - \ket{2d2u} - \ket{2u2d} + \ket{d2u2} - \ket{du22} + \ket{u2d2} - \ket{ud22} \right) \nonumber 
\eeq
{\subsubsection{\boldmath Eigenvectors for ${\rm  N_e}=6$ and   ${\rm m_s}$= $1$.}
\beq
\ket{\Psi_{242}} & = & \ket{6,1,2,\Gamma_{4,1}} \nonumber \\ 
&=& \frac{1}{2}
 \left ( \ket{2u2u} + \ket{2uu2} + \ket{u22u} + \ket{u2u2} \right) \nonumber 
\eeq
\beq
\ket{\Psi_{243}} & = & \ket{6,1,2,\Gamma_{4,2}} \nonumber \\ 
&=& \frac{1}{2}
 \left ( \ket{22uu} + \ket{2u2u} - \ket{u2u2} - \ket{uu22} \right) \nonumber 
\eeq
\beq
\ket{\Psi_{244}} & = & \ket{6,1,2,\Gamma_{4,3}} \nonumber \\ 
&=& \frac{1}{2}
 \left ( \ket{22uu} - \ket{2uu2} + \ket{u22u} + \ket{uu22} \right) \nonumber 
\eeq
\beq
\ket{\Psi_{245}} & = & \ket{6,1,2,\Gamma_{5,1}} \nonumber \\ 
&=& \frac{1}{2}
 \left ( \ket{22uu} + \ket{2uu2} - \ket{u22u} + \ket{uu22} \right) \nonumber 
\eeq
\beq
\ket{\Psi_{246}} & = & \ket{6,1,2,\Gamma_{5,2}} \nonumber \\ 
&=& \frac{1}{2}
 \left ( \ket{2u2u} - \ket{2uu2} - \ket{u22u} + \ket{u2u2} \right) \nonumber 
\eeq
\beq
\ket{\Psi_{247}} & = & \ket{6,1,2,\Gamma_{5,3}} \nonumber \\ 
&=& \frac{1}{2}
 \left ( \ket{22uu} - \ket{2u2u} + \ket{u2u2} - \ket{uu22} \right) \nonumber 
\eeq
{\subsubsection{\boldmath Eigenvectors for ${\rm  N_e}=7$ and   ${\rm m_s}$= $- \frac{1}{2} $.}
\beq
\ket{\Psi_{248}} & = & \ket{7,- \frac{1}{2} , \frac{3}{4} ,\Gamma_1} \nonumber \\ 
&=& \frac{1}{2}
 \left ( \ket{222d} + \ket{22d2} + \ket{2d22} + \ket{d222} \right) \nonumber 
\eeq
\beq
\ket{\Psi_{249}} & = & \ket{7,- \frac{1}{2} , \frac{3}{4} ,\Gamma_{4,1}} \nonumber \\ 
&=& \frac{1}{2}
 \left ( \ket{222d} + \ket{22d2} - \ket{2d22} - \ket{d222} \right) \nonumber 
\eeq
\beq
\ket{\Psi_{250}} & = & \ket{7,- \frac{1}{2} , \frac{3}{4} ,\Gamma_{4,2}} \nonumber \\ 
&=& \frac{1}{2}
 \left ( \ket{222d} - \ket{22d2} - \ket{2d22} + \ket{d222} \right) \nonumber 
\eeq
\beq
\ket{\Psi_{251}} & = & \ket{7,- \frac{1}{2} , \frac{3}{4} ,\Gamma_{4,3}} \nonumber \\ 
&=& \frac{1}{2}
 \left ( \ket{222d} - \ket{22d2} + \ket{2d22} - \ket{d222} \right) \nonumber 
\eeq
{\subsubsection{\boldmath Eigenvectors for ${\rm  N_e}=7$ and   ${\rm m_s}$= $\frac{1}{2} $.}
\beq
\ket{\Psi_{252}} & = & \ket{7,\frac{1}{2} , \frac{3}{4} ,\Gamma_1} \nonumber \\ 
&=& \frac{1}{2}
 \left ( \ket{222u} + \ket{22u2} + \ket{2u22} + \ket{u222} \right) \nonumber 
\eeq
\beq
\ket{\Psi_{253}} & = & \ket{7,\frac{1}{2} , \frac{3}{4} ,\Gamma_{4,1}} \nonumber \\ 
&=& \frac{1}{2}
 \left ( \ket{222u} + \ket{22u2} - \ket{2u22} - \ket{u222} \right) \nonumber 
\eeq
\beq
\ket{\Psi_{254}} & = & \ket{7,\frac{1}{2} , \frac{3}{4} ,\Gamma_{4,2}} \nonumber \\ 
&=& \frac{1}{2}
 \left ( \ket{222u} - \ket{22u2} - \ket{2u22} + \ket{u222} \right) \nonumber 
\eeq
\beq
\ket{\Psi_{255}} & = & \ket{7,\frac{1}{2} , \frac{3}{4} ,\Gamma_{4,3}} \nonumber \\ 
&=& \frac{1}{2}
 \left ( \ket{222u} - \ket{22u2} + \ket{2u22} - \ket{u222} \right) \nonumber 
\eeq
{\subsubsection{\boldmath Eigenvectors for ${\rm  N_e}=8$ and   ${\rm m_s}$= $0$.}
\beq
\ket{\Psi_{256}} & = & \ket{8,0,0,\Gamma_1} \nonumber \\ 
&=& 1
 \left ( \ket{2222} \right) \nonumber 
\eeq
}



\begin{thebibliography}{99}
\bibitem{Schumann08}R. Schumann, Ann. Phys. (Leipzig) {\bf 17} (2008) 221-254, cond-mat 0701060
\bibitem{CornwellBook}J.F. Cornwell, Group Theory in Physics (Academic Press, London, 1984).
\bibitem{Schumann02} R. Schumann, Ann. Phys. (Leipzig) \textbf{11}, 49 (2002)
\bibitem{webpage} schumann@theory.phy.tu-dresden.de
\end{thebibliography}
\end{document}